%% file: main.tex
\documentclass[a4paper,12pt,twoside,openany]{book}

\input{preambule_these.tex}

\title{Semiclassical Analysis of Wave Dynamics via Wigner Transforms: Applications to the Quantum Harmonic, Anharmonic, and Dirac Oscillator}
\author{August Burg}
\date\today

\begin{document}

\frontmatter
\input{frontmatter/titlepage}

\newpage
\input{chapters/abstract}
\newpage

\input{frontmatter/note_on_origin}
\newpage
\input{frontmatter/Acknowledgements}
\newpage
\newpage
\setcounter{page}{1}
\tableofcontents
\mainmatter
 \input{chapters/Introduction}

\input{chapters/Introduction_to_Wigner-Weyl_phase_space_representaion_of_quantum_mechanics}
 \input{chapters/Quantum_Harmonic_Oscillator}
\input{chapters/Wigner-Weyl_formalism_for_Multiband_systems}

 \input{chapters/Anharmonic_Oscillator}
 \input{chapters/Dirac_Oscillator} 
 \input{chapters/Lagranigan_Ansatz_for_wave_packet_center}
  \input{chapters/Conclusion}

\backmatter
\bibliographystyle{ieeetr}
\bibliography{literature}

\end{document}

%% file: preambule_these.tex
\usepackage[T1]{fontenc}
\usepackage[utf8]{inputenc}
\usepackage{lmodern}
\usepackage[english]{babel} 

\usepackage{geometry} 
\usepackage{setspace}        
\usepackage{fancyhdr}        
\usepackage{emptypage} 
 
\fancypagestyle{plain}{
\fancyhf{}
\fancyfoot[RO,RE]{\thepage}

}

\usepackage[Lenny]{fncychap}

\usepackage[nottoc]{tocbibind} 
\usepackage{tocloft}

\usepackage{amsmath,amssymb,amsthm,mathtools} 
\usepackage{amsfonts}        
\usepackage{mathrsfs}        
\usepackage{bbm}             
\usepackage{bbold}           
\usepackage{physics}         
\usepackage{braket}          
\usepackage{bm}              
\usepackage{booktabs}        

\usepackage{graphicx}        
\usepackage{tikz}            
\usetikzlibrary{arrows.meta,calc,positioning,decorations.pathreplacing}
\usepackage{standalone}      
\usepackage{float}           
\usepackage{placeins} 

\usepackage{here}            
\usepackage{wrapfig}         
\usepackage[font=small,labelfont=bf,width=0.995\linewidth]{caption}
\usepackage{subcaption}      
\usepackage{sidecap}         

\usepackage[locale=DE,range-phrase={{ bis }}]{siunitx} 
\DeclareSIUnit\permille{\text{\textperthousand}}

\usepackage{booktabs}
\usepackage{array}
\usepackage{paralist} 

\usepackage{cite}         
\usepackage{hyperref} 
\hypersetup{
    bookmarks=true,
    colorlinks=false, 
    pdfborder={0 0 0} }

\let\orgautoref\autoref

\renewcommand{\autoref}[1]{\def\equationautorefname{Eq.}\orgautoref{#1}}
\usepackage{doi} 

\usepackage{verbatim} 
\usepackage{textcomp} 
\usepackage{empheq}   
\usepackage{dirtytalk} 

\usepackage{textcomp} 
\usepackage{xcolor} 
\usepackage{epigraph} 
\usepackage{cite} 


%% file: frontmatter/titlepage.tex
\begin{titlepage}
\thispagestyle{empty} 

\vspace*{-1.2cm} 
\noindent
\centerline{%
\begin{minipage}[t]{0.78\paperwidth}
\includegraphics[height=25mm]{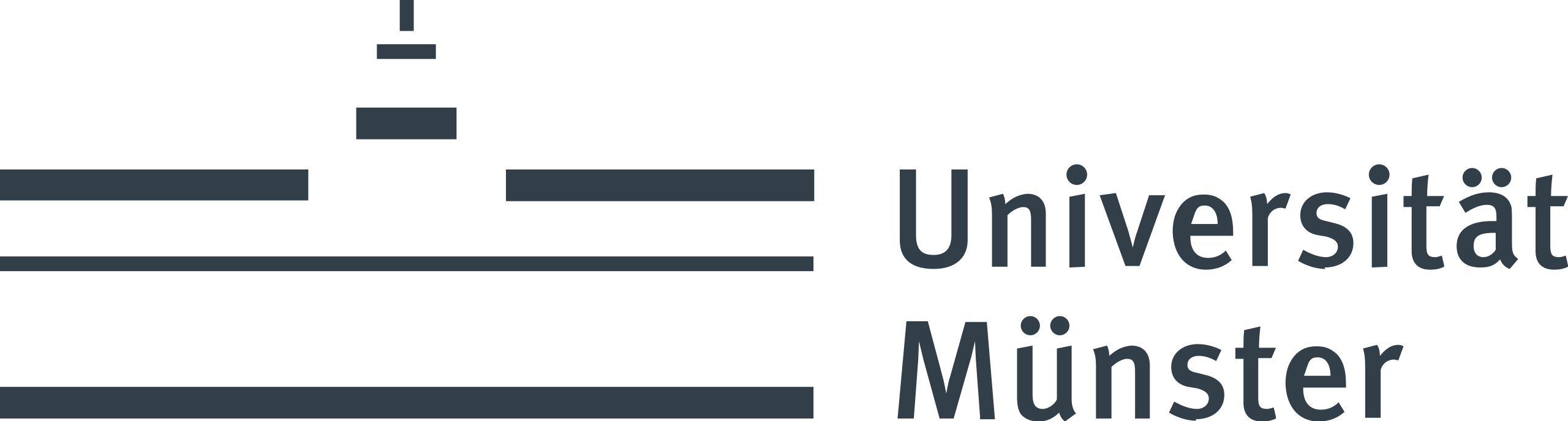}\hfill%
\includegraphics[height=25mm]{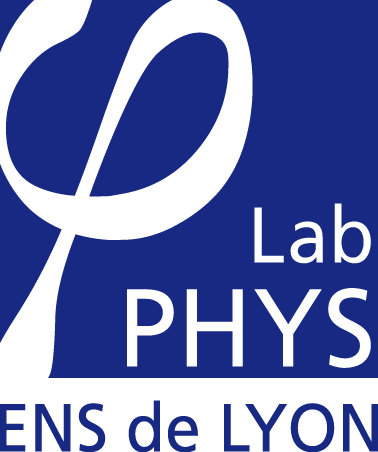}
\end{minipage}}

\centering
\vspace{10mm} 
\par
\vspace{0.2cm}

\newcommand{\HRule}{\rule{\linewidth}{0.5mm}}

\HRule\\[0.1cm]

{\LARGE \bfseries
\begin{spacing}{1.05}
Semiclassical Analysis of Wave Dynamics via Wigner Transforms:\\ 
Applications to the Quantum Harmonic, Anharmonic, and Dirac Oscillator
\end{spacing}
\par}

\HRule\\[1.5cm]
\centerline{%
\begin{minipage}[t]{0.75\paperwidth}
\includegraphics[height=51mm]{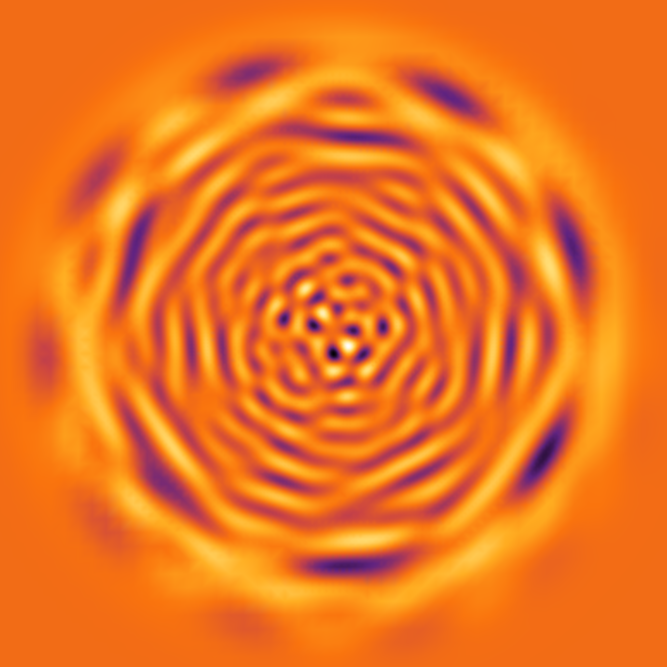}
\hfill%
\includegraphics[height=51mm]{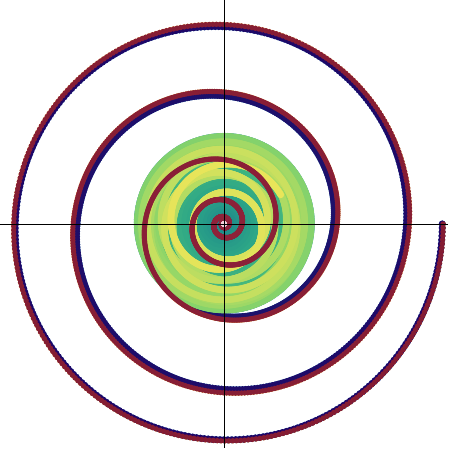}
\hfill%
\includegraphics[height=51mm]{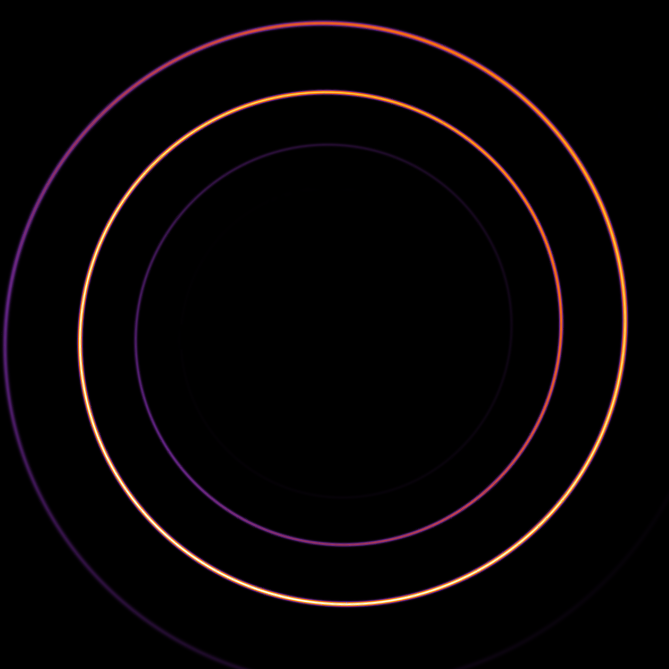}
\end{minipage}}
\vspace{4cm}
{\Large \bfseries \scshape August Burg\par}



\vspace{1.5cm}

{\large \bfseries Universität Münster, Institut für Festkörpertheorie}\par
\vspace{0.05cm}
{\large  in cooperation with}\par
{\large \bfseries ENS de Lyon, Laboratoire de Physique}\par

\end{titlepage}














%% file: chapters/abstract.tex
\chapter*{Abstract}
\markboth{Abstract}{Abstract}
In this Master’s thesis, we give a pedagogical introduction to the semiclassical analysis of wave dynamics in single-band and multi-band systems. We show how the semiclassical evolutions of Wigner functions can approximate the dynamics of wave packet centers and wave packet deformations. Three systems of increasing complexity are considered: the quantum harmonic oscillator and the quantum anharmonic oscillator as single-band systems, and the Dirac oscillator as a two-band system. We explain the origins of the first-order semiclassical corrections (Moyal, Littlejohn–Flynn, and Berry) in the equations of motion and examine their effects on predictive accuracy.\\
Another focus is set on the conditions for validity of the semiclassical analysis. For both the quantum anharmonic and the Dirac oscillators, a dimensionless parameter controlling the accuracy of the approximation is identified. In multi-band systems, the formalism applies only to states within a single adiabatic subspace, which, for the Dirac oscillator, corresponds to an energy branch. This motivates the construction of single-branch almost coherent states: approximately Gaussian wave packets formed by superpositions of states from a single energy branch. Their centers undergo oscillations with an aperiodic envelope. These states enable comparison between semiclassical Wigner function evolution and the commonly used Sundaram–Niu formalism. We show that semiclassical Wigner function evolution captures deformation-related dynamics, including the initial decay of oscillations, which the Sundaram–Niu formalism cannot describe by construction.\\
Additionally, exact quantum evolution reveals partial relocalization of these single-branch almost coherent states, a phenomenon that cannot be captured by the semiclassical evolution.

%% file: frontmatter/note_on_origin.tex
\chapter*{Note on the origin of this work}
This work is a revised and polished version of my Master’s thesis entitled \textit{''Semiclassical Analysis of Wave Dynamics via Wigner Transforms: A Study of Multiband Oscillators''}.\\
The thesis was carried out as a collaboration between the Universität Münster and the ENS de Lyon within the framework of the Erasmus exchange program.\\
The research was supervised by Prof. Pierre Delplace (Laboratoire de Physique, ENS de Lyon) and Prof. Michael Rohlfing (Institut für Festkörperphysik, Universität Münster), who also served as examiners for the Master’s thesis.

%% file: frontmatter/Acknowledgements.tex
\chapter*{Acknowledgments}
\markboth{Acknowledgments}{Acknowledgments}
First of all, I would like to thank Pierre for giving me the opportunity to come to Lyon and work on this exciting topic, for his competent support and for sharing his enthusiasm during our frequent and long sessions at the blackboard, without which this work would not have been possible in this form. 

I also thank Prof. Rohlfing very much for making my research stay in France possible by supervising this work from Münster.

I thank my friends Valentin, Rémy and Maria-Myrto for the great time together in Lyon and for the valuable and intensive feedback on this manuscript, which helped me a lot. Further, I want to thank my friends, parents and grandparents, with whom I stayed during the period of writing this Master’s thesis, for their constant support and company.

%% file: chapters/Introduction.tex
\chapter*{Introduction}
\addcontentsline{toc}{chapter}{Introduction}
In most quantum systems, no exact analytical solutions are known for the time evolution of quantum states. Often, solving the Schrödinger equation numerically  is computationally costly and leads to a loss of intuition  about the system at hand.  For systems with a Hamiltonian  $\hat{\mathcal{H}}$ acting on a product  of slow and fast degrees of freedom,  for example position and the spin or nuclei position and electron position in the Born-Oppenheimer approximation,  it is possible to approximately simplify the problem of solving the Schrödinger equation to solving the semiclassical equations of motion, which are Hamilton's equations from classical mechanics with additional correction terms. \\
For this purpose, it is convenient to express quantum mechanics in the Wigner-Weyl phase space representation: Operators $\hat{o}$ get mapped on the Weyl symbols $ \sigma _\hbar[\hat{o}]=o(x,p)$, which  are functions of scalars $x,\,p$.  The noncommutativity of $[\hat{x},\hat{p}]=\mathrm{i}\hbar$ is taken into account in the symbol $o(x,p)$ via terms in powers of the commutator $\mathrm{i}\hbar$, for example $\sigma _\hbar[\hat{x}^2\hat{p}^2]=x^2p^2+2\mathrm{i}\hbar xp-\hbar^2/2 $. The semiclassical  approximation then consists in neglecting all terms of $\mathcal{O}(\hbar^2)$ in the limit of  ``small $\hbar$``. \\
For a Hamiltonian $\hat{\mathcal{H}}$ acting on both the slow degrees of freedom and the fast degrees of freedom ,  the symbol Hamiltonian $H(x,p)=H_0(x,p)+\hbar H_1(x,p)+\mathcal{O}(\hbar^2)$  depends on the scalars $x,p$, i.e. it does not act on $\ket{x}$.  But it is still an operator acting on the fast degrees of freedom~\cite{Teufel2003-AdiabaticPTQM,Luneau:phd,Stiepan_2012}. However, compared to diagonalizing  the operator Hamiltonian $\hat{\mathcal{H}}$,  it is relatively easy to find the eigenvectors $\ket{\Psi_0^\nu}$ of the lowest order symbol $H_0(x,p)$ 
\begin{equation*}
    H_0\ket{\Psi_0^\nu}=e_0^{(\nu)}\ket{\Psi_0^\nu}.
\end{equation*}
Most commonly in the literature, in the derivation of the semiclassical equations of motion, states are assumed to be wave packets, associated with a band $e_0^{(\nu)}(x,p)$,  propagating  ``like a classical particle with a well defined group velocity``~\cite{wimmerExperimentalMeasurementBerry2017ExampleNiu}. Using this assumption, a Lagrangian ansatz~\cite{SundaramNiu,perez:tel-03922294}  yields the approximate evolution  for the real- and momentum space wave packet centers  $x_c$ and $p_c$ 
\begin{equation*}
\begin{aligned}
\dot{x}_{c}=&+\partial_{p_c} h^{\nu} - \hbar\left(F_{p_{c}x_c}^{\nu} \dot{x}_c+F_{p_c p_{c}}^{\nu} \dot{p}_{c}\right)+\mathcal{O}(\hbar^2) \\
 \dot{p_c}=&-\partial_{x_c} h^{\nu}+\hbar \left( F_{x_cx_c}^{\nu} \dot{x}_c+F_{x_c p_{c}}^{\nu} \dot{p}_{c}\right) +\mathcal{O}(\hbar^2) \\[2ex]
h^{\nu}=&e_0^{(\nu)}+\hbar\left(\braket{H_1}^{(\nu)}+M^{(\nu)}\right).
\end{aligned}
\end{equation*}
At 0th order, these are the Hamilton's equations of motion for a classical system with energy $e_0^{(\nu)}(x_c,p_c)$.  At first order, the noncommutativity  of $\hat{x}$ and $\hat{p}$ manifests itself in the Berry curvature $F^\nu$ and a scalar correction to $e_0^{(\nu)}$ by $\braket{H_1}^{(\nu)}=\braket{\Psi_0^\nu|H_1|\Psi_0^\nu}$ and  $M^{(\nu)}$, the Littlejohn-Flynn term (also called ``no name term``) .\\
In the derivation of these semiclassical equations of motion, only $x_c$ and $p_c$ are  included as dynamical degrees of freedom of the wave packet: it is assumed that the wave packet moves in phase space but never changes form. As we show in \autoref{chapter:Niu}, this leads to substantial deviations from the real evolution since wave packet deformations play an important role in wave dynamics. \\
This motivates the use of a less common but much more powerful semiclassical formalism introduced in~\cite{Stiepan_2012}, which includes in the approximated dynamics, not just the wave centers $x_c$ and $p_c$,  but the whole Wigner function $W_\psi(x,p)$, which is a phase space analog of the density matrix $\hat{\rho}=\ket{\psi}\bra{\psi}$ and can be seen as a quasi phase space probability distribution for a state $\ket{\psi}$. In this formalism, the semiclassical equations of motion are defined in adiabatic subspaces $\mathcal{H}^\nu_\text{adiabatic}$ corresponding to the bands $e_0^{(\nu)}(x,p)$ of the symbol $H_0(x,p)$. This makes it possible to approximate the evolution of a state $\ket{\psi_\nu}$ associated with  $\mathcal{H}^\nu_\text{adiabatic}$ by the semiclassical evolution of its Wigner function.  \\

The first main goal of this Master's thesis is to use illustrative examples to provide a practical guide to wave dynamics in the semiclassical limit via Wigner function evolution. The focus is on the following questions:
\begin{itemize}
    \item \textbf{What parameter controls the semiclassical approximation?}  
    Since $\hbar$ is a natural constant, the limit $\hbar\to 0$ has no physical meaning. Instead, one must identify a dimensionless parameter that characterizes the degree to which the noncommutativity of  $\hat{x}$ and $\hat{p}$ is important in the phase space representation of quantum mechanics. 
    \item \textbf{Which states are associated with the adiabatic subspaces $\mathcal{H}^\nu_{\mathrm{adiabatic}}$?}  
    As the semiclassical equations are defined separately for each adiabatic subspace, it is essential to determine which states belong to a given adiabatic subspace. 
    \item \textbf{What is the origin and role of the first-order correction terms $\braket{H_1}^{(\nu)}$, $M^{(\nu)}$, and $F^{(\nu)}$?}  
    The correction terms are the first order manifestation of the noncommutativity of $\hat{x}$ and $\hat{p}$. While  $\braket{H_1}^{(\nu)}$  originates from $H_1(x,p)$, the first-order term in the Hamiltonian symbol, the Littlejohn–Flynn term $M^{(\nu)}$ and the Berry curvature $F^{(\nu)}$ arise from the projection on adiabatic subspaces. Their effects on the prediction of wave dynamics are analyzed explicitly. 
\end{itemize}
The semiclassical formalism is applied to 3 example systems with known analytical solutions, so that the quality of the semiclassical predictions is evaluated by comparison to the exact solutions. The studied systems are oscillators with different levels of complexity:
\begin{enumerate}
    \item the Quantum Harmonic Oscillator,  a rare case where the semiclassical and exact evolution coincide, and thus a simple pedagogical example to introduce quantum dynamics in phase space
    \item the (Quantum) Anharmonic Oscillator~\cite{proppDecoherenceLimitsCost}, a single-band system with nontrivial semiclassical first order and higher order corrections, allowing us to show how the semiclassical Wigner function evolution can capture key features of wave dynamics.
    \item the Dirac Oscillator~\cite{moshinsky_dirac_1989},   a simple and widely studied two-band system that requires the identification of adiabatic subspaces for an approximate semiclassical description.
\end{enumerate}
In both the Anharmonic Oscillator and the Dirac Oscillator, wave packets exhibit rich and complex dynamics, which themselves are worth investigating~\cite{banerjiNonlinearWavePacket2001,NogamiCoherentStateDO}. The second main goal of this master thesis is to gain a global understanding of the dynamics of initially localized wave packets. For both systems the approximate evolution of the Wigner function allows one to predict characteristic quantities of the dynamics, such as the period $T_\text{oscill}$ of the oscillations of wave packet centers $\braket{\hat{x}}(t)$ and $\braket{\hat{p}}(t)$ and $\tau_\text{dec}$ the typical time it takes the envelope of the oscillations to decay to $\sqrt{\braket{\hat{x}}^2+\braket{\hat{p}}^2}\approx 0$. 
Although exact solutions are known, they can be too complex to directly extract how these quantities depend on the initial conditions and the control parameters.\\
For simplicity, this Master's thesis deals with semiclassical equations of motion in only one spatial dimension. For a discussion of the semiclassical evolution of the Wigner function in higher spatial dimensions, the reader is referred to~\cite{Stiepan_2012,Luneau:phd}.

%% file: chapters/Introduction_to_Wigner-Weyl_phase_space_representaion_of_quantum_mechanics.tex
\chapter{Wigner-Weyl Phase Space Representation of Quantum Mechanics}
To study the semiclassical limit of a quantum system, it is desirable to represent quantum mechanical operators  in the $(x,p)$-phase space, the natural representation space in classical physics~\cite{Zachos-QuantumMechanicsinPS}. Such a phase space representation of quantum mechanics was already introduced in the early days of quantum mechanics (in the first half of the 20th century) and is today known as the Wigner-Weyl formalism, (see for example ~\cite{Weyl1927},\cite{vNeumann1931},\cite{Wigner-QuantumCorrections},\cite{GROENEWOLD1946405},\cite{Moyal_1949} all reprinted in~\cite{Zachos-QMPS:OverviewwithPapers}). A pedagogical overview is given in \cite{Zachos-QuantumMechanicsinPS},\cite{Case-WFandWTforpedestrinas},\cite{chaturvedi-PSdescriptionsoperators} and Chapter 2 of~\cite{Luneau:phd}. 
The Wigner-Weyl formalism is a completely equivalent formulation of quantum mechanics, in the same way as it is equivalent to represent quantum mechanics  in Dirac notation  $\ket{n}$ or as wave functions $\psi_n(x)$ in $x$ space-representation. Given the commutator of  a hermitian cartesian pair $[\hat{x},\hat{p}]=\mathrm{i}\hbar$, a quantum mechanical operator $\hat{o}$ can be represented in phase space by the Wigner transform
\begin{equation}\label{eq:WignertransfromScalar}
    \sigma_\hbar[\hat{o}]=o(x,p)=\int\mathrm{d}y\mathrm{e}^{-\mathrm{i}\frac{py}{\hbar}}\braket{x+y/2|\hat{o}|x-y/2}.
\end{equation}
$o(x,p)$ is called the Weyl-symbol of $\hat{o}$ and is a function of the scalars $x$ and $p$, i.e. it does not act on $\ket{x}$ or $\ket{p}$. Its inverse is the Weyl quantization of an operator
\begin{equation}
    \hat{o}=\text{Op}[o(x,p)]=\frac{1}{2\pi\hbar}\int\mathrm{d}x\int\mathrm{d}p\int\mathrm{d}y\mathrm{e}^{\mathrm{i}\frac{p(x-y)}{\hbar}}\ket{x}\bra{y}o(x+y,p).
\end{equation}

For the moment, only operators $\hat{o}$ are considered that are purely built up from $\hat{x}\equiv\,x $ and $\hat{p}\equiv-\mathrm{i}\hbar\partial_x$, i.e. operators which that act only on $L^2(\mathbb{R})$, the Hilbert space of wave function $\psi(x)$ with 
\begin{equation}
    \int_{-\infty}^{+\infty}|\psi(x)|^2 \mathrm{d} x<\infty.
\end{equation}
The Wigner transform makes possible to calculate the trace of such operators $\hat{o}$ as a phase space integral
\begin{equation}
    \text{Tr}[\hat{o}]=\frac{1}{2\pi\hbar}\int\int\mathrm{d}x\mathrm{d}po(x,p),
\end{equation}
as can be shown directly by inserting~\autoref{eq:WignertransfromScalar} on the right-hand side. Furthermore, the Wigner transform is constructed such that the trace of a product of operators $\hat{a}\hat{b}$ is simply given by the phase space integral of the product of their symbols
\begin{equation}
     \text{Tr}[\hat{a}\hat{b}]=\frac{1}{2\pi\hbar}\int\int\mathrm{d}x\mathrm{d}p\,a(x,p)b(x,p).
\end{equation} 
This allows to express $\braket{\hat{o}}=\text{Tr}[\hat{o}\hat{\rho}]$, the expectation value given a density operator $\hat{\rho}=\ket{\psi}\bra{\psi}$, as a phase space integral
\begin{equation}\label{eq:IntroWW-into}
\braket{\hat{o}}=\int\mathrm{d}x\mathrm{d}p\frac{\rho(x,p)}{2\pi\hbar}o(x,p).
\end{equation}
This form is analogous to the calculation of a classical expectation value of a quantity $a(x,p)$ given a phase space probability density  $\braket{o}=\int \int \mathrm{d}x\mathrm{d}p P(x,p)a(x,p)$.  One can identify 
\begin{equation}
    w(x,p)=\frac{\rho(x,p)}{2\pi\hbar}
\end{equation}  
as the probability weight of the state $\ket{\psi}$. $w(x,p)$ is called the Wigner function and is normed for pure states $\ket{\psi}$ 
\begin{equation}
    \int \mathrm{d}x\mathrm{d}pw(x,p)=1.
\end{equation} 
But the Wigner function is rather a quasiprobability than a true probability as it can take negative values. Apart from its use in phase-space analysis of quantum mechanics, the Wigner function is a practical tool to graphically illustrate quantum states. For operators $\hat{o}$  purely acting on $L^2(\mathbb{R})$, the symbol $o(x,p)$ is a scalar and scalars commute. Therefore, in general, the symbol of a product of 2 noncommuting operators $\hat{a}\hat{b}$ cannot just be given by the product of the symbol of each:  $\sigma_\hbar[\hat{a}\hat{b}]\neq\sigma_\hbar[\hat{a}]\sigma_\hbar[\hat{b}]$. And indeed, the symbol of a product of operators is calculated from a noncommuting product, called Moyal-product or $\star$-product
\begin{equation}
\begin{aligned}
\sigma_\hbar[\hat{a}\hat{b}]=(a \star b)_{\hbar}(x, p) &=\sum_{n=0}^{\infty} \frac{1}{n!}\left(\frac{\mathrm{i} \hbar}{2}\right)^n \sum_{k=0}^n(-1)^k\binom{n}{k} \frac{\partial^n a}{\partial p^k \partial x^{n-k}} \frac{\partial^n b}{\partial x^k \partial p^{n-k}} \\
& =a(x,p) b(x, p)+\frac{\mathrm{i} \hbar}{2}\{a(x,p),b(x,p)\}+\mathcal{O}\left(\hbar^2\right)
\end{aligned}
\end{equation}
with the Poisson-brackets 
\begin{equation}\label{eq:Poisson}
   \{a,b\}= \left(\frac{\partial a}{\partial x} \frac{\partial b}{\partial p}-\frac{\partial a}{\partial p} \frac{\partial b}{\partial x}\right)
\end{equation}
not to be confused with the anticommutator. The $\star$-product makes possible to formulate fundamental quantum mechanical relations in phase space like the idempotency condition $\hat{\rho}\hat{\rho}=\hat{\rho}$ for pure states
\begin{equation}
    2\pi\hbar\, w\star w=w
\end{equation}
and the energy eigenequation $\hat{\mathcal{H}}\hat{\rho}_n=E_n\hat{\rho}_n=\hat{\rho}_n\hat{\mathcal{H}}$ as $\star$-genvalue equation of the symbol $\sigma_\hbar[\hat{\mathcal{H}}]=H$
\begin{equation}
    H\star w_n=E_n w_n=w_n\star H.
\end{equation}
Furthermore, the Liouville-von Neumann equation $ \frac{\partial \hat{\rho}}{\partial t}=-\frac{\mathrm{i}}{ \hbar}[\hat{\mathcal{H}},\hat{\rho}]$ can be expressed as
\begin{equation}
   \frac{\partial w}{\partial t}=-\frac{\mathrm{i}}{ \hbar}[H,w]_\star
\end{equation} 
and the Heisenberg equation  $ \frac{\partial \hat{o}}{\partial t}=\frac{\mathrm{i}}{ \hbar}[\hat{\mathcal{H}},\hat{o}]$ as
\begin{equation}\label{eq:partialto scalar exact}
   \frac{\partial o}{\partial t}=\frac{\mathrm{i}}{ \hbar}[H,o]_\star\,,
\end{equation} 
which correspond to phase space representation of quantum dynamics in the Schrödinger and Heisenberg picture respectively. $[\cdot,\cdot]_\star$ is the Moyal commutator defined as 
\begin{equation}
  [a,b]_\star=a\star b-b\star a . 
\end{equation}
\section{Semiclassical equations of motion for scalar valued symbols}\label{sec:scalarsemiclass} 
\begin{figure}[t]
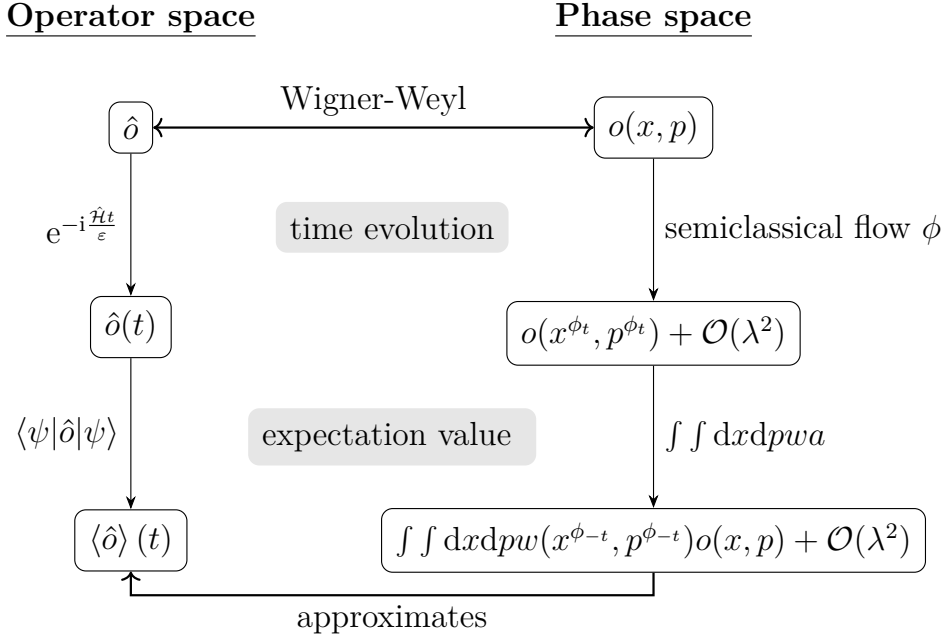

    \centering
    \includestandalone[width=0.8\linewidth]{graphics/sketches/semiclass_approximation_single_band_H}
    \caption{Sketch of steps when performing the semiclassical approximation for the expectation value of an operator $\hat{o}$ for an Hamiltonian $\hat{\mathcal{H}}$ acting purely on a Hilbert space $L^2(\mathbb{R})$.}
    \label{fig:sketch semiclass approx single band}
\end{figure}
In the limit of ``small $\hbar$``, truncation of all $\mathcal{O}(\hbar^2)$-terms at the right hand side of~\autoref{eq:partialto scalar exact} gives Hamilton's equations of motion $\partial_to=\{o,H\}$, used in classical physics. This way, the Wigner-Weyl representation makes it possible to approximate quantum dynamics by a semiclassical motion in phase space. In the following, the semiclassical evolution of symbols given a Hamiltonian $\hat{\mathcal{H}}$ acting purely on $L^2(\mathbb{R})$ is shown as a special case of arguments in chapter 2 of~\cite{Luneau:phd},\cite{Stiepan_2012} and chapter 3 of~\cite{Teufel2003-AdiabaticPTQM}. First of all, it is to emphasize that  $\hbar=6.626 070 15  \cdot 10^{-34}\text{Js}/2\pi$ is a natural constant~\cite{BIPM2025} and therefore cannot tend to zero. What is referred to as ``the limit of $\hbar \rightarrow0$ `` is $\hbar$ being relatively small compared to the typical quantities of a system. To identify the condition of `` small $\hbar$`` one can introduce a tracing parameter $\lambda=1$ as a prefactor in the commutator $[\hat{x},\hat{p}]=\mathrm{i}\lambda\hbar$. Then, $\lambda$ traces the order of the commutator $\lambda\hbar$ in the equation of motion
\begin{equation}\label{eq:partialt o scalar approx}
\begin{aligned}
\partial_t o
&= \frac{\mathrm{i}}{\lambda\hbar}[H,o]_\star \\
&= \sum_{l=0}^{\infty} 
\frac{(-1)^{l+1}}{2(2l+1)!}
\left(\frac{\lambda\hbar}{2}\right)^{2l}
\sum_{k=0}^{2l+1}
(-1)^k\binom{2l+1}{k}
\bigl(
\partial_p^k\partial_x^{2l+1-k}H
\;\partial_x^k\partial_p^{2l+1-k}o
-(p\leftrightarrow x)
\bigr)
\\[0.5ex]
&= \{o,H\} + \mathcal{O}(\lambda^2)
\end{aligned}
\end{equation}


where all even terms of $[H,o]_\star$ have canceled out due to the symbols $H$ and $o$ and all their derivatives being scalars therefore commuting.   

If the $(2l+1)$th derivatives  in $x $ and $p $ of $H$ and $o$ times $\hbar^{2l}$ are much smaller than $\{o,H\}$, it is possible to neglect all the order $\mathcal{O}(\lambda^2)$-terms. This is referred to as the semiclassical approximation. Neglecting even all $\mathcal{O}(\lambda)$-terms in  $H=\sigma[\hat{H}]$ and $o=\sigma[\hat{o}]$ means not taking into account the noncommutativity of $[\hat{x},\hat{p}]=\lambda\hbar$ at any order and thus corresponds to the classical approximation $[\hat{x},\hat{p}]=0+\mathcal{O}(\lambda)$. After neglecting all $\mathcal{O}(\lambda^2)$-terms, \autoref{eq:partialt o scalar approx} is approximately solved by  $o(x^{\phi_t},p^{\phi_t})$ where $\phi_t$ denotes the evolution under the semiclassical Hamiltonian flow \cite{Arnold1989} defined by the solution of 
\begin{equation}\label{eq:Hamoltonian eq of motion} 
    \begin{aligned}
        \dot{x}=&\partial_{p}H\\
        \dot{p}=&-\partial_xH\,.
    \end{aligned}
\end{equation}
It follows directly to approximate the expectation value of
$\hat{o}$ under quantum time evolution
\begin{equation}
    \braket{\hat{o}}(t)=\braket{\mathrm{e}^{\mathrm{i}\frac{\hat{\mathcal{H}}t}{\hbar}}\hat{o}\mathrm{e}^{-\mathrm{i}\frac{\hat{\mathcal{H}}t}{\hbar}}}
\end{equation} 
by 
\begin{equation}
    \braket{\hat{o}}(t)=\int\int\mathrm{d}x\mathrm{d}pw(x,p)o(x^{\phi_t},p^{\phi_t})+\mathcal{O}(\lambda^2),
\end{equation}
where $\mathcal{O}(\lambda^2)$ is the order of magnitude of the neglected terms. One can change into the Schrödiger picture by pulling the time dependence into the Wigner function via the coordinate transformation $(x',p')=(x^{\phi_{-t}},p^{\phi_{-t}})$ 
\begin{equation}\label{eq:<o>(t) semicl scalar wigner}
\begin{aligned}
        \braket{\hat{o}}(t)=&\int\int \underbrace{\mathrm{d}x'^{\phi_{-t}}\mathrm{d}p'^{\phi_{-t}}}_{\mathrm{d}x'\mathrm{d}p'}w(x'^{\phi_{-t}},p'^{\phi_{-t}})o(x',p')+\mathcal{O}(\lambda^2)\\
        =&\int\int\mathrm{d}x\mathrm{d}pw(x^{\phi_{-t}},p^{\phi_{-t}})o(x,p)+\mathcal{O}(\lambda^2)\\
\end{aligned}
\end{equation}
where in the last line the prime $'$ was lifted, and it was used that volume element $\mathrm{d}x\mathrm{d}p$ is preserved under the Hamiltonian flow as follows directly from
\begin{equation}
    \text{div}\left(\begin{array}{cc}
     \dot{x}^{\phi_t} \\
     \dot{p}^{\phi_t}
\end{array}\right)=\partial_x\partial_pH-\partial_p\partial_xH=0.
\end{equation}
Figure~\ref{fig:sketch semiclass approx single band} summarizes the most important steps to arrive at~\autoref{eq:<o>(t) semicl scalar wigner}.

The approximated evolution $w(x^{\phi_{-t}},p^{\phi_{-t}})$ can be a useful mathematical tool when exact evolutions of the density operator $\hat{\rho}$ are not known or too complex to give a physical intuition. Furthermore, the study of $w(x^{\phi_{-t}},p^{\phi_{-t}})$ can be insightful when investigating classical-quantum correspondence, as will be shown in the next chapter by the example of the Harmonic Oscillator. 

%% file: graphics/sketches/semiclass_approximation_single_band_H.tex
\begin{tikzpicture}[node distance=1.8cm and 4cm, every node/.style={align=center}]

    \node (operator) [text=black] {\underline{\textbf{Operator space}}};
    \node (phase space) [right=3.5cm of operator, text=black] {\underline{\textbf{Phase space}}};

    \node (asymbol) [below=of phase space, yshift=1.2cm]  [draw,  rounded corners, text=black,  inner sep=5pt, anchor=north]{$o(x,p)$};
 
     \node (asymbol evolution)[below=of asymbol] [draw,  rounded corners, text=black,  inner sep=5pt, anchor=north]{$ o(x^{\phi_t},p^{\phi_t})+\mathcal{O}(\lambda^2)$};
   \node (asymbol expectation) [below=of asymbol evolution] [draw,  rounded corners, text=black,  inner sep=5pt, anchor=north]{$\int\int\mathrm{d}x\mathrm{d}p w(x^{\phi_{-t}},p^{\phi_{-t}}) o(x,p)+\mathcal{O}(\lambda^2)$};
 \node (aoperator) 
    [draw, rounded corners, text=black, inner sep=5pt, anchor=center] 
     at (operator |- asymbol) 
    {$\hat{o}$};

    \node (aoperator evolution)  [below=of aoperator] [draw,  rounded corners, text=black,  inner sep=5pt, anchor=north] {$\hat{o}(t)$};
    \node (aoperator expectation) [draw, rounded corners, text=black, inner sep=5pt] 
    at (aoperator evolution |- asymbol expectation) {$\braket{\hat{o}}(t)$};

   \node (time evolution)  at ($ (aoperator evolution)!0.5!(asymbol) $)  [ rounded corners, text=black,fill=gray!20,  inner sep=5pt]  {time evolution};
   
 \node (expectation value)  at ($ (aoperator expectation)!0.5!(asymbol evolution) $) [ rounded corners, text=black,fill=gray!20,  inner sep=5pt]  {expectation value  }; 

\draw[<->, thick] (aoperator.east) -- node[above] {Wigner-Weyl} (asymbol.west);

\draw[-{Stealth}] (asymbol.south) -- (asymbol evolution.north)
    node(semicl flow)[midway, right] {semiclassical flow $\phi$ 
    };
 \draw[-{Stealth}] (aoperator.south) -- (aoperator evolution.north) node(semicl flow)[midway, left] {$\mathrm{e}^{-\mathrm{i}\frac{\hat{\mathcal{H}}t}{\varepsilon}}$};
  \draw[-{Stealth}] (aoperator evolution.south) -- (aoperator expectation.north) node(semicl flow)[midway, left] {$\braket{\psi|\hat{o}|\psi}$};
\draw[-{Stealth}] (asymbol evolution.south) -- (asymbol expectation.north)
    node(semicl flow)[midway, right] { $\int\int\mathrm{d}x\mathrm{d}p wa$ 
    };

   \draw[<-, thick] 
    (aoperator expectation.south) 
        -- ++(0,-0.3cm) coordinate(corner)  
        -| (asymbol expectation.south);     


\node[below, yshift=-0.1cm] (apprimates) at ($(corner)!0.5!(asymbol expectation.south)$) [ rounded corners, text=black, inner sep=5pt] {approximates};




\end{tikzpicture}

%% file: chapters/Quantum_Harmonic_Oscillator.tex
\chapter{Wigner Functions of the Quantum Harmonic Oscillator and its classical limit}\label{chapter:HO}
The best way to get familiar with the Wigner-Weyl phase space representation is by studying its application to a technically uncomplicated example: the Quantum Harmonic Oscillator. The Quantum Harmonic Oscillator is a special case, where  the semiclassical phase space flow $\phi$ describes the exact quantum dynamics, because all $\mathcal{O}(\lambda^2)$-terms in~\autoref{eq:partialt o scalar approx} are zero. This makes it relatively easy to show the reader how the Wigner-Weyl phase space representation can reproduce an equivalent result to the standard Hilbert space formulation. Furthermore, the Wigner-Weyl formalism will be employed to let the limit of a classical dynamics arise for coherent states in a natural way. The content of this chapter was mainly adapted from~\cite{Mostowski}. 
\section{The Quantum Harmonic Oscillator}
The Harmonic Oscillator in quantum dynamics~\cite{Munster2020}  is given by the Hamiltonian  
\begin{equation}
    \hat{\mathcal{H}}=\frac{\hat{p}^2}{2m}+\frac{m\omega^2}{2}\hat{x}^2,
\end{equation}
which is exactly the Weyl quantization $\hat{\mathcal{H}}=\text{Op}[H(x,p)]$ of the Harmonic Oscillator in classical mechanics 
\begin{equation}\label{eq:HO symbolH}
    H=\frac{p^2}{2m}+\frac{m\omega^2}{2}x^2.
\end{equation}

One can use the annihilation and creation operators of Fock states $\ket{n},\, n \in \mathbb{N}$
\begin{equation}
    \begin{aligned}
        \hat{a}\ket{n}=&\sqrt{n}\ket{n-1}\\
        \hat{a}^\dagger\ket{n}=&\sqrt{n+1}\ket{n+1}
    \end{aligned}
\end{equation}
given by
\begin{equation}\label{eq:HO a adagger}
\begin{aligned}
   \hat{a}=\frac{1}{\sqrt{2}}\left(\frac{\hat{x}}{l}+\mathrm{i} \frac{l}{\hbar} \hat{p}\right)\\ 
      \hat{a}^\dagger=\frac{1}{\sqrt{2}}\left(\frac{\hat{x}}{l}-\mathrm{i} \frac{l}{\hbar} \hat{p}\right),\\ 
\end{aligned}
\end{equation}
 with $l=\sqrt{\frac{\hbar}{m\omega}}$ the characteristic length scale, to express the Hamiltonian as
 \begin{equation}
     \hat{\mathcal{H}}=\hbar\omega\left(\hat{a}^\dagger\hat{a}+\frac{1}{2}\right).
 \end{equation} 
 This way, it becomes clear that the energy eigenstates of $\hat{\mathcal{H}}$ are Fock states $\ket{n}$  such that:
\begin{equation}
\begin{aligned}
        \hat{\mathcal{H}}\ket{n}=&E_n\ket{n}\\
        E_n=&\hbar \omega\left(n+\frac{1}{2}\right).\\
\end{aligned}
\end{equation}
The  $x$-space representation of $\ket{n}$ is given by
 \begin{equation}\label{eq:phin(x)}
    \braket{x|n}=\psi_n(x)=\frac{1}{\sqrt{2^n n!l \sqrt{\pi}}} e^{-x^2 / 2 l^2} H_n(x/l),
\end{equation}
where $H_n(x) \equiv(-1)^n e^{x^2} \frac{d^n}{d x^n} e^{-x^2}$ denotes the  Hermite Polynomials.  Writing $\hat{x}$ and $\hat{p}$ in terms of $\hat{a}$ and $\hat{a}^\dagger$, one can directly show that $\ket{n}$ is centered around $0$ in real and momentum space
\begin{equation}
    \begin{aligned}
        \braket{n|\hat{x}|n}=&0\\
        \braket{n|\hat{p}|n}=&0,\\
    \end{aligned}
\end{equation}
while a superposition of the form  \begin{equation}\label{eq:HO |n,n+1>}
    \ket{n,(n+1)_{\varphi_{n+1}}}=\frac{\ket{n}+\mathrm{e}^{\mathrm{i}\left(\varphi_{n+1}\right)}\ket{n+1}}{\sqrt{2}}
\end{equation}
has  nonzero expectation values for the operators  $\hat{x}$ and $\hat{p}$  
\begin{equation}
\begin{aligned}
 \langle\hat{x}\rangle_{n,n+1_{\varphi_{n+1}}}&=\frac{l}{\sqrt{2}}\sqrt{n+1}\cos{\varphi_{n+1}} \\
\langle\hat{p}\rangle_{n,n+1_{\varphi_{n+1}}}&=\frac{lm\omega}{\sqrt{2}}\sqrt{n+1} \sin{\varphi_{n+1}} .\\
\end{aligned}
\end{equation}

In Hilbert space representation of quantum mechanics, time evolution is generated by the time evolution operator, $\hat{U}(t)=\mathrm{e}^{-\mathrm{i}\frac{\hat{\mathcal{H}}t}{\hbar}}$, leading to the evolution of $\ket{n,(n+1)_{\varphi_{n+1}}}$ in the Schrödinger picture: 
\begin{equation}\label{eq:HO |n,n+1>(t)}
    \ket{n,(n+1)_{\varphi_{n+1}}}(t)=\frac{\ket{n}+\mathrm{e}^{\mathrm{i}\left(\varphi_{n+1}-\omega t\right)}\ket{n+1}}{\sqrt{2}}
\end{equation}
modulo a general phase $\mathrm{e}^{-\mathrm{i}(n+1/2)\omega t}$.  As a consequence, its  expectation values oscillate with an angular velocity $\omega$
\begin{equation}
\begin{aligned}
 \langle\hat{x}\rangle_{n,n+1_{\varphi_{n+1}}}(t)&=\frac{l}{\sqrt{2}}\sqrt{n+1}\cos{(\varphi_{n+1}-\omega t)} \\
 \langle\hat{p}\rangle_{n,n+1_{\varphi_{n+1}}}(t)&=\frac{\hbar}{l}\frac{1}{\sqrt{2}}\sqrt{n+1} \sin{(\varphi_{n+1}-\omega t)} .\\
\end{aligned}
\end{equation}

 \section{Wigner functions of Fock states }
 \begin{figure}[h]
     \centering
     \includegraphics[width=0.3\linewidth]{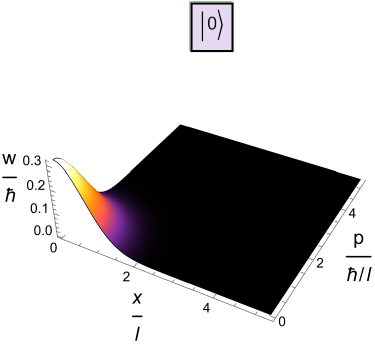}
     \includegraphics[width=0.3\linewidth]{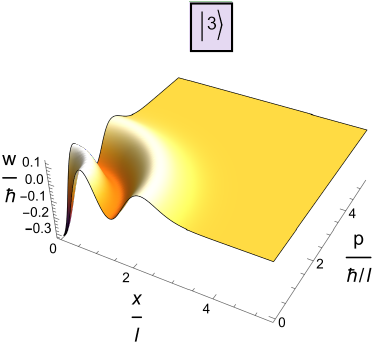}
     \includegraphics[width=0.3\linewidth]{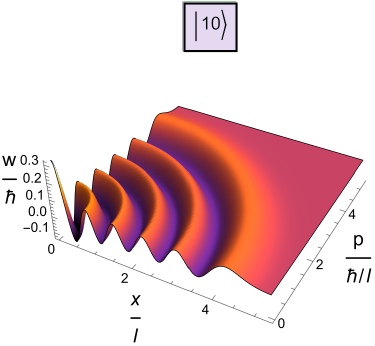}
    \includegraphics[width=0.3\linewidth]{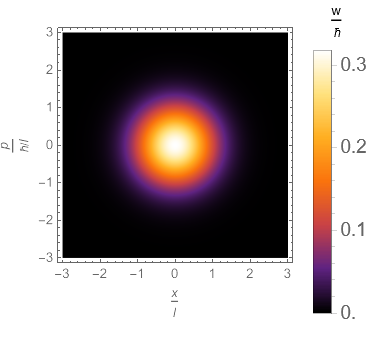}
    \includegraphics[width=0.3\linewidth]{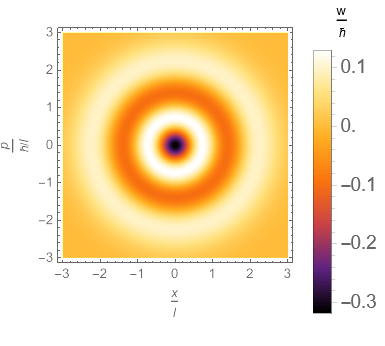}
    \includegraphics[width=0.3\linewidth]{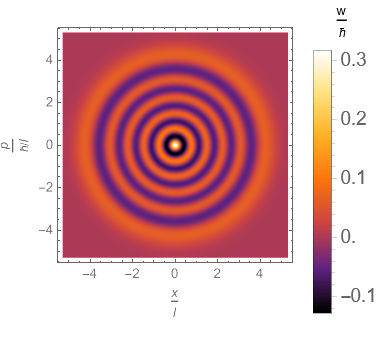}
     \caption{Graphical representation (3D-plot and map) of  $w_n(x,p)$, the Wigner functions of  the Fock state $\ket{n}$, Left: $n=0$, Middle: $n=3$, Right: $n=10$. All $w_n(x,p)$ are centered around the origin. For larger $n$, the Wigner function becomes increasingly oscillatory. Figure inspired by~\cite{Mostowski}.}
     \label{fig:plot Wignerfunctions wn}
 \end{figure}
 
  \begin{figure}[]
     \centering
     \includegraphics[width=0.3\linewidth]{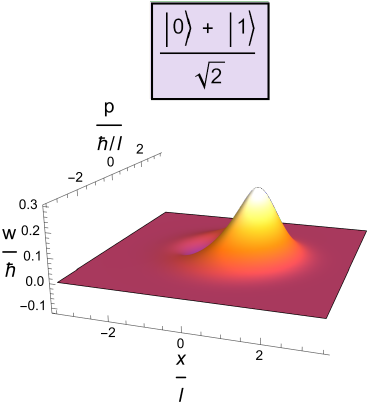}
     \includegraphics[width=0.3\linewidth]{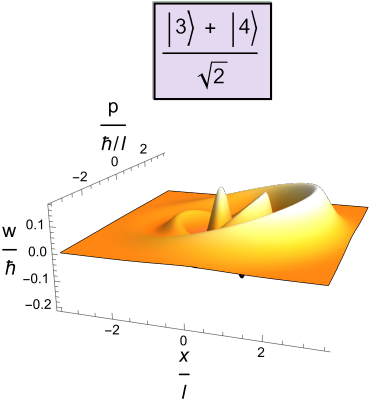}
     \includegraphics[width=0.3\linewidth]{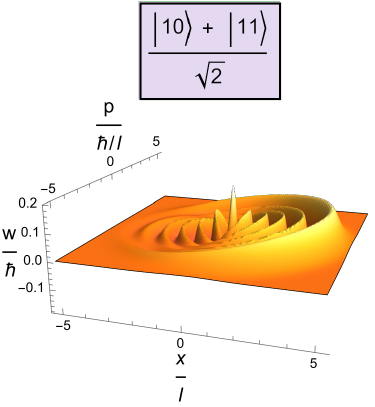}
    \includegraphics[width=0.3\linewidth]{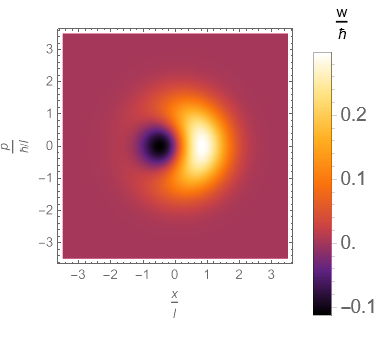}
    \includegraphics[width=0.3\linewidth]{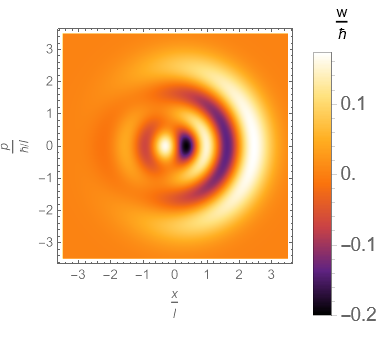}
    \includegraphics[width=0.3\linewidth]{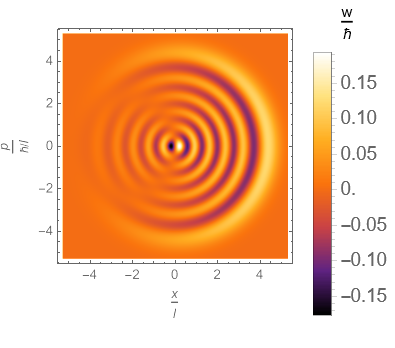}
       \caption{Graphical representation of  $w_{n,n+1}(x,p)$ the Wigner functions of a superposition of Fock states $(\ket{n}+\ket{n+1})/{\sqrt{2}}$.  Left: $n=0$, Middle: $n=3$, Right: $n=10$ . All $w_{n,n+1}(x,p)$ have a average weight  $\braket{\hat{p}}=0$ in $p$-direction and have a average weight of $\braket{\hat{x}}=l\sqrt{(n+1)/2}$ in $x$-direction.}
     \label{fig:plot Wignerfunctions wn,n+1}
 \end{figure}

The same results can be derived in phase space representation, which makes it possible to visually  gain an intuition on the  quantum states.  The Wigner function $w_n(x,p)$ of a Fock state $\ket{n}$ is given by 
\begin{equation}\label{eq:Wignerfunction wn}
\begin{aligned}
    w_n(x,p)=&\frac{(-1)^n}{\hbar\pi}L_n\left(2 \xi(x,p)\right) 
e^{-\xi(x,p)},\\[1ex]
\xi(x,p)=&\left(\frac{x}{l}\right)^2+\left(\frac{pl}{\hbar}\right)^2
\end{aligned}
\end{equation}

where $L_n(x)$ are the Laguerre Polynomials defined as $L_n(x)=\sum_{k=0}^n \frac{(-1)^k}{k!}\binom{n}{k} x^k$~\cite{WeissteinLaguerrePolynomial}.  $w_n(x,p)$ can be calculated via the $x$-space representation of a Fock state, \autoref{eq:phin(x)}, \cite{Mostowski} 
\begin{equation}
\sigma_{\hbar}[|n\rangle\langle n|]=\int d y \exp \left(-\mathrm{i} \frac{p y}{\hbar}\right)\braket{x+y/2|n}\braket{n|x-y/2}
\end{equation}
or alternatively $w_n(x,p)$  derived directly in phase space representation~\cite{Zachos-QuantumMechanicsinPS} from the symbol $H$, as normalized solution of 
\begin{equation}
 H \star w_n=E_n w_n=w_n\star H.
\end{equation}
In \autoref{fig:plot Wignerfunctions wn}, one can see that $w_n(x,p)$ becomes more oscillatory for larger the larger values of $n$.  For all possible $n$, $w_n(x,p)$ is centered around the origin. This visualizes directly that  one obtains
\begin{equation}
    \begin{aligned}
        \braket{n|\hat{x}|n}=&0\\
        \braket{n|\hat{p}|n}=&0,\\
    \end{aligned}
\end{equation}
when integrating $\sigma_\hbar[\hat{x}]=x$ and $\sigma_\hbar[\hat{p}]=p$ weighted by  $w_n(x,p)$.   The Wigner function of a superposition state $\ket{n,m_{\varphi_m}}=\frac{\ket{n}+\mathrm{e}^{\mathrm{i}\varphi_m}\ket{m}}{\sqrt{2}}$ is given by 
\begin{equation}
\begin{aligned}
       w_{n,m_{\varphi_m}}(x,p)=\frac{w_n(x,p)+w_m(x,p)}{2}&+\frac{\sigma_\hbar\left[\mathrm{e}^{\mathrm{i}\varphi_m}\ket{m}\bra{n}+\text{h.c.}\right]}{2\pi\hbar}
\end{aligned}
\end{equation}
with 
\begin{equation}\label{eq:symbol ketm bran plus hc}
    \frac{\sigma_\hbar\left[\mathrm{e}^{\mathrm{i}\varphi_m}\ket{m}\bra{n}+\text{h.c.}\right]}{2\pi\hbar}=\frac{(-1)^n}{\hbar \pi}2^{\frac{m-n}{2}}\sqrt{\dfrac{n!}{m!}} \Re\left[\mathrm{e}^{-\mathrm{i}\varphi_m}\left(\frac{x}{l}+\mathrm{i}\frac{pl}{\hbar}\right)^{m-n}\right] \ L_n^{(m-n)}\left(2 \xi\right)   e^{-\xi}\,,
\end{equation}
where the associated Laguerre Polynomials are given by~\cite{WeissteinAssociatedLaguerrePolynomial}
\begin{equation}
    L_n^{(k)}(x)=\sum_{m=0}^n(-1)^m \frac{(n+k)!}{(n-m)!(k+m)!m!} x^m.
\end{equation}
$w_{n,m_{\varphi_m}}$ can be derived straightforwardly from~\autoref{eq:WignertransfromScalar},~\autoref{eq:phin(x)} using the relation \cite{Feldheim1942}
\begin{equation}\label{eq:Feldheim}
\begin{aligned}
\text{for }& m\geq n:\\
& 2^m n!(v+\mathrm{i} u)^{m-n} L_n^{(m-n)}\left(2 u^2+2 v^2\right)
=\frac{1}{\sqrt{\pi}} \int_{-\infty}^{+\infty} \mathrm{d} xe^{-(v-\mathrm{i} u)^2} H_m(x+v) H_n(x-v). 
\end{aligned}
\end{equation}
\autoref{fig:plot Wignerfunctions wn,n+1} shows the Wigner functions of superpositions $\ket{n,n+1}=\frac{\ket{n}+\ket{n+1}}{\sqrt{2}}$ .  One can directly observe, that the Wigner functions are centered around $0$ in $p$-direction but not in $x$-direction illustrating visually  $\braket{\hat{p}}=0$ but $\braket{\hat{x}}\neq 0$. Again, 
\begin{equation}
\begin{aligned}
 \langle\hat{x}\rangle_{n,n+1}&=\frac{l}{\sqrt{2}}\sqrt{n+1} \\
\langle\hat{p}\rangle_{n,n+1}&=0\\
\end{aligned}
\end{equation}
can be calculated via $w_{n,n+1}(x,p)$ using~\autoref{eq:IntroWW-into}.  
\subsubsection{Time evolution in phase space representation}
\begin{figure}[h]
     \centering
    \includegraphics[width=0.32\linewidth]{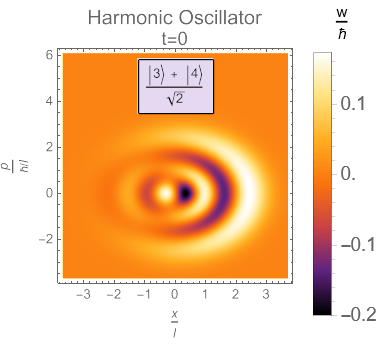}
    \includegraphics[width=0.32\linewidth]{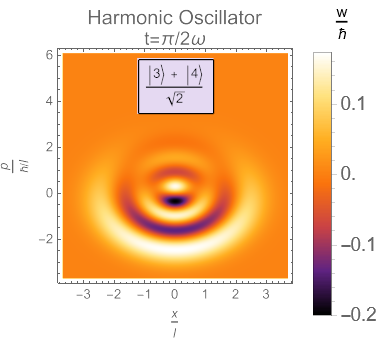}
    \includegraphics[width=0.32\linewidth]{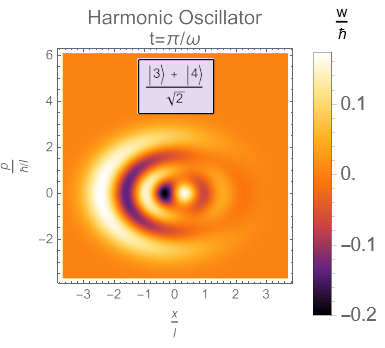}
       \caption{Time evolution of the initial Wigner function $w_{3,4}(x,p)$ illustrated at the times $t=0$, $t=\pi/2\omega$ and $t=\pi/\omega$. $w_{3,4}$ is rotating clockwise with an angular velocity $\omega$.}
     \label{fig:HO wnn+1 evolution}
\end{figure}
For the Quantum Harmonic Oscillator, the right hand side of ~\autoref{eq:partialt o scalar approx} is exactly given by the Poisson bracket
\begin{equation}
    \partial_to=\{o,H\},
\end{equation}
since all derivatives in $x$ and $p$ of the 3rd order and beyond of the symbol $H(x,p)$ are zero. Therefore, the solution to the equation of motion is exactly given by  $o(t)=o(x^{\phi_{t}},p^{\phi_{t}})$, the evolution under the Hamiltonian flow $\phi$ of the classical Harmonic oscillator 
\begin{equation}\label{eq:HO x,p classical}
    \begin{aligned}
        x^{\phi_{t}}=&l\cdot r_0\cos(\varphi_0-\omega t)\\
          p^{\phi_{t}}=&lm\omega\cdot  r_0\sin(\varphi_0-\omega t)\\
    \end{aligned}
\end{equation}
with $r_0=\sqrt{(x/l)^2+(p/lm\omega)^2}$ and $\varphi_0=\arctan{\left(\frac{p/lm\omega}{x/l}\right)}$  the dimensionless cylindrical coordinates of the initial conditions  $x^{\phi_{t=0}}=x$ and $p^{\phi_{t=0}}=p$. After pulling the time dependence into the Wigner function as $w(x^{\phi_{-t}},p^{\phi_{-t}})$, the Wigner function is rotating  around the phase space origin with an angular velocity $-\omega$.  From~\autoref{fig:plot Wignerfunctions wn} one can see that $w_n(x,p)$ is invariant under such a rotation, which illustrates how time  invariance is reproduced in phase space representation. For the dynamical state $\ket{n,n+1}$ the evolution of  
\begin{equation}\label{eq:HO <x>nn+1(t) <p>nn+1(t) }
\begin{aligned}
 \langle  \hat{x}\rangle_{n.n+1}(t)&=\frac{l}{\sqrt{2}}\sqrt{n+1}\cos{(\omega t)} \\
 \langle \hat{p}\rangle_{n,n+1}(t)&=-\frac{l m \omega}{\sqrt{2}} \sqrt{n+1}\sin{(\omega t) }, \\
\end{aligned}
\end{equation}
follows directly from the rotation of the Wigner function as illustrated in figure~\ref{fig:HO wnn+1 evolution}.
\section{The coherent state $\ket{\alpha}$ and its  classical limit}

 \begin{figure}[h]
     \centering
    \includegraphics[width=0.3\linewidth]{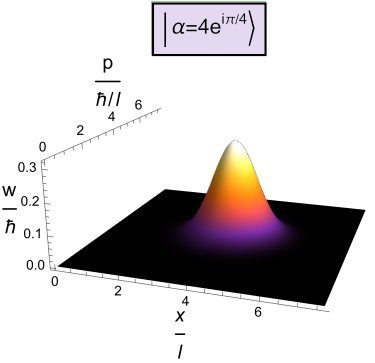}
    \includegraphics[width=0.3\linewidth]{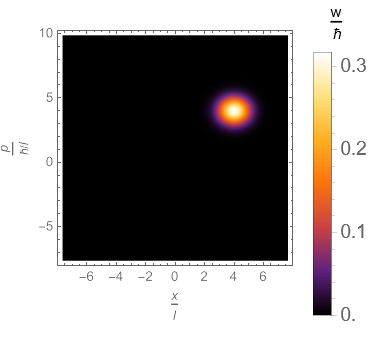}
       \caption{Graphical representation of the Wigner function $w_{\alpha}(x,p)$ of a coherent state with $\alpha=4\mathrm{e}^{\mathrm{i}\pi/4}$. $w_{\alpha}(x,p)$ does not take negative values and is well localized around its maximum value.}
     \label{fig:HO walpha}
\end{figure}

In this section,  the classical limit of the Quantum  Harmonic Oscillator is studied, following mainly the results of~\cite{Mostowski}. In \autoref{fig:plot Wignerfunctions wn}  and \autoref{fig:plot Wignerfunctions wn,n+1}, one can see that the Wigner functions of a single Fock state $\ket{n}$ and a superposition $\ket{n,n+1}$ can  be highly non-localized and take negative values. These features are in contrast to the nature of the point-like probability density of a classical particle $P(x,p)=\delta (x_c-x)\delta(p_c-p)$ . Therefore,  the states  $\ket{n}$ and  $\ket{n,n+1}$  are in general not suitable for obtaining a classical limit of quantum dynamics. Only for the ground state $\ket{0}$, $w(x,p)$ is taking only positive values and has minimal delocalization $\Delta x \Delta p =\frac{\hbar}{2}$.  A common choice for studying quantum-classical correspondence of the Quantum Harmonic Oscillator is a displacement of the ground state
   \begin{equation}
   \hat{D}(\alpha)\ket{0}=\exp\left({\alpha \hat{a}^\dagger-\bar{\alpha}\hat{a}}\right)\ket{0}=\ket{\alpha}.
\end{equation}
The obtained state ~
\begin{equation}\label{eq:coherent state superposition}
    |\alpha\rangle=\exp \left(-|\alpha|^2 / 2\right) \sum_n \frac{\alpha^n}{\sqrt{n!}}|n\rangle,
\end{equation}
is called the coherent state~\cite{Mostowski,Munster2020}  and is said to be the ``most classical`` state, because it has minimal uncertainty $\Delta x\Delta p=\frac{\hbar}{2}$.   The coherent state $\ket{\alpha}$ can also be found as the eigenvector of the annihilation operator $\hat{a}$ 
\begin{equation}
    \hat{a}\ket{\alpha}=\alpha\ket{\alpha}
\end{equation}
with eigenvalue $\alpha=|\alpha|\mathrm{e}^{\mathrm{i}\varphi_\alpha}$.   Its $x$-space representation has the form of a plane wave modulated with a Gaussian 
\begin{equation}
    \psi_\alpha(x)=\mathrm{e}^{\mathrm{-i}\frac{p_\alpha x_\alpha}{2\hbar}}\frac{1}{\sqrt{l\sqrt{\pi}}}\mathrm{e}^{-\frac{(x-x_\alpha)^2}{2l^2}}\mathrm{e}^{\mathrm{i}\frac{p_\alpha x}{\hbar}},
\end{equation}
where  $x_\alpha $ and $p_\alpha$  are the wave packet centers in real and momentum space 
\begin{equation}
     \begin{aligned}
         \braket{\hat{x}}_\alpha=&x_\alpha =l\sqrt{2}|\alpha|\cos{\varphi_\alpha}\\
         \braket{\hat{p}}_\alpha=&p_\alpha=l m \omega \sqrt{2}|\alpha|\sin{\varphi_\alpha}.\\
     \end{aligned}
\end{equation}
Under quantum time evolution, all terms of~\ref{eq:coherent state superposition}  are in phase
\begin{equation}\label{eq:coherent state superposition}
    |\alpha\rangle(t)=\exp \left(-|\alpha|^2 / 2\right) \sum_n \frac{\left(|\alpha|\mathrm{e}^{\mathrm{i}(\varphi_\alpha-\omega t)}\right)^n}{\sqrt{n!}}|n\rangle.
\end{equation}
As a consequence, the evolution of the wave packet centers
\begin{equation}\label{eq:HO <x>(t) <p>(t) coherent}
    \begin{aligned}
           \braket{\hat{x}}_\alpha(t)&=l\sqrt{2}|\alpha|\cos{\left(\varphi_\alpha -\omega t\right)}\\
\braket{\hat{p}}_\alpha(t)&=lm\omega\sqrt{2}|\alpha|\sin{\left(\varphi_\alpha -\omega t\right)} 
    \end{aligned}
\end{equation}
follow exactly the trajectories of a classical Harmonic Oscillator, which is another reason why the coherent state is called the most classical state of the Quantum Harmonic Oscillator. The uncertainties $\Delta x=\sqrt{\braket{\hat{x}^2}-\braket{\hat{x}}^2}$ and $\Delta p=\sqrt{\braket{\hat{p}^2}-\braket{\hat{p}}^2}$ remain minimal under time evolution
\begin{equation}
    \begin{aligned}
        \Delta x (t)=&\frac{l}{\sqrt{2}}\\
        \Delta p (t)=&\frac{\hbar}{l\sqrt{2}}.\\
    \end{aligned}
\end{equation}
 \begin{figure}[t]
     \centering
    \includegraphics[width=0.3\linewidth]{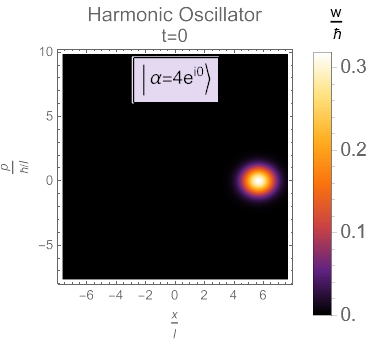}
    \includegraphics[width=0.3\linewidth]{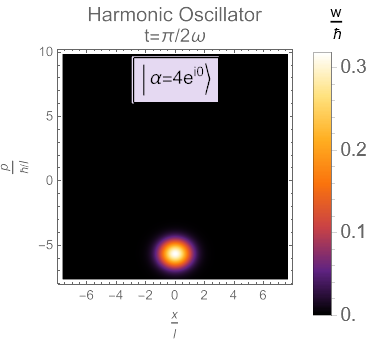}
    \includegraphics[width=0.3\linewidth]{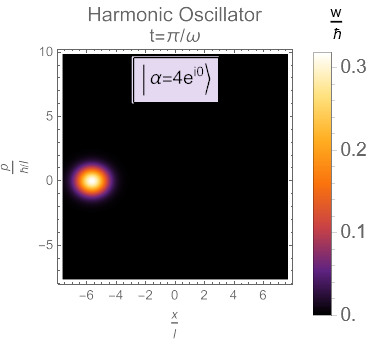}
       \caption{Time evolution of an initial Wigner function $w_\alpha(x,p)$ with $\alpha=4$. The center of the Wigner function follows the phase space trajectory of a classical particle in the classical Harmonic Oscillator.}
     \label{fig:HO walpha evolution}
\end{figure}
Now, one can ultimately study the classical limit of the coherent state via its Wigner function~\cite{Mostowski} 
\begin{equation}\label{eq:Wignerfunction_alpha}
    \begin{aligned}
        w_\alpha(x,p)=&\frac{1}{\pi\hbar}\mathrm{e}^{-\frac{(x_\alpha-x)^2}{l^2}}\mathrm{e}^{-\frac{(p_\alpha-p)^2}{(\hbar/l)^2}},
    \end{aligned}
\end{equation}
illustrated in \autoref{fig:HO walpha}. One can see that $w_\alpha(x,p)$ does not take negative values and is localized around $x_\alpha$ and $p_\alpha $ with minimal width $\Delta x=\frac{l}{\sqrt{2}}$ and $\Delta p =\frac{\hbar}{l\sqrt{2}}$. \autoref{fig:HO walpha evolution}  shows $w_\alpha(x,p)$ under time evolution:  The Hamiltonian flow $\phi$~(\autoref{eq:HO x,p classical} ) generates a rotation of $w_\alpha (x,p)$ such that the center of $w_\alpha (x,p)$ follows the phase space trajectory of a classical particle.  The classical limit then arises when measuring the state with some device unable to resolve a precision of  $\Delta x=\frac{l}{\sqrt{2}}$ and $\Delta p =\frac{\hbar}{l\sqrt{2}}$. One would then detect a point-like probability distribution in phase space evolving under Hamiltonian flow, which is exactly the dynamics of a classical particle. 

%% file: chapters/Wigner-Weyl_formalism_for_Multiband_systems.tex
\chapter{Semiclassical Analysis of Multiband Systems via Wigner Function Evolution}\label{chapter:operatorvauledsemiclass}

So far, we have only considered Hamiltonians $\hat{\mathcal{H}}$ built up from $\hat{x}$ and $\hat{p}$, i.e. Hamiltonians acting purely on $L^2(\mathbb{R})$.  For these Hamiltonians, the symbol $H( x,p)$ is just a scalar-valued function $f( x,p)$. In contrast, a Hamiltonian $\hat{\mathcal{H}}$ that also acts on other (fast) degrees of freedom, like for example the spin of an electron or the sub-lattice site of a crystal, has a symbol $H(x,p)$ which is still operator valued. Therefore, $H (x,p)$ has a band-spectrum $e^\nu(x,p) $ of eigenvalues.  In this chapter, it is presented  how  the semiclassical equations of motion can be derived for such a  multiband symbol $H (x,p)$ via adiabatic subspaces $\mathcal{H}^\nu _\text{adiabatic}$. The aim is to address the questions \say{Under what conditions can the semiclassical equations of motion be applied?} and \say{What are the origins of the different first order correction terms in the semiclassical equations of motion?}. A reader primarily interested in applying the semiclassical Wigner function  evolution to predict wave dynamics, but not concerned with multiband systems, may  skip this chapter and directly move ahead to~\autoref{chapter:AO}.       \\
For an operator $\hat{A}$ that acts on both $L^2(\mathbb{R})$ and $\mathcal{H}_\text{f}$, the Hilbert space of fast degrees of freedom,   the symbol
\begin{equation}\label{eq:WignertransfromOperator}
\sigma_\hbar[\hat{A}]=A(x,p)=\int\mathrm{d}y\mathrm{e}^{-\mathrm{i}\frac{py}{\hbar}}\braket{x+y/2|\hat{A}|x-y/2}.
\end{equation}
still acts on $\mathcal{H}_\text{f}$ and thus in general $[A(x,p),B(x,p)]\neq0$. $\hat{x}$ and $\hat{p}$ will be referred to as slow degrees of freedom and we choose lowercase letters to designate slow operators, i.e. operators that act purely on $L^2(\mathbb{R})$. Historically, one of the first examples of a slow-fast decomposition of Hilbert spaces is the Born-Oppenheimer approximation for a molecule: the electrons are assumed to move much faster than the nuclei such that the Hilbert space of the electrons can be diagonalized by the electron energy eigenvectors $\ket{\psi^\nu}$ for fixed nuclei position.  The motion of the nuclei then has the electron energy as an effective potential, i.e. $\hat{\mathcal{H}}$ acts on a product of nuclei position $\ket{x_\text{nuclei}}$ and electron wave functions $\{\ket{\psi^\mu}\}$. \\
Given a Hamiltonian acting on a product of slow and fast Hilbert spaces $L^2(\mathbb{R}) \otimes \mathcal{H}_{\text {f}}$, the goal is now to approximate the dynamics of some slow observable $\hat{o}$   by a scalar equation of motion for its symbol $o(x,p)$.  In the following, we will present our understanding of the work of Stiepan and Teufel \cite{Stiepan_2012}, and of parts of chapter 3 in~\cite{Teufel2003-AdiabaticPTQM} and chapter 8 in~\cite{Lein2010}, in which the semiclassical phase space flow is defined individually for adiabatic subspaces associated with bands of the symbol $H(x,p)$. The focus will be on showing the reader the origin of the different first order correction terms in the semiclassical equations of motion, rather than on a precise mathematical formulation of theorems. Please note that in \cite{Stiepan_2012}, chapter 3 of~\cite{Teufel2003-AdiabaticPTQM}, the Poisson bracket $\{,\}$ is defined with a opposite sign than in~\autoref{eq:Poisson}.\\
Even for an operator $\hat{o}=\hat{o}\,\mathbb{1}_{\mathcal{H}_\text{f}}$, that acts purely on $L^2(\mathbb{R})$, the Heisenberg equation in phase space representation
\begin{equation}
    \frac{\partial o}{\partial t}=\frac{\mathrm{i}}{\lambda\hbar}[H,o]_\star
\end{equation}
is operator valued in the right-hand side if $\hat{\mathcal{H}}$ acts on $\mathcal{H}_\text{f}$.  As a consequence, the equation of motion at first order
\begin{equation}
     \frac{\partial o}{\partial t}=\{o,H\}+\mathcal{O}(\lambda^2 )
\end{equation}
cannot  be directly solved by a Hamiltonian flow $\phi$, because the Hamiltonian flow is defined by a scalar differential equation. Therefore, it is necessary to replace the operator valued symbol $H(x,p)$ with a scalar symbol. Firstly,  note that the eigenvalues $e_0^{(\nu)}$ of $H(x,p)$ are scalar symbols.  The strategy is now to  associate  an adiabatic subspace $\mathcal{H}^\nu_\text{adiabatic}$ to each eigenvalue  $e_0^{(\nu)}$, so that for each subspace the operator valued symbol $H(x,p)$ can be approximated by a scalar symbol $h^{(\nu)}(x,p)$. The $ 0$th order of  $h^{(\nu)}(x,p)$ will just be $e_0^{(\nu)}$. \\
For this purpose, one has to project the phase space dynamics on adiabatic subspaces  $\mathcal{H}^\nu_\text{adiabatic}$ via an adiabatic projector $\pi^\nu(x,p)$. In order to approximate the projector $ \pi^\nu(x,p)$, a  control parameter $\varepsilon$ has to be introduced that allows to simplify $\pi^\nu$ in the limit $\varepsilon\ll1$.  This can be done by introducing new dimensionless coordinates for  $x,p,t$ such that $[\hat{x},\hat{p}]\rightarrow\mathrm{i}\varepsilon$, $\mathrm{i}\hbar\partial_t\rightarrow\mathrm{i}\varepsilon\partial_t$ and that the band gap  $\Delta e_0^{}$ separating $e_0^{(\nu)}$ from other bands becomes $\Delta e_0\gtrsim1$.   Expressing the operator Hamiltonian in  the new dimensionless $x$-space representation, $\hat{\mathcal{H}}(x,-\mathrm{i}\varepsilon\partial_x )$,  it becomes clear that the limit $\varepsilon\ll1$  corresponds to a system that is slowly varying compared to the band gap. Note that $L^2(\mathbb{R})$ is actually a ``slow`` Hilbert space only for $\varepsilon\ll1$. After nondimensionalization, $\mathrm{i}\varepsilon$ is the commutator in Moyal product 
\begin{equation}
    (A\star B)_{\varepsilon} =\sum_{n=0}^{\infty} \frac{1}{n!}\left(\frac{\mathrm{i} \varepsilon}{2}\right)^n \sum_{k=0}^n(-1)^k\binom{n}{k} \frac{\partial^n A}{\partial p^k \partial x^{n-k}} \frac{\partial^n B}{\partial x^k \partial p^{n-k}}
\end{equation}
and the symbol Hamiltonian can be expressed as 
\begin{equation}
    H(x,p)=H_0(x,p)+\varepsilon H_1(x,p)+\mathcal{O}(\varepsilon^2).
\end{equation}
The projector on the adiabatic subspace $\pi^\nu$ is now assumed to be a development in $\varepsilon$ 
\begin{equation}
    \pi^\nu=\sum_i\varepsilon^i\pi_i^\nu. 
\end{equation}
An adiabatic subspace $\mathcal{H}^\nu_\text{adiabatic}$ is defined as being invariant under time evolution up to non-perturbative effects denoted with  $\mathcal{O}(\varepsilon^\infty)$.  Non-perturbative effects are mentioned for completeness, but will not explicitly be studied in this Master's thesis.
This expresses a condition on the adiabatic projector 
\begin{equation}\label{eq:Teuf pinu-invariant}
    \begin{aligned}
        \pi^\nu\star H-H\star\pi^\nu=&\mathcal{O}(\varepsilon^\infty).\\
    \end{aligned}
\end{equation}
Further, the condition of $\pi^\nu$ being a true projector is 
\begin{equation}\label{eq:Teuf pinu-idem}
     \pi^\nu\star\pi^\nu=\pi^\nu.\\
\end{equation}
These two conditions give at $0$th order that the adiabatic  projector is the projector on $\ket{\Psi_0^\nu}$, the eigenvector of $H_0(x,p)$ with the associated eigenvalue $e_0^{(\nu)}$ 
\begin{equation}
    \pi_0^\nu=\ket{\Psi_0^\nu}\bra{\Psi_0^\nu}.
\end{equation}
We can see that the evolution of this projector
\begin{equation}
\begin{aligned}
    \partial_t\pi_0^\nu&=\frac{\mathrm{i}}{\varepsilon}[H,\pi_0^\nu]_\star\\
    &=\frac{\mathrm{i}}{\varepsilon}\underbrace{[H_0,\pi_0^\nu]}_0+\mathcal{O}(1)
\end{aligned}
\end{equation}
is slow compared to the evolution of a projector $\pi_0^{\nu\mu}$ on a superposition of eigenvectors $(\ket{\Psi_0^\nu}+\ket{\Psi_0^\mu})/\sqrt{2}$
\begin{equation}
\begin{aligned}
    \partial_t\pi_0^{\nu\mu}&=\frac{\mathrm{i}}{\varepsilon}[H,\pi_0^{\nu\mu}]_\star\\
    &=\frac{\mathrm{i}}{\varepsilon}\underbrace{[H_0,\pi_0^{\nu\mu}]}_{\sim(\Delta e_0)}+\mathcal{O}(1)\sim\frac{\Delta e_0}{\varepsilon},
\end{aligned}
\end{equation}
if $\varepsilon\ll1$.
Using again the conditions \autoref{eq:Teuf pinu-invariant} and \autoref{eq:Teuf pinu-idem} one can show the following relation for first order term of $\pi^\nu(x,p)$  
\begin{equation}\label{eq:pi0pi1pi0}
    \pi_0^\nu\pi_1^\nu\pi_0^\nu=-\frac{\mathrm{i}}{2} \pi_0^\nu\{ \pi_0^\nu, \pi_0^\nu\} \pi_0^\nu.
\end{equation} 
Let us now consider a state $\ket{\psi}\in L^2(\mathbb{R})\otimes \mathcal{H}_\text{f}$ that can be associated to the adiabatic subspace $\mathcal{H}^\nu_\text{adiabatic}$, e.g. a state in the range of $\hat{\pi}^\nu=\text{Op}[\pi^\nu]$,   
\begin{equation}
\begin{aligned}
\hat{\pi}^\nu\ket{\psi}=&\ket{\psi}+\mathcal{O}(\varepsilon^\infty),\\
\end{aligned}
\end{equation}
so that its expectation value of an observable $\hat{o}$ is invariant under a projection of $\hat{o}$ on  $\mathcal{H}^\nu_\text{adiabatic}$
\begin{equation}
    \begin{aligned}
        \braket{\hat{o}}=&\braket{\psi|\hat{o}|\psi}\\
        =&\braket{\psi|\hat{\pi}^\nu\hat{o}\hat{\pi}^\nu|\psi}+\mathcal{O}(\varepsilon^\infty).
    \end{aligned}
\end{equation}
The same holds for the time evolution of $\hat{o}(t)$ in the Heisenberg picture and the equation of motion can be expressed in phase space 
\begin{equation}
\begin{aligned}
        \pi^\nu\star\partial_to\star\pi^\nu = &\frac{\mathrm{i}}{\varepsilon}\pi^\nu\star[H,o]_\star\star\pi^\nu\\
        =&\frac{\mathrm{i}}{\varepsilon}\pi^\nu\star[\pi^\nu\star H\star \pi^\nu,o]_\star\star\pi^\nu.\\
        \end{aligned}
\end{equation}
\begin{figure}[t]
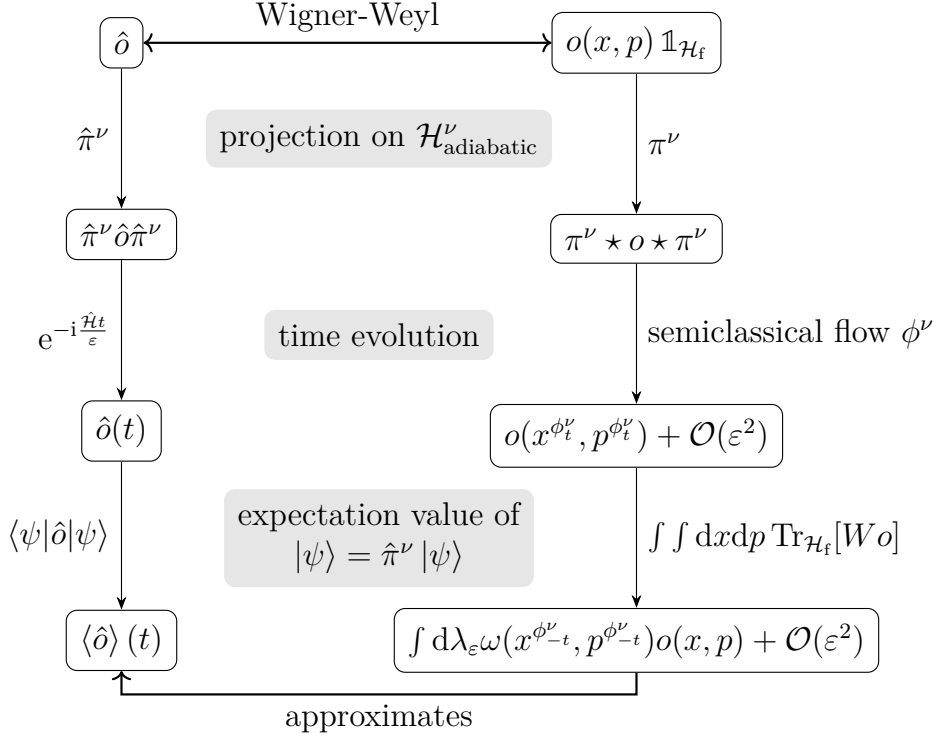

    \centering
   \includestandalone[width=0.8\linewidth]{graphics/sketches/semiclassical_approximation}
    \caption{Sketch of the steps when performing the semiclassical approximation for the evolution of the expectation value of  $\hat{o}$ for a state lying in $\mathcal{H}_\text{adiabatic}^\nu$.}
    \label{fig:sketch semiclass approx multiband}
\end{figure}
Ultimately, the  projected operator valued symbol $H(x,p)$ can be approximated with the projected scalar Hamiltonian symbol $h^{(\nu)}(x,p)$
\begin{equation}\label{eq:Teuf projected H}
    \begin{aligned}
        \pi^\nu\star H\star\pi^\nu=&\pi^\nu\star h^{(\nu)}\star\pi^\nu+\mathcal{O}(\varepsilon^2)
    \end{aligned}
\end{equation}
with 
\begin{equation}\label{eq:Teuf projected H}
    \begin{aligned}      h^{(\nu)}=&e_0^{(\nu)}+\varepsilon(M^{(\nu)}+\text{Tr}_{\mathcal{H}_\text{f}}[H_1\pi_0^\nu])\\[0.4ex]
        M^\nu=&\frac{\mathrm{i}}{2}\text{Tr}_{\mathcal{H}_\text{f}}[\pi_0^\nu \{\pi_0^\nu,H_0-e^\nu_0\}]
    \end{aligned}
\end{equation}

as can be shown by calculating the Moyal products (the $\star$-products) up to order $\mathcal{O}(\varepsilon)$ and using the relation for $\pi_1^\nu$,~\autoref{eq:pi0pi1pi0}. As announced previously, the 0th order of $h^{(\nu)}$ is nothing but the eigenvalue $e_0^{(\nu)}(x,p)$ of $H_0(x,p)$.  The term 
\begin{equation}
    \varepsilon\text{Tr}_{\mathcal{H}_\text{f}} [H_1\pi_0^\nu]=\varepsilon\braket{\Psi_0^\nu|H_1|\Psi_0^\nu}
\end{equation} 
is the correction by the first order term of the  symbol Hamiltonian $H=H_0+\varepsilon H_1+\mathcal{O}(\varepsilon^2)$. 
There are two possible origins for $ H_1$: either already the operator  $\hat{\mathcal{H}}$ is decomposed into $\hat{\mathcal{H}}=\hat{\mathcal{H}}_0+\varepsilon\hat{\mathcal{H}}_1+\mathcal{O}(\varepsilon^2)$ or the Wigner transform of $\sigma_\varepsilon[\hat{\mathcal{H}}_0]$ gives $\mathcal{O}(\varepsilon)$-terms. Depending on which is the case, this term will be called the `perturbation correction` or  the `Moyal term`. The term $\varepsilon M^\nu$ is necessary when  replacing the projected $H_0$ with a projected scalar $h^{(\nu)}$ up to errors $\mathcal{O}(\varepsilon^2)$, because the  first order projector $\varepsilon \pi ^\nu_1$ and the first order terms in the $\star$-product both have a nontrivial  contribution.  $\varepsilon M^\nu$ is the term that Littlejohn and Flynn called `no-name term` in their multicomponent WKB
 approximation~\cite{Littlejohn1992}.  We will refer to it as the `Littlejohn-Flynn term` or `scalarisation term`. The projection of the time derivative of the scalar symbol $o(x,p)\mathbb{1}_{\mathcal{H}_\text{t}}$ can now be approximated by a scalar equation
 \begin{equation}
\begin{aligned}
        \pi^\nu\star\partial_to\star\pi^\nu   =&\frac{\mathrm{i}}{\varepsilon}\pi^\nu\star\underbrace{[\pi^\nu\star h^{(\nu)}\star \pi^\nu,o]_\star}_{\neq\pi^\nu\star[ h^{(\nu)} ,o]_\star\star\pi^\nu}\star\pi^\nu+\mathcal{O}(\varepsilon^2),\\
\end{aligned}
\end{equation}
but now $\frac{\mathrm{i}}{\varepsilon}[h^{(\nu)},\pi^\nu]_\star=\{\pi^\nu_0,h^{(\nu)}\}+\mathcal{O}(\varepsilon)\neq\mathcal{O}(\varepsilon^2)$. So that $\pi^\nu$ cannot be brought outside the Moyal bracket by swapping $h^{(\nu)}\star\pi^\nu$. This leads after rearranging to the semiclassical equations of motion
\begin{equation}
     \pi^\nu\star\partial_to\star\pi^\nu= \pi^\nu\star\left(\{o,h^{(\nu)}\}+\mathrm{i}\varepsilon\operatorname{Tr}_{\mathcal{H}_{\text{f}}}[\,\pi_0^\nu\left[\{h^{(\nu)}, \pi_0^\nu\},\left\{o, \pi_0^\nu\right\}\right]\,]\right)\star\pi_0^\nu+\mathcal{O}(\varepsilon^2)
\end{equation}
where $\mathrm{i}\varepsilon\operatorname{Tr}_{\mathcal{H}_{\text{f}}}[\,\pi_0\left[\left\{h, \pi_0\right\},\left\{a, \pi_0\right\}\right]\,]$ can be seen as a geometric correction due to the projection of the dynamics on $\mathcal{H}^\nu_\text{adiabatic}$: it is the manifestation of the Berry curvature dating back to~\cite{Berry1984}. More commonly, the semiclassical equations of motion are expressed for $\dot{x}$ and $\dot{p}$ as \cite{Luneau:phd,perez:tel-03922294}
\begin{equation}\label{eq:Semiclass eq of motion}
\boxed{
\begin{aligned}
& \dot{x}=+\frac{\partial h^{\nu}}{\partial p} - \varepsilon\left(F_{p_{}x}^{\nu} \dot{x}+F_{p p_{}}^{\nu} \dot{p}_{}\right)+\mathcal{O}(\varepsilon^2) \\
& \dot{p}=-\frac{\partial h^{\nu}}{\partial x}+\varepsilon \left( F_{xx}^{\nu} \dot{x}+F_{x p_{}}^{\nu} \dot{p}_{}\right) +\mathcal{O}(\varepsilon^2) \\
\end{aligned}
}
\end{equation}
with  
\begin{equation}\label{Berry:curvature}
 F_{ab}^\nu=-2 \Im\left(\sum_{\mu \neq \nu} \frac{\bra{\Psi^\nu}\partial_a H_0\ket{\Psi^\mu}\bra{ \Psi^\mu}\partial_b H_0 \ket{\Psi^\nu}}{\left(e_0^{(\nu)}-e_0^{(\mu)}\right)^2}\right) \quad \text { (Berry curvature) }
\end{equation}
and
\begin{equation}
h^{(\nu)}=e_0^{(\nu)}+\varepsilon\left(M^{(\nu)}+\braket{H_1}^{(\nu)}\right),\quad   \quad \text { (scalar symbol Hamiltonian) }
\end{equation}
where
\begin{equation}\label{eq:no name term}
    \begin{aligned}
        M^{(\nu)}=&- \Im\left(\sum_{\mu \neq \nu} \frac{\bra{\Psi^\nu}\partial_x H_0\ket{\Psi^\mu}\bra{ \Psi^\mu}\partial_p H_0 \ket{\Psi^\nu}}{e_0^{(\nu)}-e_0^{(\mu)}}\right) \quad \text { (Littlejohn-Flynn term) }&\\[4ex]
        \braket{H_1}^{(\nu)}=&\braket{\Psi_0^\nu|H_1|\Psi_0^\nu} \quad \quad \quad \quad \quad \quad \quad \quad \text { (perturbation correction/Moyal term).}&
    \end{aligned}
\end{equation}
It now becomes apparent that the condition on the gap $\Delta e_0\gtrsim 1 $ is necessary for $\varepsilon M$ and $\varepsilon F$ to be a correction of the order $\mathcal{O}(\varepsilon)$. Note that  in one spatial dimension the $xx$ and $pp$ components are always $F^\nu_{xx}=F^\nu_{pp}=0$, but in higher spatial dimensions one can have $F_{x_i,x_j}^\nu\neq0$ and $F_{p_i,p_j}^\nu\neq0$.

Finally, given a state $\ket{\psi}$ in the range of $\hat{\pi}^\nu$, the evolution of $\braket{\hat{o}}(t)$ can be approximated by the semiclassical phase space flow $\phi^\nu$, that solves the semiclassical equations of motion associated to $\mathcal{H}^\nu_\text{adiabatic}$, \autoref{eq:Semiclass eq of motion},
        \begin{equation}
    \braket{\hat{o}}(t)=\int\int\mathrm{d}x\mathrm{d}p\operatorname{Tr}_{\mathcal{H}_\text{f}}[\,W(x,p)]o(x^{\phi^\nu_t},p^{\phi^\nu_t})+\mathcal{O}(\varepsilon^2),
\end{equation}
where the Wigner function $W=\frac{\sigma_\varepsilon[\ket{\psi}\bra{\psi}]}{2\pi}$ is an operator acting on $\mathcal{H}_\text{f}$. 
The semiclassical flow $\phi^\nu$ does not preserve the phase space volume element $\mathrm{d}x\mathrm{d}p$ anymore. Instead, the phase space measure $\mathrm{d}\lambda_\varepsilon=\mathrm{d}x\mathrm{d}p(1+\varepsilon F^\nu_{xp})$
is approximately invariant under semiclassical flow
\begin{equation}
    \mathrm{d}x^{\phi^\nu_t}\mathrm{d}p^{\phi^\nu_t}\left(1+\varepsilon F^\nu_{x^{\phi^\nu_t}p^{\phi^\nu_t}}\right)=\mathrm{d}x\mathrm{d}p(1+\varepsilon F^\nu_{xp})+\mathcal{O}(\varepsilon^2).
\end{equation}
Using this phase space measure, the time dependence can be pulled into a measurement-corrected scalar Wigner function $\omega(x,p)$
    \begin{equation}\label{eq:<o>(t) semiclass including phasespacemeasure}
    \begin{aligned}
          \braket{\hat{o}}(t)=&\int\int\mathrm{d}x\mathrm{d}p(1+\varepsilon F^\nu_{xp})\omega(x^{\phi^\nu_{-t}},p^{\phi^\nu_{-t}}) o(x,p)+\mathcal{O}(\varepsilon^2)\\
    \end{aligned}
\end{equation}
with 
    \begin{equation}\label{eq:scalarWig}
    \begin{aligned}
          \omega(x,p) =&\frac{\operatorname{Tr}_{\mathcal{H}_\text{f}}[\,W(x,p)]}{1+\varepsilon F^\nu_{x,p}}\\
    \end{aligned}
\end{equation}
Figure~\ref{fig:sketch semiclass approx multiband} summarizes the most important steps to arrive at~\autoref{eq:<o>(t) semiclass including phasespacemeasure}. For the special case where $F^\nu_{x,p}$ is invariant under $\phi^\nu$,  the volume element $\mathrm{d}x\mathrm{d}p$ is approximately preserved under $\phi^\nu$ and  it is sufficient  to calculate the evolution of $w=\operatorname{Tr}[W]$
\begin{equation}
\begin{aligned}
\text{for } F^\nu_{x^{\phi^\nu_t}p^{\phi^\nu_t}}=F^\nu_{xp}: \quad 
   &\langle\hat{o}\rangle(t)
     = \iint \mathrm{d}x\,\mathrm{d}p\;
       w(x^{\phi^\nu_{-t}},p^{\phi^\nu_{-t}})\,o(x,p)
       + \mathcal{O}(\varepsilon^2),\quad \quad \\[1ex]
   &w(x,p)
     = \operatorname{Tr}_{\mathcal{H}_\text{f}}[W(x,p)].
\end{aligned}
\end{equation}

%% file: graphics/sketches/semiclassical_approximation.tex
\begin{tikzpicture}[node distance=1.8cm and 4cm, every node/.style={align=center}]

    \node (operator) [text=black] {\underline{\textbf{Operator space}}};
    \node (phase space) [right=3.5cm of operator, text=black] {\underline{\textbf{Phase space}}};

    \node (asymbol) [below=of phase space, yshift=1.2cm]  [draw,  rounded corners, text=black,  inner sep=5pt, anchor=north]{$o(x,p)\,\mathbb{1}_{\mathcal{H}_\text{f}}$};
   \node (projected asymbol) [below=of asymbol] [draw,  rounded corners, text=black,  inner sep=5pt, anchor=north]{$\pi^\nu\star o\star \pi^\nu$};
     \node (asymbol evolution) [below=of projected asymbol] [draw,  rounded corners, text=black,  inner sep=5pt, anchor=north]{$o(x^{\phi_t^\nu},p^{\phi_t^\nu})+\mathcal{O}(\varepsilon^2)$};
   \node (asymbol expectation) [below=of asymbol evolution] [draw,  rounded corners, text=black,  inner sep=5pt, anchor=north]{$\int\mathrm{d}\lambda_\varepsilon \omega(x^{\phi_{-t}^\nu},p^{\phi_{-t}^\nu}) o(x,p)+\mathcal{O}(\varepsilon^2)$};

 \node (aoperator) 
    [draw, rounded corners, text=black, inner sep=5pt, anchor=center] 
     at (operator |- asymbol) 
    {$\hat{o}$};

   \node (projected aoperator) [below=of aoperator] [draw,  rounded corners, text=black,  inner sep=5pt, anchor=north] {$\hat{\pi}^\nu\hat{o}\hat{\pi}^\nu$};
    \node (aoperator evolution) [below=of projected aoperator] [draw,  rounded corners, text=black,  inner sep=5pt, anchor=north] {$\hat{o}(t)$};
    \node (aoperator expectation) [draw, rounded corners, text=black, inner sep=5pt] 
    at (aoperator evolution |- asymbol expectation) {$\braket{\hat{o}}(t)$};




    \node (orangebox)  at ($ (projected aoperator)!0.5!(asymbol) $) [rounded corners,fill=gray!20, text=black,  inner sep=5pt]
        {projection on $\mathcal{H}_\text{adiabatic}^\nu$};
    \node (time evolution)  at ($ (aoperator evolution)!0.5!(projected asymbol) $) [ rounded corners, text=black,fill=gray!20,  inner sep=5pt]  {time evolution};
     \node (expectation value)  at ($ (aoperator expectation)!0.5!(asymbol evolution) $) [ rounded corners, text=black,fill=gray!20,  inner sep=5pt]  {expectation value of \\$\ket{\psi}=\hat{\pi}^\nu\ket{\psi}$};

\draw[<->, thick] (aoperator.east) -- node[above] {Wigner-Weyl} (asymbol.west);
    
\draw[-{Stealth}] (aoperator.south) -- (projected aoperator.north) node(semicl flow)[midway, left] {$\hat{\pi}^\nu$};
\draw[-{Stealth}] (asymbol.south) -- (projected asymbol.north) node(semicl flow)[midway, right] {$\pi^\nu$};
\draw[-{Stealth}] (projected asymbol.south) -- (asymbol evolution.north)
    node(semicl flow)[midway, right] {semiclassical flow $\phi^{\nu}$ 
    };
 \draw[-{Stealth}] (projected aoperator.south) -- (aoperator evolution.north) node(semicl flow)[midway, left] {$\mathrm{e}^{-\mathrm{i}\frac{\hat{\mathcal{H}}t}{\varepsilon}}$};
  \draw[-{Stealth}] (aoperator evolution.south) -- (aoperator expectation.north) node(semicl flow)[midway, left] {$\braket{\psi|\hat{o}|\psi}$};
\draw[-{Stealth}] (asymbol evolution.south) -- (asymbol expectation.north)
    node(semicl flow)[midway, right] { $\int\int\mathrm{d}x\mathrm{d}p \operatorname{Tr}_{\mathcal{H}_\text{f}}[Wo]$ 
    };

   \draw[<-, thick] 
    (aoperator expectation.south) 
        -- ++(0,-0.3cm) coordinate(corner)  
        -| (asymbol expectation.south);     


\node[below, yshift=-0.1cm] (apprimates) at ($(corner)!0.5!(asymbol expectation.south)$) [ rounded corners, text=black, inner sep=5pt] {approximates};

\end{tikzpicture}

%% file: chapters/Anharmonic_Oscillator.tex
\chapter{Wave Dynamics of the Anharmonic Oscillator}\label{chapter:AO}
After having introduced a formalism for approximating the evolution of expectation values $\braket{\hat{o}}(t)$ by the semiclassical evolution of the Wigner function, we ultimately apply this formalism to study the dynamics of wave packets in the (Quantum) Anharmonic Oscillator in this chapter,  and in the Dirac Oscillator in the following chapter. \\
There are 2 main goals. Firstly, we want to give the reader an example of how this formalism can be used and what its limitations are. A focus is also set on showing the effect of including or omitting the different first order corrections in the semiclassical equations of motion.
The other goal is to gain a global understanding of wave packet dynamics in each system. For this purpose, the semiclassical formalism is utilized  to predict characteristics of the dynamics, such as the oscillation period of the wave packet center and the typical delocalization time.  The exact solutions, though known, can be too complex to extract directly these quantities. \\     
As a first example, the (Quantum) Anharmonic Oscillator is chosen as a single band system (no fast degrees of freedom)  with nonzero Moyal term and nonzero higher-order terms in the equation of motion $\partial_to=\frac{\mathrm{i}}{\lambda \hbar}[H,o]_\star\,$. This allows us to study the effect of the Moyal term and the limitations of the semiclassical predictions without the additional complexity of a multiband system. Furthermore,  wave packets in the Anharmonic Oscillator exhibit a rich dynamics~\cite{sudheeshManifestationsWavePacket2004}, which themselves are worth investigating.  The reader concerned mainly with the application of the semiclassical formalism to multiband systems, rather than wave packet dynamics in the Anharmonic Oscillator, may  jump directly to~\autoref{chapter:DO}  and return to this chapter for the prediction of characteristic quantities via the semiclassical flow.  
\subsection{The Anharmonic Oscillator}
Adding a term quadratic in the number operator $\hat{n}=\hat{a}^\dagger \hat{a}$ to the Harmonic Oscillator (modulo $\hbar\omega\frac{1}{2}$), gives the Hamiltonian of an Anharmonic Oscillator~\cite{milburnQuantumClassicalLiouville1986} 
\begin{equation}\label{eq:AO hamiltonian with dimensions}
    \hat{\mathcal{H}}=\hbar\omega\hat{a}^\dagger\hat{a}-\hbar K(\hat{a}^\dagger\hat{a})^2\,,
\end{equation}
which models light propagation in a Kerr-like nonlinear media~\cite{drummondQuantumTheoryOptical1980}. Apart from optics, the (Quantum) Anharmonic Oscillator has applications in various fields, such as dynamics of Bose-Einstein condensates~\cite{greinerCollapseRevivalBEC2002} or quantum information~\cite{ReagorCricuitQED2016}. \\
Its eigenenergies $E_n$ are not linear in $n$
\begin{equation}
\hat{\mathcal{H}}\ket{n}=\hbar\left(\omega n-Kn^2\right)\ket{n},
\end{equation}
what breaks the equidistance between neighboring energy levels
\begin{equation}
    E_{n+1}-E_n=\hbar\omega-\hbar K(2n+1).
\end{equation}
For the Anharmonic Oscillator, there are now 2 intrinsic timescales: $1/\omega$ of the harmonic term and $1/K$ of the anharmonic term.

\section{Nondimensionalization}
Anticipating  the semiclassical analysis, it would be useful to nondimensionalize the Schrödinger equation such that the Planck constant is replaced by a small parameter $\hbar\rightarrow\varepsilon$, which also gives the commutator of some rescaled position and momentum operators   $[\hat{X},\hat{P}_X]=\mathrm{i}\varepsilon$. However, 
when using the characteristic length scale $l$ for introducing a nondimensional space coordinate $\tilde{x}=\frac{x}{l}$, the creation  and annihilation operator $\hat{a}^\dagger$ and $\hat{a}$, given in~\ref{eq:HO a adagger},  become
\begin{equation}
\hat{a}\equiv\frac{\tilde{x}+\partial_{\tilde{x}}}{\sqrt{2}},
\end{equation}
which is purely dependent on operators in nondimensional coordinates $\hat{\tilde{x}}\equiv \tilde{x}$, $\hat{\tilde{p}}\equiv-\mathrm{i}\partial_{\tilde{x}}$. There are no prefactors of $\hat{\tilde{x}}$ or $\hat{\tilde{p}}$ carrying a dimension that could be compared to other quantities of the system. 
The only two quantities of the problem that carry a dimension are $\hbar \omega$, the typical energy of the harmonic term, and $\hbar K$,  the typical energy of the anharmonic term. Choosing the typical time scale of the harmonic term $\tau_\text{harmonic}=1/\omega$ to introduce a dimensionless time $\tilde{t}=\frac{t}{1/\omega}$,  the Schrödinger equation can be expressed as
\begin{equation}    \mathrm{i}\partial_{\tilde{t}}\ket{\psi}=\hat{\tilde{\mathcal{H}}}\ket{\psi},
\end{equation}
with the dimensionless Hamiltonian 
\begin{equation}
    \hat{\tilde{\mathcal{H}}}=\hat{a}^\dagger\hat{a}-\zeta(\hat{a}^\dagger\hat{a})^2\,,
\end{equation}
where
\begin{equation}
    \zeta=\frac{K}{\omega}
\end{equation}
is the dimensionless parameter that compares the typical time scales of the harmonic -and anharmonic term.  In the limit $\zeta\ll1$, the contributions of the anharmonic term are happening so slowly that they can be neglected and one obtains the limit of a Harmonic Oscillator (modulo $\frac{1}{2}$). In the following, we will only work in dimensionless coordinates and drop the tildes  $\tilde{}$  to lighten the notation.
\begin{figure}
    \centering
    \includegraphics[width=0.5\linewidth]{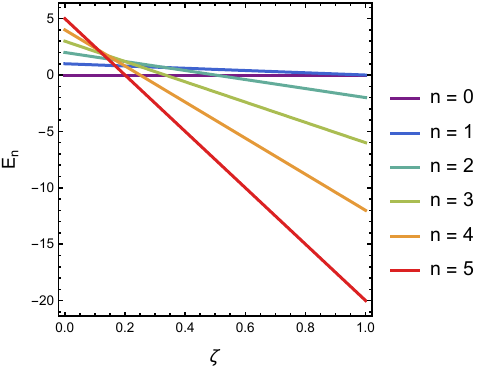}
    \caption{Plot of nondimensional eigenenergies $E_n$ of the dimensionless Anharmonic Oscillator. $\zeta$ interpolates between a Harmonic Oscillator $E_{n+1}-E_n=1$ for $\zeta=0$ and a simplified Transmon qubit $E_{n+1}-E_n=-2n$ for $\zeta=1$.} 
    \label{fig:SpectrumEnAO}
\end{figure}
The nondimensional eigenenergies 
\begin{equation}\label{eq:AO En}
    \begin{aligned}
        E_n=&n-\zeta n^2
    \end{aligned}
\end{equation}
are plotted in~\autoref{fig:SpectrumEnAO} as a function of $\zeta$. In the following, we choose $\zeta=1$, which is the special case of  degeneracy of the lowest energy levels $E_0=E_1=0$. For this choice, the Hamiltonian can be expressed as 
\begin{equation}
    \hat{\mathcal{H}}=-\hat{a}^\dagger\hat{a}^\dagger\hat{a}\hat{a},
\end{equation}
which is a simplified description of the transmon qubit~\cite{ReagorCricuitQED2016}.

\section{Quantum time evolution of an initially coherent state $\ket{\alpha}$}
\subsection{Decoherence}
\begin{figure}
    \centering
    \includegraphics[width=0.45\linewidth]{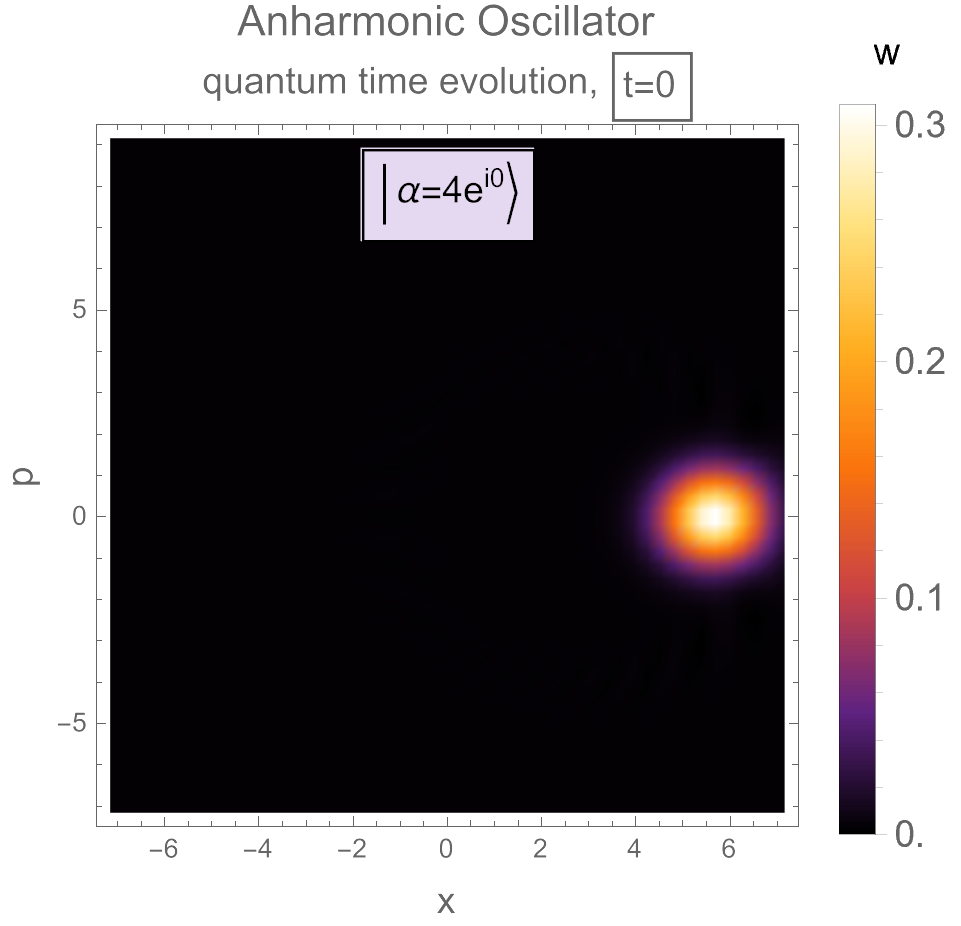}
    \includegraphics[width=0.45\linewidth]{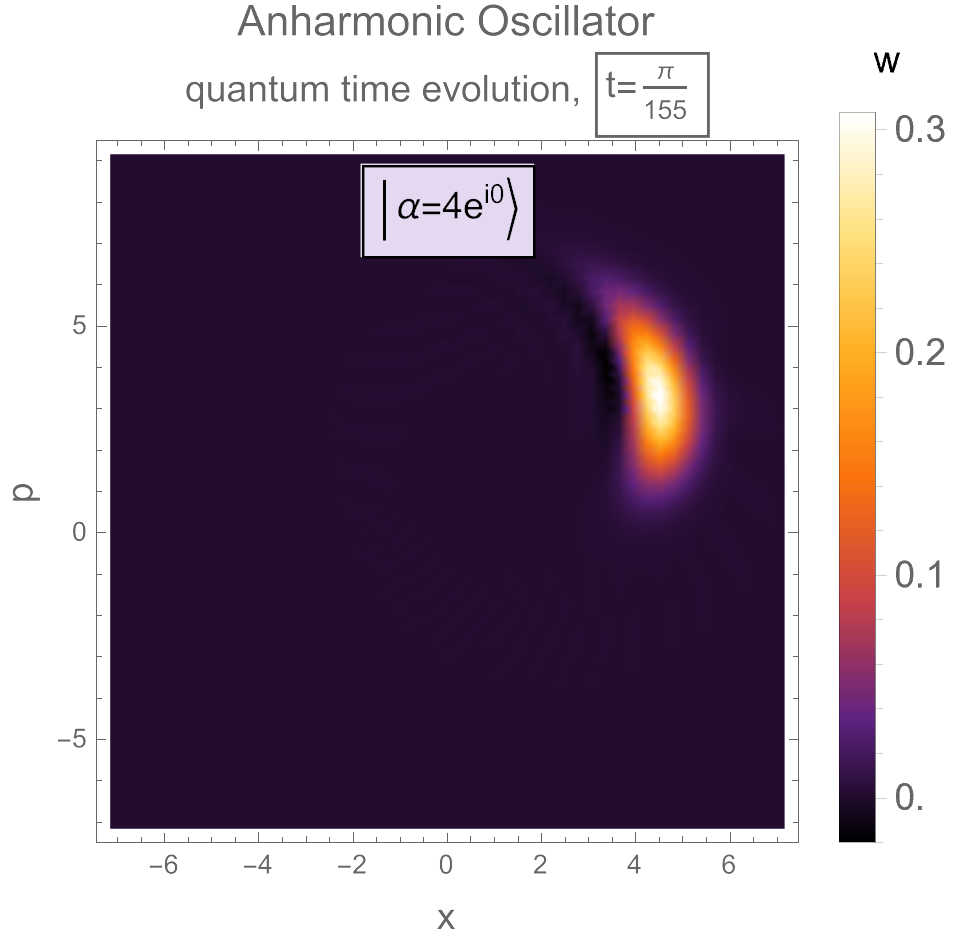}
     \includegraphics[width=0.45\linewidth]{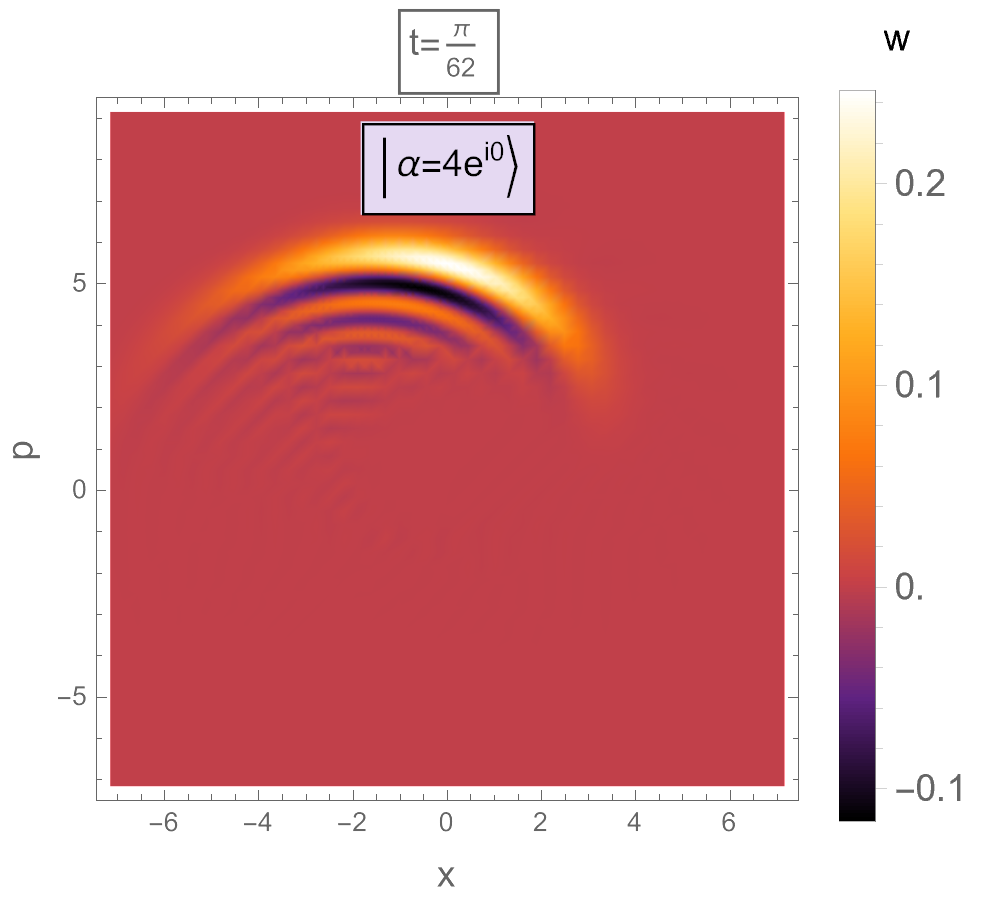}
     \includegraphics[width=0.45\linewidth]{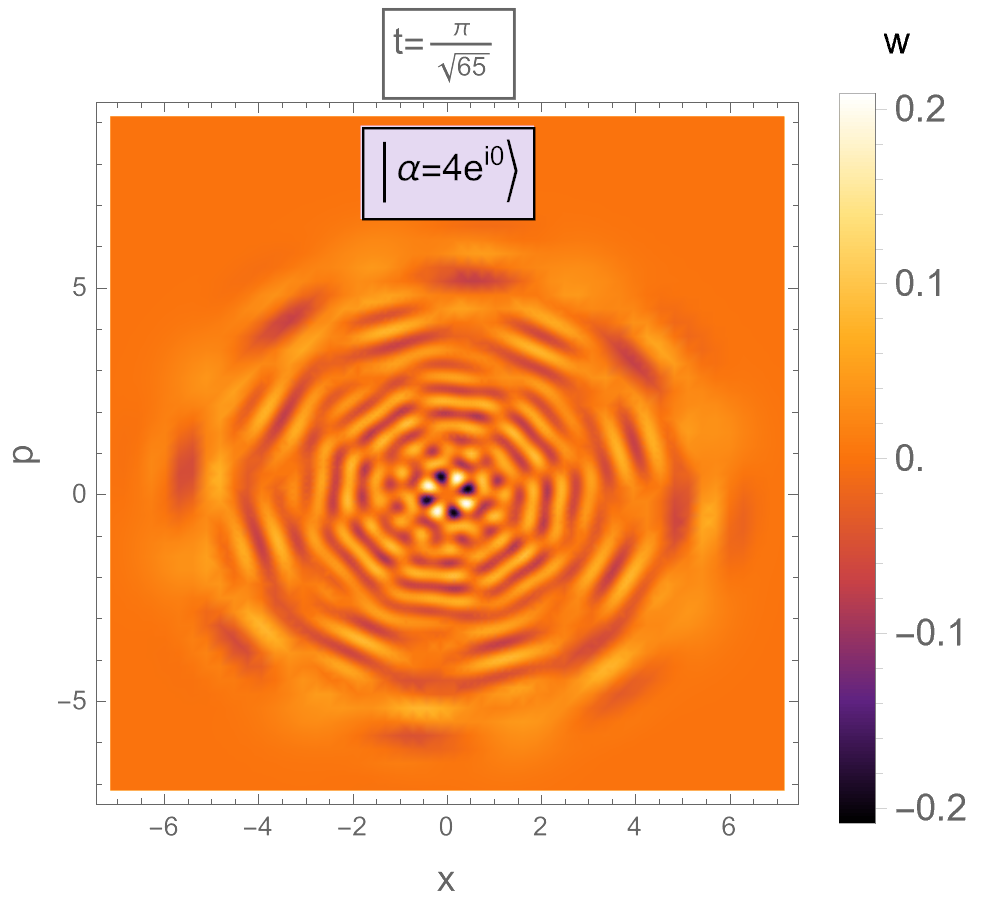}
   \includegraphics[width=0.45\linewidth]{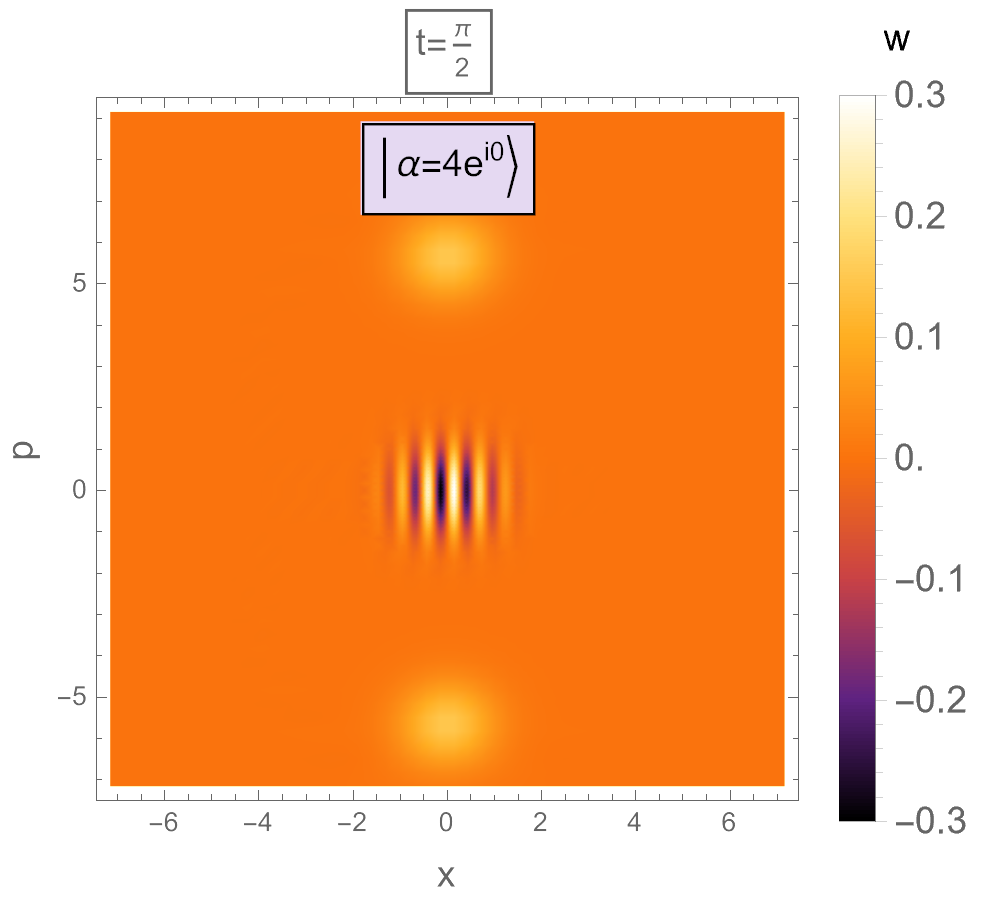}
\includegraphics[width=0.45\linewidth]{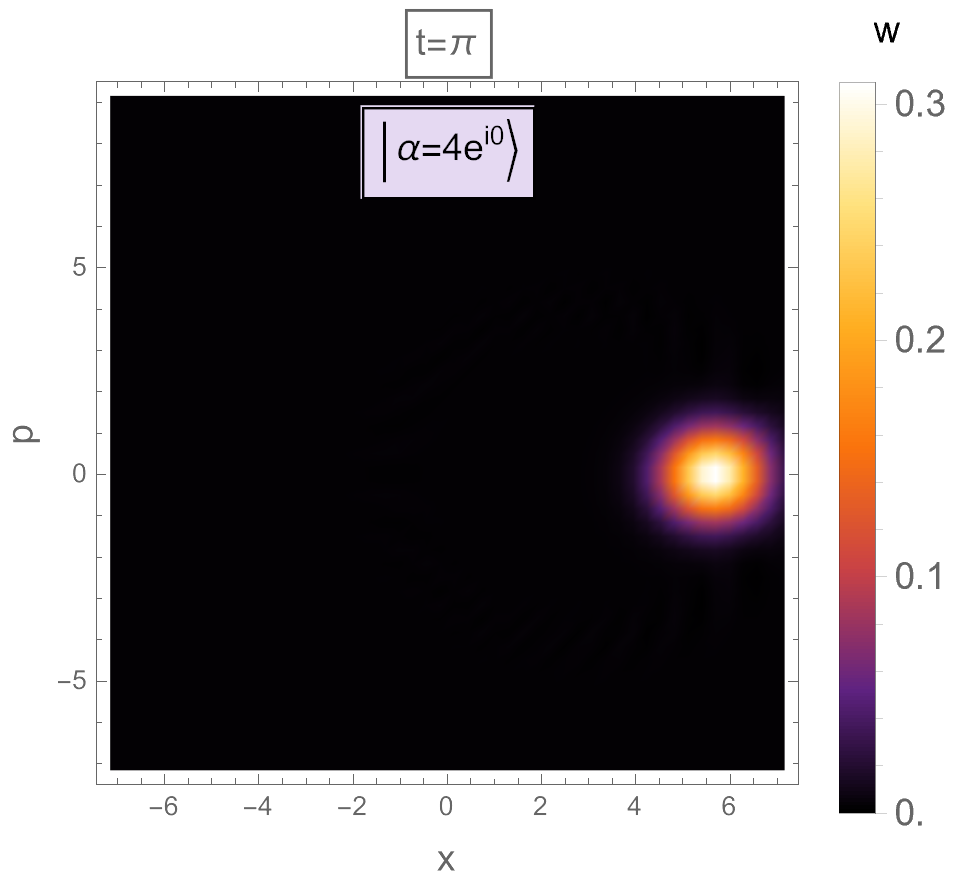}
    \caption{Quantum time evolution of the Wigner function $w_\alpha(t)$ of an initially coherent state $\ket{\alpha}(t)$ with $\alpha=4$.  Initially, $w_\alpha(t)$ loses its coherent form, but after a time $T_\text{rev}=\pi$, the coherent form is fully recovered again.  For $t=\pi/2$, $w_\alpha(t)$ becomes  the Wigner function of a superposition of two coherent states $\ket{\mathrm{i}\alpha}$ and $\ket{-\mathrm{i}\alpha}$.  All plots are calculated with a numerical truncation at $n_\text{max}=32$ in \autoref{eq:AO Wigalpha(t)}. Figure inspired by~\cite{Wigner-QuantumCorrections}.}
    \label{fig:AO PlotWig quantum time evolution}
\end{figure}
The breaking of equidistance between energy levels gives rise to a dynamic of an initially coherent state $\ket{\alpha}$, which is much more complex than for the Harmonic Oscillator~\cite{sudheeshManifestationsWavePacket2004,rohithFractionalRevivalsSuperposed2014,proppDecoherenceLimitsCost,milburnQuantumClassicalLiouville1986}: Under time evolution $\hat{U}(t)=\exp(-\mathrm{i}\hat{\mathcal{H}}t)$,
\begin{equation}\label{eq:AO |alpha>(t)}
    \ket{\alpha}(t)=\mathrm{e}^{-|\alpha|^2/2}\sum_{n=0}^\infty\frac{ \mathrm{e}^{\mathrm{in(n-1)t}}\alpha^n}{\sqrt{n!}}\ket{n}
\end{equation}
the different terms in the sum are in general not in phase, which leads to a decoherence of the wave packet. A vivid way to illustrate this decoherence is the quantum time evolution of the Wigner function  \cite{stobinskaWignerFunctionEvolution2008}  
\begin{equation}\label{eq:AO Wigalpha(t)}
\begin{aligned}
        w_\alpha(x,p)(t)=&\frac{\sigma[\ket{\alpha}\bra{\alpha}](t)}{2\pi}\\[1.5ex]
=&\mathrm{e}^{-|\alpha|^2}
\sum_{n=0}^{\infty}
\left[
  \frac{|\alpha|^{2n}}{n!}\, w_{n}(x,p)\right.\\
 &\left.
   \quad + \sum_{m=n+1}^{\infty}
    \frac{|\alpha|^{n+m}}{\sqrt{n!\,m!}}\,\frac{ \sigma\left[\mathrm{e}^{\mathrm{i}\varphi_{m,n,\alpha}(t)}\ket{m}\bra{n}+\text{h.c.}\right]}{2\pi}
\right].
\end{aligned}
\end{equation}
with 
\begin{equation}\label{eq:AO phi-m,n,alpha}
      \varphi_{m,n,\alpha}(t)=(m-n)\varphi_\alpha+(m-n)(m+n-1)t
\end{equation}
and $w_n$ and $ \sigma\left[\mathrm{e}^{\mathrm{i}\varphi}\ket{m}\bra{n}+\text{h.c.}\right]$ defined in~\autoref{eq:Wignerfunction wn} and \autoref{eq:symbol ketm bran plus hc}, but in dimensionless coordinates $\hbar=l=1$.  \autoref{fig:AO PlotWig quantum time evolution}  shows $w_\alpha(x,p)(t) $ at different times. One can see that the Wigner function  $w_\alpha (x,p)(t)$  gets deformed and spreads out, but after a time $T_\text{rev}=\pi$ the initial form of $w_\alpha (x,p)$ is recovered. This is the revival of the initial state: all terms of $\ket{\alpha}(t)$ in~\autoref{eq:AO |alpha>(t)} are in phase again~\cite{sudheeshManifestationsWavePacket2004}. For times $t=\pi/k,\,k\in \mathbb{Z}$, $\ket{\alpha}(t)$  becomes a finite superposition of coherent states with equal $|\alpha|$ and a rotated phase $\varphi_{\alpha'}$~\cite{rohithFractionalRevivalsSuperposed2014}.  At a time $t=\pi/2$, $\ket{\alpha}(t)$ becomes a superposition of $ \ket{\mathrm{i}\alpha}$  and $\ket{-\mathrm{i}\alpha}$. This behavior is called fractional revival.  

\subsection{Evolution of wave packet centers $\braket{\hat{x}}$ and $\braket{\hat{p}}$}\label{sec:AO <x> <p> quantum}
\begin{figure}[h]
    \centering
    \includegraphics[width=0.32\linewidth]{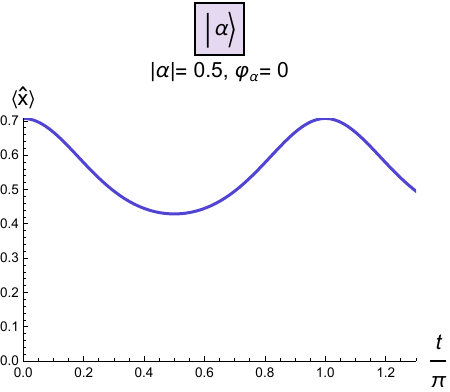}
    \includegraphics[width=0.32\linewidth]{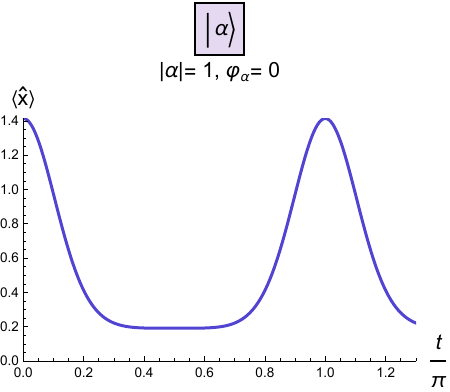}
    \includegraphics[width=0.32\linewidth]{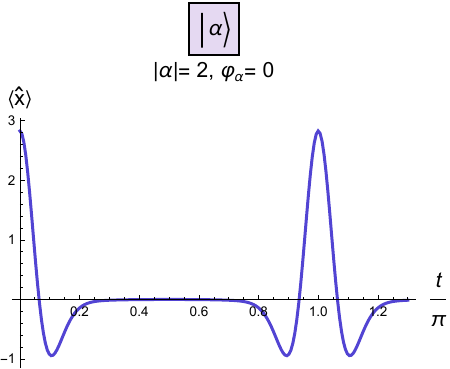}
    \includegraphics[width=0.32\linewidth]{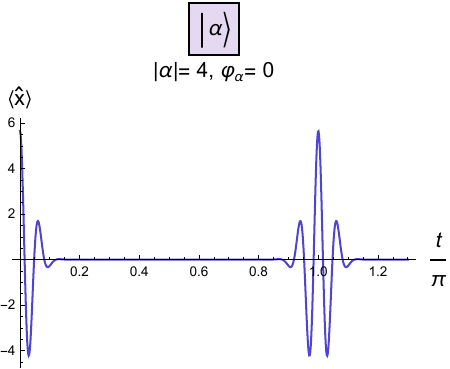}
    \includegraphics[width=0.32\linewidth]{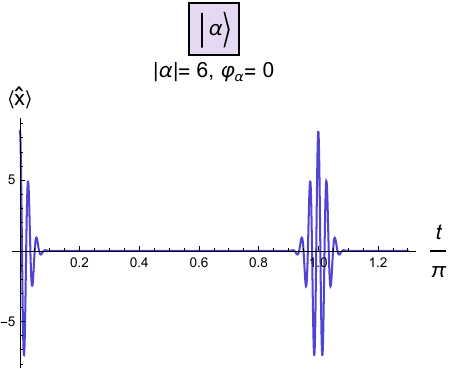}
    \includegraphics[width=0.32\linewidth]{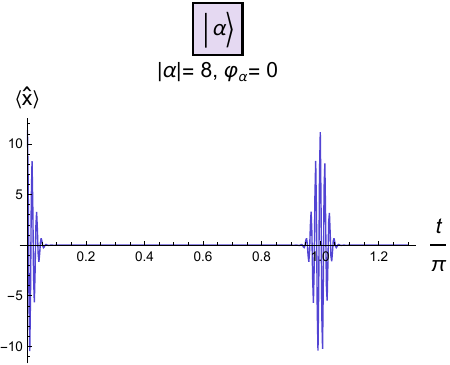}
    \includegraphics[height=0.7cm]{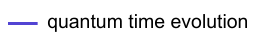}
    \caption{Quantum time evolution of $\braket{\hat{x}}$ for initially coherent states with   fixed $\varphi_\alpha=0$ and varying $|\alpha|$. After the onset of motion, $\braket{\hat{x}}(t)$ decays, eventually remain at $\braket{\hat{x}}\approx0$ and later grows again before reaching the initial maximum at $t=\pi$. The initial decay and later growth of $\braket{x}(t)$ become more oscillatory for larger $|\alpha|$. Figure inspired by~\cite{sudheeshManifestationsWavePacket2004}}
    \label{fig:AO coherent state <x>(t)}
\end{figure}
\begin{figure}[h]
    \centering    
    \includegraphics[width=0.3\linewidth]{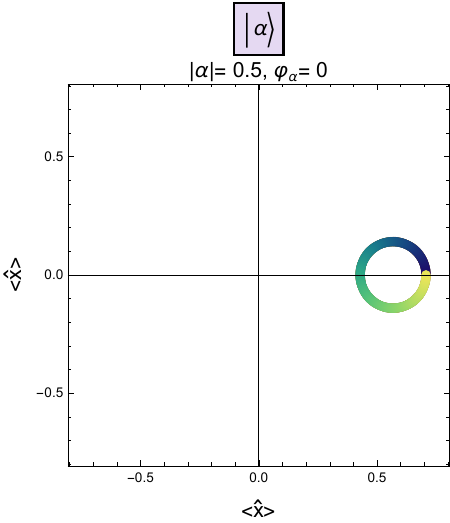}
    \includegraphics[width=0.3\linewidth]{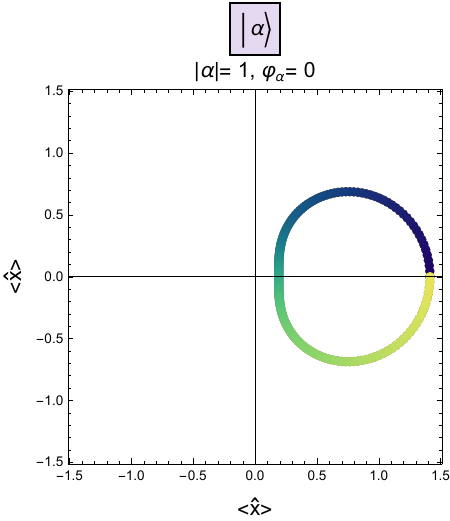}
    \includegraphics[width=0.3\linewidth]{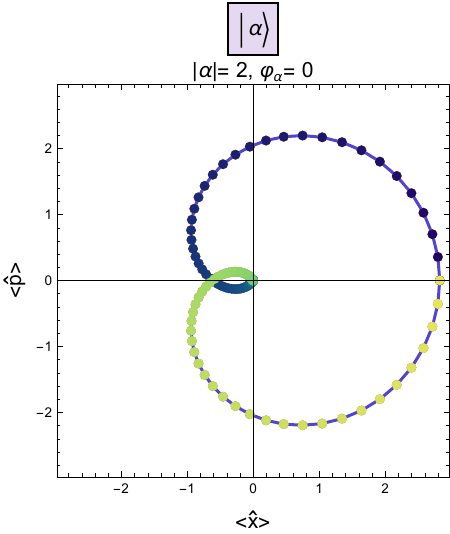}
    \includegraphics[width=0.3\linewidth]{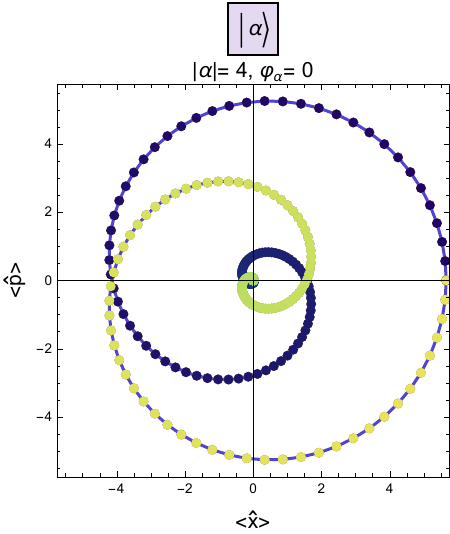}
    \includegraphics[width=0.3\linewidth]{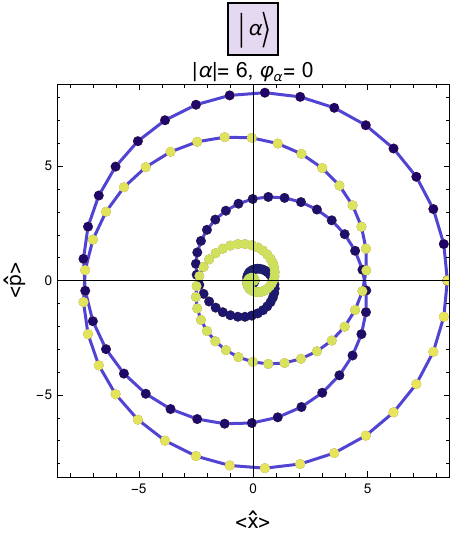}
    \includegraphics[width=0.3\linewidth]{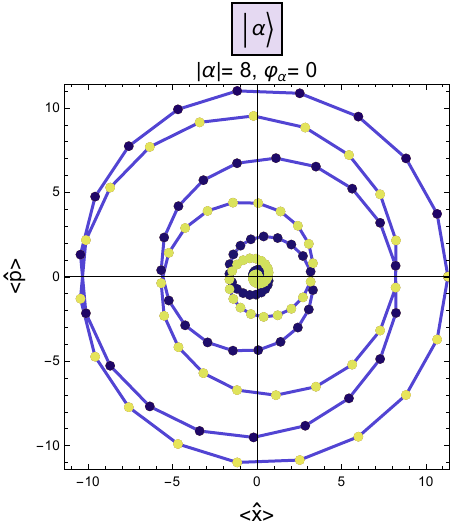}
    \includegraphics[height=1.3cm]{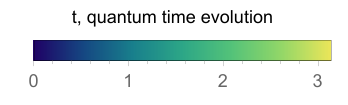}
    \caption{Quantum time evolution of $\braket{\hat{x}}$ and $\braket{\hat{p}}$ for initially coherent states with $\varphi_\alpha=0$. For small $|\alpha|$,  the evolution of  $\braket{\hat{x}}$ and $\braket{\hat{p}}$ resemble a classical harmonic oscillation around an offset. For larger $|\alpha|$, the dynamics resemble an (over)damped oscillation mirrored after a time $t=\pi/2$. Figure inspired by~\cite{sudheeshManifestationsWavePacket2004}}
    \label{fig:AN coherent state <x>,<p>}
\end{figure}
For an initially coherent state, the evolution of the wave packet centers in real space and momentum space is given by~\cite{sudheeshManifestationsWavePacket2004} 
\begin{equation}\label{eq:AO coherent state <x> <p>}
\begin{aligned}
\braket{\alpha|\hat{x}|\alpha}(t)
&= \sqrt{2}|\alpha|e^{-|\alpha|^2}
   \sum_{n=0}^{\infty} \frac{|\alpha|^{2n}}{n!}
   \cos(\phi_\alpha + 2nt) \\[1ex]
&= e^{|\alpha|^2(1-\cos{2t})}\sqrt{2}
   \left[\Re[\alpha]\cos(|\alpha|^2\sin{2t})
   + \Im[\alpha]\sin(|\alpha|^2\sin{2t})\right],
\\[2ex]
\braket{\alpha|\hat{p}|\alpha}(t)
&= \sqrt{2}|\alpha|e^{-|\alpha|^2}
   \sum_{n=0}^{\infty} \frac{|\alpha|^{2n}}{n!}
   \sin(\phi_\alpha + 2nt) \\[1ex]
&= e^{|\alpha|^2(1-\cos{2t})}\sqrt{2}
   \left[-\Re[\alpha]\sin(|\alpha|^2\sin{2t})
   + \Im[\alpha]\cos(|\alpha|^2\sin{2t})\right].
\end{aligned}
\end{equation}

\autoref{fig:AO coherent state <x>(t)} and \autoref{fig:AN coherent state <x>,<p>} show their evolution for fixed $\varphi_\alpha =0$ and varying $|\alpha|$. For all $|\alpha|$, the initial values of $\braket{\hat{x}}$ and $\braket{\hat{p}}$ are restored after a time $T_\text{rev}=\pi$ corresponding to the revival of the initial $\ket{\alpha}$. For $|\alpha|=1/2$, the dynamics of the wave packet centers resembles a harmonic oscillation around an offset. By increasing the value of $|\alpha|$ a transition takes place towards a damped oscillation mirrored at $t=\pi/2$. One can observe that $\tau_\text{dec}$,  the typical time it takes for the amplitude of the oscillations to decay towards $\braket{\hat{x}}\approx\braket{\hat{p}}\approx 0$, decreases with $|\alpha|$. Nevertheless, the oscillations become more and more pronounced with increasing $|\alpha|$, since the oscillation period $T_\text{oscill}$ decreases even more quickly.\\ 
The dynamics for low  values of $|\alpha|$ can  be understood directly from the development in~\autoref{eq:AO coherent state <x> <p>}: for small $|\alpha|$  all terms of $n\geq2$ become negligible so that $\braket{\hat{x}}(t)\approx\cos(\varphi_\alpha+2t)+\text{const.}$ The dependence of $\tau_\text{dec}$ on $|\alpha|$ can be extracted from the prefactor $\mathrm{e}^{|\alpha|^2(1-\cos{2t})}$ in the compact version of~\ref{eq:AO coherent state <x> <p>}. The dependence of the oscillation period $T_\text{oscill}$ on $|\alpha|$ on the other hand can only  be approximately deduced as $T_\text{oscill}=2\pi/2|\alpha|^2$  from~\autoref{eq:AO coherent state <x> <p>} when approximating $\sin{2t}\approx 2t$. This approximation quickly loses its validity. In the following sections, we will present how $T_\text{oscill}$ and $\tau_\text{dec}$ can be obtained using a semiclassical phase space analysis, which can in general still be valid at times $t\sim 1$.

\FloatBarrier
\section{Semiclassical analysis of the Anharmonic Oscillator}\label{sec:AO semiclass eq of motion}
\begin{figure}[h]
    \centering
    \includegraphics[width=0.5\linewidth]{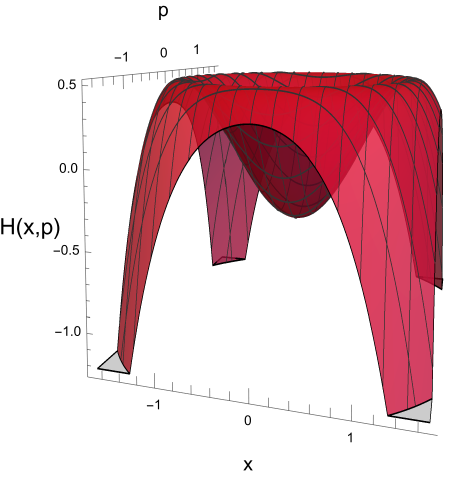}
    \caption{Plot of the symbol Hamiltonian $H(x,p)$ of the Anharmonic Oscillator for $\zeta=1$} 
    \label{fig:SpectrumE0AO}
\end{figure}
By nondimensionalizing the Anharmonic Oscillator, the commutation relation of the (new) dimensionless conjugated operators became $[\hat{x},\hat{p}]=\mathrm{i}$: In order to still keep track of the order in the commutator $[\hat{x},\hat{p}]$ a tracking parameter $\lambda=1$ is introduced as
\begin{equation}
    [\hat{x},\hat{p}]=\mathrm{i}\lambda.
\end{equation}
 Later, we will show that for an initially coherent state $\ket{\alpha}$, the semiclassical approximation for the initial dynamics is controlled  by the eigenvalue $\alpha$ via the condition $\epsilon=\frac{1}{2|\alpha|^2}\ll1$. Furthermore, we will reveal how the semiclassical analysis can help to obtain a better qualitative understanding of the evolution of $\braket{\hat{x}}(t)$  and $\braket{\hat{p}}(t)$.\\
The symbol of $\hat{\mathcal{H}}$ is given by the Moyal product with commutator $\mathrm{i}\lambda$ instead of $\mathrm{i}\hbar$ \cite{Polkovnikov_PSRoQD_2010}
\begin{equation}\label{eq:AO symbolH}
\begin{aligned}
        \sigma_\lambda[\hat{\mathcal{H}}]=H(x,p)=&-\frac{(x-\mathrm{i}p)\star(x-\mathrm{i}p)\star (x+\mathrm{i}p)\star(x+\mathrm{i}p)}{4}\\
        =&-\frac{(x^2+p^2)^2}{4}+\lambda(x^2+p^2)-\frac{\lambda^2}{2}.\\
\end{aligned}
\end{equation}
A plot of $H(x,p) $ can be seen in \autoref{fig:SpectrumE0AO}. Note that the symbol $H$ has a non-trivial  $\mathcal{O}(\lambda)$-term. The importance of this first-order Moyal-term for approximating the dynamics will be investigated in~\autoref{sec:Moyalterm}.

In the Heisenberg picture, the equation of motion for an operator $\hat{o}$  in phase space representation is given by
\begin{equation}
    \begin{aligned}
    \partial_to=&\frac{\mathrm{i}}{\lambda}[H,o]_\star\\
    =&-\{H,o\}+\lambda^2\frac{1}{16} \sum_{k=0}^3(-1)^k\binom{3}{k}\left(\partial_p^k \partial_x^{3-k} H \partial_x^k \partial_p^{3-k}o-(H \leftrightarrow o)\right) 
    \end{aligned}
\end{equation}
The semiclassical approximation now consists in demanding that all terms of $\mathcal{O}(\lambda^2)$ are negligible. This requires to identify a control parameter $\epsilon^2\ll1$ ensuring this request. Expressing the derivatives in cylindrical coordinates $\varphi=\arctan\left(\frac{p}{x}\right)$ and $r=\sqrt{x^2+p^2}$
\begin{equation}
\begin{aligned}
        \partial_x=\cos \varphi \partial_r-\frac{\sin \varphi}{r} \partial_\varphi,\\
         \partial_p=\sin \varphi \partial_r-\frac{\cos \phi}{r} \partial_\varphi,       
\end{aligned}
\end{equation}
it becomes clear that for any $o(x,p)$ that is a finite polynomial in $x$ and $p$ the $\mathcal{O}(\lambda^2)$-term in the equations of motion is of the order 

\begin{equation}
    \lambda^2\cdot\partial_p^k\partial_x^{3-k} H \partial_x^k \partial_p^{3-k}o=\lambda^2\cdot\mathcal{O}\left(\frac{o}{r^2}\right),
\end{equation}
while the Poisson bracket is
\begin{equation}\label{eq:AO possionbraket}
    \{o,H\}=\lambda^0\cdot\mathcal{O}\left(r^2o\right)+\lambda^1\cdot\mathcal{O}\left(o\right).
\end{equation}

For a coherent state $\ket{\alpha}$, the Wigner function
\begin{equation}
    w_\alpha(x,p)=\frac{1}{\pi}\mathrm{e}^{-(x-x_\alpha)^2}\mathrm{e}^{-(p-p_\alpha)^2}
\end{equation} 
is a product of Gaussians with width $\Delta x =\Delta p=1/\sqrt{2}$ centered around
\begin{equation}
\begin{aligned}
      x_\alpha=r_\alpha\cos\varphi_\alpha\\
      p_\alpha=r_\alpha\sin\varphi_\alpha,
\end{aligned}
\end{equation}
where 
\begin{equation}
    r_\alpha=\sqrt{2}|\alpha|
\end{equation}
Due to the Gaussian form, phase space regions where $r\sim r_\alpha$ give the main contribution when calculating 
\begin{equation}
    \braket{\hat{o}}=\int_0^\infty \mathrm{d} r\,r \int_0^{2\pi} \mathrm{d}\varphi\, o(r,\varphi)\,w_\alpha(r,\varphi).
\end{equation}
\begin{figure}
    \centering
    \includegraphics[width=0.3\linewidth]{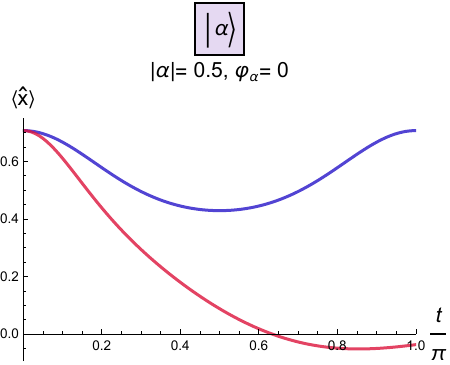}
    \includegraphics[width=0.3\linewidth]{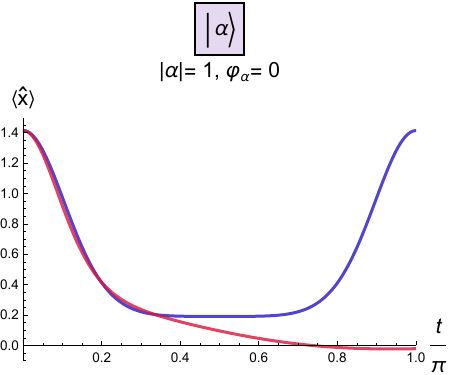}
    \includegraphics[width=0.3\linewidth]{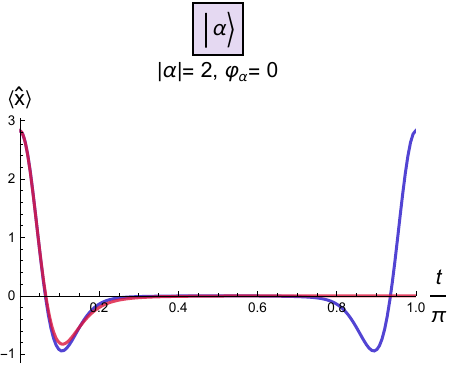}
    \includegraphics[width=0.3\linewidth]{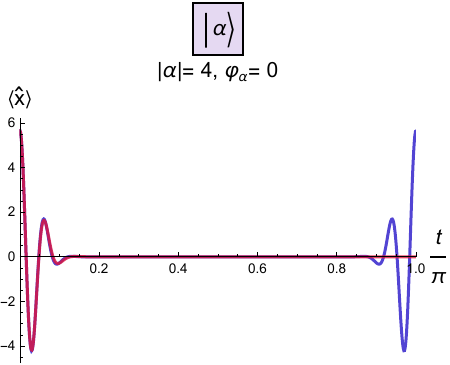}
    \includegraphics[width=0.3\linewidth]{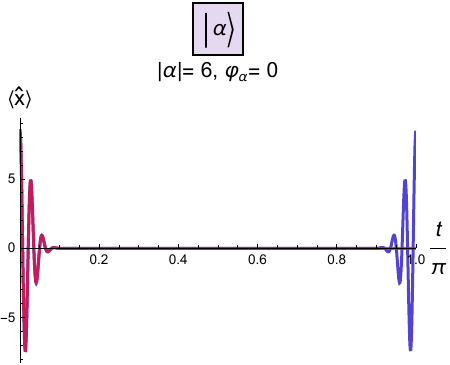}
    \includegraphics[width=0.3\linewidth]{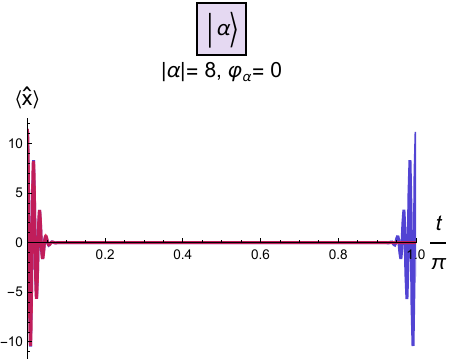}
    \includegraphics[height=0.7cm]{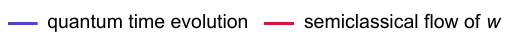}
    \caption{Comparison of $\braket{\hat{x}}(t)$ under quantum time evolution and under semiclassical flow $\phi$. For larger values of $|\alpha|$, the semiclassical prediction becomes better, but it always fails to predict the onset of motions after the initial decay of the amplitude of $\braket{\hat{x}}(t)$. Figure inspired by~\cite{Polkovnikov_PSRoQD_2010} }
    \label{fig:AN coherent state <x>(t) semicl}
\end{figure}
Under quantum time evolution, $w_\alpha(t)$ can differ drastically from $0$ in phase space regions $r\ll r_\alpha$ as can be seen in~\autoref{fig:AO PlotWig quantum time evolution}. However, in the regions of $r\ll r_\alpha$ the Wigner function is highly oscillatory such that the contributions of these phase space regions mainly cancel out when integrating over $o(x,p)$. Under the assumption
that only phase space regions of $r\sim r_\alpha$ contribute to the dynamics, the condition for the $\mathcal{O}(\lambda^2)$-term to be small compared to $\{o,H\}$ can be expressed as 
\begin{equation}
    \epsilon^2=\frac{1}{4|\alpha|^4}\ll1.
\end{equation}
Then, the evolution of an expectation value $\braket{\alpha|\hat{o}|\alpha}(t)$ can be approximated by
\begin{equation}\label{eq:osemicl alpha}
    \braket{\hat{o}}_\text{semicl}(t)=\int\mathrm{d}x\mathrm{d}pw_\alpha(x^{\phi_{-t}},p^{\phi_{-t}})o(x,p)
\end{equation}
where the semiclassical flow $\phi_t$ is given by the solution of 
\begin{equation}\label{eq:AO semiclEoM}
\begin{aligned}
        \dot{x}=&\Omega(r)p+\mathcal{O}(\lambda^2)\\
          \dot{p}=&-\Omega(r)x+\mathcal{O}(\lambda^2),\\
\end{aligned}
\end{equation}
 which are the differential equations of an oscillator with an $r$-dependent angular velocity 
\begin{equation}
     \Omega(r)=- r^2+2\lambda.
\end{equation}
Using $\dot{\Omega}(r)=0+\mathcal{O}(\lambda^2)$, it is straightforward to show that 
\begin{equation}
    \begin{aligned}
        x^{\phi_t}=&r_0\cos(\varphi_0-\Omega(r_0)t)\\
         p^{\phi_t}=&r_0\sin(\varphi_0-\Omega(r_0)t)
    \end{aligned}
\end{equation}
with  $r_0=\sqrt{x_0^2+p_0^2}$ and $\varphi_0=\arctan\left(\frac{p_0}{x_0}\right)$ as initial conditions.

\subsection{Dynamics of wave packet centers $\braket{\hat{x}}$, $\braket{\hat{p}}$ via semiclassical flow $\phi$}

\begin{figure}[h]
    \centering    
    \includegraphics[width=0.3\linewidth]{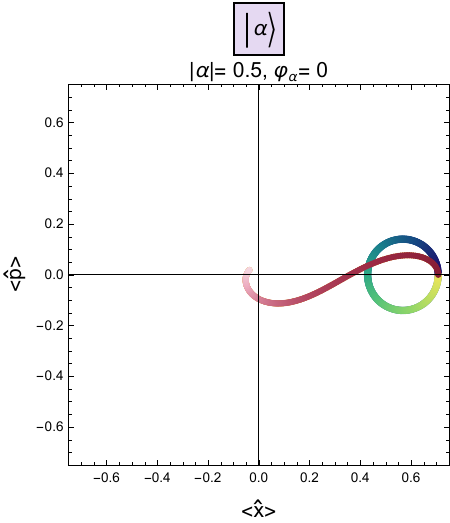}
    \includegraphics[width=0.3\linewidth]{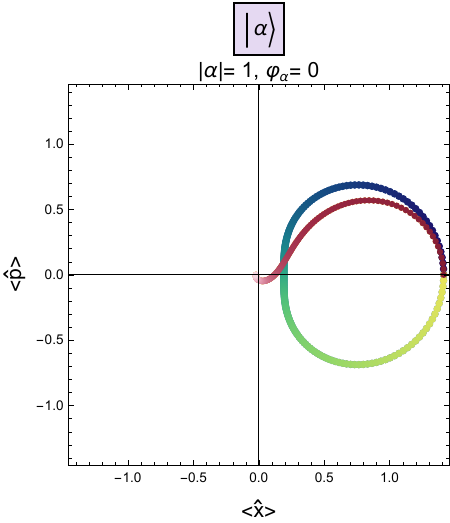}
    \includegraphics[width=0.3\linewidth]{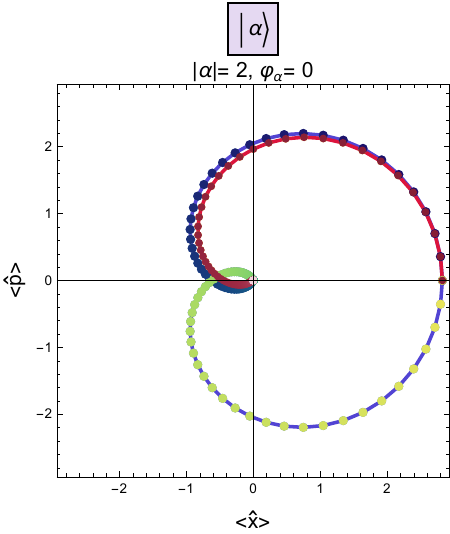}
    \includegraphics[width=0.3\linewidth]{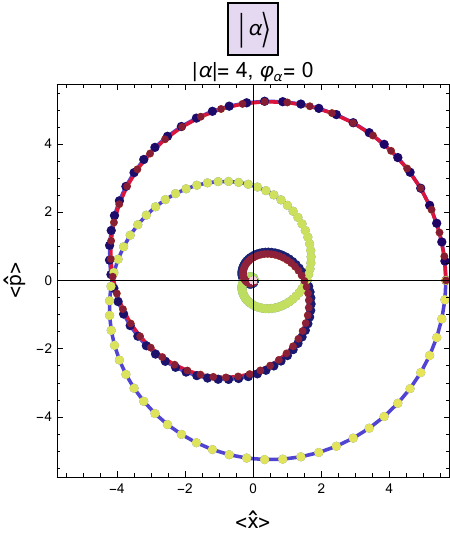}
    \includegraphics[width=0.3\linewidth]{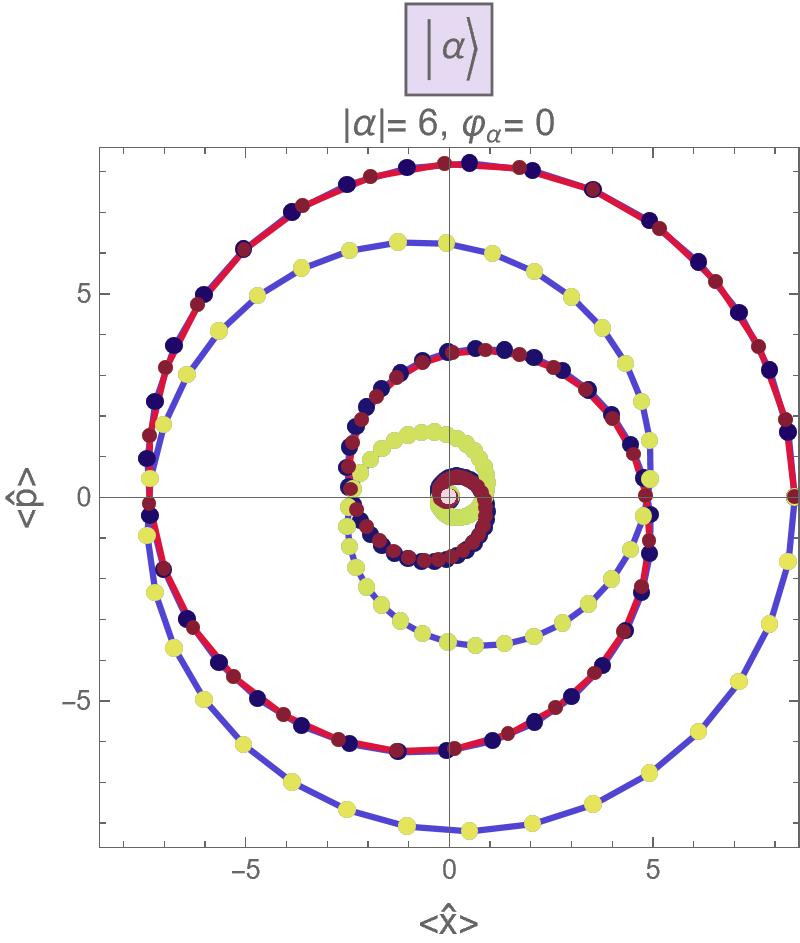}
    \includegraphics[width=0.3\linewidth]{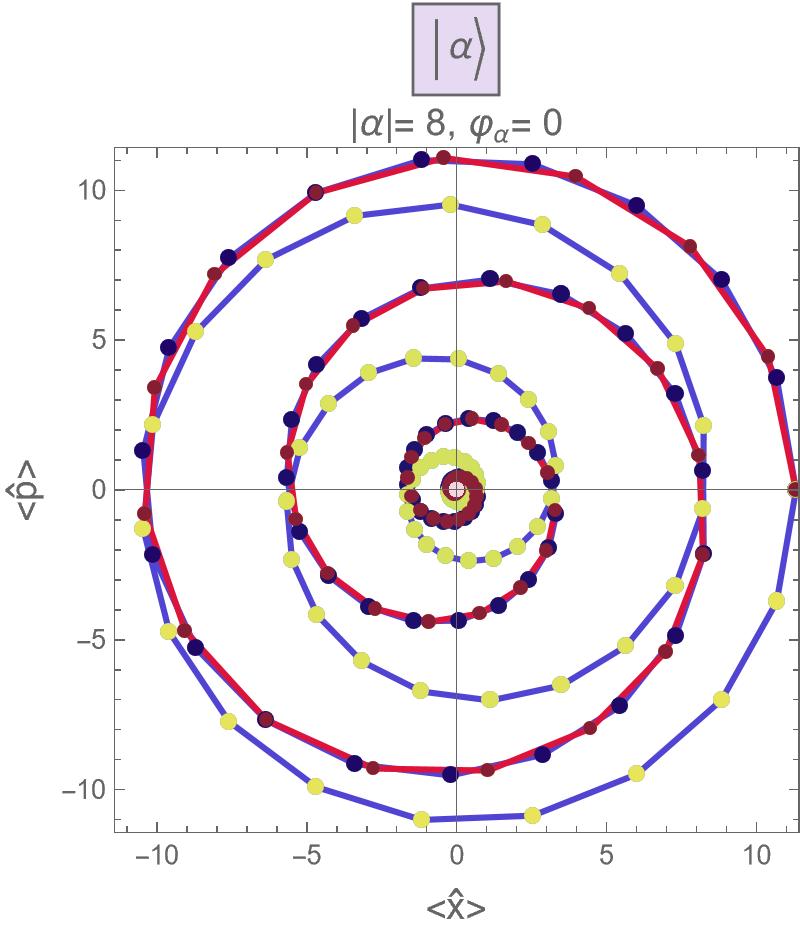}
     \includegraphics[height=1.3cm]{graphics/Anharmonic_Oscillator/Phasespace/AO_phasespace-xp-exact/tbar-quantumAO.pdf}
      \includegraphics[height=1.3cm]{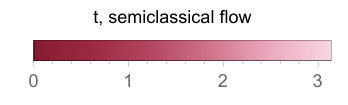}
        \caption{Comparison of the evolution of $\braket{\hat{x}}$ and $\braket{\hat{p}}$ under quantum time evolution and semiclassical flow $\phi$. Once $\braket{\hat{x}}\approx\braket{\hat{p}}\approx0$  is reached under semiclassical flow, the semiclassical dynamics become  become stationary.}
    \label{fig:AO ocoherent state <x>,<p> semicl}
\end{figure}
The approximate $\braket{\hat{x}}_\text{semicl}(t)$ is compared to the quantum time evolution $\braket{\hat{x}}(t)$ in ~\autoref{fig:AN coherent state <x>(t) semicl}. One can qualitatively observe that $\epsilon^2=1/4|\alpha|^2$ controls indeed the quality of the prediction: while for $|\alpha|=\frac{1}{2}$, the 2 graphs do not match at all, the semiclassical evolution describes much more accurately the ``damped oscillation`` for larger values of $|\alpha|$. The ``driven oscillation`` of $\braket{\hat{x}}(t)$ and $\braket{\hat{p}}(t)$ however, i.e. the revival of the initial state, cannot be reproduced by the semiclassical flow~\cite{proppDecoherenceLimitsCost}, as can be seen in~\autoref{fig:AO ocoherent state <x>,<p> semicl}.

\subsection{Importance of the Moyal term}\label{sec:Moyalterm}
\begin{figure}[h]
    \centering
    \includegraphics[width=0.3\linewidth]{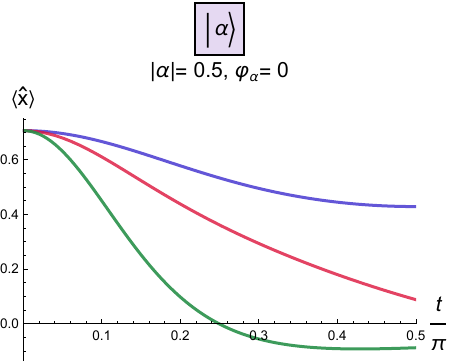}
    \includegraphics[width=0.3\linewidth]{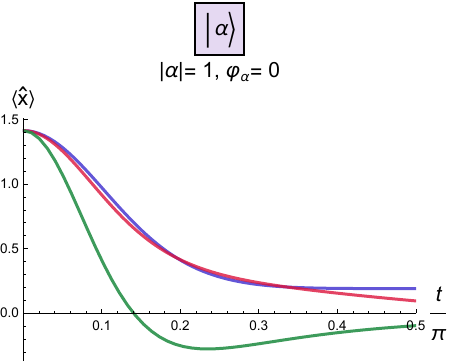}
    \includegraphics[width=0.3\linewidth]{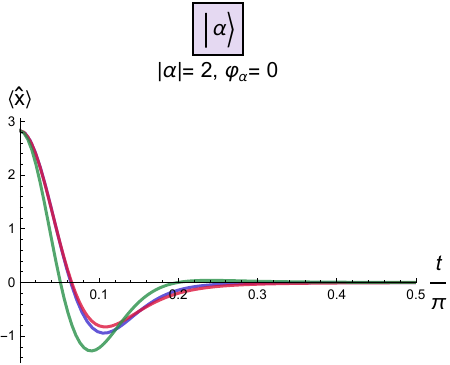}
    \includegraphics[width=0.3\linewidth]{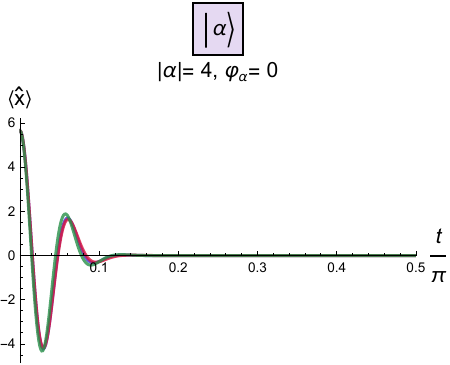}
    \includegraphics[width=0.3\linewidth]{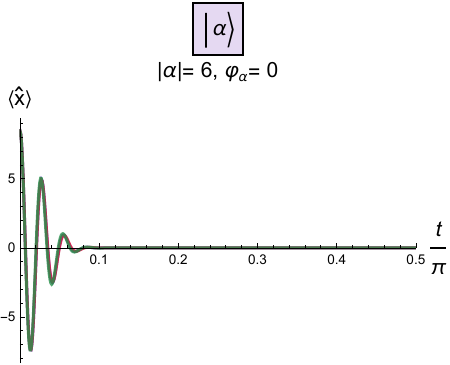}
    \includegraphics[width=0.3\linewidth]{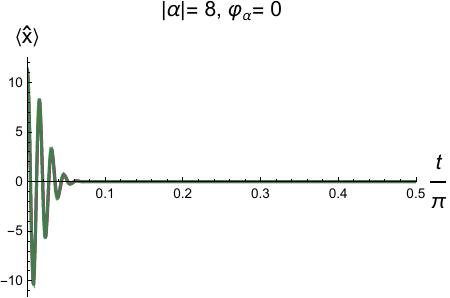}
   \includegraphics[height=0.7cm]{graphics/Legends/plotlegend-exact-semicl.pdf}
   \includegraphics[height=0.7cm]{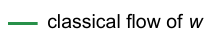}
    \caption{Comparison of  the evolution of $\braket{\hat{x}}$ under classical and semiclassical flow with the exact quantum evolution. For $|\alpha|<4$, the classical prediction is clearly worse than semiclassical, while for larger values of $|\alpha|$ both the classical predictions are very close to the exact result.}
    \label{fig:AO coherent state <x>(t) semicl cl}
\end{figure}
In the following, we will compare the quality of the classical and semiclassical predictions of $\braket{\hat{x}}(t)$, i.e. we will study the difference between  neglecting or including  the $\mathcal{O}(\lambda)$-term when calculating $H(x,p)=\sigma_\lambda[\hat{\mathcal{H}}]$ via the Moyal-product, which we refer to as the Moyal term. When neglecting the Moyal term, the symbol Hamiltonian reads  
\begin{equation}
    H(x,p)=-\frac{(x^2+p^2)^2}{4}+\mathcal{O}(\lambda),
\end{equation}
which directly implies that the classical phase space flow is an oscillation with angular velocity  $\Omega(r)_\text{class}=- r^2$  
\begin{equation}
    \begin{aligned}
        x^{\phi_{\text{class }t}}(t)=&r_0\cos(\varphi_0+r^2t)\\
         p^{\phi_{\text{class }t}}(t)=&r_0\sin(\varphi_0+r^2t).\\ 
    \end{aligned}
\end{equation}

Figure~\ref{fig:AO coherent state <x>(t) semicl cl} shows $\braket{\hat{x}}_\text{class}(t)$, the expectation valued obtained by the evolution of $w_\alpha$ under $\phi_{\text{class}}$, together with $\braket{\hat{x}}_\text{semicl}(t)$ and the exact evolution of $\braket{\hat{x}}(t)$. One can see that neglecting the Moyal term for $H(x,p)$ clearly gives a worse prediction.  For  $|\alpha|\gtrsim4$ however, the classical flow also provides a relatively good prediction.  This is not surprising, as in~\autoref{eq:AO possionbraket}, the $\mathcal{O}(\lambda)$-term and the $\mathcal{O}(\lambda^0)$-term differ by a factor $1/r^2$. Assuming again that only the phase space region of $r\sim \sqrt{2}|\alpha|$ contribute gives 
\begin{equation}
    \epsilon=\frac{1}{2|\alpha|^2}\ll1
\end{equation}
as control parameter of the classical approximation. Now that we have seen that the Moyal term indeed improves the prediction, it will always be included in the approximate dynamics for the rest of this chapter.

\subsection{Semiclassical prediction for the period $T_\text{oscill}$ and decay time $\tau_\text{dec}$ of wave packet center oscillations}\label{sec:AO Toscill and taudecay}
\begin{figure}[h] 
    \centering
    \includegraphics[width=0.45\linewidth]{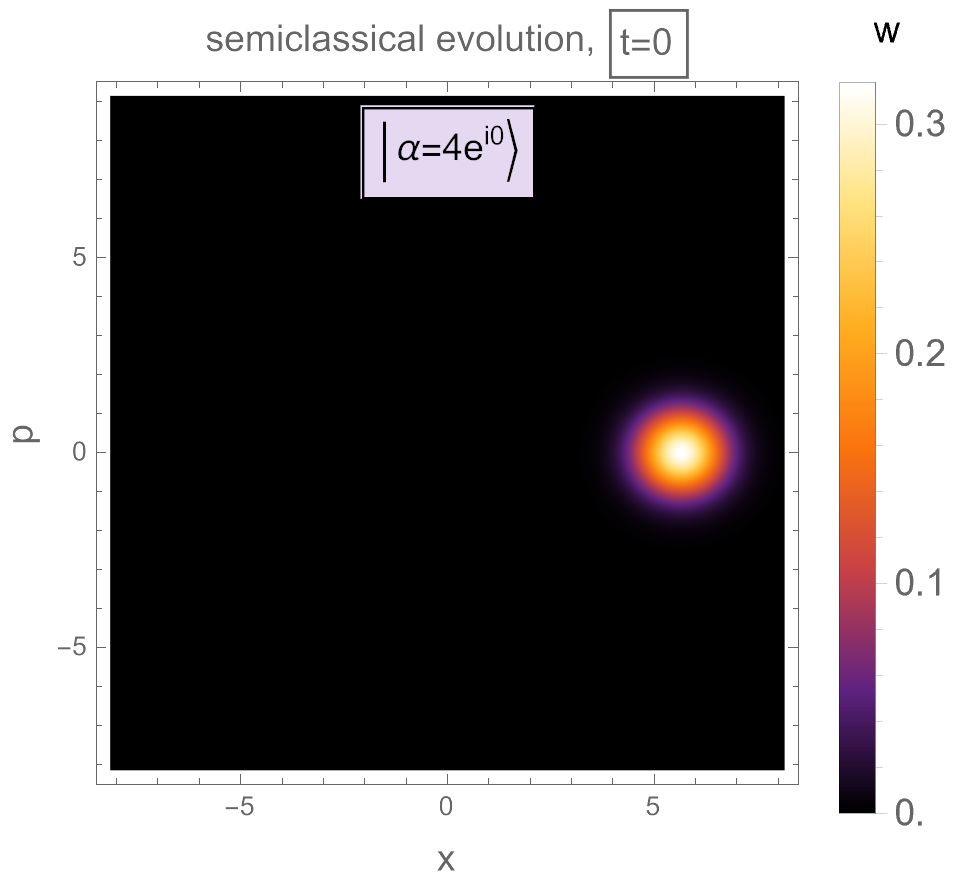}
    \includegraphics[width=0.45\linewidth]{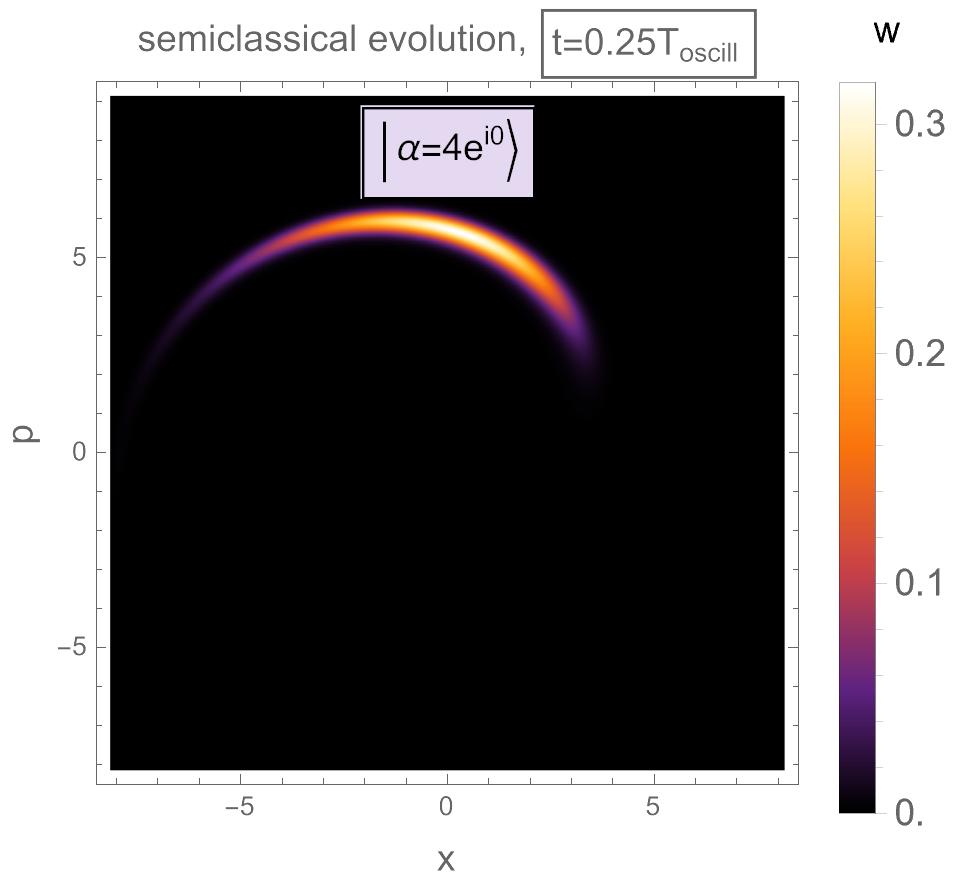}
    \caption{Evolution of  $w_\alpha(x^{\phi_{-t}},p^{\phi_{-t}})$, the Wigner function for an initial state $\ket{\alpha}$ with $\alpha=4$ under semiclassical flow $\phi$. After a time  $t=\frac{T_\text{oscill}}{4}=\frac{\pi}{62}$, $w_\alpha$ has made an average phase space rotation of $\pi/2$ and has been pulled in a banana-like shape. Comparison to~\autoref{fig:AO PlotWig quantum time evolution} shows that for $t=\pi/62$,  the form of $w_\alpha(x^{\phi_{-t}},p^{\phi_{-t}})$  mainly resembles  $w_\alpha(t)$ under the quantum evolution, but the semiclassical evolution misses the oscillator region for $r<r_\alpha$.}
    \label{fig:AO 2DWig alpha=2 t=0 t=T/4 semicl}
\end{figure}
The aim of this section is to predict the period $T_\text{oscill}$ and the typical decay time $\tau_\text{dec}$ of the oscillations of $\braket{\hat{x}}(t)$ and   $\braket{\hat{p}}(t)$ using the semiclassical evolution of $w_\alpha(x,p)$. This will allow a classification of the  semiclassical dynamics in terms of $|\alpha|$. Please note that the Anharmonic Oscillator is a special case where the evolution of $\braket{\hat{x}}$ and $\braket{\hat{p}}$ for an initially coherent state can be written in a compact form. The method that will be presented would allow to approximately predict $T_\text{oscill}$ and $\tau_\text{dec}$ also for oscillator systems where no compact form or not even an analytical solution is known. 

\begin{figure}
    \centering
    \includegraphics[width=0.48\linewidth]{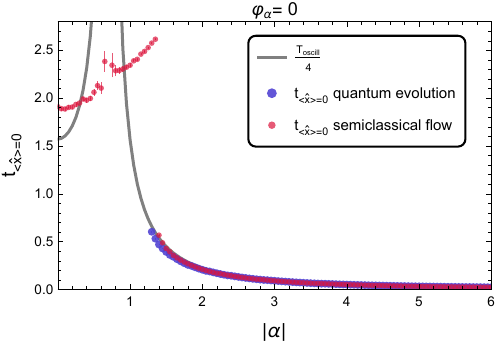}
    \includegraphics[width=0.48\linewidth]{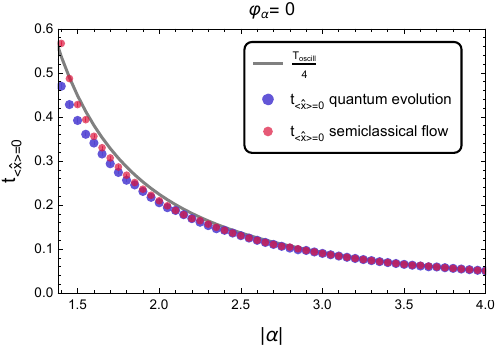}
    \caption{Comparison of $T_\text{oscill}/4$, as in~\autoref{eq:AO Toscill}, with the times of first $\braket{\hat{x}}=0$ under quantum time evolution (blue) and  under semiclassical flow (red) for an initial state $\ket{\alpha}$ with $\varphi_\alpha=0$. The graph on the right side is a zoom of the left side.  $T_\text{oscill}/4$ gives a good prediction of $t_{\braket{\hat{x}}=0}$ for larger  $|\alpha|$, but it fails for smaller $|\alpha|$.  The values of  $t_{\braket{\hat{x}}=0}$ have been found by iteration in $t$. 
    }
    \label{fig:AO xeq0 Toscill/4}
\end{figure}
We predict $T_\text{oscill}$  assuming that the angular velocity of $\braket{\hat{x}}_\text{semicl}$ and $\braket{\hat{p}}_\text{semicl}$ is given by the Wigner function average of  the angular velocity of the semiclassical flow  $\Omega(r)$,
\begin{equation}
\begin{aligned}
     \braket{\Omega}=&\int\mathrm{d}r r\int\mathrm{d}\varphi w_\alpha(r,\varphi) \Omega(r)\\
     =&1-r_\alpha^2\\
\end{aligned}
\end{equation}
and then 
\begin{equation}\label{eq:AO Toscill}
    \boxed{T_\text{oscill}=\frac{2\pi}{|\braket{\Omega}|}=\frac{2\pi}{|1-2|\alpha|^2|}}.
\end{equation}
In \cite{proppDecoherenceLimitsCost}, the limit of the period time $T_\text{oscill}\rightarrow\frac{\pi}{|\alpha|^2}$ for $|\alpha|\gg1$ is mentioned as ``the expected form of the Kerr effect``. This limit is the same expression as can be found for $t\ll1$ from the exact solutions~\ref{eq:AO coherent state <x> <p>}. Figure~\ref{fig:AO 2DWig alpha=2 t=0 t=T/4 semicl} qualitatively confirms~\autoref{eq:AO Toscill}: We can see that  after $t=\frac{T_\text{oscill}}{4}$, $w_\alpha(x,p)$ has performed, as  expected, a mean phase space rotation of $\pi/2$ under the semiclassical flow. It is worth stressing that $T_\text{oscill}$ in  \autoref{eq:AO Toscill} is a prediction for the semiclassical flow and thus only (approximately) describes a physical quantity if the semiclassical approximation is valid. 

For fixed  $\varphi_\alpha=0$, we expect  $t_{\braket{\hat{x}=0}}$, the first time when $\braket{\hat{x}}=0$, to occur at $T_\text{oscill}/4$. This prediction is tested in \autoref{fig:AO xeq0 Toscill/4} for different values of $|\alpha|$: For $|\alpha|\geq 1.4$, the prediction matches well with the semiclassical $t_{\braket{\hat{x}}=0}$, but for $|\alpha|<1.4$ it does not match at all. An explanation for this failure could be that for $|\alpha|<1.4$ phase space regions around  $r=\frac{1}{\sqrt{2}}$ where $\Omega\approx0$ become important: a part of the Wigner functions rests centered around $x=0$ so that $\braket{\Omega}$ alone is not sufficient to describe the dynamics. This can be illustrated by an exaggerated example: If 95$\%$ of the weight of a Wigner function remain still standing and are centered around a $x=r`$ then even if the remaining $5\%$-weight becomes centered around $x=-r`$ one never obtains $\braket{\hat{x}}=0$.  Under quantum time evolution  $\braket{\hat{x}}=0$ is not reached for any   $|\alpha|<1.3$,  which is in agreement with the observations in~\autoref{sec:AO <x> <p> quantum}.  For $|\alpha|\geq1.4$,  $t_{\braket{\hat{x}}=0}$ of quantum time evolution is well predicted by $T_\text{oscill}/4$.
\\Previously, we saw that the semiclassical evolution cannot  predict the revival of the initial state.  Therefore $t=2\tau_\text{dec}$, the approximate time when $\braket{\hat{x}}_\text{semicl}(t)$ and   $\braket{\hat{p}}_\text{semicl}(t)$  reach the origin $\braket{\hat{x}}_\text{semicl}\approx\braket{\hat{p}}_\text{semicl}\approx0$, becomes the limit for the validity of the semiclassical approximation, no matter how small $\epsilon$.  An explanation might be given by the following argument: Even though the oscillatory parts at $r<r_\alpha$ in the exact Wigner function (see \autoref{fig:AO PlotWig quantum time evolution}) can be neglected when integrating over some symbol $o(x,p)$, the oscillatory parts still contain physical information that gives rise to the revivals~\cite{proppDecoherenceLimitsCost}. Therefore, when dealing with wave packet revivals, the previous assumption of only including phase space regions where $r\sim r_\alpha $ is not valid, showing that $\epsilon^2=1/4|\alpha|^4$ is not the right control parameter anymore.               

\begin{figure}
    \centering
    \begin{subfigure}{0.55\linewidth}
       \includegraphics[width=0.99\linewidth]{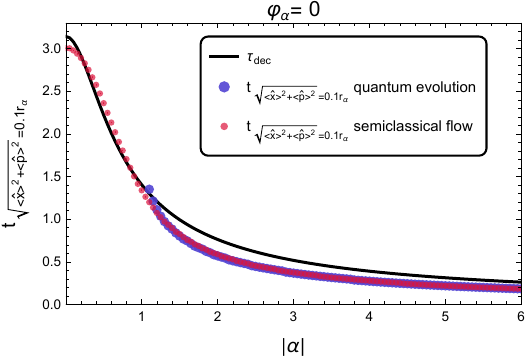} 
       \caption{}
    \end{subfigure}
    \begin{subfigure}{0.44\linewidth}
            \includegraphics[width=0.99\linewidth]{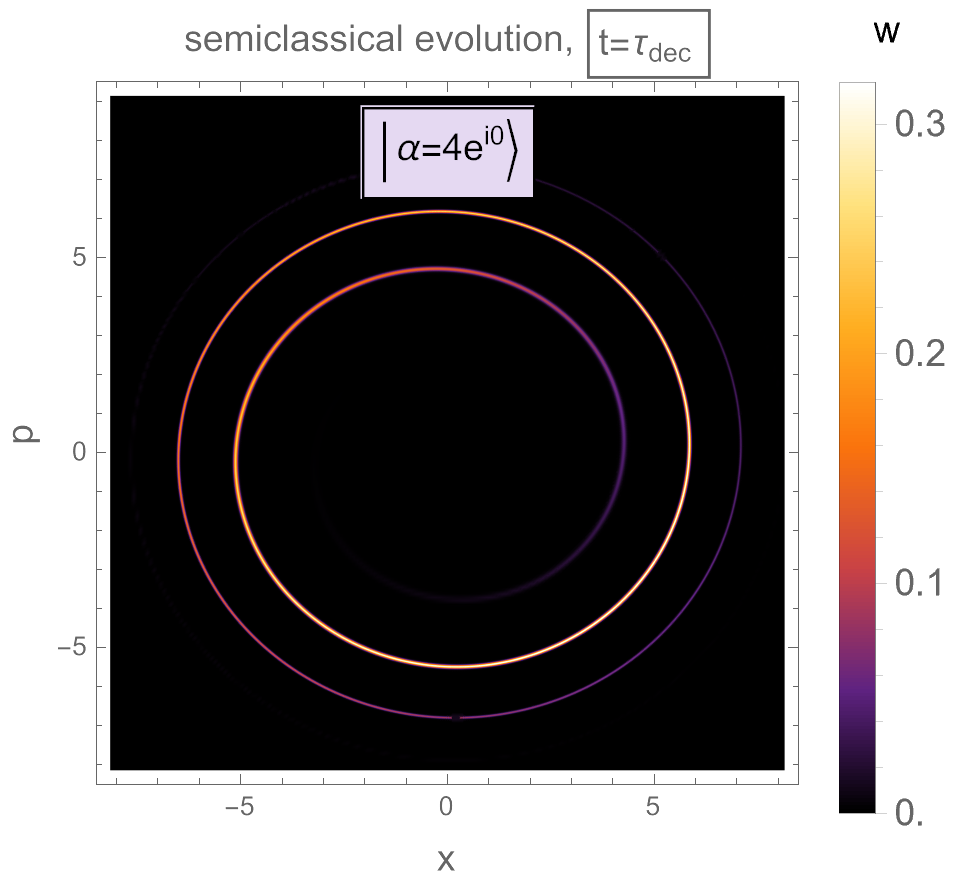} 
            \caption{}
    \end{subfigure}    
    \caption{a) Comparison of $\tau_\text{dec}$, as in~\autoref{eq:AO taudecay},  with the time when the envelope of $\braket{\hat{x}}$ and $\braket{\hat{p}}$ for $\ket{\alpha}(t)$ has decayed under a threshold $\sqrt{\braket{\hat{x}}^2+\braket{\hat{p}}^2}=0.1r_\alpha$.  $\tau_\text{dec}$ describes well the order of the values for the semiclassical flow (red), but cannot capture that for  $|\alpha|\leq1$,   $\sqrt{\braket{\hat{x}}^2+\braket{\hat{p}}^2}$  never reaches $0.1r_\alpha$ under quantum time evolution (blue).  The values for the semiclassical flow and quantum time evolution have been found iteratively. b) Semiclassical evolution of  the Wigner function at $t=\tau_\text{dec}=\pi/\sqrt{65}$ for an initial state $\ket{\alpha}$ with $\alpha =4$. Figure b) inspired by~\cite{proppDecoherenceLimitsCost} }
    \label{fig:AO taudec}
\end{figure}

As visible in~\autoref{fig:AO 2DWig alpha=2 t=0 t=T/4 semicl}, $w_\alpha(x,p)$ spreads out under semiclassical evolution, because the angular velocity $\Omega(r)$ depends on the radial phase space position. This mechanism can explain the decay of the oscillation amplitude: as the Wigner function spreads out along a disk around $r=r_\alpha$~\cite{proppDecoherenceLimitsCost} it becomes increasingly centered around $(x=0,p=0)$. We expect $w_\alpha(x,p)$ to be well centered around the origin when 2 neighboring parts of the Wigner function say at $(r=r_0,\varphi=\varphi_0)$ and $(r=r_0+\Delta r,\varphi=\varphi_0)$  have moved to positions $(r=r_0,\varphi=\varphi_1)$ and $(r=r_0+\Delta r,\varphi=\varphi_1+\pi)$ with a relative phase $\pi$. More precisely, we expect  $\tau_\text{dec}$,  the typical decay time of the oscillations, to be when the Wigner function has accumulated a standard phase deviation $\Delta \Omega\, t =\pi$.  Thus, $\tau_\text{dec}$ is also the typical time of decoherence of the Wigner function. With  
\begin{equation}
\begin{aligned}
        \langle\Omega^2\rangle=&\int \mathrm{d} r r \int \mathrm{~d} \phi w_\alpha(r, \phi) \Omega(r)^2\\
        =&2+r_\alpha^4,\\
        \Delta\Omega=&\sqrt{\braket{\Omega^2}-\braket{\Omega}^2}\\
        =&\sqrt{1+2r_\alpha^2}\\
\end{aligned}
\end{equation}
we obtain
\begin{equation}\label{eq:AO taudecay}
    \boxed{\tau_\text{dec}=\frac{\pi}{\Delta\Omega}=\frac{\pi}{\sqrt{1+4|\alpha|^2}}}.
\end{equation}
\autoref{fig:AO taudec} b) shows that at $t=\tau_\text{dec}$, $w_\alpha(x^{\phi_{-t}},p^{\phi_{-t}})$  has indeed evolved into a spiral well centered around the origin. In \autoref{fig:AO taudec} a),   $\tau_\text{dec}$ is compared with $t_{\sqrt{\braket{\hat{x}}^2+\braket{\hat{p}}^2}=0.1r_\alpha}$, the time when the dynamics reach $\sqrt{\braket{\hat{x}}^2+\braket{\hat{p}}^2}=\frac{r_\alpha}{10}$. One can see that under semiclassical flow $t_{\sqrt{\braket{\hat{x}}^2+\braket{\hat{p}}^2}=0.1r_\alpha}$ is truly at the order of $\tau_\text{dec}$. For $|\alpha|\leq1$, under quantum time evolution $\sqrt{\braket{\hat{x}}^2+\braket{\hat{p}}^2}$ does not become smaller than $\frac{r_\alpha}{10}$. This is the regime where the semiclassical prediction quickly deviates from the exact solution and that $\tau_\text{dec}$ is not giving the correct prediction. For $|\alpha|>1$ on the other hand,  $\tau_\text{dec}$ describes well also the  order of the values obtained by quantum time evolution.\\
\begin{figure}[t]
    \centering
    \includegraphics[width=0.6\linewidth]{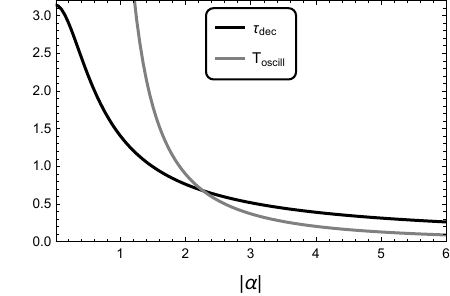}
    \caption{Comparison of $\tau_\text{dec}$ and $T_\text{oscill}$. For small $|\alpha|$,   $T_\text{oscill}\gg\tau_\text{dec}$, while around $|\alpha|=2$ ,  $T_\text{oscill}\sim\tau_\text{dec}$. $T_\text{oscill}$ further decreases faster  with $|\alpha|$  than $\tau_\text{dec}$. }
    \label{fig:AO taudecay vs Toscill}
\end{figure}
Finally, one can explain why the onset of the oscillations during the initial decay of $\braket{\hat{x}}_\text{semicl}(t)$ occurs around $|\alpha|=2$, by comparing $\tau_\text{dec}$ to $T_\text{oscill}$ in~\autoref{fig:AO taudecay vs Toscill}. At $|\alpha|=1$, the decay happens much faster than one period of an oscillation takes, resulting in an ``overdamped oscillation``,   while for $|\alpha|=2$,    $\tau_\text{dec}$ and $T_\text{oscill}$ are already comparable, so that one observes more than half of an oscillation  (see figure~\ref{fig:AN coherent state <x>(t) semicl}). Furthermore, $T_\text{oscill}$ decreases faster with $|\alpha|$ than with $\tau_\text{dec}$: the number of clearly visible oscillations increases with $|\alpha|$.
\FloatBarrier
\section{Evolution of wave packet width}\label{sec:AO DeltaxDeltap}

\begin{figure}
    \centering
  
    \begin{subfigure}{0.99\linewidth}
        \includegraphics[width=0.3\linewidth]{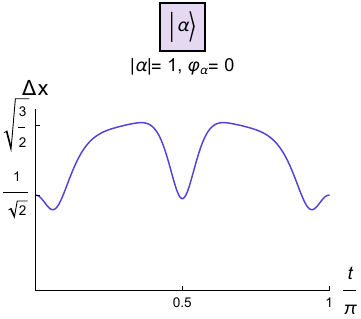}
        \includegraphics[width=0.3\linewidth]{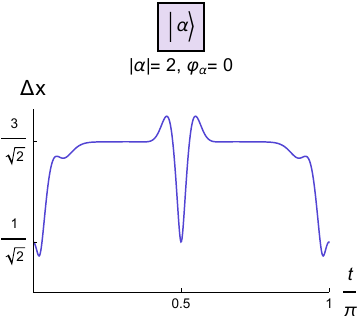}
        \includegraphics[width=0.3\linewidth]{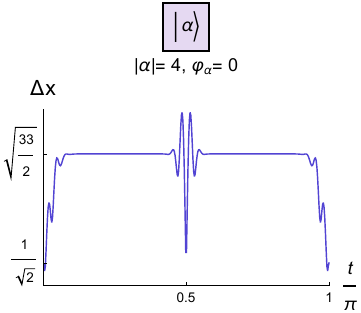}
    \end{subfigure}

    \begin{subfigure}{0.99\linewidth}
        \includegraphics[width=0.3\linewidth]{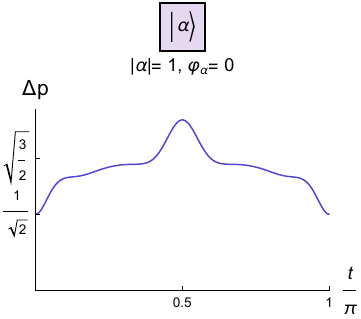}  
        \includegraphics[width=0.3\linewidth]{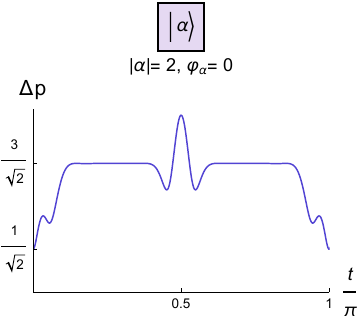}
        \includegraphics[width=0.3\linewidth]{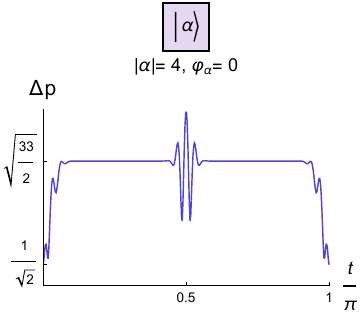}
    \end{subfigure}
   
    \begin{subfigure}{0.99\linewidth}
        \includegraphics[width=0.3\linewidth]{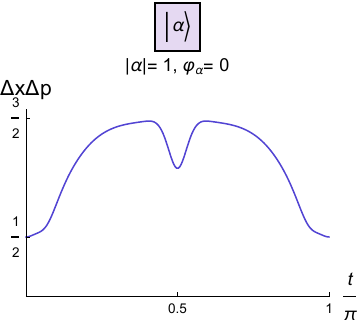}
        \includegraphics[width=0.3\linewidth]{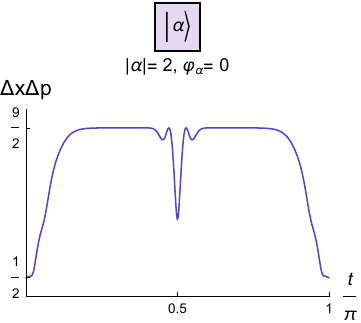}
        \includegraphics[width=0.3\linewidth]{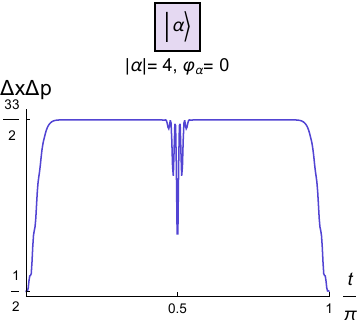} 
    \end{subfigure}
    \includegraphics[height=0.7cm]{graphics/Legends/plotlegend-exact.pdf}
    \caption{Evolution of $\Delta x$ and $\Delta p$ for $\ket{\alpha}(t)$. Initially,  $\Delta x$ and $\Delta p$ grow towards $\sqrt{|\alpha|^2+1/2}$ and then remain approximately constant until close to $t=\pi/2$ there is again an onset of motion. For larger $|\alpha|$, the grow happens faster and becomes stronger modulated with oscillations. The uncertainty product $\Delta x \Delta p$  has an upper bound of $|\alpha|^2+1/2$. Figure inspired by~\cite{sudheeshManifestationsWavePacket2004}}
    \label{fig:AO Heisenberg}
\end{figure}

In the Anharmonic Oscillator, an initially coherent state does not saturate the Heisenberg inequality under time evolution: $\Delta x$ and $\Delta p$ become dynamical quantities 
\begin{equation}\label{eq:AO Deltax and Deltap}
    \begin{aligned}
                 \Delta x=\sqrt{\frac{1}{2}+|\alpha|^2\left(1+\exp(-|\alpha|^2)\sum_{n=0}^\infty\frac{|\alpha|^{2n}}{n!}\cos{\varphi_{n+2,n,\alpha}(t)} \right.}\quad \quad \quad \quad\quad \quad \quad &\\%
                \quad \quad \space\overline{\left. -2\left(\exp \left(-|\alpha|^2\right) \sum_{n=0}^{\infty} \frac{|\alpha|^{2 n}}{n!} \cos \varphi_{n+1,n,\alpha}(t)\right)^2\right)}&\\
                \Delta p=\sqrt{\frac{1}{2}+|\alpha|^2\left(1-\exp(-|\alpha|^2)\sum_{n=0}^\infty\frac{|\alpha|^{2n}}{n!}\cos{\varphi_{n+2,n,\alpha}(t)} \right.}\quad \quad \quad \quad\quad \quad \quad &\\%
                \quad \quad \space\overline{\left. -2\left(\exp \left(-|\alpha|^2\right) \sum_{n=0}^{\infty} \frac{|\alpha|^{2 n}}{n!} \sin \varphi_{n+1,n,\alpha}(t)\right)^2\right)}&
    \end{aligned}
\end{equation}
with $\varphi_{m,n,\alpha}(t)$ as in \autoref{eq:AO phi-m,n,alpha}. The evolution of $\Delta x$ and $\Delta p$ is shown in  \autoref{fig:AO Heisenberg} for $\varphi_\alpha=0$ and different values of $|\alpha|$: 
After the onset of motion,  $\Delta x$ and $\Delta p$ grow towards $\sqrt{\frac{2|\alpha|^2+1}{2}}$~\cite{sudheeshManifestationsWavePacket2004}. During this growth, the initially coherent state becomes squeezed  such that $\Delta x <\frac{1}{\sqrt{2}}$~\cite{sudheeshManifestationsWavePacket2004}. When increasing $|\alpha|$, the growth happens more rapidly and becomes more oscillatory. After $\Delta x$ and $\Delta p$ have remained for a while at $\sqrt{\frac{2|\alpha|^2+1}{2}}$, there is an onset of an oscillation close before $t=\pi/2$. At $t=\pi/2$, $\Delta x$  and $\Delta p$  become a local minimum and maximum, which corresponds to a fractional revival~\cite{sudheeshManifestationsWavePacket2004}, illustrated via the exact Wigner function evolution in~\autoref{fig:AO PlotWig quantum time evolution} bottom left. Further, one can see that the uncertainty product $\Delta x\Delta p$ never exceeds $|\alpha|^2+1/2$. This is indeed the maximal uncertainty of any initially coherent state
\begin{equation}\label{eq:AO maxiDeltaxDelta}
    \begin{aligned}
        \Delta x\Delta p\leq |\alpha|^2+\frac{1}{2}.
    \end{aligned}
\end{equation}
\begin{proof}
    \begin{equation}
    \begin{aligned}        
       \Delta x\Delta p&= \sqrt{\braket{\hat{x}^2}-\braket{\hat{x}}^2}\sqrt{\braket{\hat{p}^2}-\braket{\hat{p}}^2}\\
       &\leq \sqrt{\braket{\hat{x}^2}\braket{\hat{p}^2}}\\
       &=\left(\frac{1}{2}+|\alpha|^2\right)\sqrt{1-\left(\exp(-|\alpha|^2)\sum_{n=0}^\infty\frac{|\alpha|^{2n}}{n!}\cos{\varphi_{n+2,n,\alpha}(t)} \right)^2}\\
       &\leq|\alpha|^2+\frac{1}{2}
    \end{aligned}
    \end{equation}
\end{proof}
\subsection{Semiclassical analysis of the wave packet width}
\begin{figure}
    \centering
    \begin{subfigure}{0.99\linewidth}
           \includegraphics[width=0.32\linewidth]{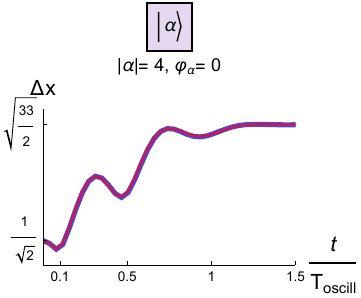}
    \includegraphics[width=0.32\linewidth]{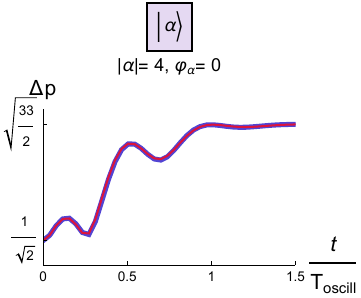}
    \includegraphics[width=0.32\linewidth]{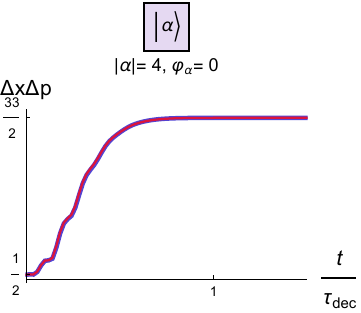}
    \end{subfigure}
     \includegraphics[height=0.7cm]{graphics/Legends/plotlegend-exact-semicl.pdf}
    \caption{The evolution of $\Delta x$, $\Delta p$ and $\Delta x\Delta p$ under quantum time evolution and under the semiclassical flow for an initial state $\ket{\alpha}$ with $|\alpha|=4$ and $\varphi_\alpha=0$.  During the initial growth the quantum and the semiclassical evolution match. Left and Middle: At  $t=T_\text{oscill}/2$,  $\Delta x$ becomes locally minimal and $\Delta p$  locally maximal.  Right: At $t=\tau_\text{dec}$, the uncertainty product $\Delta x \Delta p$ has already reached its maximal value.}
    \label{fig:AO Deltax Deltap semicl}
\end{figure}
\begin{figure}
    \centering
    \begin{subfigure}{0.49\linewidth}
        \includegraphics[width=0.99\linewidth]{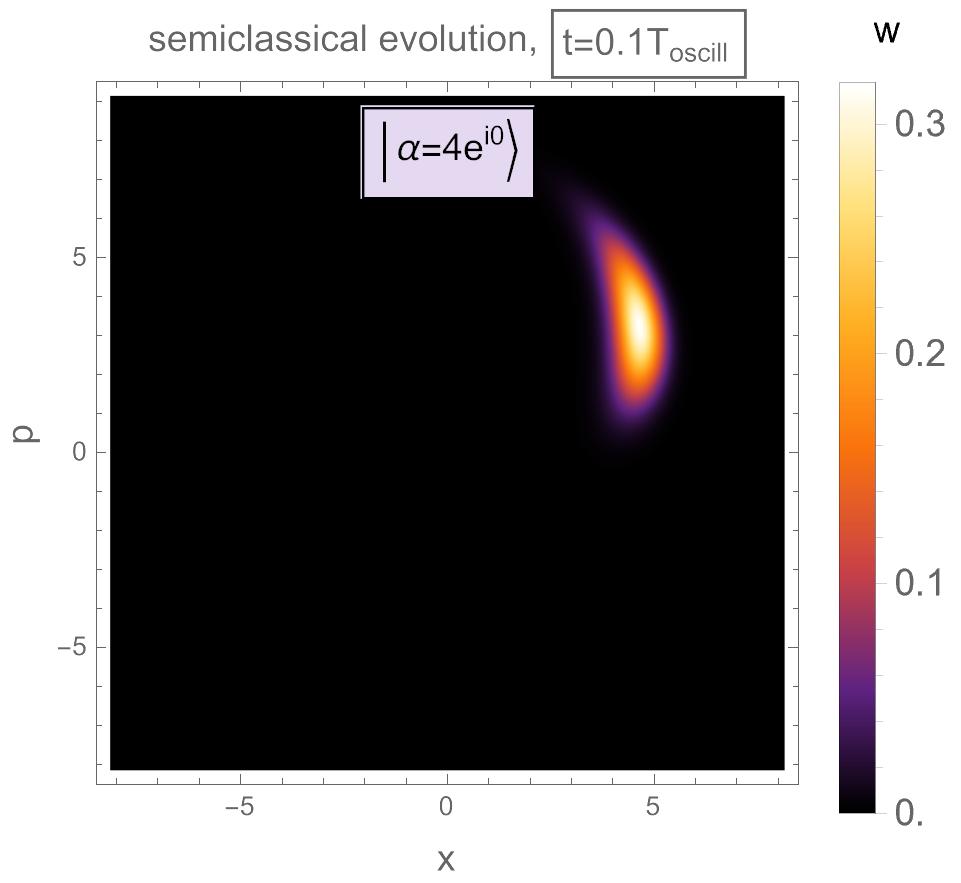}
    \end{subfigure}
    \begin{subfigure}{0.49\linewidth}
        \includegraphics[width=0.99\linewidth]{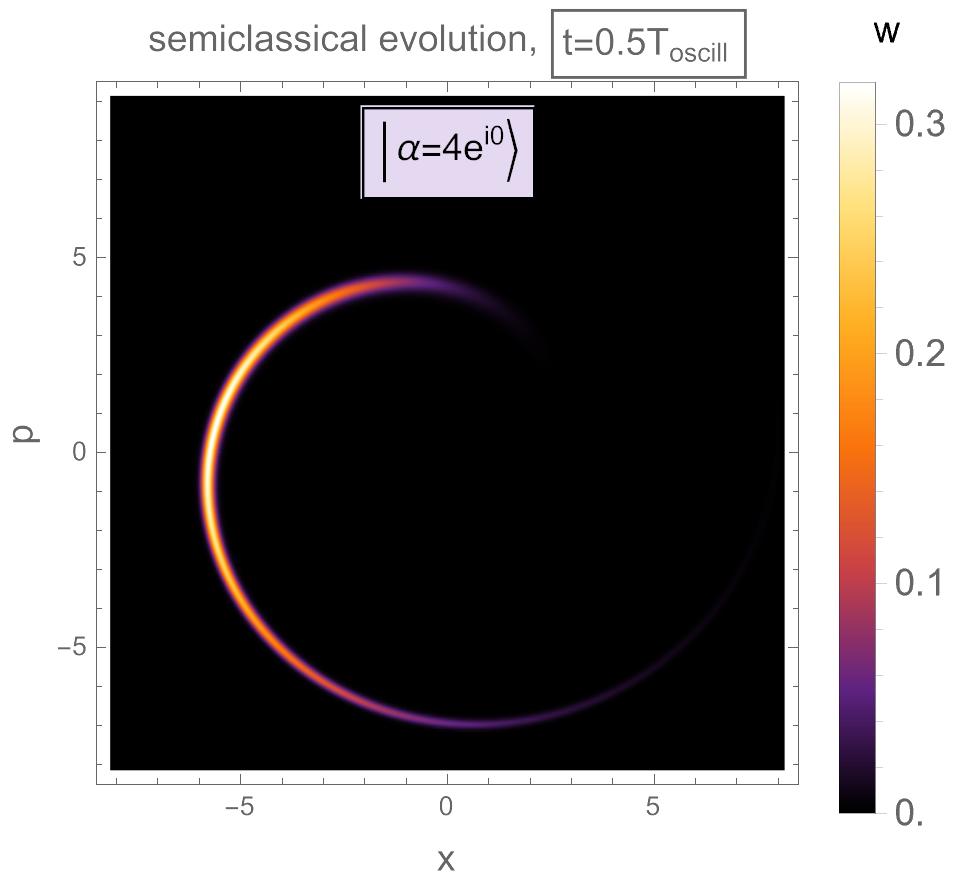}
    \end{subfigure}
    \caption{Left: After a time  $t=T_\text{oscill}/10=\pi/155$, the Wigner function $w_\alpha$ is squeezed into $\Delta x<\sqrt{\frac{1}{2}}$ by the semiclassical flow $\phi$.  Right: After a time $t=T_\text{oscill}/2=\pi/32$,  the Wigner function has the form of a crescent moon centered around the $x$-axis.}
    \label{fig:AO 2DWig alpha=4 t=0.1 0.5Toscill}
\end{figure}
To gain a qualitative understanding of the oscillatory growth of $\Delta x$ and $\Delta p$ it is insightful to study the evolution of $w_\alpha(x,p)$ under semiclassical flow. \autoref{fig:AO Deltax Deltap semicl} shows exemplarily that the semiclassical prediction can describe the initial dynamics  of $\Delta x$ and $\Delta p$ for  not too small $|\alpha|$.  $\Delta x $ getting squeezed below $\frac{1}{\sqrt{2}}$ can be understood qualitatively in the following way: Parts of $w_\alpha(x,p)$ with different values of $r$ move with different angular velocities $\Omega(r)$, this way the initially round form of $w_\alpha(x,p)$ gets  pulled into a banana-like shape~\cite{proppDecoherenceLimitsCost} as illustrated in \autoref{fig:AO 2DWig alpha=4 t=0.1 0.5Toscill} on the left. One can see that the width of the banana-shaped Wigner function is smaller in $\vec{e}_r$-direction than in $\vec{e}_\varphi$-direction. Therefore, $\Delta x$ becomes  a local maximum at around $T_\text{oscill}/4$ (\autoref{fig:AO 2DWig alpha=2 t=0 t=T/4 semicl}) and a local minimum at around $T_\text{oscill}/2$ (\autoref{fig:AO 2DWig alpha=4 t=0.1 0.5Toscill} on the right). For $\alpha=4$, $T_\text{oscill}$, as in \autoref{eq:AO Toscill}, gives a rather accurate prediction of the oscillation period for $\Delta x$ and $\Delta p$ as shown in~\autoref{fig:AO Deltax Deltap semicl}. Under semiclassical flow, the Wigner function is constantly stretched in direction $\vec{e}_\varphi$. When at a time around $t=\tau_\text{dec}$, it has expanded over the whole disk around $r=r_\alpha$, (\autoref{fig:AO taudec} on the right),  we expect the uncertainty product $\Delta x \Delta p$  to  have reached its maximal value, which confirmed for $\alpha=4$ in \autoref{fig:AO Deltax Deltap semicl}. The argument of $T_\text{oscill}$ decreasing  much faster with $|\alpha|$ than $\tau_\text{dec}$ (see \autoref{fig:AO taudecay vs Toscill}) can now explain that the growth of $\Delta x$ and $\Delta p$ becomes increasingly oscillatory with larger values of $|\alpha|$. The semiclassical Wigner function evolution is incapable of describing the onset of oscillation close to $t=\pi/2$  related to fractional revival.          


%% file: chapters/Dirac_Oscillator.tex
\chapter{Wave Dynamics of the Dirac Oscillator}\label{chapter:DO}
The semiclassical formalism for multiband systems presented in~\autoref{chapter:operatorvauledsemiclass}  is ultimately used in this chapter to study the dynamics of the Dirac Oscillator, a two-band system with know analytic solutions. One aim is to give a proof of concept of the presented methods via  validation through the exact solution. A focus is set on the identification of the adiabatic subspaces.  The Dirac Oscillator has nontrivial  Berry curvature and Littlejohn-Flynn term, but a zero Moyal term. This allows to complete the study of the effects of the different first-order semiclassical correction terms.
The other main goal of this chapter is to gain new understandings on wave dynamics in the Dirac Oscillator. For this purpose the  semiclassical formalism is exploited to predict characteristic quantities in a similar way as done for the Anharmonic Oscillator.         
\section{The Dirac Oscillator}
The Dirac Oscillator~\cite{moshinsky_dirac_1989} is a relativistic generalization of the Quantum Harmonic Oscillator. In the (1+1) dimensional case~\cite{NogamiCoherentStateDO}, its Hamiltonian is expressed by the $(2\times2)$ Pauli matrices $\sigma_i$ acting on (pseuod)spin 1/2  particles
\begin{equation}
 \hat{\mathcal{H}}=\sigma_y(c\hat{p}-\mathrm{i} \sigma_z cm \omega \hat{x})+\sigma_z mc^2\,
\end{equation}
or in matrix form
\begin{equation}\label{eq:DO operatorH-HO style}
    \hat{\mathcal{H}}=c\left(\begin{array}{cc} 
mc & \sqrt{2 m\hbar \omega} \hat{a}^{\dagger} \\
\sqrt{2 m \hbar\omega} \hat{a} & -mc
\end{array}\right).
\end{equation}
In the literature, $\hat{\mathcal{H}}$ is also  commonly expressed with $\hat{a}$ and $\hat{a}^\dagger$ interchanged \cite{torres_dirac-moshinsky_2010,sadurni_dirac-moshinsky_2011,franco-MicrowaveDO_2013} corresponding to an interchanging of the (pseudo)spins $\ket{\uparrow}\leftrightarrow \ket{\downarrow}$ and an opposite sign of $m$. The energy spectrum of the Dirac Oscillator 
\begin{equation}
\begin{aligned}
    E_{n\pm}=&\pm c\sqrt{|m|^2c^2+2\hbar \omega m(n+1) }=\pm E_n\\
    E_{(-1)\pm}=&\pm|m|c^2\\
\end{aligned}
\end{equation}
consists of 2 branches `$+$` and `$-$` corresponding to particles and antiparticles. For an infinite-size system, one of the modes $E_{(-1)\pm}$,  has a non-normalizable wave function and is therefore not physical~\cite{raghu_analogs_2008}. Here, these modes are abusively label with ``$n=-1$`` to stress the conceptual difference to $E_{n\pm}$ modes. Usually the ``$n=-1$`` modes are called ``zero-modes`` in the literature~\cite{raghu_analogs_2008}. In the nonrelativistic limit 
$\hbar \omega\ll m c^2$, 
\begin{equation}
   E_n= mc^2+(n+1)\hbar \omega +\mathcal{O}\left(\left(\frac{\hbar\omega}{mc^2}\right)^2\right) 
\end{equation}
the spectrum of the Quantum Harmonic Oscillator is recovered  (modulo the energy offset) for the particle branch.\\ 
\subsection{Physical realizations}
\begin{figure}
    \centering
    
    \centering
    \begin{subfigure}[b]{0.85\textwidth}
    \centering
    \includestandalone[height=4cm]{graphics/sketches/2DCrystal/HexagonalLattice}
     \hspace{2cm}
      \includestandalone[height=4cm]{graphics/sketches/2DCrystal/Brillouinzone}
        \caption{Left: Sketch of a tight-binding 2D hexagonal crystal with opposite onsite energy $\Delta$. Right: the corresponding first Brillouin zone}
        \label{fig:DO 2DCrystalsketch}
    \end{subfigure}
    \hfill
    \begin{subfigure}[b]{0.85\textwidth}
        \includegraphics[width=0.32\linewidth]{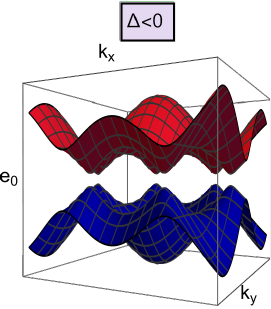}
        \includegraphics[width=0.32\linewidth]{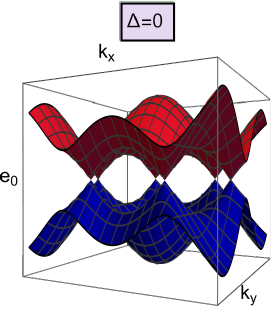}
       \includegraphics[width=0.32\linewidth]{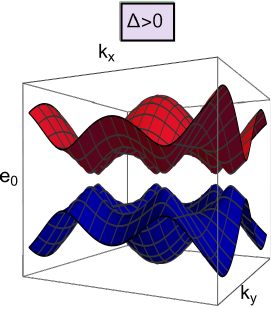}
        \caption{Band structure of the hexagonal crystal: For $\Delta=0$ the bands touch at the corners $K$ and $K'$ of the first Brillouin zone. A nonzero relative onsite energy opens a gap between the bands.}
        \label{fig:DO 2DCrystalbands}
    \end{subfigure}

    \caption{Hexagonal lattice and its band structures. Figure inspired by \cite{LectNotesDelplace}}
    \label{fig:2Dcrystal}
\end{figure}

The Dirac Oscillator is a model with application to various fields of physics. In quantum optics for instance~\cite{torres_dirac-moshinsky_2010,bermudez-Relativistic_Landau_Levels_2007,sanchez-jaynes-cummings-in-PSQM_2025}, the Dirac Oscillator can be mapped on the Jaynes-Cummings model~\cite{jaynes_JC_model_1963} describing the interaction between a two level atom and a cavity mode. A photonic realization is proposed in~\cite{longhi_photonicDO_2010}. Experimentally in microwave physics, the Dirac oscillator has been realized as a tight-binding system of a chain of coupled dielectric disks~\cite{franco-MicrowaveDO_2013}. In the following, we will focus on still another realization: the Dirac oscillator as the Hamiltonian of a tight-binding hexagonal 2D-crystal in the vicinity of band crossing points~\cite{raghu_analogs_2008,LectNotesDelplace}. \autoref{fig:DO 2DCrystalbands} shows the bands of the Bloch Hamiltonian (the symbol Hamiltonian) of a tight-binding hexagonal crystal. For equal on side energies of the two sublattice sites A and B shown in \autoref{fig:DO 2DCrystalsketch}, the 2 bands touch at the corners of the first Brillouin zone $K$ and $K'$. For a fermionic crystal, this degeneracy happens at the Fermi energy. Altering the relative onsite energy of the 2 sublattice positions to $2\Delta$ opens a gap at the degeneracy points. In the vicinity of $K$ ($K'$) the Bloch Hamiltonian can be linearized \cite{xiao_valley-contrasting_2007} to
\begin{equation}
    H=\left(\begin{array}{cc}
       \Delta  & \pm cp_x-\mathrm{i}cp_y \\
         \pm cp_x+\mathrm{i}cp_y & -\Delta
    \end{array}\right)
\end{equation} 
acting on the wave functions of the sublattice sites $A$ and $B$, here represented as pseudospin $\ket{\uparrow},\ket{\downarrow}$.  $c$ is the fermi velocity,  and the different signs  $+$ ($-$) correspond to a linearization around $K$ ($K'$), where in new coordinates $p_x=p_y=0$ indicates the position of the corner of the Brilloun zone. Now, a slow variation of the relative onsite energy in one spatial direction is considered
\begin{equation}
    \Delta(x) =x\beta c^2.
\end{equation}
such that the Bloch Hamiltonian acquires a real space dependence and can be interpreted as a symbol Hamiltonian. $x=0$ indicates the spatial position where the bands 
\begin{equation}
    e_{0\pm}=\pm c\sqrt{x^2\beta^2 c^2+p_x^2+p_y^2}=\pm e_0
\end{equation} 
touch for $p_x=p_y=0$. It is convenient to perform a unitary rotation of the Pauli-matrices such that for linearization around $K$, the Weyl-quantized Hamiltonian becomes 
\begin{equation}
       \hat{\mathcal{H}}=c\left(\begin{array}{cc}
       \hat{p}_y & \sqrt{2\hbar \beta}c\hat{a}^\dagger \\
         \sqrt{2\hbar \beta}c\hat{a} & -\hat{p}_y
    \end{array}\right)
\end{equation}
 where $\hat{a}=\frac{1}{\sqrt{2}}\left(\frac{\hat{x}}{l}+\mathrm{i}\frac{\hat{p}_x}{\hbar/l}\right)$ with the characteristic length of the Dirac Oscillator
 \begin{equation}
     l=\sqrt{\frac{\hbar}{\beta c}}.
 \end{equation}
 If we had chosen to linearize around $K'$, $\hat{a}$ and $\hat{a}^\dagger$ would be interchanged corresponding to an interchange of the rotated (pseudo)spins $\ket{\uparrow}\leftrightarrow\ket{\downarrow}$ and a sign flip of $\hat{p}_y\rightarrow-\hat{p}_y$. \\
With the ansatz of a monochromatic wave in $y$-direction $\Psi=\mathrm{e}^{\mathrm{i}\frac{p_y y}{\hbar}}\ket{\psi}$ the equation of motion becomes
\begin{equation}
       \mathrm{i}\hbar \partial_t \ket{\psi}=c\left(\begin{array}{cc}
       p_y & \sqrt{2\hbar \beta}c\hat{a}^\dagger \\
         \sqrt{2\hbar \beta}c\hat{a} & -p_y
    \end{array}\right) \ket{\psi}\,.
\end{equation}
$p_y$ is now an external control parameter taking the role of $m$ in~\autoref{eq:DO operatorH-HO style}. For 3D Weyl semi-metals subjected to a magnetic field~\cite{yang_Weyl_semimetal_2011,son_chiral_anomaly_Weylmetals_2013}, the Dirac Oscillator can be realized in a similar way by a plane wave ansatz in direction of the magnetic field~\cite{LectNotesDelplace}.
\section{Nondimensionalization}
In order to apply the semiclassical formalism for multiband systems, a nondimensionalization needs to be performed such that the energy gap between the bands of the symbol $H(x,p)$  becomes  $\Delta \tilde{e}_0\gtrsim 1$  and that $\hbar$ is replaced by a dimensionless parameter $\varepsilon$ controlling the semiclassical approximation. 
As a first step, we introduce dimensionless coordinates $x=l \tilde{x}$, $t=\frac{l}{c}\tilde{t}$ and divide by half the minimal band gap $\Delta e_{0}/2$. Then, the Schrödinger equation can be expressed as 
\begin{equation}
      \mathrm{i}\sqrt{\varepsilon} \partial_{\tilde{t}}\ket{\psi}=\left(\begin{array}{cc}
       \mu & \sqrt{\varepsilon}\tilde{x}-\sqrt{\varepsilon}\partial_{\tilde{x}} \\
        \sqrt{\varepsilon}\tilde{x}+\sqrt{\varepsilon}\partial_{\tilde{x}} & -\mu
    \end{array}\right)\ket{\psi},
\end{equation}
where 
\begin{equation}\label{eq:DO varepsilon mu}
\begin{aligned}
        \sqrt{\varepsilon}&=\frac{\hbar c}{l \Delta e_{0}/2 }\\[1ex]
     \mu &=\frac{c p_y}{ \Delta e_{0}/2} \,.
\end{aligned}
\end{equation}

In order to have a form where $\varepsilon $ is at the original position of $\hbar$, slow variables $\tau= \sqrt{\varepsilon}\tilde{t}$ and $X=\sqrt{\varepsilon}\tilde{x}$ are introduced 
\begin{equation}
\begin{aligned}
    \mathrm{i} \varepsilon\partial_\tau|\psi\rangle&=
\left(\begin{array}{cc}
       \mu & X-\mathrm{i}\varepsilon\partial_{X} \\
        X+\mathrm{i}\varepsilon\partial_{X} & -\mu
    \end{array}\right)\ket{\psi}. 
\end{aligned}
\end{equation}
The Hamiltonian now depends on the operators  $\hat{X}\equiv X$ and $\hat{P}_X\equiv-\mathrm{i}\varepsilon\partial_X$ which have indeed a rescaled commutator
\begin{equation}
   [X,-\mathrm{i}\varepsilon\partial_X]=\mathrm{i}\varepsilon.
\end{equation}
For a monochromatic wave in $y$-direction and no restriction on the $(x,p_x)$-plane, the minimal band gap is given by
\begin{equation}
\Delta e_{0}=2 c |p_y|.
\end{equation}
Thus
\begin{equation}\label{eq:mu-pm1}
   \boxed{
   \mu=\frac{p_y}{|p_y|}
}
\end{equation}
and 
\begin{equation}
\boxed{
    \varepsilon=\frac{\hbar \beta c}{p_y^2},}
\end{equation}
showing that there are ``four ways`` towards the semiclassical limit $\varepsilon\ll1$:
\begin{align}
   \text{relatively small $\hbar$ } \quad & \Rightarrow  \hbar \ll \frac{\beta c}{p_y^2}  \\
     \text{ slowly varying $\Delta(x)$} \quad & \Rightarrow    \beta \ll \frac{\hbar c}{p_y^2} \\ \text{relatively small  $c$ }\quad & \Rightarrow c \ll \frac{\hbar \beta}{p_y^2} \\
      \text{large band gap $\Delta e_0$}        \quad & \Rightarrow p_y^2 \gg \hbar \beta c \,.
\end{align}
From now on, we will work in nondimensional slow coordinates and rename $\tau \rightarrow t, \,X \rightarrow x, \,P_X\rightarrow p $ to lighten the notation. In order to describe a large range of physical realizations of the Dirac oscillator we treat $\mu$ in the following as an external control parameter instead of using directly \autoref{eq:mu-pm1}. We will later make use of \autoref{eq:mu-pm1} to justify the assumption that $|\mu|\gtrsim1$. 
\section{Energy spectrum of $\hat{\mathcal{H}}$}
\begin{figure}
    \centering
    \includegraphics[width=0.5\linewidth]{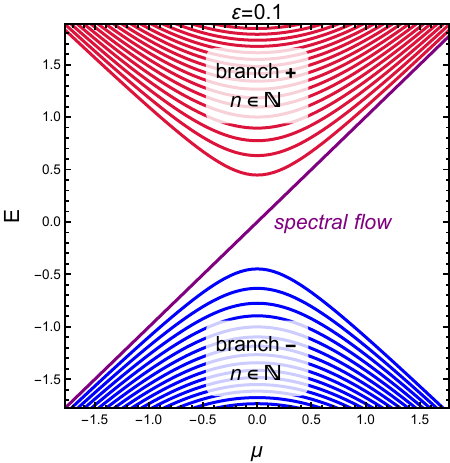}
    \caption{Spectrum of the dimensionless operator Hamiltonian $\hat{\mathcal{H}}$ of Dirac oscillator for $\varepsilon=0.1$. Figure inspired by~\cite{LectNotesDelplace}}
    \label{fig:DO SpectrumEn}
\end{figure}

In dimensionless slow coordinates, the annihilation operator is given by
\begin{equation}
    \hat{a}=\frac{\hat{x}+\mathrm{i}\hat{p}}{\sqrt{2\varepsilon}}\equiv\frac{1}{\sqrt{2}}\left(\frac{x}{\sqrt{\varepsilon}}+\sqrt{\varepsilon}\partial_x\right),
    \end{equation} 
which clearly satisfies $[\hat{a},\hat{a}^\dagger]=1$. This corresponds to a replacement  $l\rightarrow \sqrt{\varepsilon}$  and $\hbar\rightarrow\varepsilon$ in \autoref{eq:HO a adagger}. As a consequence, the dimensionless Hamiltonian can be expressed as 
\begin{equation}
    \hat{\mathcal{H}}=\left(\begin{array}{cc}
       \mu & \sqrt{2\varepsilon} \hat{a}^\dagger \\
         \sqrt{2\varepsilon} \hat{a} & -\mu
    \end{array}\right).
\end{equation}
 Its eigenvectors can be constructed from Fock states

\begin{equation}\label{eq:DO |psi_npm>}
\ket{\psi_{ n\pm}}=\frac{1}{\sqrt{2 E_n(E_n \mp \mu)}} \binom{\sqrt{2\varepsilon}\sqrt{n+1}|n+1\rangle}{( \pm E_n-\mu)|n\rangle},
\end{equation}
which have two branches of eigenenergys
\begin{equation}\label{eq:DO E_npm}
    E_{n\pm}=\pm\sqrt{\mu^2+2\varepsilon(n+1)}=\pm E_n
\end{equation}
For $n=-1$, there is only one possible eigenenergy 
\begin{equation}\label{eq:DO E_-1}
    E_{-1}=\mu
\end{equation}
with a corresponding eigenstate
\begin{equation}
    \ket{\psi_{-1}}=\left(\begin{array}{cc}
       \ket{0}\\
       0
    \end{array}\right).
\end{equation}
The formal solution $E=-\mu$ would correspond to a state~\cite{Haldane}
\begin{equation}
    \ket{\psi_{E=-\mu}}\propto \left(\begin{array}{cc}
        0\\
       \exp\left(\frac{x^2}{2\varepsilon}\right)
    \end{array}\right)
\end{equation}
which is not normalizable. $\ket{\psi_{-1}}$ is also called spectral flow state, because $E_{-1}$ ``flows`` from the branch `$-$` to the branch `$+$` when increasing the value of $\mu$, as shown in \autoref{fig:DO SpectrumEn}. As a consequence of \autoref{eq:DO E_-1}, $\ket{\psi_{-1}}$ is moving unidirectionally towards $+y$. For the Hamiltonian of the valley $K'$, $\ket{\psi_{-1}}$ is a state moving towards $-y$. 
\section{Adiabatic subspaces}\label{sec:DO Adiabatic subspaces}
\begin{figure}[!h]
        \includegraphics[width=0.32\textwidth]{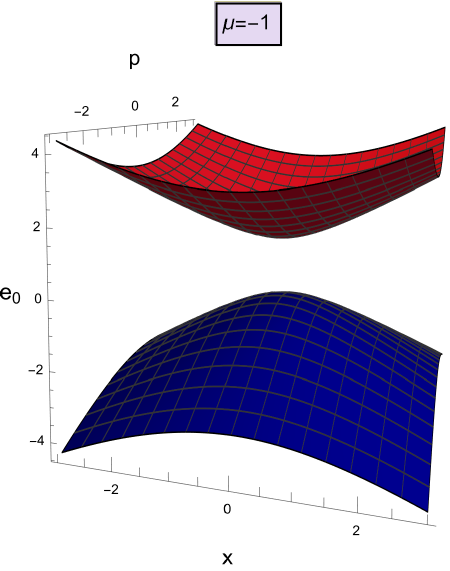}
        \includegraphics[width=0.32\textwidth]{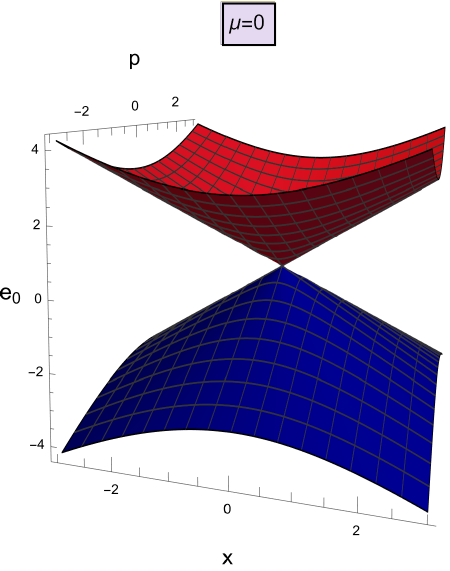}
        \includegraphics[width=0.32\textwidth]{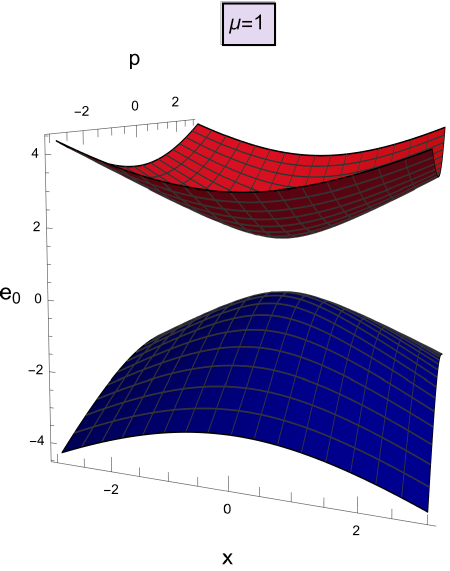}
        
        \caption{Plot of the bands $e_0^\pm (x,p)$ of the symbol Hamiltonian $H(x,p)$ of the Dirac Oscillator for different values of $\mu$}
        \label{fig:DO SpectrumSymbolHpy-101}
\end{figure}
The semiclassical equations of motion are defined individually for each adiabatic subspace $\mathcal{H}^\nu_\text{adiabatic}$ associated with the bands $e^{(\nu)}_0(x,p)$ of the lowest order symbol Hamiltonian $H_0(x,p)$. Therefore, to make use of the multiband semiclassical formalism, one needs to construct the different $\mathcal{H}^\nu_\text{adiabatic}$   from eigenvectors $\ket{\Psi_0}$ of $H_0(x,p)$ and has to identify the states contained in each subspace. The symbol Hamiltonian of the Dirac Oscillator is given by
\begin{equation}\label{eq:DO symbolH}
    \begin{aligned}
        H(x,p)=&\left(\begin{array}{cc}
        \mu & x-\mathrm{i}p \\
        x+\mathrm{i}p & -\mu
        \end{array}\right),
    \end{aligned}
\end{equation}
which does not have any terms of $\mathcal{O}(\varepsilon)$, i.e. $H=H_0$. Its two-level eigenspectrum 
\begin{equation}
    e^{\pm}_0=\pm\sqrt{\mu^2+x^2+p^2}=\pm e_0
\end{equation}
remains gapped along the whole phase space as long as $\mu\neq 0$ (see \autoref{fig:DO SpectrumSymbolHpy-101}). The nondimenionalisation was preformed such that $|\mu|=1$, ensuring that the minimal gap width $\Delta e_{0}=2\mu$ is at the order $\Delta e_{0}\sim 1$. Under this condition, the projectors $\pi^\pm=\sum_j\varepsilon^j\pi^\pm_j$ on the adiabatic subspaces $\mathcal{H}^\pm_\text{adiabatic}$ can be approximated by their lowest order terms for $\varepsilon\ll1$. The lowest order $\pi^{\pm}_0$ is given by the projector on the eigenvector $\ket{\Psi^\pm_0}$ of $H(x,p)$ 
\begin{equation}
    \begin{aligned}
        \pi^{\pm}_0=\ket{\Psi^\pm_0}\bra{\Psi^\pm_0}=&\frac{1}{2e_0\left(e_0 \mp \mu\right)}\left(\begin{array}{ll}
|z|^2 & ( \pm e_0-\mu) \bar{z} \\
( \pm e_0-\mu) z & (\pm e_0-\mu)^2
\end{array}\right)
    \end{aligned}
\end{equation}
where $z=x+\mathrm{i}p$. Note that $\ket{\Psi^\pm_0}\bra{\Psi^\pm_0}$ is not a Wigner function of any physical state, since
\begin{equation}
    \int\mathrm{d}x\int\mathrm{d}p\operatorname{Tr}_{\mathcal{H}_\text{f}}[\ket{\Psi^\pm_0}\bra{\Psi^\pm_0}]\rightarrow \infty\,.
\end{equation}
This is because in  Wigner-Weyl representation, invariant states are found as solutions of the $\star$-gen value equation $H\star W=EW$ and not by the eigenvalue equation $H\ket{\Psi_0}=e_0\ket{\Psi_0}$.
After having identified $\pi^\pm_0$, the remaining question is \say{Which physical states are contained in the adiabatic subspaces $\mathcal{H}^\pm_\text{adiabatic}$?} Comparing the spectrum of $\hat{\mathcal{H}}$ (the actual physical energy spectrum), shown in \autoref{fig:DO SpectrumEn}, to the spectrum of the symbol $H(x,p)$, shown in ~\autoref{fig:DO SpectrumSymbolHpy-101}, one intuitively associates the upper branch of eigenenergies $E_{n+}$ to the band $e_0^+$ and the lower branch $E_{n-}$ to $e_0^-$. Furthermore, the subspaces spanned by the branches $\{\ket{\psi_{n+}}\}$ and $\{\ket{\psi_{n-}}\}$  can be seen as time invariant subspaces: A superposition of states from a single branch might acquire a dynamical phase under time evolution, but will always remain in the initial branch. Expressed differently, the projectors $\sum_n\ket{\psi_{n\pm}}\bra{\psi_{n\pm}}$ on each branch are invariant under time evolution $\left[\sum_n\ket{\psi_{n\pm}}\bra{\psi_{n\pm}},\hat{\mathcal{H}}\right]=0$. This suggests a correspondence between the subspaces spanned by the branches $\pm$  and $\mathcal{H}^\pm_\text{adiabatic}$. For this correspondence to be true, each branch has to be in the range of its corresponding (Weyl-quantized) adiabatic projector 
\begin{equation}
    \hat{\pi}^\pm\ket{\psi_{n\pm}}=\ket{\psi_{n\pm}}+\mathcal{O}(\varepsilon^\infty).
\end{equation}
At lowest order, one can confirm this correspondence, by acting on $\ket{\psi_{n\pm}}$ with the Weyl-quantization of the zero-order projector
\begin{equation}
\begin{aligned}
\hat{\pi}^\pm_0=&\left(\begin{array}{lc}
2\varepsilon\frac{\hat{a} \hat{a}^{\dagger}+\hat{a}^{\dagger} \hat{a}}{2 \hat{A}} & \frac{\pm \hat{e}_0-\mu}{2 \hat{A}}\sqrt{2\varepsilon}\hat{a}^{\dagger}+\sqrt{2\varepsilon}\hat{a}^{\dagger}\frac{ \pm \hat{e}_0-\mu}{2 \hat{A}} \\
\frac{\pm \hat{e}_0-\mu}{2 \hat{A}}\sqrt{2\varepsilon}\hat{a}+\sqrt{2\varepsilon}\hat{a}\frac{\pm \hat{e}_0-\mu}{2 \hat{A}} & \frac{\left( \pm \hat{e}_0-\mu\right)^2}{\hat{A}}
\end{array}\right)+\mathcal{O}(\varepsilon^2),\\
\end{aligned}
\end{equation}
where 
\begin{equation}
\begin{aligned}
\hat{A}=&2 \hat{e}_0\left(\hat{e}_0 \mp \mu\right)\\
\hat{e}_0=&\sqrt{\mu^2+2\varepsilon(\hat{a}\hat{a}^\dagger+\hat{a}^\dagger\hat{a})}+\mathcal{O}(\varepsilon^2).
\end{aligned}
\end{equation}
It is straightforward to show that 
\begin{equation}
    \hat{\pi}^\pm_0\ket{\psi_{n\pm}}=\ket{\psi_{n\pm}}+\mathcal{O}\left(\frac{\varepsilon}{E_n}\ket{\psi_{n\pm}}\right).
\end{equation}
Indeed, $\hat{\pi}^\pm_0$ leaves $\ket{\psi_{n\pm}}$ unchanged up to an error of $\mathcal{O}(\varepsilon)$ if  $|\mu|	\gtrsim1  $. The role of the spectral flow state $\ket{\psi_{-1}}$ is more subtle. As one can see in \autoref{fig:DO SpectrumEn}, $\ket{\psi_{n-1}}$ can be associated to the branch $-$ for $\mu\lesssim-1$ and to the branch $+$ for $\mu\gtrsim1$. This is the condition that determines in which adiabatic subspace $\ket{\psi_{-1}}$ is included

\begin{equation}
\begin{aligned}
    \hat{\pi}^+_0 \ket{\psi_{-1}}
      &= \begin{cases}
            \ket{\psi_{-1}} + \mathcal{O}(\varepsilon^2 \ket{\psi_{-1}}), & \text{for } \mu \gtrsim 1, \\[1ex]
            0 + \mathcal{O}(\varepsilon \ket{\psi_{-1}}), & \text{for } \mu \lesssim -1,
         \end{cases} \\[2ex]
    \hat{\pi}^-_0 \ket{\psi_{-1}}
      &= \begin{cases}
            0 + \mathcal{O}(\varepsilon \ket{\psi_{-1}}), & \text{for } \mu \gtrsim 1, \\[1ex]
            \ket{\psi_{-1}} + \mathcal{O}(\varepsilon^2 \ket{\psi_{-1}}), & \text{for } \mu \lesssim -1 .
         \end{cases}
\end{aligned}
\end{equation}

Up to $\mathcal{O}(\varepsilon)$-corrections,  $\ket{\psi_{-1}}$ is in the range of $\hat{\pi}^-_0$ for $\mu\lesssim-1$ and in the range of $\hat{\pi}^+_0$ for $\mu\gtrsim1$. 
\subsubsection{Semiclassical equations of motion}
\begin{figure}
    \centering
    \includegraphics[width=0.45\linewidth]{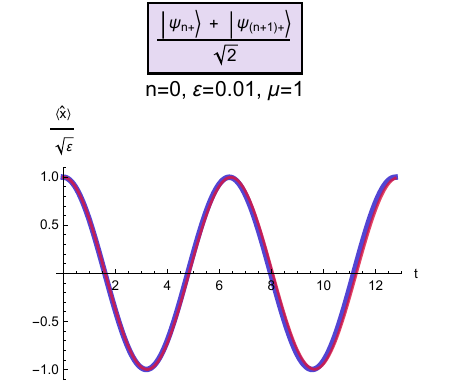}
    \includegraphics[width=0.45\linewidth]{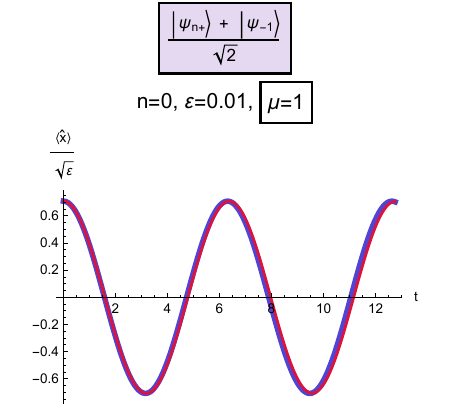}
    \hspace{0.5cm}
    \includegraphics[width=0.45\linewidth]{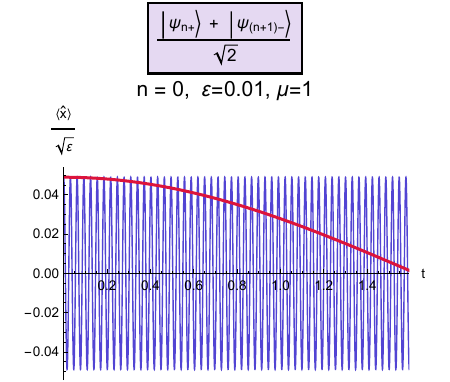}
      \includegraphics[width=0.45\linewidth]{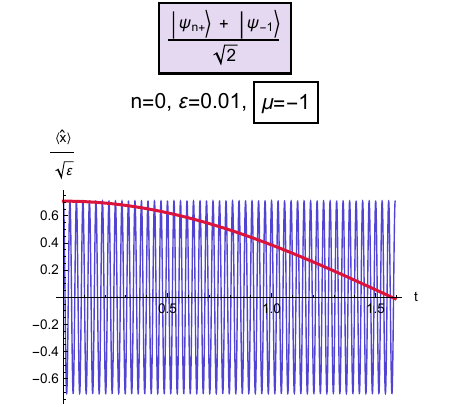}
      \includegraphics[height=0.7cm]{graphics/Legends/plotlegend-exact-semicl.pdf}
    \caption{Comparison of the exact $\braket{\hat{x}}(t)$  obtained via quantum evolution with the approximation $\braket{\hat{x}}_\text{semicl}(t)$ calculated by evolution of $w(x,p)$ under the semiclassical flow $\phi^+$ . Only for states that are purely composed from the branch +, the evolution can be approximated by  $\phi^+$.}
    \label{fig:DO x semicl Schroedingwe cats}
\end{figure}
The calculation of the Berry curvature~\cite{Berry1984} and the Littlejohn-Flynn term follows straightforwardly from \autoref{Berry:curvature} and \autoref{eq:no name term}
\begin{equation}
    \begin{aligned}
        M^\pm&=\frac{\mu}{2(\mu^2+x^2+p^2)}\\
       F^\pm_{px}&=\mp\frac{\mu}{2(\mu^2+x^2+p^2)^\frac{3}{2}}.\\
    \end{aligned}
\end{equation}
 The other components of the Berry curvature are given implicitly by 
 \begin{equation}
     \begin{aligned}
         F_{xp}=&-F_{px}\\
         F_{xx}=&F_{pp}=0.\\
     \end{aligned}
 \end{equation}
 The semiclassical equations of motion,~\autoref{eq:Semiclass eq of motion}, then become the equations of motion of an oscillator
    \begin{equation}
    \begin{aligned}
     \dot{x}&=\Omega^\pm(r) p +\mathcal{O}(\varepsilon^2) \\
       \dot{p}&=-\Omega^\pm(r) x+\mathcal{O}(\varepsilon^2), \\[1ex]
    \end{aligned}
    \end{equation}
with an angular velocity $\Omega^\pm(r)$ that depends on $r=\sqrt{x^2+p^2}$ 
\begin{equation}\label{eq:DO Omega(r)}
    \begin{aligned}
        \Omega^\pm(r)&=\left(\pm\frac{1}{e_0}-\varepsilon\frac{\mu}{e_0^4}\right)\frac{1}{1\mp\varepsilon\frac{\mu}{2e_0^3}}\\&=
      \pm\frac{1}{\sqrt{\mu^2+r^2}} +\varepsilon\frac{ \mu}{2(\mu^2+r^2)^2}+\mathcal{O}(\varepsilon^2).\\
    \end{aligned}
\end{equation}
Neglecting the $\mathcal{O}(\varepsilon^2)$-terms, the semiclassical equations of motion are solved by the semiclassical phase space flow $\phi^\pm$
\begin{equation}
    \begin{aligned}
        x^{\phi^\pm_t}&=r_0\cos(\varphi_0-\Omega^\pm(r_0)t)\\
        p^{\phi^\pm_t}&=r_0\sin(\varphi_0-\Omega^\pm(r_0)t),\\
    \end{aligned}
\end{equation} with initial conditions $r_0=\sqrt{x_0^2  +p_0^2}$ and $\varphi_0=\arctan{\frac{p_0}{x_0}}$: The semiclassical flow $\phi^\pm$ moves along lines of constant $r$ and thus preserves the phase space measure $\mathrm{d}x\mathrm{d}p$. Therefore, the correction to phase space measurement $\mathrm{d}\lambda_\varepsilon=\mathrm{d}x\mathrm{d}p\left(1+\varepsilon F^\pm_{xp}\right)$ simplifies in \autoref{eq:<o>(t) semiclass including phasespacemeasure}, such that quantum dynamics in phase space can be approximated by the semiclassical flow of $w=\operatorname{Tr}[W]$. For an energy eigenstate $\ket{\psi_{n\pm}}$, the trace of the Wigner function  
\begin{equation}
    w_{n\pm}=\operatorname{Tr}[W_{n\pm}]=\frac{2\varepsilon(n+1)w_{n+1}+(\pm E_n-\mu)^2w_n}{2E_n(E_n\mp\mu)}
\end{equation} 
is a sum of Wigner functions of single Fock states 
\begin{equation}
     w_n=\frac{(-1)^n}{\varepsilon\pi}L_n\left(\frac{2r^2}{\varepsilon}\right)
e^{-\frac{r^2}{\varepsilon}}
\end{equation}
which are invariant under $\phi^\pm$. Thus, time invariance of an energy eigenstates is reproduced by $\phi^\pm$ without any error
\begin{equation}
    \braket{\psi_{n\pm}|\hat{o}|\psi_{n\pm}}_\text{semicl}(t)=\iint \mathrm{d} x \mathrm{d} p w_{n\pm}(x^{\phi^\pm_{-t}}, p^{\phi^\pm_{-t}}) o(x, p) = \text{const.}
\end{equation}
For a superposition of states from the same branch $(\ket{\psi_{n\pm}}+\ket{\psi_{m\pm}})/\sqrt{2}$, quantum time evolution can be approximated by the corresponding semiclassical flow $\phi^\pm$ of its Wigner function trace 
\begin{equation}
    \begin{aligned}
        w_{n\pm,m\pm}=\frac{1}{2}\left(w_{n\pm}+w_{m\pm} +
        \frac{2\varepsilon(n+1)(m+1)}{\sqrt{2E_n(E_n\mp\mu){2E_m(E_m\mp\mu)}}}\frac{\sigma[\ket{m+1}\bra{n+1}+\text{h.c}]}{2\pi\varepsilon}\right.\\
        \left.+\frac{(\pm E_n-\mu)(\pm E_m-\mu)}{\sqrt{2E_n(E_n\mp\mu){2E_m(E_m\mp\mu)}}}\frac{\sigma[\ket{m}\bra{n}+\text{h.c}]}{2\pi\varepsilon}\right)
        \end{aligned}
\end{equation}
with 
\begin{equation}\label{eq:symbol ketm bran plus hc}
\begin{aligned}
     \frac{\sigma\left[\ket{m}\bra{n}+\text{h.c.}\right]}{2\pi\varepsilon}=&\\
     \frac{(-1)^n}{\varepsilon \pi}&2^{\frac{m-n}{2}}\sqrt{\dfrac{n!}{m!}} \Re\left[r\left(\frac{\cos{\varphi}+\mathrm{i}\sin{\varphi}}{\sqrt{\varepsilon}}\right)^{m-n}\right] \ L_n^{(m-n)}\left(2\frac{r^2}{\varepsilon}\right)   e^{-\frac{r^2}{\varepsilon}}.
\end{aligned}
\end{equation}
But the same is not possible for a superposition of states from different branches $(\ket{\psi_{n-}}+\ket{\psi_{m+}})/\sqrt{2}$ by the evolution of $w_{n-,m+}$ under $\phi^+$ or $\phi^-$. This is illustrated in \autoref{fig:DO x semicl Schroedingwe cats} by the comparison between 
\begin{equation}
    \braket{\hat{x}}_\text{semicl}(t)=\iint \mathrm{d} x \mathrm{d} p\, w(x^{\phi^\pm_{-t}}, p^{\phi^\pm_{-t}}) x
\end{equation}
and the exact evolution $\braket{\hat{x}}(t)$ for superposition states. It also shows that $\ket{\psi_{-1}}$  belongs to the branch $+$ for $\mu=1$ but does not for $\mu=-1$. Furthermore,  \autoref{fig:DO x semicl Schroedingwe cats} illustrates that there are 2 possible different time scales $\tau$ for the oscillation of superposition states:  Under the condition $\varepsilon\ll|\mu|$, the angular velocity of the oscillation is much faster for a state composed from different branches
\begin{equation}
    \frac{E_{n+}-E_{m-}}{\varepsilon}=\frac{2|\mu|}{\varepsilon} +\mathcal{O}\left(|\mu|\right)
\end{equation}
than for a superposition from the same branch 
\begin{equation}
   \frac{E_{n+}-E_{m+}}{\varepsilon}=\frac{(n-m)}{|\mu|} +\mathcal{O}\left(\left(\frac{\varepsilon}{|\mu|^2}\right)\right)\,.
  \end{equation}
It is this separation between 
\begin{equation}
    \tau_\text{single branch}=|\mu|\gg\tau_\text{mixed branches}=\frac{\varepsilon}{|\mu|}
\end{equation}
 that controls the approximation of the adiabatic projector as $\hat{\pi}^\pm\approx\hat{\pi}^\pm_0+\varepsilon\hat{\pi}^\pm_1$. 
 
\section{Coherent states of Dirac Oscillator}
In experiments, one often does not excite just one or two modes of a system but rather a whole collection of modes eg. a wave packet, see for example~\cite{greinerCollapseRevivalBEC2002,leibfriedTwoIonWavePacket,maternyWavePacketDynamics2000,Feldmann-CoherentDynamacisWavePackets}. In quantum physics, the best known example of a wave packet is the coherent state $\ket{\alpha}$ of the harmonic Oscillator, which is a wave packet centered around the expectation values $\braket{\hat{x}}=x_c$ and $\braket{\hat{p}}=p_c$ with minimal uncertainty $\Delta x=\Delta p=\sqrt{\frac{\varepsilon}{2}}$ (see \autoref{chapter:HO}  and \autoref{chapter:AO}). To gain insights into wave packet propagation of the Dirac Oscillator, we are interested in generalizing the concept of the coherent state $\ket{\alpha}$ of the Harmonic Oscillator to the Dirac Oscillator. There are different ways to define the coherent state, namely as the eigenstate of the annihilation operator $\hat{a}\ket{\alpha}=\alpha\ket{\alpha}$, as the displacement of the ground state by the displacement operator $\hat{D}\ket{0}=\exp \left(\alpha \hat{a}^{\dagger}-\alpha^* \hat{a}\right)\ket{0}=\ket{\alpha}$, and as a minimum uncertainty state, which for the Harmonic Oscillator all give the same state $\ket{\alpha}$~\cite{dodonovSqueezedRewiew2002}. For other quantum systems, this definitions might not coincide~\cite{Ghosh2021QuasicoherentstateDO}.\\ In \cite{NogamiCoherentStateDO}, the authors have studied the evolution of a state of the Dirac Oscillator initially prepared at  
\begin{equation}
    \binom{\ket{\alpha}}{0},
\end{equation}
which can be created by acting with the displacement operator $\hat{D}(\alpha)$ on the spectral flow state 
\begin{equation}
    \hat{D}(\alpha)\ket{\psi_{-1}}=\binom{\exp \left(\alpha \hat{a}^{\dagger}-\alpha^* \hat{a}\right)\ket{0}}{0}=\binom{\ket{\alpha}}{0},
\end{equation} clearly this is an eigenstate of the annihilation operator $\hat{a}\mathbb{1}_{\mathcal{H}_{\mathrm{f}}}$ 
\begin{equation}
    \left(\begin{array}{cc}\hat{a} & 0 \\0 & \hat{a}\end{array}\right) \binom{\ket{\alpha}}{0}=\alpha\binom{\ket{\alpha}}{0}\end{equation}
and it obviously has minimal uncertainty. For these reasons, it seems valid for the authors to  name it `the coherent state of the Dirac Oscillator`. However, one can argue that $\ket{\psi_{-1}}$ cannot be identified with the ground state as the spectrum of the Dirac Oscillator~\ref{eq:DO E_npm}  ranges from $-\infty$ to $\infty$. Furthermore, the authors point out that the eigenequation
\begin{equation}\label{eq:DO apsialpha=alpha}
    \hat{a}\mathbb{1}_{\mathcal{H}_{\mathrm{f}}}\ket{\psi_\alpha}=\alpha \ket{\psi_\alpha}
\end{equation} 
is fulfilled by any state
\begin{equation}\label{eq:DO general coherent state}
    \ket{\psi_\alpha}=\frac{1}{\sqrt{c_\uparrow^2+c_\downarrow^2}}\binom{c_\uparrow\ket{\alpha}}{c_\downarrow\ket{\alpha}},
\end{equation}
which also saturates the Heisenberg inequality.
Therefore, $\binom{\ket{\alpha}}{0}$  is rather  `a coherent state of the Dirac oscillator` among other.

$\binom{\ket{\alpha}}{0}$  is constructed from states of all  branches~\cite{NogamiCoherentStateDO}. As argued in the previous section, it is therefore not possible to approximate the evolution of expectation values for $\binom{\ket{0}}{0}$ with the semiclassical formalism presented in \autoref{chapter:DO}. In~\cite{Ghosh2021QuasicoherentstateDO}, `quasi coherent states` of the Dirac Oscillator were constructed from a single branch as a solution of a generalized annihilation operator. In the context of examining the form and localization of wave packet however, it is natural to choose a minimal uncertainty $\Delta x\Delta p=\varepsilon/2$ as the generalized definition of a coherent state, which is not given for these  `quasi coherent states`. In the following, we will use states of only one branch to  construct wave packets  $\ket{\psi_{\alpha\pm}}$ with almost minimal uncertainty. This states can be approximated in the limit of $\varepsilon\ll1$ as states  $\ket{\psi_\alpha} $ with the form of \autoref{eq:DO general coherent state}. Such single-branch almost coherent states then make a semiclassical analysis of the dynamic of wave packets possible. Furthermore, in the model of a 2D-crystal with varying on site energy, such states are wave packets moving unidirectionally in $y$-direction, what might have interesting experimental applications. \\ 
\section{The single-branch almost coherent state}
\subsection{Constructing $\ket{\psi_{\alpha\pm}}$}

The coherent state of the Quantum Harmonic Oscillator $\ket{\alpha}$ is built by a superposition of Fock states $\ket{n}$ including the ground state $\ket{0}$, as one can see in \autoref{eq:coherent state superposition}. For the Dirac Oscillator on the other side, note that the state $n=0$ of each band 
\begin{equation}
\ket{\psi_{0\pm}}=\frac{1}{\sqrt{2 E_0\left(E_0 \mp \mu\right)}} \cdot\binom{\sqrt{2 \varepsilon} |1\rangle}{\left( \pm E_0-\mu\right)|0\rangle}
\end{equation}
is proportional to $\ket{1}$ for $\uparrow$-polarization. Thus, to construct a state that is for both $\uparrow$-and $\downarrow$-polarization proportional to $\ket{\alpha}$, one needs to include the spectral flow state $\ket{\psi_{-1}}=\binom{\ket{0}}{0}$. As shown in section~\ref{sec:DO Adiabatic subspaces}, the condition to include $\ket{\psi_{-1}}$ into a branch with a corresponding adiabatic subspace is
\begin{equation}\label{eq:DO condition on mu}
    \begin{aligned}
        \text{`branch +`:}\quad & \mu	\gtrsim1\\
        \text{`branch -`:}\quad & \mu	\lesssim-1.\\
    \end{aligned}
\end{equation}

This is also the condition for approximating the almost coherent single branch superposition

\begin{equation}\label{eq:DO-pis-alpha-exact}
\boxed{
\begin{aligned}
\ket{\psi_{\alpha_\pm}} 
   &= \frac{1}{\sqrt{ \tilde{N}}
      }
   \left(
        \sqrt{2 \varepsilon}\ket{\psi_{-1}}
        + \displaystyle\sum_{n=0}^{\infty} 
          \sqrt{\tfrac{2 E_n(E_n \mp \mu)}{n+1}} 
          \tfrac{\alpha^{n+1}}{\sqrt{(n+1)!}} 
          \ket{\psi_{n \pm}}
      \right)
         \\[1.5ex]
   &= \frac{1}{ \mathrm{e}^{-\frac{|\alpha|^2}{2}}\sqrt{\tilde{N}}
      }  \left(\begin{array}{cc}\sqrt{2\varepsilon}\ket{\alpha}  \\ \mathrm{e}^{-\frac{|\alpha|^2}{2}}\sum_{n=0}^\infty\frac{(\pm E_n-\mu)\alpha^{n+1}}{(n+1)\sqrt{n!}}\ket{n}\end{array}\right), \\
\end{aligned}
}
\end{equation}
 where 
 \begin{equation}
     \tilde{N}=2 \varepsilon
        + \displaystyle\sum_{n=0}^\infty 
          \tfrac{2E_n(E_n\pm\mu)|\alpha|^{2(n+1)}}{(n+1)(n+1)!},
\end{equation}
to be proportional to $\ket{\alpha}$ in both pseudospin polarizations $\uparrow$ and $\downarrow$
\begin{equation}\label{eq:DO |psi_alpha> approx}
    \boxed{
    \begin{aligned}
         \ket{\psi_{\alpha+}} =\frac{\ket{\alpha} }{\sqrt{2 \varepsilon + \frac{\varepsilon^2}{|\mu|^2}|\alpha|^2}}\left(\begin{array}{cc}\sqrt{2 \varepsilon} \\+ \frac{\alpha\varepsilon}{|\mu|} \end{array}\right)+\mathcal{O}\left(\frac{\varepsilon^\frac{3}{2}\left(1+|\alpha|^3\right)}{|\mu|^2}\right), \quad \text{for } \mu \gtrsim1 \\[1.5ex]
       \ket{\psi_{\alpha-}} =\frac{\ket{\alpha}}{\sqrt{2 \varepsilon + \frac{\varepsilon^2}{|\mu|^2}|\alpha|^2}}\left(\begin{array}{cc}\sqrt{2 \varepsilon}  \\-\frac{\alpha\varepsilon}{|\mu|}\end{array}\right)+\mathcal{O}\left(\frac{\varepsilon^\frac{3}{2}\left(1+|\alpha|^3\right)\ket{\alpha}}{|\mu|^2}\right), \quad \text{for } \mu \lesssim-1 .
    \end{aligned}
    }
\end{equation}

\begin{proof}
    Firstly, we need the condition we imposed for $\ket{\psi_{-1}}$ ``to belong to the right band``, \autoref{eq:DO condition on mu}, to develop 
 \begin{equation}
     \begin{aligned}
        -\mu\pm E_n =&-\mu \pm \sqrt{\mu^2+2 \varepsilon(n+1)}\\
         =&\pm \frac{\varepsilon(n+1)}{|\mu|}+\mathcal{O}\left(\frac{(\varepsilon(n+1))^2}{|\mu|^3}\right).\\
     \end{aligned}
 \end{equation}
 Now, we can approximate the $\downarrow$-component by 
 \begin{equation}\label{eq:DO develop psi_alpha_down}
 \begin{aligned}
          \mathrm{e}^{-\frac{|\alpha|^2}{2}}\sum_{n=0}^\infty\frac{(\pm E_n-\mu)\alpha^{n+1}}{(n+1)\sqrt{n!}}\ket{n}=&\pm\frac{\alpha\varepsilon}{|\mu|}\mathrm{e}^{-\frac{|\alpha|^2}{2}}\sum_{n=0}^\infty\frac{\alpha^{n}}{\sqrt{n!}}\ket{n}\\&\quad \quad \quad +\mathcal{O}\left(\frac{\alpha\varepsilon^2}{|\mu|^3}\mathrm{e}^{-\frac{|\alpha|^2}{2}}\sum_{n=0}^\infty\frac{\alpha^{n}}{\sqrt{n!}}(n+1)\ket{n}\right)\\[1ex]
          =&\pm\frac{\alpha\varepsilon}{|\mu|}\ket{\alpha}+\mathcal{O}\left(\frac{\alpha\varepsilon^2}{|\mu|^3}\hat{a}\hat{a}^\dagger\ket{\alpha}\right)\\
            =&\pm\frac{\alpha\varepsilon}{|\mu|}\ket{\alpha}+\mathcal{O}\left(\frac{\alpha\varepsilon^2}{|\mu|^3}\sqrt{1+|\alpha|^4}\ket{\alpha}\right)
 \end{aligned}
 \end{equation}
 where in the last line we estimated that for any operator $\hat{o}$ that can be expressed as a finite products of $\hat{a}$ and $\hat{a}^\dagger$ the error of an expectation values
 \begin{equation}
     \braket{\alpha|\hat{a}\hat{a}^\dagger\hat{o}\hat{a}\hat{a}^\dagger|\alpha}=(|\alpha|^4+\mathcal{O}(1))\braket{\alpha|\hat{o}|\alpha}
 \end{equation}
 and thus $\mathcal{O}(\hat{a}\hat{a}^\dagger\ket{\alpha})=\mathcal{O}\left(\sqrt{|\alpha|^4+1}\ket{\alpha}\right)$. In order for the approximated state to be normed, we reexpress the normalization factor as 
 \begin{equation}\label{eq:DO normdevellopment}
     \begin{aligned}
         \frac {1}{\mathrm{e}^{-\frac{|\alpha|^2}{2}}\sqrt{2 \varepsilon+\sum_{n=0}^\infty\frac{2E_n(E_n\pm\mu)|\alpha|^{2(n+1)}}{(n+1)(n+1)!}}}=&\frac{1}{||\left(\begin{array}{cc}\sqrt{2 \varepsilon} \ket{\alpha} \\\left(\pm \frac{\alpha\varepsilon}{|\mu|} + \mathcal{O}\left(\frac{\alpha\varepsilon^2}{|\mu|^3}\sqrt{1+|\alpha|^4}\right)\right)\ket{\alpha}\end{array}\right)||}\\
         =&\frac{1}{\sqrt{2 \varepsilon + \frac{\varepsilon^2}{|\mu|^2}|\alpha|^2}}\left(1+\mathcal{O}\left(\varepsilon^{\frac{3}{2}}\frac{|\alpha|^3+1}{\mu}\right)\right).
     \end{aligned}
 \end{equation}
 Inserting \autoref{eq:DO normdevellopment} and ~\autoref{eq:DO develop psi_alpha_down} into \autoref{eq:DO-pis-alpha-exact} gives \autoref{eq:DO |psi_alpha> approx}. The order of the correction term is estimated in a way that it holds in the limits $|\alpha|\ll1$ and $|\alpha|\gg1$ and $\mu \gg1$.  
\end{proof}
\begin{figure}[H]
    \centering
    \includegraphics[width=0.45\linewidth]{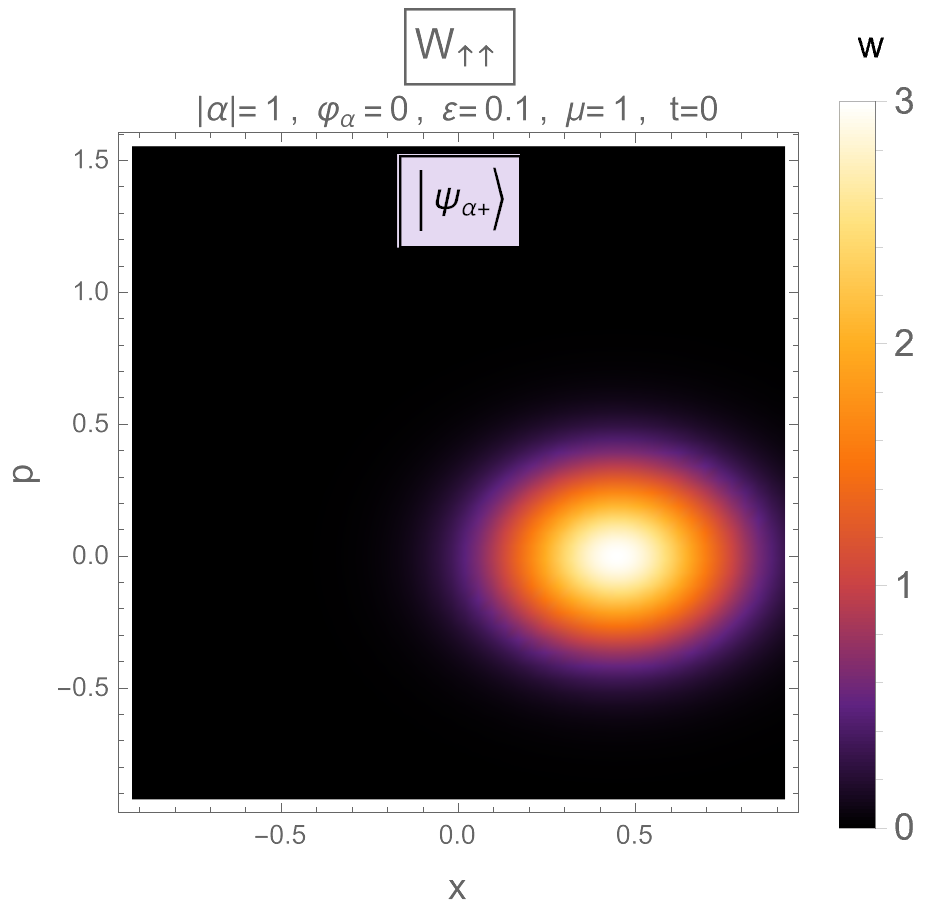}
    \includegraphics[width=0.45\linewidth]{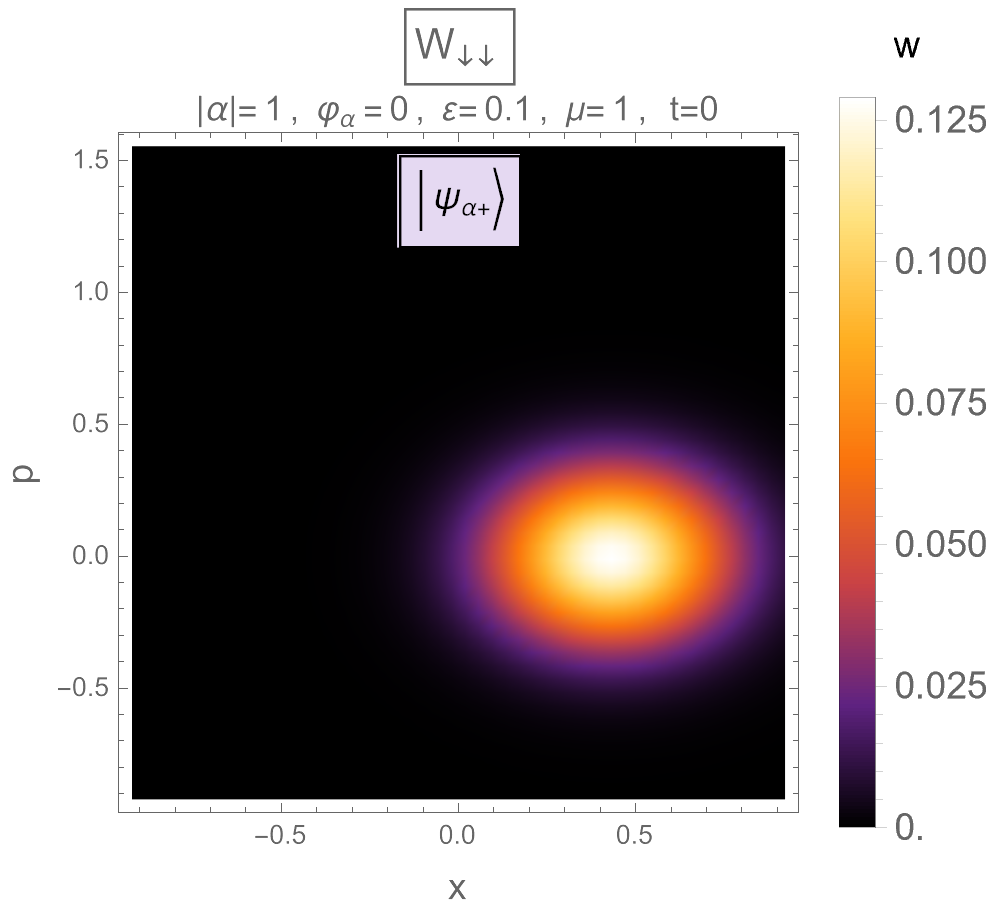}
    \caption{Plot of the diagonal components of the Wigner function of $\ket{\psi_{\alpha+}}$. $W_{\uparrow\uparrow}$ is exactly given by $w_\alpha (x,p)$ modulo a normalization factor. For the chosen parameters, $W_{\downarrow\downarrow}$ has a form that is very close to $w_\alpha(x,p)$. Both plots were calculated for a numerical truncation of $\ket{\psi_{\alpha+}}$  at $n_{\text{max}}=32$  in~\autoref{eq:DO-pis-alpha-exact}}
    \label{fig:DO W alpha+ exact}
\end{figure}
It might look like we are neglecting terms lower than $\mathcal{O}(\varepsilon^2)$. However, when calculating the expectation values of a operator $\hat{o}$, only the diagonal terms of the density matrix  come into play
\begin{equation}
    \boxed{\text{Tr}_{\mathcal{H}_f}[\ket{\psi_{\alpha\pm}}\bra{\psi_{\alpha_\pm}}]=\ket{\alpha}\bra{\alpha}+\mathcal{O}\left(\frac{\varepsilon^2|\alpha|^2\sqrt{1+2|\alpha|^2+|\alpha|^4}}{|\mu|^4}\right),}
\end{equation}

which is exact up to order $\mathcal{O}(\varepsilon^2)$ if $|\alpha|\sim1$ and $|\mu|\gtrsim 1$, which will be the assumed for the rest of this chapter. 
Note that for $\alpha \rightarrow0$, $\ket{\psi_{\alpha_\pm}}\rightarrow\ket{\psi_{-1}}$, showing that there is still a link between $\ket{\psi_{-1}}$ and $\ket{\psi_{\alpha\pm}}$.  To illustrate that $\ket{\psi_{\alpha+}}$ is indeed very close to a coherent state for $\mu=1$, the diagonal components of its Wigner function $W_{\alpha+}$ are shown in~\autoref{fig:DO W alpha+ exact} .  The $\uparrow$-polarization (on the left side) is exactly proportional to $\ket{\alpha}$ and therefore
\begin{equation}
    W_{\uparrow\uparrow}\propto w_\alpha.
\end{equation}
The $\downarrow$-polarization (on the right side) is not exactly proportional to $\ket{\alpha}$
\begin{equation}
\begin{aligned}
        W_{\downarrow\downarrow}\propto\,|\alpha|^2&
\sum_{n=0}^{\infty}
\left[
  \frac{(E_n-\mu)^2|\alpha|^{2n}}{(n+1)^2n!}\, w_{n}(x,p)\right.\\
 \quad &\left.
  + \sum_{m=n+1}^{\infty}
    \frac{(E_n-\mu)(E_m-\mu)|\alpha|^{n+m}}{(m+1)(n+1)\sqrt{n!\,m!}}\,\frac{ \sigma_\varepsilon\left[\mathrm{e}^{\mathrm{i}\varphi_{\alpha}(m-n)}\ket{m}\bra{n}+\text{h.c.}\right]}{2\pi\varepsilon}
\right],
\end{aligned}
\end{equation}
 but still one can see that the shape of  $W_{\downarrow\downarrow}$ closely resembles that of $w_\alpha$ for $\varepsilon=0.1$.

\begin{figure}
    \centering
    \begin{subfigure}{0.48\linewidth}
        \includegraphics[width=0.99\linewidth]{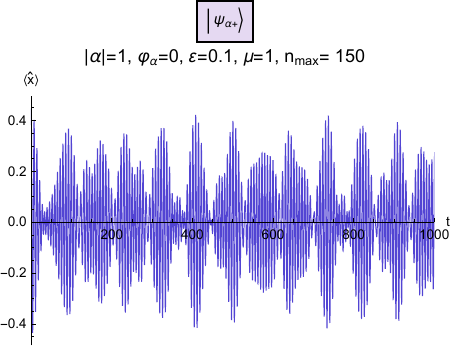}
        \includegraphics[width=0.99\linewidth]{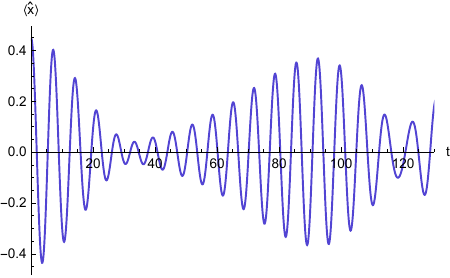}
    \end{subfigure}
    \begin{subfigure}{0.48\linewidth}
        \includegraphics[width=0.99\linewidth]{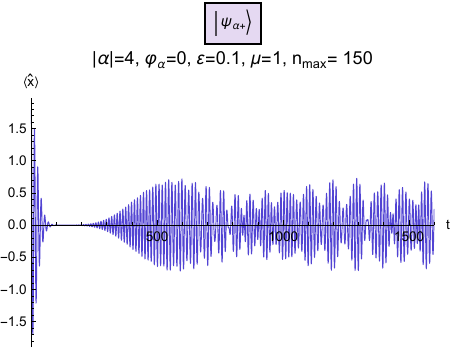}
        \includegraphics[width=0.99\linewidth]{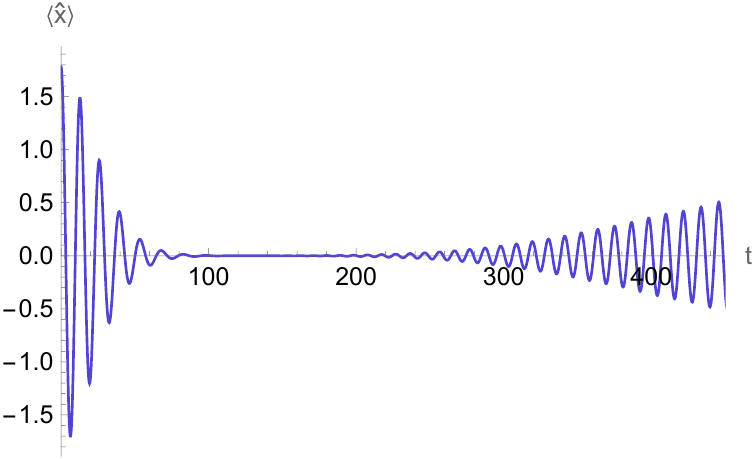}
    \end{subfigure}
\includegraphics[height=0.7cm]{graphics/Legends/plotlegend-exact.pdf}
    
    \caption{Quantum evolution of $\braket{\hat{x}}$ for a initial state $\ket{\psi_{\alpha_+}}$, calculated via~\autoref{eq:DO <x> <p>}   with truncation at term $n_\text{max}=150$. The oscillation amplitude  of $\braket{\hat{x}}(t)$ initially decays and increases again later, but never reaches its initial value again.  For $\alpha=1$, the amplitude take values after the decay much closer to its initial value than for $\alpha=4$. The second row is a zoom out of the first row.}
    \label{fig:DO <x> large times}
\end{figure}

\subsection{Quantum time evolution of $\ket{\psi_{\alpha\pm}}$}
\begin{figure}
    \centering
        \includegraphics[width=0.48\linewidth]{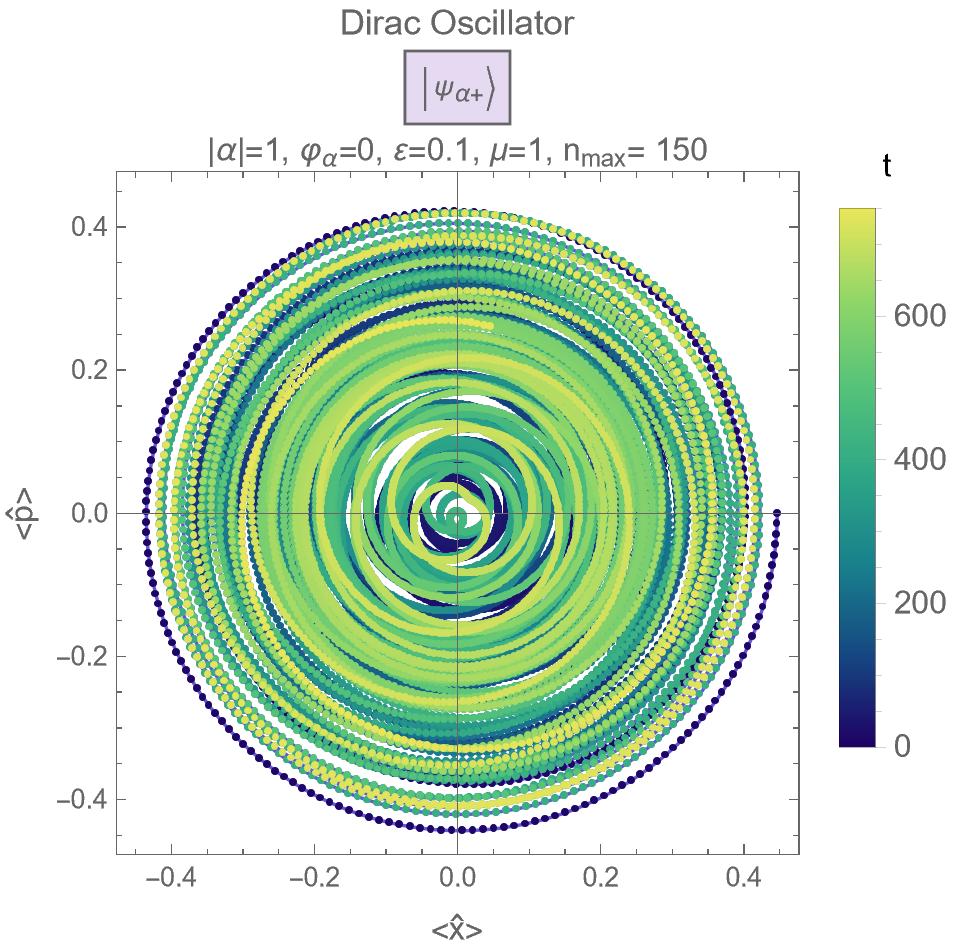}
        \includegraphics[width=0.48\linewidth]{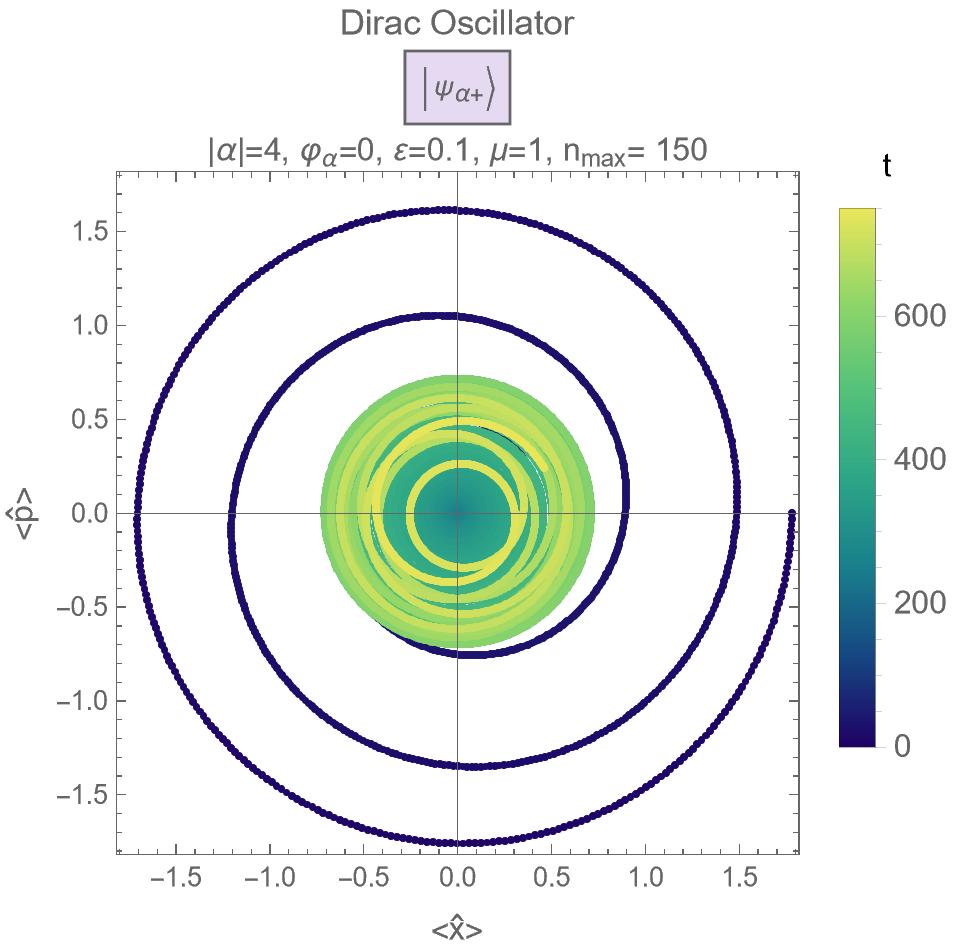}
    \caption{Quantum evolution of $\braket{\hat{x}}$ and $\braket{\hat{p}}$ for a initial state $\ket{\psi_{\alpha_+}}$. Almost the whole close neighborhood around the origin $\braket{\hat{x}}=\braket{\hat{p}}=0$ is reached by the dynamics.}
    \label{fig:DO <x>-<p>}
\end{figure}
\begin{figure}[h]
    \centering
    \includegraphics[height=5.8cm]{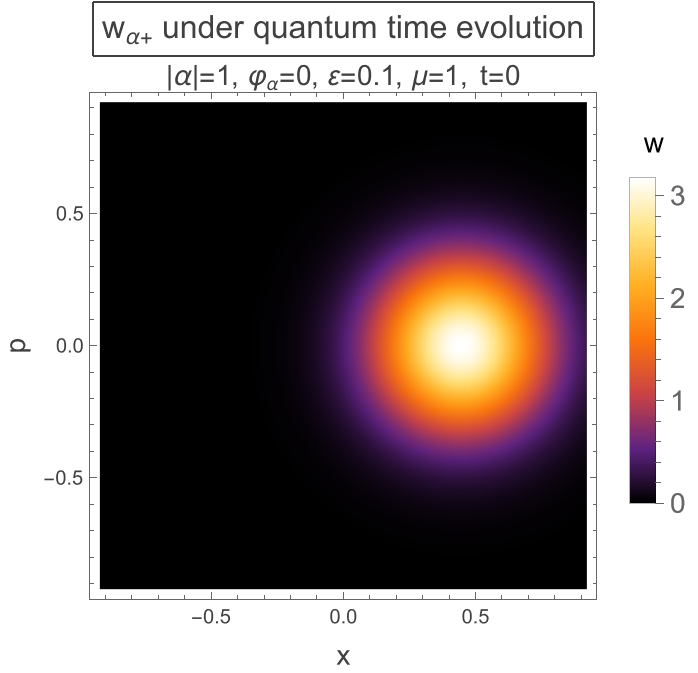}
    \includegraphics[height=5.8cm]{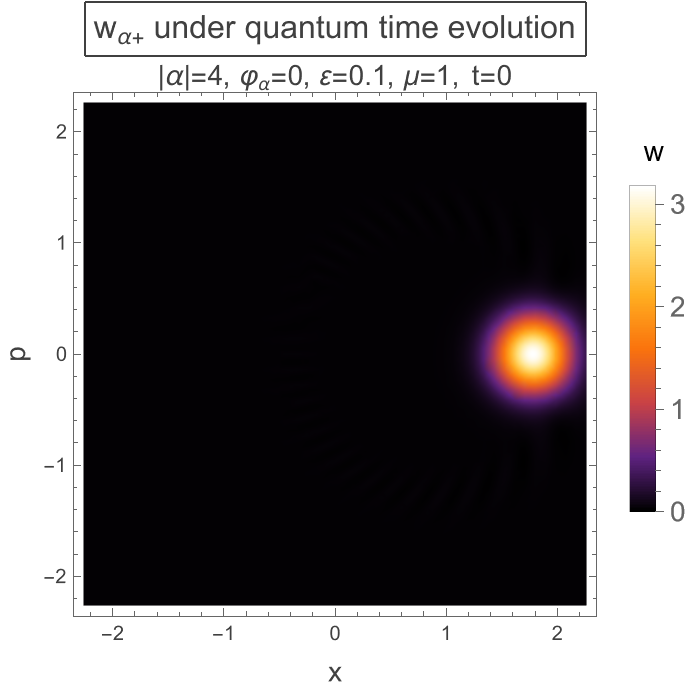}
    \includegraphics[height=5.2cm]{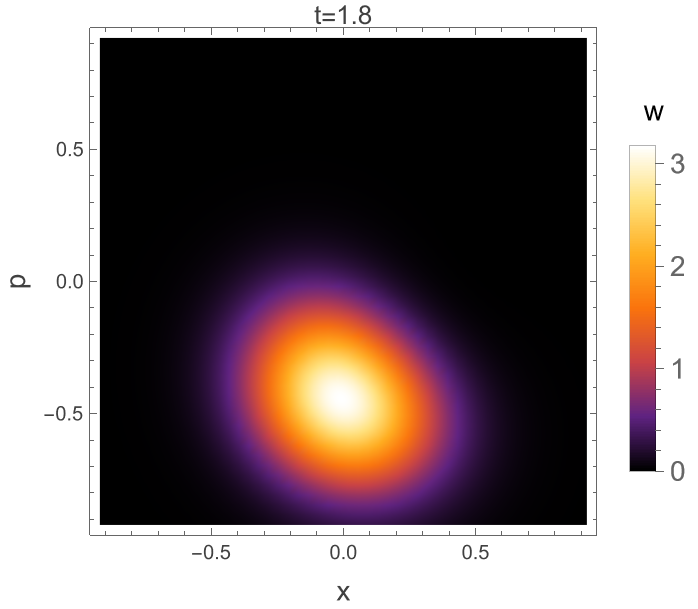}
       \includegraphics[height=5.2cm]{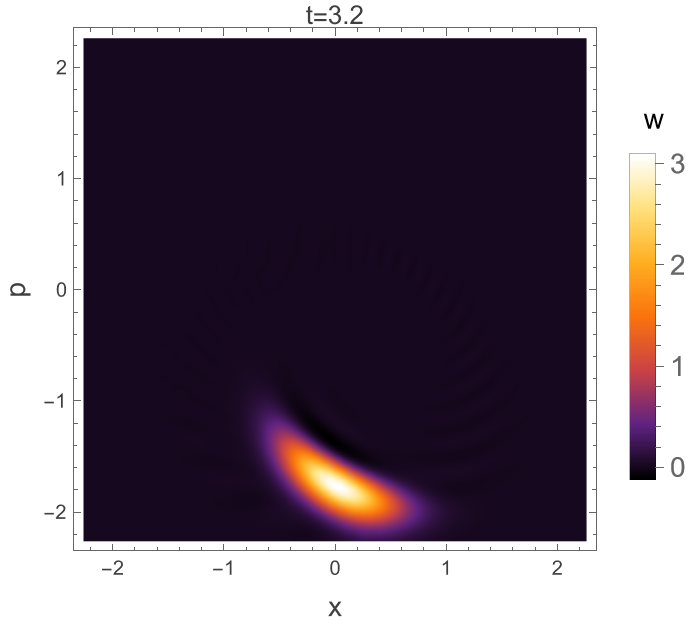}
    \includegraphics[height=5.2cm]{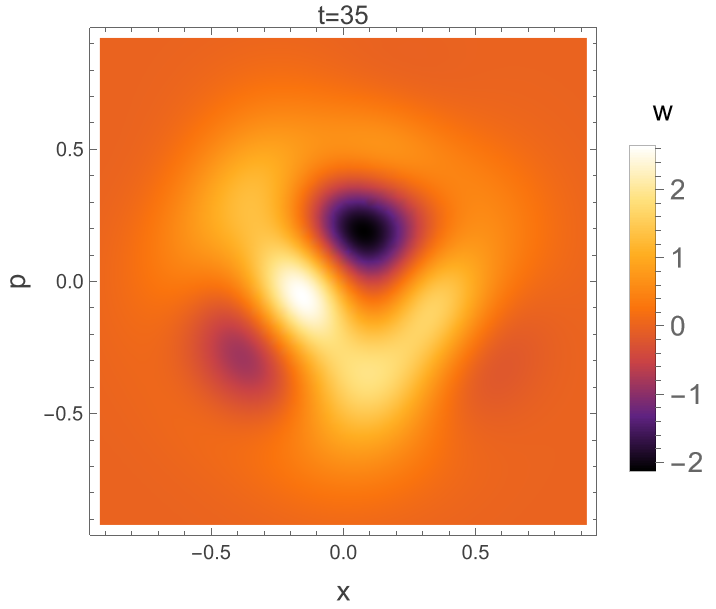}
    \includegraphics[height=5.2cm]{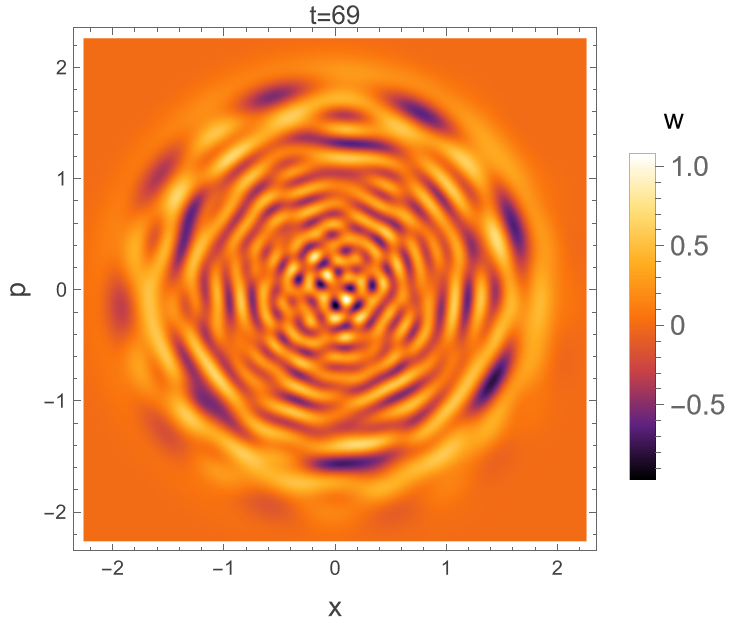}
    \includegraphics[height=5.2cm]{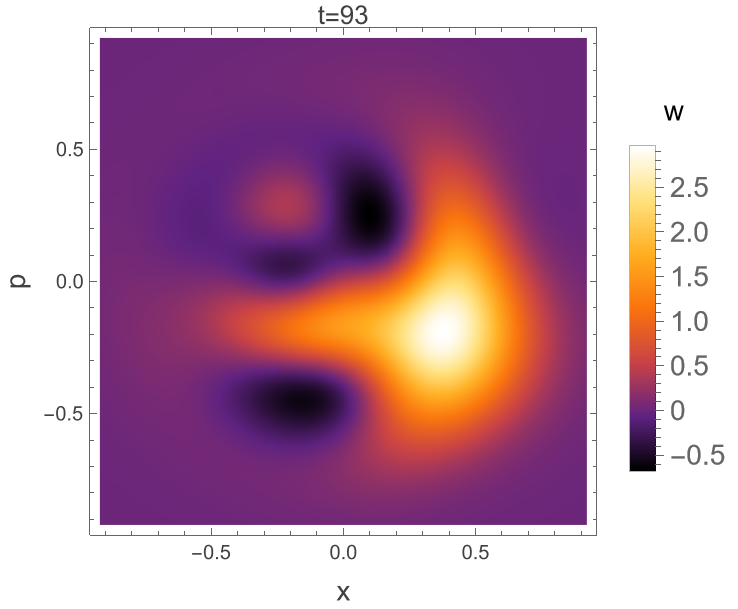}
       \includegraphics[height=5.2cm]{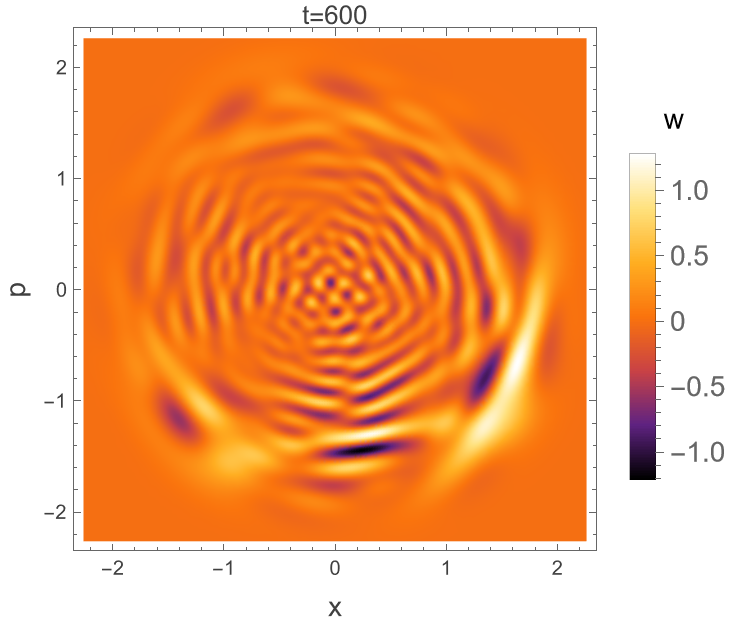}
    \caption{Plots of $w_{\alpha +}=\operatorname{Tr}[W_{\alpha+}]$ under quantum time evolution for $|\alpha|=1$ (left) and $|\alpha|=4$ (right), calculated for a $\ket{\psi_{\alpha +}}$ truncated at $n_\text{max}=32$ in~\autoref{eq:DO-pis-alpha-exact}. $w_{\alpha+}$ delocalizes under quantum time evolution (2nd and 3rd row). The initial form is never restored, but $w_{\alpha+}$ can become relatively well localized again (4th row). }
    \label{fig:DO PlotWig(t) exact}
\end{figure}
The quantum time evolution of a state initially prepared as the single-branch almost coherent state $\ket{\psi_{\alpha+}}$ is given by 
\begin{equation}\label{eq:DO psi_alpha(t)}
  \ket{\psi_{\alpha+}}(t)=  \frac{1}{ \sqrt{\tilde{N}}
      } \sum_{n=0}^\infty\frac{\alpha^{n}}{\sqrt{n!}} \left(\begin{array}{cc}\sqrt{2\varepsilon}\,\mathrm{e}^{\mathrm{-i}\frac{E_{n-1}}{\varepsilon}t}\ket{n}  \\ \frac{(\pm E_n-\mu)\alpha}{(n+1)}\,\mathrm{e}^{\mathrm{-i}\frac{E_{n}}{\varepsilon}t}\ket{n}\end{array}\right) .
\end{equation}
In the Dirac Oscillator revival of initial state does not accrue, as the eigenenergies $E_n=\sqrt{\mu^2+2\varepsilon(n+1)}$ are different irrational numbers~\cite{NogamiCoherentStateDO}, what makes it impossible for all terms of the sum to be in phase at any time other than $t=0$.  
As a consequence the evolution of the wave packet centers of $\ket{\psi_{\alpha+}}$
 
          
\begin{equation}\label{eq:DO <x> <p>}
\begin{aligned}
\braket{\hat{x}}
&= \frac{\sqrt{2\varepsilon}\,|\alpha|}{\tilde{N}}
  \Biggl(
    2\varepsilon \sum_{n=0}^{\infty} \frac{|\alpha|^{2n}}{n!}\,
      \cos\!\bigl(\varphi_{n,n-1,\alpha}(t)\bigr)\\
&\qquad\qquad\qquad
    + |\alpha|^2 \sum_{n=0}^{\infty}
      \frac{(\pm E_{n+1}-\mu)(\pm E_n-\mu)}{(n+2)(n+1)}
      \frac{|\alpha|^{2n}}{n!}\,
      \cos\!\bigl(\varphi_{n+1,n,\alpha}(t)\bigr)
  \Biggr),\\[2ex]
\braket{\hat{p}}
&= \frac{\sqrt{2\varepsilon}\,|\alpha|}{\tilde{N}}
  \Biggl(
    2\varepsilon \sum_{n=0}^{\infty} \frac{|\alpha|^{2n}}{n!}\,
      \sin\!\bigl(\varphi_{n,n-1,\alpha}(t)\bigr)\\
&\qquad\qquad\qquad
    + |\alpha|^2 \sum_{n=0}^{\infty}
      \frac{(\pm E_{n+1}-\mu)(\pm E_n-\mu)}{(n+2)(n+1)}
      \frac{|\alpha|^{2n}}{n!}\,
      \sin\!\bigl(\varphi_{n+1,n,\alpha}(t)\bigr)
  \Biggr).
\end{aligned}
\end{equation}

with
\begin{equation}
    \begin{aligned}
        \varphi_{m,n,\alpha}(t)=&(m-n)\varphi_\alpha-\frac{E_m-E_n}{\varepsilon}t\,,\\
    \end{aligned}
\end{equation}
is an aperiodic signal. The aperiodicity can be observed in \autoref{fig:DO <x> large times} by variations of the oscillation amplitude of $\braket{\hat{x}}(t)$.   Additionally, \autoref{fig:DO <x>-<p>} shows that in the $(\braket{\hat{x}},\braket{\hat{p}})$-phase space, the dynamics reach almost the entire close neighborhood of the origin, another manifestation of the aperiodicity. Despite this aperiodicity, there are similarities to the dynamics  of the Anharmonic Oscillator presented in~ \autoref{chapter:AO}: $\braket{\hat{x}}(t)$ and $\braket{\hat{p}}(t)$ are modulated oscillations that decay  after $t=0$, eventually remain at $\braket{\hat{x}}\approx\ \braket{\hat{p}}\approx0$ and grow again. But other than for the Anharmonic Oscillator, the maximal oscillation amplitude is not reached again. \autoref{fig:DO PlotWig(t) exact} visualizes these characteristics of the dynamics via $w_{\alpha+}(t)=\operatorname{Tr}[W_{\alpha+}(t)]$,  trace of the Wigner function under quantum time evolution: $w_{\alpha+}(t)$  gets pulled into a banana shaped form corresponding to the decay of the oscillation amplitude (2nd row),  becomes largely spread around the origin  (3rd row) corresponding to $\braket{\hat{x}}\approx\ \braket{\hat{p}}\approx0$ and later partially localized again (4th row) corresponding to a growth of the oscillation amplitude.  The partial localization occurs when the most important terms of~\autoref{eq:DO psi_alpha(t)} are approximately in phase. For higher values of $|\alpha|$,  more terms in \autoref{eq:DO psi_alpha(t)}  become important, which makes it less probable for terms to be in phase: For $|\alpha|=1$, the oscillation amplitude becomes much closer to its initial value than for $|\alpha|=4$, as can be seen in \autoref{fig:DO <x> large times} and \autoref{fig:DO <x>-<p>}. \\
 The angular velocity of the oscillations can be estimated from~\ref{eq:DO <x> <p>}
  by the average $(E_n-E_{n-1})/\varepsilon$ of the most important terms. By construction, the typical time of the oscillations is $\tau_\text{single branch}=|\mu|$. In contrast to $\ket{\psi_{\alpha+}}$, the state $\binom{\ket{0}}{0}$ is built up with states of all branches leading to additional oscillations at a time scale~$\tau_\text{mixed branches}=\varepsilon/|\mu|$, which the authors of~\cite{NogamiCoherentStateDO}  called a `Zitterbewegung`.\\
  Writing~\ref{eq:DO <x> <p>} in a compact form is, to our knowledge, not possible. Therefore, it is rather difficult to have a global understanding on how the signals of $\braket{\hat{x}}(t)$ and  $\braket{\hat{p}}(t)$ depend on $\alpha$ and $\varepsilon$. For instance, if one wants to know \say{Given $\mu,\alpha,\varepsilon$, how many oscillations of $\braket{\hat{x}}(t)$ does it takes until the amplitude has decay to $\approx0$?}, all one can do is to plot $\braket{\hat{x}}(t)$ and `count by hand``. Later in \autoref{sec:DO Toscill taudeay}, we will use the methods developed in \autoref{sec:AO Toscill and taudecay} to predict $T_\text{oscill}$, the period time, and $\tau_\text{dec}$, the typical decay time, via the semiclassical evolution of the Wigner function.

\subsection{Semiclassical evolution of wave packet centers $\braket{\hat{x}}$, $\braket{\hat{p}}$}\label{sec:Semicl <x> <p>}
Firstly, we  investigate how the quality of the semiclassical prediction  depends on $\varepsilon$ and $|\alpha|$ by comparing the semiclassical prediction for the evolution of $\ket{\psi_{\alpha+}}$ with the exact evolution. The scalar Wigner function of $\ket{\psi_{\alpha_+}}$ can be approximated by the Wigner function of $\ket{\alpha}$,  given that $\mu\gtrsim1$, 
\begin{equation}\label{eq:DO Wig |alphapm>}
    \begin{aligned}
        w_{\alpha_+}(x,p)=&w_\alpha(x,p)+\mathcal{O}\left(\varepsilon^2\right),
    \end{aligned}
\end{equation}
where in slow dimensionless coordinates 
\begin{equation}\label{eq:DO Wig_alpha}
    \begin{aligned}
        w_\alpha(x,p)=&\frac{1}{\pi\varepsilon}\mathrm{e}^{-\frac{(x_\alpha-x)^2}{\varepsilon}}\mathrm{e}^{-\frac{(p_\alpha-p)^2}{\varepsilon}},
    \end{aligned}
\end{equation}
 with $x_\alpha=r_\alpha\cos{\varphi_\alpha}$ and  $p_\alpha=r_\alpha\sin{\varphi_\alpha}$, where 
 \begin{equation}
     r_\alpha=\sqrt{2\varepsilon}|\alpha|.
 \end{equation}
 This makes it possible to approximate an expectation value $\braket{\psi_{\alpha+}|\hat{o}|\psi_{\alpha +}}(t)$ up to an error $\mathcal{O}(\varepsilon^2)$  by 
\begin{equation}\label{eq:DO o(t) semicl}
    \braket{\hat{o}}_\text{semicl}(t)=\int\int\mathrm{d}x\mathrm{d}po(x,p)w_\alpha(x^{\phi^+_{-t}},p^{\phi^+_{-t}})
\end{equation}

\begin{figure}
    \centering
       \includegraphics[width=0.3\linewidth]{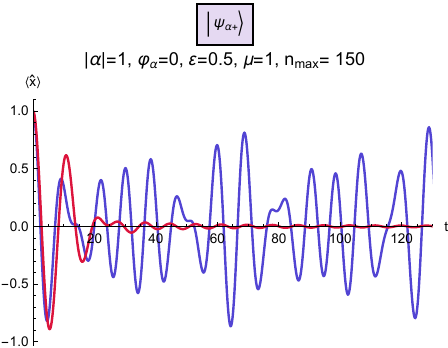}
        \includegraphics[width=0.3\linewidth]{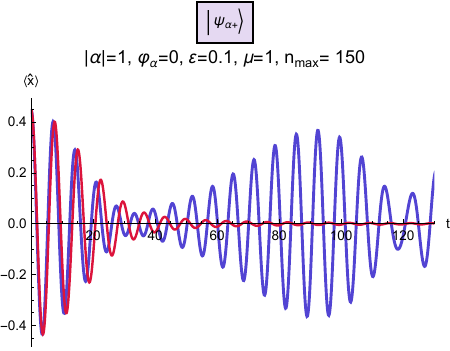}
        \includegraphics[width=0.3\linewidth]{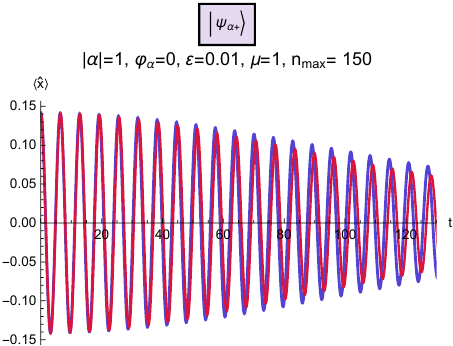}
        \includegraphics[width=0.3\linewidth]{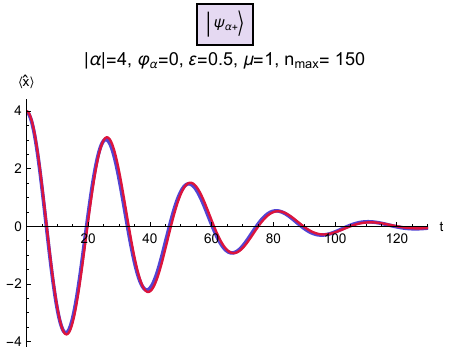}
        \includegraphics[width=0.3\linewidth]{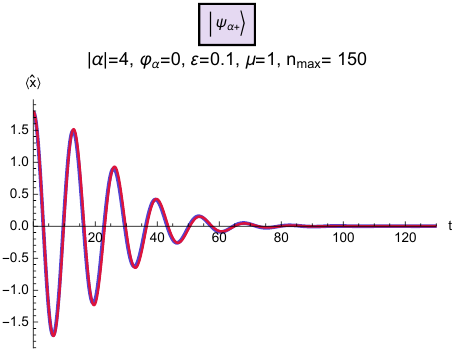}
        \includegraphics[width=0.3\linewidth]{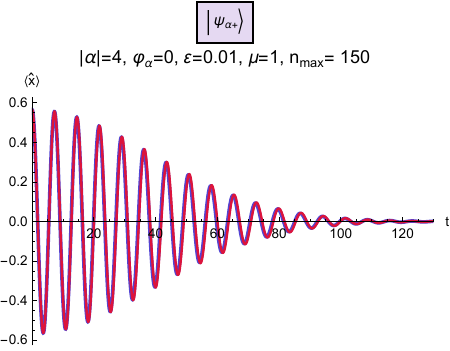}
        \includegraphics[height=0.7cm]{graphics/Legends/plotlegend-exact.pdf}
        \includegraphics[height=0.7cm]{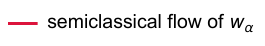}
    \caption{Comparison of the quantum time evolution of $\braket{\hat{x}}$ with $\braket{\hat{x}}_\text{semicl}(t)$. For $|\alpha|$ fixed, the semiclassical prediction becomes better by decreasing  $\varepsilon$ from $0.5$  to $0.01$ (left to right),  but varying $|\alpha|$  from $1$ to $4$ (top row and bottom row) has an even stronger effect on the quality of the semiclassical predictions.}
    \label{fig:DO x semiclflow}
\end{figure}
\begin{figure}
    \centering
        \includegraphics[width=0.4\linewidth]{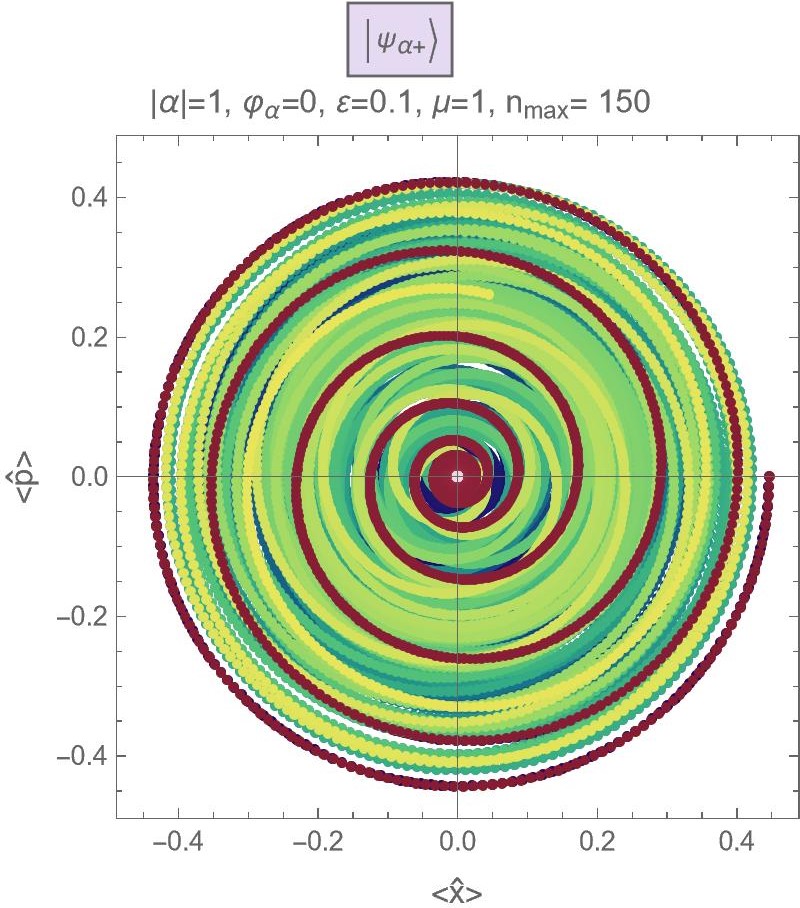}
        \includegraphics[width=0.4\linewidth]{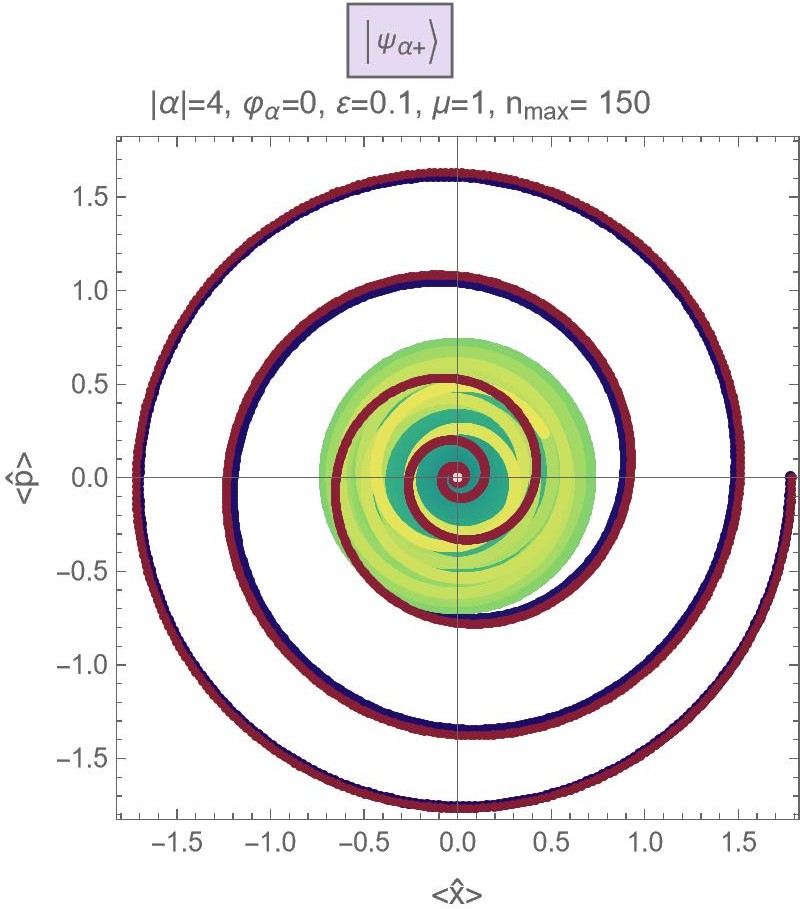}
         \includegraphics[height=1.5cm]{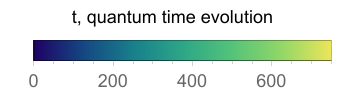}
          \includegraphics[height=1.5cm]{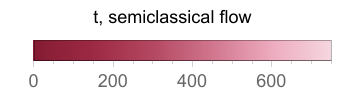}
    \caption{Comparison of the quantum time evolution of $\braket{\hat{x}}$ and $\braket{\hat{p}}$ with $\braket{\hat{x}}_\text{semicl}(t)$  and $\braket{\hat{p}}_\text{semicl}(t)$ . Once $\braket{\hat{x}}\approx\braket{\hat{p}}\approx0$  is reached under semiclassical flow the semiclassical dynamics become  become stationary.}
    \label{fig:DO <x>-<p> semicl}
\end{figure}
\autoref{fig:DO x semiclflow} compares the evolution of $\braket{\hat{x}}_\text{semicl}(t)$ with the exact quantum time evolution of $\braket{\hat{x}}$. As expected  $\braket{\hat{x}}_\text{semicl}(t)$ is close to $\braket{\hat{x}}$ for a longer time for smaller $\varepsilon$. Furthermore, one can see that the quality of the semiclassical approximation is much better for $\alpha=4$ than $\alpha=1$. This is not surprising: following the same arguments as in~\autoref{sec:AO semiclass eq of motion} mainly the phase space region where $r\sim\sqrt{2\varepsilon}|\alpha|$  contributes to the dynamics of $w_{\alpha+} (t)$: Therefore, the effective gap becomes $\Delta e_0^\prime=2\sqrt{\mu^2+2\varepsilon|\alpha|^2}$ instead of $\Delta e_0=2|\mu|$ leading to a smaller effective control parameter $\varepsilon^\prime<\varepsilon$ as one can see from ~\autoref{eq:DO varepsilon mu}. In \autoref{fig:DO <x>-<p> semicl}, one can observe that the semiclassical flow is incapable of reproducing the growth of the oscillation amplitude after the dynamics have reached $\braket{\hat{x}}\approx\braket{\hat{p}}\approx0$.
\FloatBarrier
\subsection{Importance of the Littlejohn-Flynn term and the Berry curvature}
\begin{figure}

  a)
  \begin{subfigure}{0.99\linewidth}
  \centering
        \includegraphics[width=0.45\linewidth]{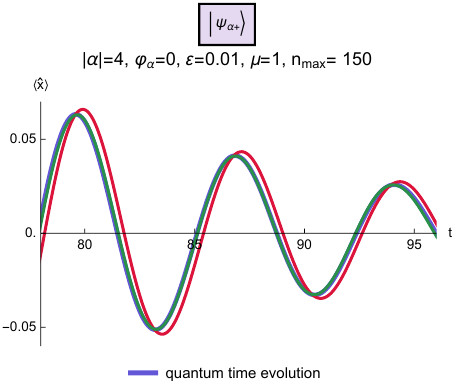}
         \includegraphics[width=0.45\linewidth]{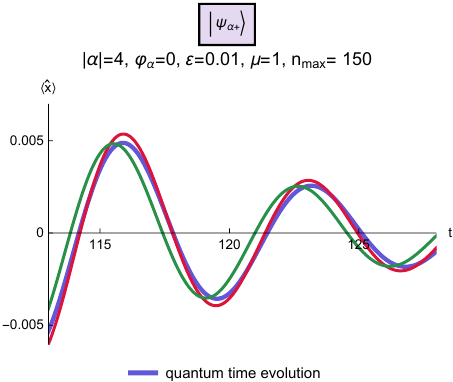}
  \end{subfigure}
  b)
  \begin{subfigure}{0.99\linewidth}
  \centering
        \includegraphics[width=0.45\linewidth]{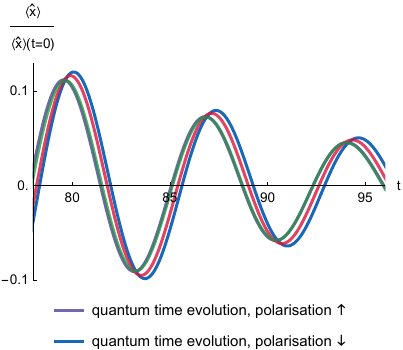}
        \includegraphics[width=0.45\linewidth]{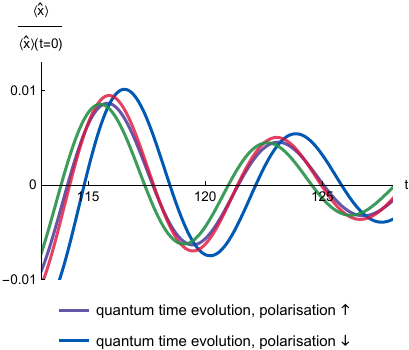} 
         \includegraphics[height=0.7cm]{graphics/Legends/linelegendwalpha-semicl.pdf}
         \includegraphics[height=0.7cm]{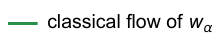}
  \end{subfigure}
  
c)
   \begin{subfigure}{\linewidth}
   \centering
\includegraphics[width=0.35\linewidth]{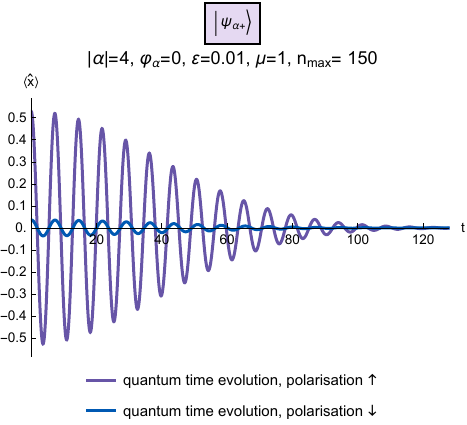}
  \end{subfigure}

\caption{a) Comparison of  $\braket{\hat{x}}_\text{class}(t)$ and $\braket{\hat{x}}_\text{semicl}(t)$ with the quantum time evolution of $\braket{\hat{x}}$ for $\varepsilon=0.01$ and $\alpha=4$. At smaller times $\braket{\hat{x}}_\text{class}(t)$ is closer to quantum evolution and only for larger times $\braket{\hat{x}}_\text{semicl}(t)$ becomes the better prediction.   b) comparison with the normalized contributions  of the $\uparrow$-polarization and $\downarrow$-polarization to $\braket{\hat{x}}(t)$.  $\braket{\hat{x}}_\text{class}(t)$ is only close to the contribution of  $\uparrow$, while the graph of $\braket{\hat{x}}_\text{semicl}(t)$ lies between graphs of both polarizations. c) Plot of the  absolute contributions  of the $\uparrow$-polarization and $\downarrow$-polarization under quantum time evolution. The $\uparrow$-polarization is much more important for the evolution of $\braket{\hat{x}}(t)$.}
    \label{fig:DO x semiclclflow}
\end{figure}
By neglecting all $\mathcal{O}(\varepsilon)$-terms, the semiclassical equations of motion reduce to 
\begin{equation}
    \begin{aligned}
     \dot{x}&=\frac{\partial e_0^{\pm}}{\partial p}+\mathcal{O}(\varepsilon) \\
     \dot{p}&=-\frac{\partial e_0^{\pm}}{\partial x}+\mathcal{O}(\varepsilon),  \\
    \end{aligned}
    \end{equation}
which is solved by the classical phase space flow $\phi_\text{class}^\pm$
\begin{equation}
    \begin{aligned}
        x^{\phi_{\text{class} t}^\pm}&=r_0\cos(\varphi_0-\Omega_\text{class}^\pm(r_0)t)\\
        p^{\phi_{\text{class} t}^\pm}&=r_0\sin(\varphi_0-\Omega_\text{class}^\pm(r_0)t)
    \end{aligned}
\end{equation}
with 
\begin{equation}
      \Omega_\text{class}^\pm=\pm\frac{1}{\sqrt{\mu^2+r^2}}.
\end{equation}
In the following, we study the evolution of $\braket{\hat{x}}$ calculated with the classical flow $\phi_{\text{class}}^\pm$ instead of the semiclassical flow in~\autoref{eq:DO o(t) semicl} to investigate how (not) including the first order corrections in the equations of motion effects the quality of the prediction. For the Dirac Oscillator, the Moyal term is $0$. The  first order corrections are thus only the Littlejohn-Flynn term (the scalarisation term) and the Berry curvature (projection term). Surprisingly, for the parameters $\alpha=4$ and $\varepsilon=0.01$, it seems like including the $\mathcal{O}(\varepsilon)$-corrections gives a worse prediction: one can see in the left plot of \autoref{fig:DO x semiclclflow}a) that the prediction by the classical flow is closer to the exact values than by the semiclassical flow. For later times (plot in the right), the semiclassical flow is better in phase than the classical flow, but still overestimates the amplitude. To gain a deeper understanding, we look at the evolution of the pseudospin polarizations of  
\begin{equation}
  \ket{\psi_{\alpha+}}= \ket{\psi_{\alpha+,\downarrow}}+ \ket{\psi_{\alpha+,\uparrow}}
\end{equation}
separately. The contribution of each to $\braket{\hat{x}}=\braket{\psi_{\alpha+,\uparrow}|\hat{x}|\psi_{\alpha+,\uparrow}}+\braket{\psi_{\alpha+,\downarrow}|\hat{x}|\psi_{\alpha+,\downarrow}}$ are shown in \autoref{fig:DO x semiclclflow}b. Now, note that for the semiclassical calculation the evolution of both polarizations depends on $ w_\alpha(x^{\phi_{-t}^+},p^{\phi_{-t}^+})$. Under quantum time evolution, on the contrary, the 2 polarizations do not evolve in the same way: a Fock-state $\ket{n}$ acquires a dynamical phase $\mathrm{e}^{-\mathrm{i}E_{n-1}t/\varepsilon}$  in the $\uparrow$-polarization but $\mathrm{e}^{-\mathrm{i}E_{n}t/\varepsilon}$ in $\downarrow$-polarization. It is impossible to predict correctly the evolution of both $\ket{\psi_{\alpha+,\uparrow}}$ and $\ket{\psi_{\alpha+,\downarrow}}$ by the evolution of $w_\alpha(x^{\phi_{-t}^+},p^{\phi_{-t}^+})$. The best that can be done is a compromise. In \autoref{fig:DO x semiclclflow}~c) one can see that the semiclassical flow predicts $\braket{\hat{x}}$ in between the normalized contributions of each polarization. The prediction by classical flow for larger times, on the other hand, lies clearly outside the curves of contributions of both polarizations. Only because the classical prediction is closer to the contribution of the $\uparrow$-polarization, which is ``by chance`` the most important when calculating $\braket{\psi_{\alpha+}|\hat{x}|\psi_{\alpha+}}$, its prediction of $\braket{\psi_{\alpha+}|\hat{x}|\psi_{\alpha+}}$ happens to be better. For a state where each component is equally important, such as a superposition of eigenstates $\ket{\psi_{n+}}$ with large $n$, the opposite would be the case. 

\FloatBarrier
\subsection{Prediction the of  oscillation period $T_\text{oscill}$ and the typical decay time $\tau_\text{dec}$ of wave packet oscillations}~\label{sec:DO Toscill taudeay}

\begin{figure}
    \centering
    \includegraphics[width=0.48\linewidth]{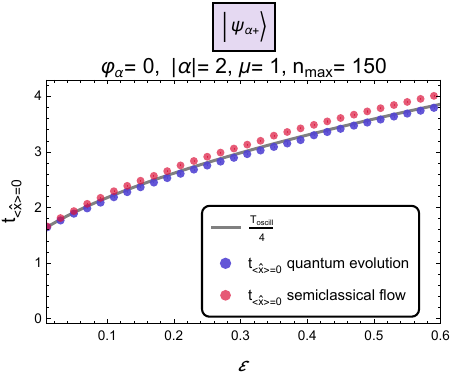}
       \includegraphics[width=0.48\linewidth]{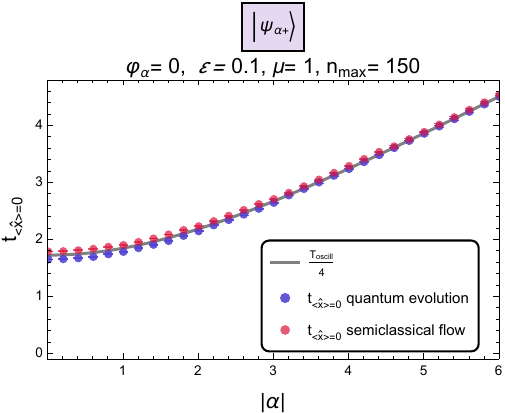}
    \caption{Comparison of $T_\text{oscill}/4=2/\pi\braket{\Omega^+}$ with $t_{\braket{\hat{x}}=0}$, the time of first $\braket{\hat{x}}=0$, under quantum time evolution (blue) and under semiclassical flow $\phi^+$ of $w_\alpha$ (red) for an initial state $\ket{\psi_{\alpha+}}$ with fixed $\varphi_\alpha=0$ and Left: fixed $|\alpha|=2$, Right: fixed $\varepsilon=0.1$. For $T_\text{oscill}/4$ gives a rather good prediction of $t_{\braket{\hat{x}}=0}$. The values of  $t_{\braket{\hat{x}}=0}$ have been found iteratively.}
    \label{fig:DO teq0 Toscill/4}
\end{figure}

\begin{figure}[h]
    \centering
    \includegraphics[width=0.48\linewidth]{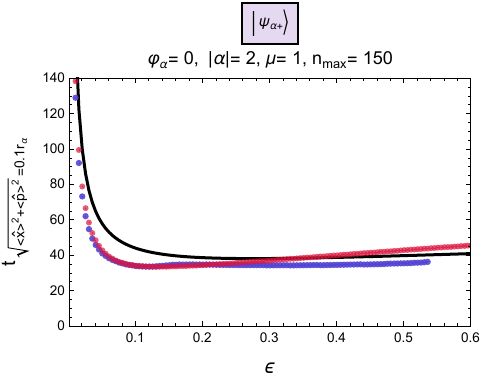}
       \includegraphics[width=0.48\linewidth]{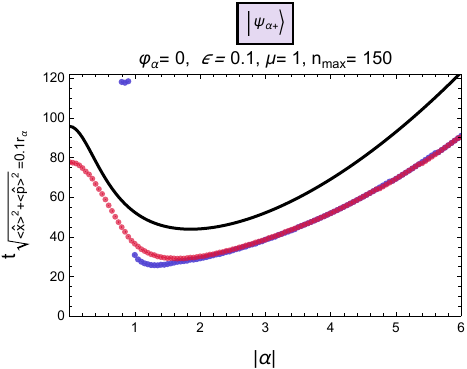}
         \includegraphics[height=0.7cm]{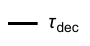}
         \includegraphics[height=0.7cm]{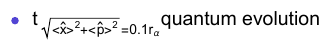}
        \includegraphics[height=0.7cm]{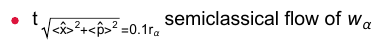}
    \caption{Comparison of $\tau_\text{dec}=\pi/\sqrt{\left\langle{\Omega^+}^2\right\rangle-\braket{\Omega^+}^2}$ with the time when the envelope of $\braket{\hat{x}}$ and $\braket{\hat{p}}$ for an initial state $\ket{\psi_{\alpha+}}$ has decayed under a threshold $\sqrt{\braket{\hat{x}}^2+\braket{\hat{p}}^2}=0.1r_\alpha$.  Left: fixed $|\alpha|=2$, Right: fixed $\varepsilon=0.1$. $\tau_\text{dec}$ describes well the order of the values for the semiclassical flow (red) for all displayed values. Under quantum time evolution  the threshold $0.1r_\alpha$ is not reached or reached much later than $\tau_\text{dec}$ for larges values of $\varepsilon$ and the smalls values $|\alpha|$ displayed.   
    }
    \label{fig:DO rleq0.1 taudecay}
\end{figure}

\begin{figure}[h]
    \centering
    \includegraphics[width=0.48\linewidth]{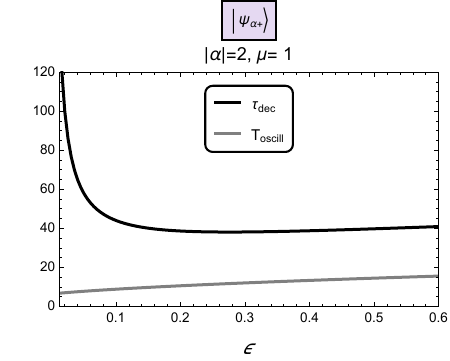}
       \includegraphics[width=0.48\linewidth]{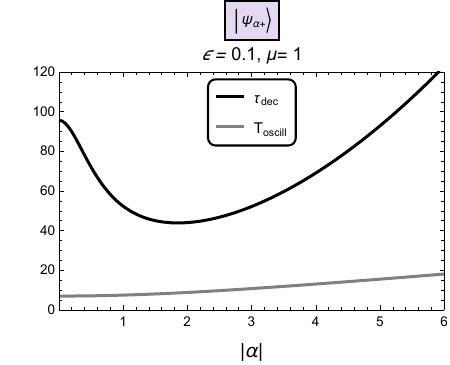}
    \caption{Comparison of $\tau_\text{dec}$ and $T_\text{oscill}$ for the branch +. For the shown parameters, $T_\text{oscill}$ is much smaller than $\tau_\text{dec}$.}
    \label{fig:DO taudecay vs Toscill}
\end{figure}

The semiclassical analysis, if valid,  now ultimately  allows to approximate period  of the oscillation of  $\braket{\hat{x}}(t)$ and $\braket{\hat{p}}(t)$  as (see \autoref{sec:AO Toscill and taudecay})
\begin{equation}
    \boxed{\begin{aligned}
    T_\text{oscill}=&\frac{2\pi}{\braket{\Omega^\pm}}\\
    \end{aligned}
    }
\end{equation}
where 
\begin{equation}
    \braket{\Omega^\pm}=\int\mathrm{d}rr\Omega^\pm(r)\int\mathrm{d}\varphi w_\alpha(r,\varphi),\\
\end{equation}
which can easily be calculated as a numerical integral. This prediction is confirmed by a numerical study shown in \autoref{fig:DO teq0 Toscill/4}: For an initial state $\ket{\psi_{\alpha+}}$ with $\varphi_\alpha=0$, the dynamics reach $\braket{\hat{x}}=0$ at a time close to $t=T_\text{oscill}/4$ for the whole presented parameter range both under the semiclassical and quantum evolution. The typical time it takes for the oscillation amplitude to decay can be predicted via 
\begin{equation}\label{eq:DO taudec}
   \boxed{ \begin{aligned}
    \tau_{\text {dec}}=&\frac{\pi}{\sqrt{\left\langle{\Omega^\pm}^2\right\rangle-\braket{\Omega^\pm}^2} }
    \end{aligned}
    }
\end{equation}
where 
\begin{equation}\label{eq:DO <Omega>}
    \begin{aligned}
    \left\langle{\Omega^\pm}^2\right\rangle =&\int \mathrm{d} r r \int \mathrm{~d} \varphi w_\alpha(r, \varphi){\Omega^\pm(r)}^2. \\
    \end{aligned}
\end{equation}
In the study shown in \autoref{fig:DO rleq0.1 taudecay}, the given formula for $\tau_\text{dec}$ predicts the time, when $\sqrt{\braket{\hat{x}}^2+\braket{\hat{p}}^2}=0.1r_\alpha$ is reached under the semiclassical flow, with an error of $30\%$ for the whole parameter range, which is quite accurate since $\tau_\text{dec}$ is only the prediction a typical time scale. For $|\alpha|=2 \& \varepsilon\leq0.52$ and  $|\alpha|\geq1\& \varepsilon=0.1$,  $\tau_\text{dec}$ predict also well the values given by the quantum evolution. 
The given formulas for $T_\text{oscill}$ and $\tau_\text{dec}$ finally can be applied to categorize the initial dynamics. Under the condition that the semiclassical approximation is valid we can expect that the initial decay of the amplitude of $\braket{\hat{x}}(t)$ and  $\braket{\hat{p}}(t)$ is modulated with clearly visible oscillations if $T_\text{oscill}<\tau_\text{dec}$.  \autoref{fig:DO taudecay vs Toscill} shows that for $\varepsilon\in[0.005,0.6]\&|\alpha|=2$ and $\varepsilon=0.1 \&|\alpha|\in[0.01,6]$ the latter is the case,  which can explain that \autoref{fig:DO x semiclflow} does not show any decay of $\braket{\hat{x}}(t)$ resembling an ``overdamped oscillation``.\\ 
\FloatBarrier
\subsection{Evolution of the wave packet width}
\begin{figure}
    \centering
    \begin{subfigure}{0.49\linewidth}
    \centering
        \includegraphics[width=0.95\linewidth]{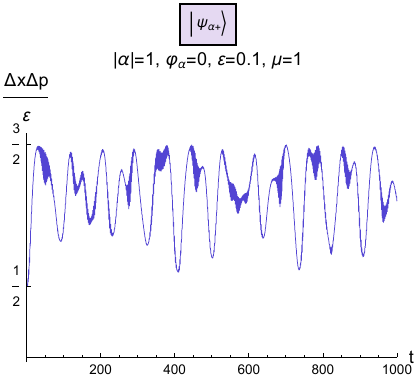}
        \includegraphics[width=0.95\linewidth]{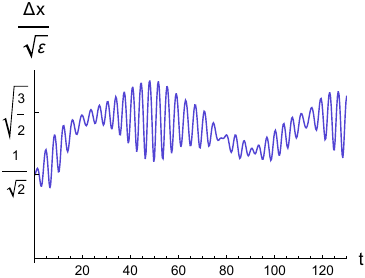}
         \includegraphics[width=0.99\linewidth]{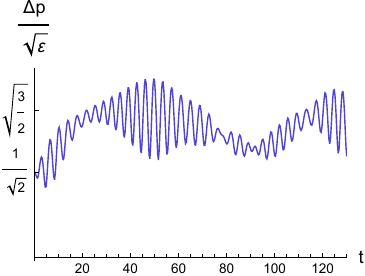}
    \end{subfigure}
      \begin{subfigure}{0.49\linewidth}
    \centering
        \includegraphics[width=0.95\linewidth]{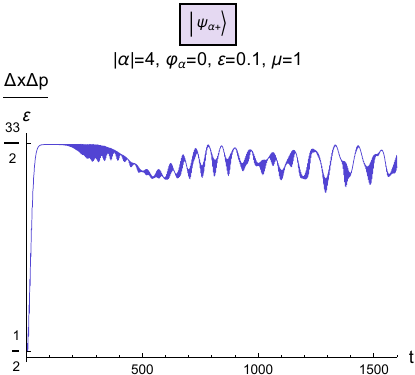}
        \includegraphics[width=0.95\linewidth]{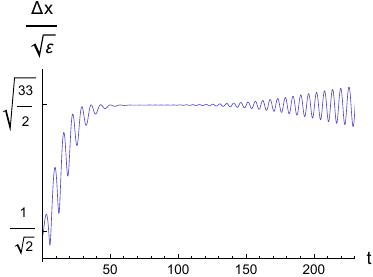}
         \includegraphics[width=0.99\linewidth]{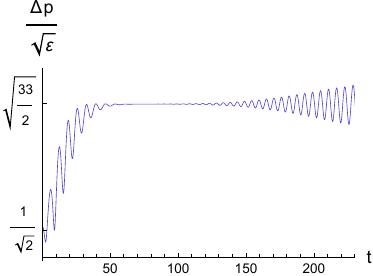}
    \end{subfigure}
    
        \includegraphics[height=0.7cm]{graphics/Legends/plotlegend-exact.pdf}      

    \caption{Evolution of the wave packet width for $\ket{\psi_{\alpha+}}$.  Top row: The uncertainty product $\Delta x\Delta p$ initially grows towards approximately $\varepsilon(|\alpha|^2 +1/2)$.  For $|\alpha|=1$ (left column) the uncertainty product quickly decays again, but for $|\alpha|=4$, it remains approximately constant until $t\approx 200$.  The uncertainty product has $\varepsilon(|\alpha|^2 +1/2)$ as an approximate upper bound. Middle and bottom row: The initial growth of $\Delta x$ and $\Delta p$ is superposed with oscillations, which have opposite phase for $\Delta x$ and $\Delta p$. During the initial growth the dynamics can reach $\Delta x <\sqrt{\varepsilon/2}$ and $\Delta p < \sqrt{\varepsilon/2}$ alternatingly .}
    \label{fig:DO Heisenberg large times}
\end{figure}

The evolution of the wave packet width of $\ket{\psi_{\alpha\pm}}$ in real space and momentum space 
\begin{equation}
    \begin{aligned}
     \Delta x =\sqrt{\braket{\hat{x}^2}-\braket{\hat{x}}^2}\\
     \Delta p=\sqrt{\braket{\hat{p}^2}-\braket{\hat{p}}^2}\\
    \end{aligned}
\end{equation}
follows from
\begin{equation}\label{eq:DO <x2> <p2>}
    \begin{aligned}
    \braket{\hat{x}^2}=&\frac{\varepsilon}{\tilde{N}}
            \left(2\varepsilon\left(\frac{\mathrm{e}^{|\alpha|^2}}{2}+|\alpha|^2\left(\mathrm{e}^{|\alpha|^2}+\sum_{n=0}^{\infty} \frac{|\alpha|^{2 n} }{n!} \cos{\varphi_{n+1,n-1,\alpha}(t)}\right)\right)\right.\\
              &\quad \quad \quad +|\alpha|^2\sum_{n=0}^{\infty}\Bigg[ \frac{(\pm E_n-\mu)^2 }{(n+1)^2} \frac{|\alpha|^{2 n}(\frac{1}{2}+n)}{n!}\\
            &\quad \quad \quad \quad \quad \quad\quad  +\left.\frac{(\pm E_{n+2}-\mu)(\pm E_n-\mu)}{(n+3)(n+1)}\frac{|\alpha|^{2 n} }{n!} \cos{\varphi_{n+2,n,\alpha}(t)}\Bigg]\right)\\[1ex]
    \braket{\hat{p}^2}=&\frac{\varepsilon}{\tilde{N}}\left(2\varepsilon\left(\frac{\mathrm{e}^{|\alpha|^2}}{2}+|\alpha|^2\left(\mathrm{e}^{|\alpha|^2}-\sum_{n=0}^{\infty} \frac{|\alpha|^{2 n} }{n!} \cos{\varphi_{n+1,n-1,\alpha}(t)}\right)\right)\right.\\
              &\quad \quad \quad +|\alpha|^2\sum_{n=0}^{\infty}\Bigg[ \frac{(\pm E_n-\mu)^2 }{(n+1)^2} \frac{|\alpha|^{2 n}(\frac{1}{2}+n)}{n!}\\
             &\quad \quad \quad \quad \quad \quad  \left.-\frac{(\pm E_{n+2}-\mu)(\pm E_n-\mu)}{(n+3)(n+1)}\frac{|\alpha|^{2 n} }{n!} \cos{\varphi_{n+2,n,\alpha}(t)}\Bigg]\right).\\
    \end{aligned}
\end{equation} 
The top row of \autoref{fig:DO Heisenberg large times}  shows the evolution of the uncertainty product $\Delta x\Delta p$ for $\alpha=1$ and $\alpha=4$ with fixed $\varepsilon=0.1$:  After the onset of motion $\Delta x\Delta p$  grows towards the approximately global maximum $\approx\varepsilon(|\alpha|^2 +1/2)$, which corresponds  to the upper limit of uncertainty for an initially coherent state $\ket{\alpha}$  (see \autoref{sec:AO DeltaxDeltap}).  At later times, one can observe partial relocalization of the wave packet as local minima of $\Delta x\Delta p$. The local minima are closer to the minimal uncertainty $\Delta x\Delta p=\frac{\varepsilon}{2}$ for $|\alpha|=1$ (left column) than for $|\alpha|=4$ (right column). This is because for larger $|\alpha|$ there are more terms in \autoref{eq:DO psi_alpha(t)} that are important, all of which have to be approximately in phase for the Wigner function to be well localized. The initial growth of $\Delta x$ and $\Delta p$ (middle and bottom row of \autoref{fig:DO Heisenberg large times} ) is superposed with oscillations such that $\ket{\psi_{\alpha+}}$ gets squeezed into $\Delta p < \sqrt{\frac{\varepsilon}{2}}$ and into $\Delta x < \sqrt{\frac{\varepsilon}{2}}$. Both squeezing and partial restoration are phenomena also observed for other types of initially coherent states of the Dirac Oscillator~\cite{NogamiCoherentStateDO,Ghosh2021QuasicoherentstateDO}.

\begin{figure}
        \centering
    \begin{subfigure}{0.45\linewidth}
       \centering
       \includegraphics[width=0.7\linewidth]{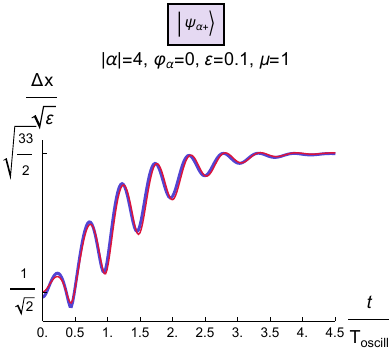}
        \includegraphics[width=0.7\linewidth]{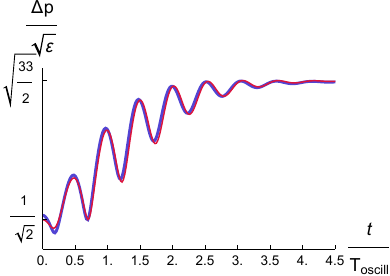}
        \includegraphics[height=0.6cm]{graphics/Legends/plotlegend-exact.pdf}
        \includegraphics[height=0.6cm]{graphics/Legends/linelegendwalpha-semicl.pdf}
    \end{subfigure}
      \begin{subfigure}{0.48\linewidth}
      \centering
        \includegraphics[width=0.99\linewidth]{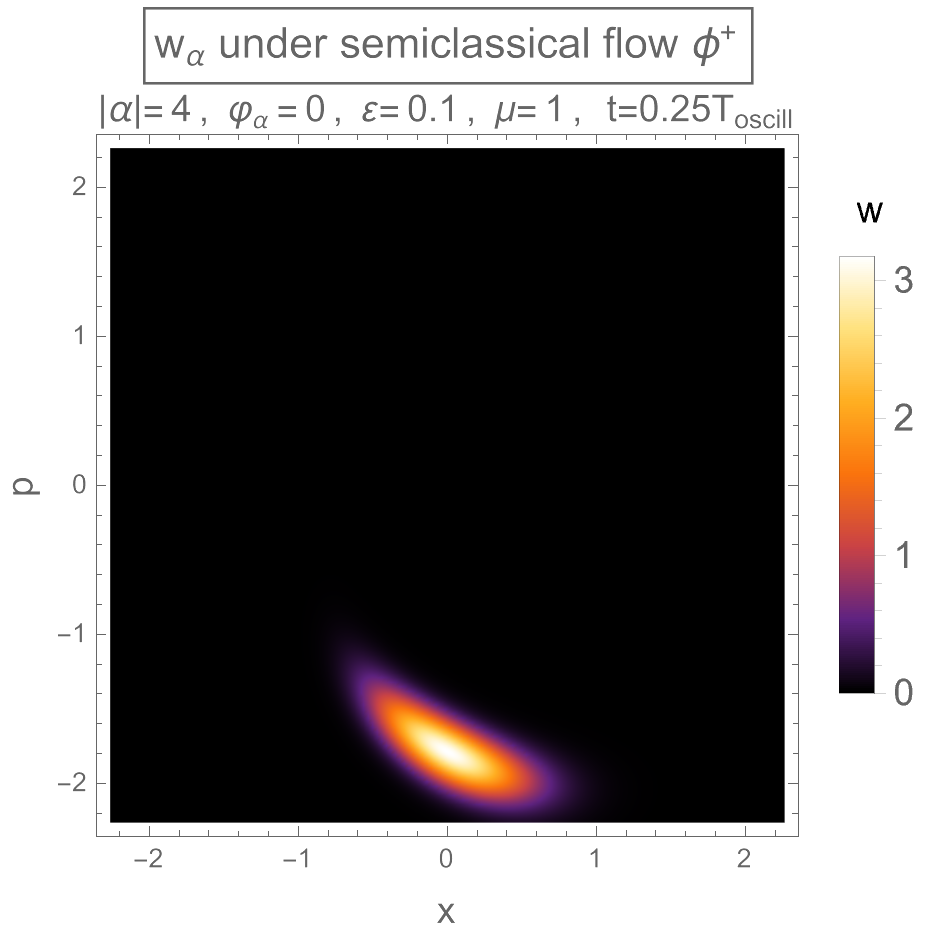}
    \end{subfigure}
    \caption{Left plot: The evolution of $\Delta x$ and $\Delta p$  for $\ket{\psi_{\alpha+}}$ with $\alpha=4$,  under quantum time evolution and under the semiclassical flow $\phi^+$ of $w_\alpha$.  During the initial growth both graphs match well.   $\Delta p$ reaches its global minimum at approximately $t=T_\text{oscill}/4$ and $\Delta x$ at approximately $t=T_\text{oscill}/2$.  Right plot:  At $t=T_\text{oscill}/2$, $w_\alpha $ has performed an average phase space rotation of $\pi/2$ under semiclassical flow $\phi^+$ and has been pulled in a banana-like shape such that it is squeezed in $p$-direction $\Delta p <\sqrt{\varepsilon/2}$.}
    \label{fig:DO Deltax Deltap semicl}
\end{figure}
\begin{figure}
    \centering
    \begin{subfigure}{0.48\linewidth}
        \includegraphics[width=0.99\linewidth]{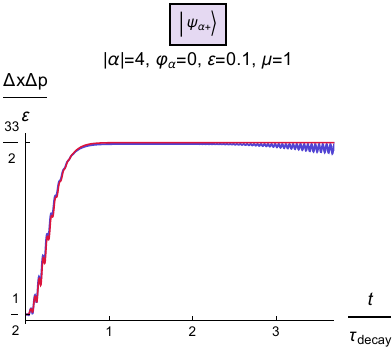}
        \includegraphics[height=0.7cm]{graphics/Legends/plotlegend-exact-semicl.pdf}
      
    \end{subfigure}
     \begin{subfigure}{0.48\linewidth}
        \includegraphics[width=0.99\linewidth]{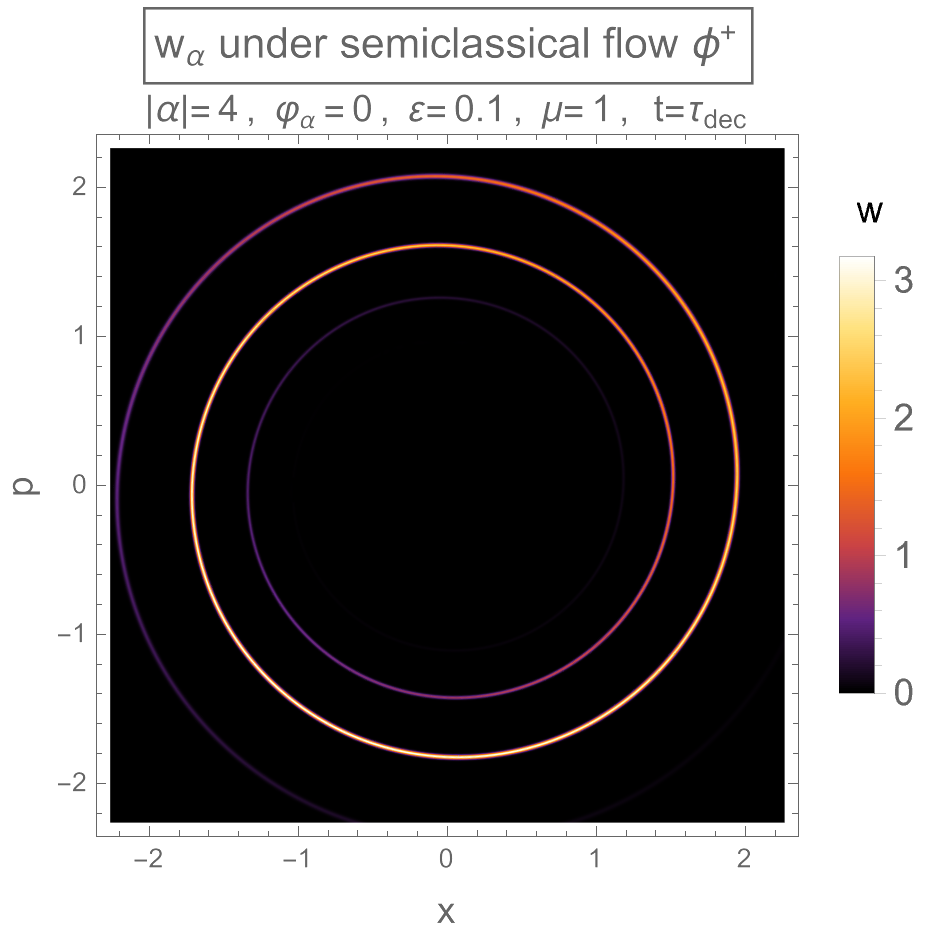}
       
    \end{subfigure}
    \caption{Left plot: The evolution of $\Delta x\Delta p$ under quantum time evolution and under the semiclassical flow $\phi^+$ of $w_\alpha$ for an initial state $\ket{\psi_{\alpha+}}$ with $|\alpha|=4$ and $\varphi_\alpha=0$ . During the initial growth the quantum and the semiclassical evolution match and at $t=\tau_\text{dec}$, the uncertainty product has already reached its approximate maximum $\Delta x\Delta p\approx(|\alpha|^2+1/2)\varepsilon$.  The semiclassical flow cannot predict the later onset of motion. Right plot: Under the semiclassical flow $\phi^+$, $w_\alpha $ has evolved into a spiral filling the disk around $r=r_\alpha$ after a time $t=\tau_\text{dec}$.}
    \label{fig:DO Heisenberg semicl}
\end{figure}
\subsubsection{Semiclassical analysis of the wave packet width}
The semiclassical analysis can be used to gain an intuition of the oscillatory growth of $\Delta x$ and $\Delta p$. Like for the Anharmonic oscillator, the semiclassical phase space angular velocity $\Omega^+(r)$ depends purely on $r$, therefore the semiclassical flow pulls $w_\alpha$ into $\vec{e}_\varphi$-direction and gives it a banana-shaped form, as can be seen in \autoref{fig:DO Deltax Deltap semicl} on the right. Due to this banana shape, the real space width $\Delta x$ (momentum space width $\Delta p$) becomes locally minimal (maximal) under $\phi^+$ at times $t$ that are multiples of $\frac{T_\text{oscill}}{2}$ for an initial state $\ket{\psi_{\alpha+}}$ with $\varphi_\alpha=0$, as illustrated in \autoref{fig:DO Deltax Deltap semicl} on the left. At a time of order $\tau_\text{dec}$, $w_\alpha$  has spread out under $\phi^+$ into a spiral on the disk around $r=r_\alpha$ in phase space (see \autoref{fig:DO Heisenberg semicl} on the right). Thus, $\tau_\text{dec}$ predicts when the semiclassical uncertainty $\Delta x\Delta p$ has approximately reached its maximal value, as illustrated in \autoref{fig:DO Heisenberg semicl} on the left. Partial relocalization, i.e. the onset of the motion  after the uncertainty product has reached  $\Delta x\Delta p\approx \varepsilon\frac{2|\alpha|^2+1}{2}$, cannot be predicted by the semiclassical flow. 

%% file: graphics/sketches/2DCrystal/HexagonalLattice.tex
\begin{tikzpicture}[scale=4, every node/.style={circle,inner sep=1pt}]

  \def\a{0.4}

  \foreach \ang in {30,90,150,210,270,330}{
    \coordinate (P\ang) at (\ang:\a);
  }

  \draw[thick,gray] (P30) -- (P90) -- (P150) -- (P210) -- (P270) -- (P330) -- cycle;

 
\coordinate (A1) at (P90);
\coordinate (B1) at (P150);
\coordinate (A2) at (P210);
\coordinate (B2) at (P270);
\coordinate (A3) at (P330);
\coordinate (B3) at (P30);

  \filldraw[blue!90] (A1) circle(0.04);
  \filldraw[red!90]  (B1) circle(0.04);
    \filldraw[blue!90] (A2) circle(0.04);
  \filldraw[red!90]  (B2) circle(0.04);
    \filldraw[blue!90] (A3) circle(0.04);
  \filldraw[red!90]  (B3) circle(0.04);

  \draw[gray, thick] (A1) -- ++(90:\a/3);
  \draw[gray, thick] (B1) -- ++(150:\a/3);
\draw[gray, thick] (A2) -- ++(-150:\a/3);
  \draw[gray, thick] (B2) -- ++(-90:\a/3);
  \draw[gray, thick] (A3) -- ++(-30:\a/3);
  \draw[gray, thick] (B3) -- ++(30:\a/3);

    \filldraw[blue!90] (A1) circle(0.04);
  \filldraw[red!90]  (B1) circle(0.04);
    \filldraw[blue!90] (A2) circle(0.04);
  \filldraw[red!90]  (B2) circle(0.04);
    \filldraw[blue!90] (A3) circle(0.04);
  \filldraw[red!90]  (B3) circle(0.04);
  \filldraw[black] (A1)  node[right=5pt] {$+\Delta$};
    \filldraw[black] (B1)  node[right=5pt] {$-\Delta$};

    \filldraw[red!90] (B2)  node[above left=5pt] {$B$};
      \filldraw[blue!90] (A3)  node[above left=5pt] {$A$};

  \fill[gray!40] (0,0) circle(0.02);
  \draw[->, thick] (-0.4,-0.5) -- ++(0:0.15) node[below] {$x$};
    \draw[->, thick] (-0.4,-0.5) -- ++(90:0.15) node[left] {$y$};

\end{tikzpicture}

%% file: graphics/sketches/2DCrystal/Brillouinzone.tex
\begin{tikzpicture}[scale=4, every node/.style={circle,inner sep=1pt}]

  \def\a{0.5}

  \foreach \ang in {0,60,120,180,240,300}{
    \coordinate (P\ang) at (\ang:\a);
  }

  \draw[thick] (P0) -- (P60) --(P120)-- (P180) -- (P240) -- (P300)  -- cycle;

 
\coordinate (K) at (P0);
\coordinate (Kprime) at (P60);
\fill[black] (P0) node[right=2pt] {$K$};
\fill[black] (P60) node[above right=2pt]{$K'$};
\fill[black] (P120) node[above right=2pt] {$K$};
\fill[black] (P180) node[right=2pt]{$K'$};
\fill[black] (P240) node[above right=2pt] {$K$};
\fill[black] (P300) node[right=2pt]{$K'$};

    \filldraw[black] (P0) circle(0.015);
  \filldraw[black]  (P60) circle(0.015);
    \filldraw[black] (P120) circle(0.015);
  \filldraw[black]  (P180) circle(0.015);
    \filldraw[black] (P240) circle(0.015);
  \filldraw[black]  (P300) circle(0.015);

  \fill[gray!40] (0,0) circle(0.02);
  \draw[->, thick] (0,0) -- ++(0:0.15) node[below] {$k_x$};
    \draw[->, thick] (0,0) -- ++(90:0.15) node[left] {$k_y$};

\end{tikzpicture}

%% file: chapters/Lagranigan_Ansatz_for_wave_packet_center.tex
\chapter{Limitations of the Lagrangian Ansatz for the Semiclassical Evolution of Wave Packet Centers}\label{chapter:Niu}    
In the literature, it is most common to derive the semiclassical equations of motion for multiband systems by an approach that consists in constructing a well localized wave packet $\psi^\nu(x)$ from an eigenvector $\ket{\Psi^\nu_0}$ of the symbol $H(x,p)$ ~(see for example  \cite{Price-BerryCurvature-OpticalLattice-NiuAnsatz,wimmerExperimentalMeasurementBerry2017ExampleNiu})  and choosing a Lagrangian ansatz to  obtain the approximate equations of motion for the wave center in real space $\braket{\hat{x}}=x_c$ and  momentum space $\braket{\hat{p}}=p_c$,  as for example in chapter 4 of~\cite{perez:tel-03922294}.  The seminal paper of this Lagrangian ansatz by Sundaram and Niu~\cite{SundaramNiu} has 1309 citations, which is 2 orders of magnitude higher than 39, the number of citations of the paper from Stiepan and Teufel~\cite{Stiepan_2012}, which derives the semiclassical evolution in multiband systems via the semiclassical flow of a scalar Wigner function (as of 13.10.2025). The aim of this chapter is to argue that the Lagrangian wave packet ansatz only takes into account the approximate dynamics at the maximum of the Wigner function but neglects the rest of the phase space. It therefore misses the characteristic of the dynamics related to the Wigner function deformation, such as decay of the oscillation amplitude of  $\braket{\hat{x}}(t)$ and $\braket{\hat{p}}(t)$ and evolution of the wave packet width $\Delta x$ and $ \Delta p$. To predict such characteristic of the dynamics via semiclassical means,  we suggest an alternative  ansatz for localized wave packets based on the work of Stiepan and Teufel.   
\section{Lagrangian approach for  semiclassical equations of motion of wave packet centers}
In the following, we will present our understanding of the semiclassical ansatz  for the evolution of wave packet centers presented in~\cite{SundaramNiu}  and chapter 4 of \cite{perez:tel-03922294}. Like in the previous chapters, the semiclassical equations of motion will be derived for a symbol Hamiltonian $H(x,p)$  that is slowly varying in $x$-direction compared to the gap of its spectrum $\Delta e_0$. Then, a nondimensionalization can be performed such that in slow variables $\hat{x} \rightarrow X$ and $\hat{P}_X\rightarrow-\mathrm{i}\varepsilon \partial_X$ the commutator becomes a small parameter $[\hat{X},\hat{P}_X]=\mathrm{i}\varepsilon$, while the gap remains  $\Delta \tilde{e}_0\gtrsim 1  $. Again, the slow variables will be denoted as $x,p$ to lighten the notation $x\equiv X$ and $p \equiv P_X$.  For simplicity, we consider a symbol with zero higher order terms $H(x,p)=H_0(x,p)$. \\
In the ansatz presented in~\cite{SundaramNiu}, an initial wave packet $\psi^\nu(x)$ associated to a band $e_0^{(\nu)}$ of the symbol Hamiltonian is constructed from the corresponding eigenvector $\ket{\Psi_0^\nu}$ as
\begin{equation}\label{eq:psi(x)nu ansatz}
    \psi^\nu(x)=a(x) \mathrm{e}^{\mathrm{i}\left(\frac{x p_c}{\varepsilon}+\varphi(x,p_c)\right)}\ket{\Psi_0^\nu(x,p_c)}
\end{equation}
where $a (x)$ is an envelope that is peaked around the real space  wave packet center $x_c$ . The complex phase  of $\psi(x)$ is chosen in different orders of $\varepsilon$ as in a WKB-framework. At lowest order, it is the phase of a plane wave with momentum $p_c$.  The wave packet $\psi^\nu(x)$ is demanded to be  centered around $p_c$ in momentum space 
\begin{equation}
\begin{aligned}
       \braket{\psi|\hat{p}|\psi}\overset{!}{=}p_c.\\
  \end{aligned}
\end{equation}
This gives an implicit condition on the phase $\varphi(x,p_c)$
\begin{equation}
     \varepsilon\int  \mathrm{d}x \,|a(x)|^2\left(\partial_x\varphi(x,p_c)+ \braket{\Psi_0^\nu|\partial_x\Psi_0^\nu}  \right) =0.
\end{equation}
Now, it is assumed that the absolute square of the envelope  $|a(x)|^2$ is sharply peaked around $x =x_c$ such that  for any function $f$ of the slow variable $x$ 
\begin{equation}\label{eq:Niu approxf(x)}
    \int  \mathrm{d}x \,|a(x)|^2 f(x)=f(x_c)+\mathcal{O}(\varepsilon^2).
\end{equation}
This implicitly fixes the phase $\varphi(x,p_c)$  at $x= x_c$
\begin{equation}\label{eq:Niu partialx phi}
    \partial_{x_c}\varphi(x_c,p_c)=-\braket{\Psi_0^\nu|\partial_{x_c}\Psi_0^\nu} +\mathcal{O}(\varepsilon^2),
\end{equation}
where $\partial_{x_c}f=\partial_{x}f(x)\big|_{x=x_c}$.  
 \\ 
To obtain the semiclassical equations of motion for $x_c$ and $p_c$, the problem of solving the Schrödinger equation 
\begin{equation}
    \mathrm{i}\varepsilon\partial_t \ket{\psi}=\hat{\mathcal{H}}\ket{\psi}, 
\end{equation}
is reformulated as minimizing the following Lagrangian   
\begin{equation}
    \mathcal{L}^\prime=\braket{\psi| \mathrm{i}\varepsilon \mathrm{d}_t-\hat{\mathcal{H}}|\psi},
\end{equation}
with $\mathrm{d}_t$ the total time derivative $\mathrm{d}_t=\dot{x}_c\partial_{x_c}+\dot{p}_c\partial_{p_c}+\partial_t$.  
Then, the goal is to approximate $\mathcal{L}'$, such that it becomes a function of $x_c,\dot{x}_c,p_c,\dot{p}_c$. Firstly, the term $\braket{\psi|\hat{\mathcal{H}}|\psi}$ is considered.   It is assumed  that each component of the Wigner function $W(x,p)$ of  $\psi(x)$ is sharply peaked around $x_c$ and $p_c$, such that in 
\begin{equation}
    \braket{\psi|\hat{\mathcal{H}}|\psi}=\int \int \mathrm{d}x\mathrm{d}p \operatorname{Tr}_{\mathcal{H}_\text{f}}\left[W(x,p) H(x,p)\right]
\end{equation}
the symbol Hamiltonian $H$ can be effectively approximated as
\begin{equation}\label{eq:Niu approx H(x,p)}
    H(x,p)=H_c+\Delta H+\mathcal{O}(\varepsilon^2)
\end{equation}
with \begin{equation}
    \begin{aligned}
        H_c=&H(x_c,p_c)\\[0.5ex]
        \Delta H=&(x-x_c)\partial_{x_c}H_c+(p-p_c)\partial_{p_c}H_c.
    \end{aligned}
\end{equation}
Weyl-quantization gives
\begin{equation}
    \hat{\mathcal{H}}_c=H_c
\end{equation} 
and 
\begin{equation}
\begin{aligned}
       \Delta \hat{\mathcal{H}}=&\frac{\hat{x}-x_c}{2}\partial_{x_c}H_c+\partial_{x_c}H_c\frac{\hat{x}-x_c}{2}\\
      & \quad \quad   +\frac{\hat{p}-p_c}{2}\partial_{p_c}H_c+\partial_{p_c}H_c\frac{\hat{p}-p_c}{2}.
\end{aligned}
\end{equation}
Using the approximated operator $\hat{\mathcal{H}}=\hat{\mathcal{H}}_c+\Delta \hat{\mathcal{H}}+\mathcal{O}(\varepsilon^2)$ and \autoref{eq:Niu approxf(x)}, \autoref{eq:Niu partialx phi} one obtains
\begin{equation}
\begin{aligned}
\braket{\psi|\hat{\mathcal{H}}|\psi}=&h^\nu(x_c,p_c)+\mathcal{O}(\varepsilon^2)\\[1ex]
  h^\nu =&  e_0^\nu+\varepsilon M^\nu, \
\end{aligned}
\end{equation}
where $M^\nu$ is the Littlejohn-Flynn term (``no name term``) defined in~\autoref{eq:no name term}. 
The term of $\mathcal{L}^\prime$ containing the total time derivative can be shown to be 
\begin{equation}
\begin{aligned}
\braket{\psi|\mathrm{i}\varepsilon\mathrm{d}_t|\psi}=
-\dot{p}_cx_c-\varepsilon\dot{p}_c\partial_{p_c}\varphi(x_c,p_c) 
+ \mathrm{i} \varepsilon \dot{p}_c \left\langle \Psi^{\nu}_0 \middle| \partial_{p_c} \Psi^{\nu}_0 \right\rangle + \mathcal{O}(\varepsilon^2).\\
\end{aligned}
\end{equation}
Adding a total time derivative to the Lagrangian $\mathcal{L^\prime}$ does not affect the variational problem. It is convenient  to work with $\mathcal{L}=\mathcal{L}^\prime+\mathrm{d}_t(x_cp_c-\varphi(x_c,p_c)$ instead of $\mathcal{L}^\prime$. This new Lagrangian can be expressed as 
\begin{equation}
    \mathcal{L}=\dot{x}_c p_c +\mathrm{i}\varepsilon\left(\dot{x}_c\braket{\Psi^\nu_0|\partial_{x_c}\Psi^\nu_0}+ \dot{p}_c\braket{\Psi^\nu_0|\partial_{p_c}\Psi^\nu_0}\right)-h_0^\nu(x_c,p_c)+\mathcal{O}(\varepsilon^2).
\end{equation}
Then, the Euler-Lagrange equations 
\begin{equation}
    \frac{\partial \mathcal{L}}{\partial q}=\frac{\mathrm{d}}{\mathrm{d}t}\left(\frac{\partial \mathcal{L}}{\partial \dot{q}}\right)
\end{equation}
with   $x_c$ or $p_c$ as degrees of freedom $q$ yield the semiclassical equations of motion for the wave packet center 
\begin{equation}\label{eq:Niu semiclass}
    \begin{aligned}
& \dot{x}_c=+\frac{\partial h^{\nu}(x_c,p_c)}{\partial p_c} - \varepsilon\left(F_{p_{c}x_c}^{\nu} \dot{x}_c+F_{p_c p_{c}}^{\nu} \dot{p}_{c}\right)+\mathcal{O}(\varepsilon^2) \\
& \dot{p}_c=-\frac{\partial h^{\nu}(x_c,p_c)}{\partial x_c}+\varepsilon \left( F_{x_cx_c}^{\nu} \dot{x}_c+F_{x_c p_{c}}^{\nu} \dot{p}_{c}\right)+\mathcal{O}(\varepsilon^2). \\
\end{aligned}
\end{equation}
\section{Limitations of the Lagrangian approach}

\begin{figure}
    \centering
    \includegraphics[height=6.5cm]{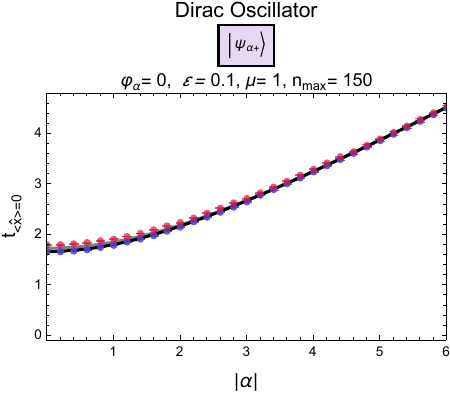}
       \includegraphics[height=6.5cm]{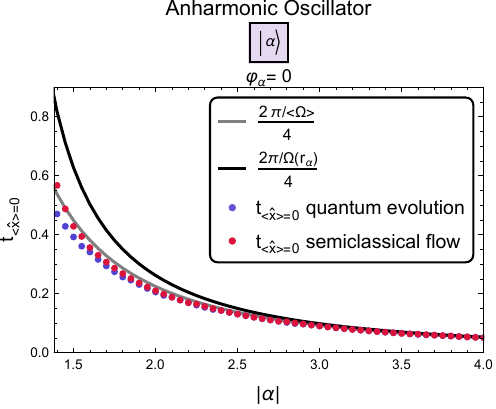} 
    \caption{Comparison of $\pi/2|\braket{\Omega}|$ and $\pi/2|\Omega(r_\alpha)|$ with $t_{\braket{\hat{x}}=0}$ the time when $\braket{\hat{x}}=0$ under quantum time evolution (blue) and under semiclassical flow of $w_\alpha(x,p)$ (red). Left: For an initial state $\ket{\psi_{\alpha+}}$ in the Dirac oscillator with $\varphi_\alpha=0$, both give a good prediction. Right: For an initial state $\ket{\alpha}$ in the Anharmonic Oscillator with $\varphi_\alpha=0$, the predictions via average semiclassical angular velocity is clearly better than via the semiclassical angular velocity at the initial maximum of $w_\alpha$.}
    \label{fig:Niu teq0 Toscill/4}
\end{figure}
\begin{figure}
    \centering
    \includegraphics[width=0.45\linewidth]{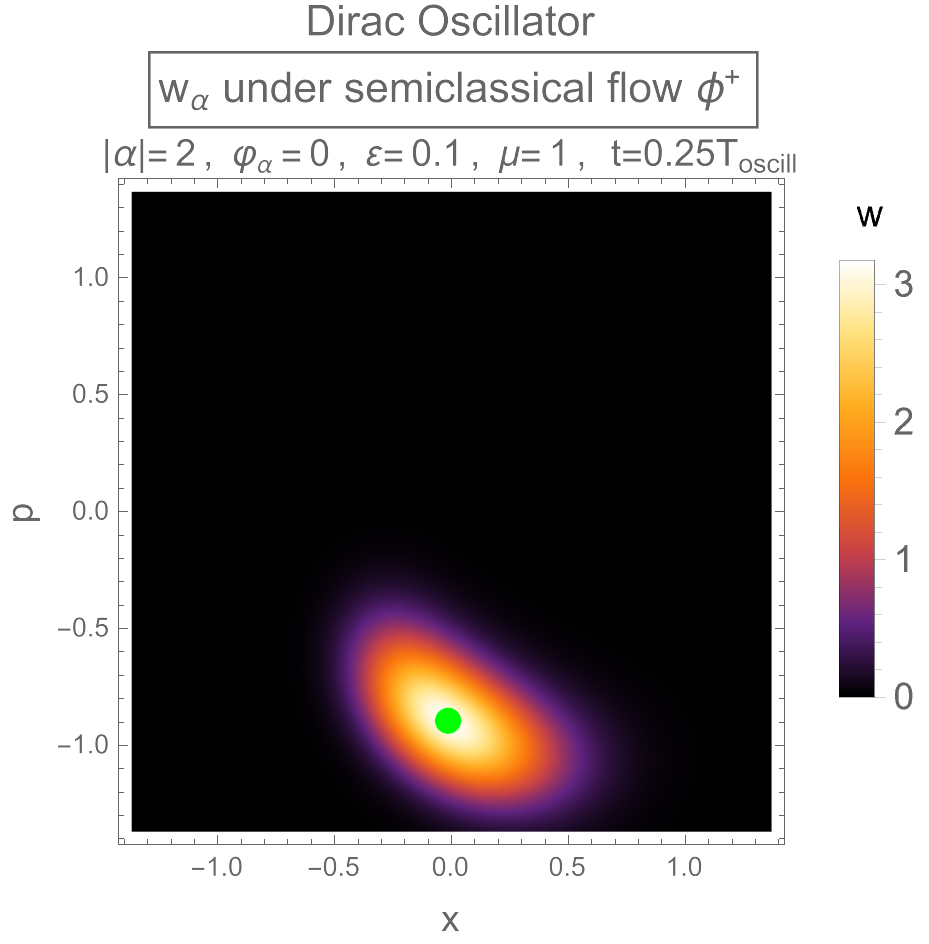}
    \includegraphics[width=0.45\linewidth]{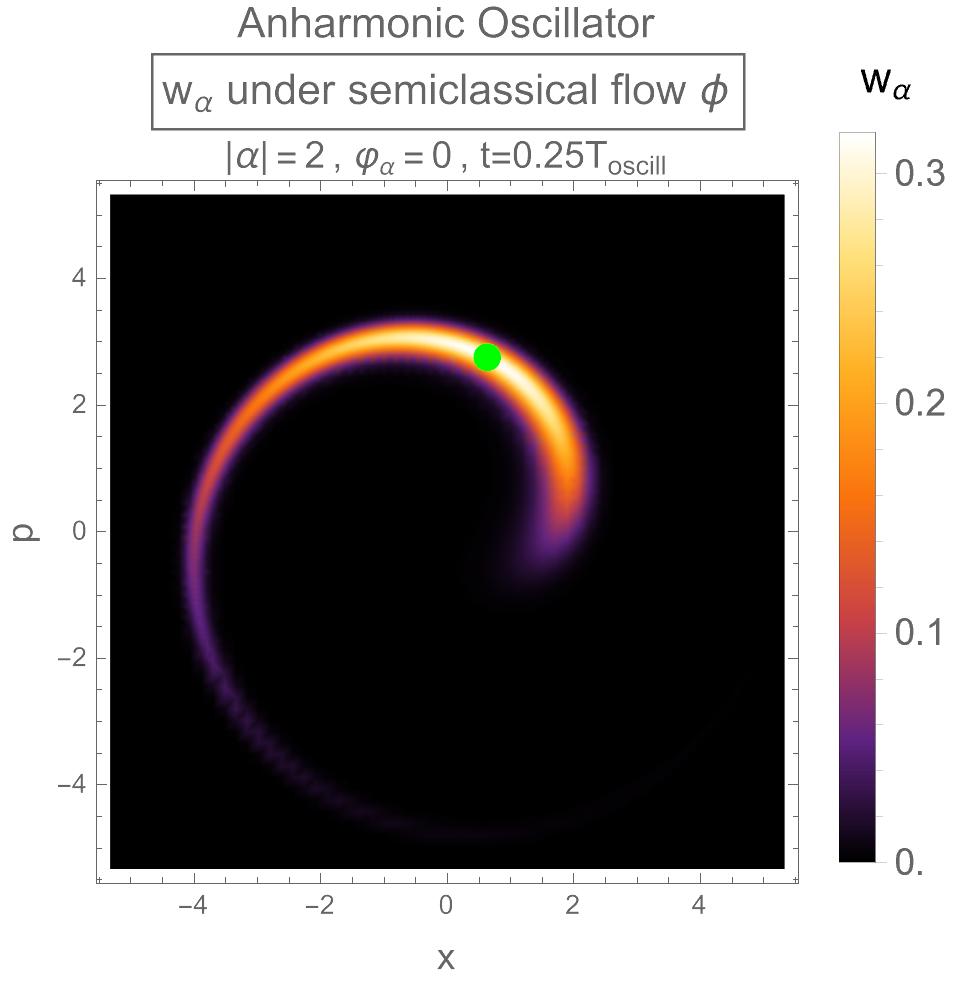}
    \includegraphics[height=0.7cm]{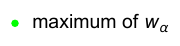}
    \caption{Evolution of $w_\alpha(x,p)$ under Left: the semiclassical flow $\phi^+$ associated to the branch + of the Dirac Oscillator Right: the semiclassical flow $\phi$ of the Anharmonic Oscillator.  After an average phase space rotation of $\pi/2$, the semiclassical $w_\alpha$ is still well centered around its maximum for the Dirac Oscillator, but the latter is not the case for a rotation of $-\pi/2$ in the Anharmonic Oscillator.}
    \label{fig:Niu PlotWig-AO}
\end{figure}

The best known example of localized wave packets in quantum physics is coherent states. In~\autoref{chapter:DO} and \autoref{chapter:AO}, we showed that the semiclassical evolution of the Wigner function allows to predict some characteristics of the dynamics of (generalized) coherent states.  In the following, we will show the limits of using instead the semiclassical wave packet Lagrangian approach for the prediction of dynamics of coherent states.\\
For the Dirac Oscillator, the almost coherent state of the branches `$+$`  and `$-$` can be approximated under the condition $\mu \gtrsim1$ or $ \mu \lesssim-1$ respectively as  
\begin{equation}\label{eq:Niu |psi_alpha> approx}
    \begin{aligned}
         \psi_{\alpha \pm}(x) =&\psi_\alpha(x)\ket{\Psi_0^\pm(x_\alpha,p_\alpha)} +\mathcal{O}(\varepsilon^2)
         \\[1.1ex]=&\mathrm{e}^{\mathrm{-i}\frac{p_\alpha x_\alpha}{2\varepsilon}}\frac{1}{(\varepsilon \pi)^{1/4}}\mathrm{e}^{-\frac{(x-x_\alpha)^2}{2\varepsilon}}\mathrm{e}^{\mathrm{i}\frac{p_\alpha x}{\varepsilon}} \ket{\Psi_0^\pm(x_\alpha,p_\alpha)} +\mathcal{O}(\varepsilon^2),\\
    \end{aligned}
\end{equation}
where $x_\alpha=x_c$ and $p_\alpha =p_c$ are the wave packet centers. This expression resembles   the ansatz for a wave packet $\psi^\pm(x)$ constructed as in \autoref{eq:psi(x)nu ansatz}  under the identification of the peaked envelope $a(x)=\mathrm{e}^{\mathrm{-i}\frac{p_\alpha x_\alpha}{2\varepsilon}}\mathrm{e}^{\frac{(x-x_\alpha)^2}{2\varepsilon}}/(\varepsilon \pi)^{1/4}$. At $x=x_c$
both differ only by a general phase 
\begin{equation}
    \mathrm{e}^{\mathrm{i}\varphi(x_c,p_c)} \psi_{\alpha \pm}(x_c) =\psi^\pm(x_c) +\mathcal{O}(\varepsilon^2).
\end{equation}
For $x\neq x_c$ however, there are differences: the complex phase of $\psi_{\alpha \pm}(x)$ is at order $\varepsilon^0$  not $x$-dependent and $\psi_{\alpha\pm}(x)$ is proportional to the eigenvector for the symbol Hamiltonian $\ket{\Psi_0^\pm(x_c,p_c)}$ with frozen $x_c$ and not proportional to $\ket{\Psi_0^\pm(x,p_c)}$. In the following, the semiclassical evolution of  $x_c(t$) and $p_c(t)$ is studied, as obtained when choosing the Lagrangian $\mathcal{L}$ to predict the dynamics of $\ket{\psi_{\alpha+ }}$ despite these differences. This can be motivated by both $\psi^+(x)$ and $\ket{\psi_{\alpha +}}$ being constructed as wave packets that are expected to belong to the branch `$+$`.  \autoref{eq:Niu semiclass} is solved for the initial conditions $x_c(0)=r_\alpha\cos{\varphi_\alpha}$ and $p_c(0)=r_\alpha\sin{\varphi_\alpha}$ as
\begin{equation}
\begin{aligned}
      x_{c}(t)&=r_\alpha\cos(\varphi_\alpha-\Omega^\pm(r_\alpha)t)\\
        p_c(t)&=r_\alpha\sin(\varphi_\alpha-\Omega^\pm(r_\alpha)t),\\
\end{aligned}
\end{equation}
corresponding to an oscillation with the angular velocity $\Omega^\pm(r)$, as defined in~\autoref{eq:DO Omega(r)} and constant amplitude. This is nothing but the semiclassical evolution of the maximum value of  $w_\alpha(x,p) \approx\operatorname{Tr}[W_{\alpha+}] $. In other words, $x_c(t)$ and $p_c(t)$ describe the semiclassical evolution that would be obtained when replacing $w_\alpha\approx\operatorname{Tr}[W_{\alpha+}]$  with a delta function $\delta(x-x_\alpha)\delta(p-p_\alpha)$. As a consequence, $x_c(t)$ and $p_c(t)$ cannot predict any characteristic of the dynamics, that is related to the deformation of the Wigner function, such as the initial decay of the oscillation amplitude of $\braket{\hat{x}}(t)$ and $\braket{\hat{p}}(t)$, shown in~\autoref{fig:DO <x>-<p>} and \autoref{fig:DO <x> large times} .  This mismatch is related to the assumption of a narrow  envelope  $a(x) $ that was made: For the coherent state $|a(x)|^2=\mathrm{e}^{\frac{(x-x_\alpha)^2}{\varepsilon}}/\sqrt{\varepsilon \pi}$, one has  
\begin{equation}
     \int  \mathrm{d}x \,|a(x)|^2 f(x)=f(x_c)+\mathcal{O}(\varepsilon)
\end{equation}

instead of an error $\mathcal{O}(\varepsilon^2)$, as can be shown from a Taylor development of $f(x)$. Also, when approximating $H(x,p)$ in~\autoref{eq:Niu approx H(x,p)} the finite width $\Delta x=\Delta p=\sqrt{\varepsilon/2}$ of the approximated $W_{\alpha +}$ leads to an error greater than $\mathcal{O}(\varepsilon^2)$
\begin{equation}
     H(x,p)=H_c+\Delta H+\mathcal{O}(\varepsilon).
\end{equation}
In order for the error to be at the order $ \mathcal{O}(\varepsilon^2)$, it is required that $\Delta x$  and $\Delta p$ are maximally $\Delta x \sim \Delta p \lesssim  \varepsilon^{1/4}$ , which would violate  the  Heisenberg uncertainty principle. This shows that the Lagrangian wave packet ansatz does not sufficiently take into account the finite width of the wave packet. \\
Reducing the semiclassical evolution of $w_\alpha(x,p)$ to the evolution of its maximum also affects the prediction of the oscillation period  $T_\text{oscill}$, since the average angular velocity of $w_\alpha(x,p)$ is in general not equal to the angular velocity at the maximum of the Wigner function trace $ \braket{\Omega}\neq \Omega(r_\alpha)$.  
 For a state $\ket{\psi_{\alpha +}}$ with $\varphi_\alpha =0$,  the first time $\braket{\hat{x}}$ reaches zero, $t_{\braket{\hat{x}=0}}$, is expected to occur after a phase space rotation of $\pm\pi/2$.  Using  the evolution  of the maximum of $w_\alpha(x,p)$,  $t_{\braket{\hat{x}=0}}$ is predicted as $\frac{\pi}{2| \Omega^+(r_\alpha )|}$, while taking into account the whole Wigner function predicts  $t_{\braket{\hat{x}=0}}$  as $\frac{\pi}{2|\braket{ \Omega^+}|}$ as in \autoref{eq:DO <Omega>}. \autoref{fig:Niu teq0 Toscill/4} on the left shows that for $\varepsilon=0.1$ and $|\alpha|\in [0.01,6]$ both make a good prediction, which is because $\Omega^+(r )$ is varying rather slowly with $r$ such that when integrating over $w_\alpha(x,p)$ 
 \begin{equation}
     \Omega(r)=\Omega(r_\alpha)+(r-r_\alpha)\partial_{r_\alpha}\Omega(r_\alpha)+\mathcal{O}((r-r_\alpha)^2)
 \end{equation}
  is well approximated by the first term: The center of mass and the maximum value of $w_\alpha$ are still close after an average phase rotation of $\pi/2$, as illustrated in \autoref{fig:Niu PlotWig-AO} on the left. For the Anharmonic Oscillator (\autoref{chapter:AO}) on the other side of \autoref{fig:Niu teq0 Toscill/4},  $\Omega(r)$ changes more rapidly with $r$ and thus using $\braket{\Omega}$ instead of  $\Omega (r_\alpha)$ clearly improves the prediction of $t_{\braket{\hat{x}}=0}$ for a state $\ket{\alpha}$ with $\varphi_\alpha=0$. In \autoref{fig:Niu PlotWig-AO}  on the right, one can see that for $\ket{\alpha}$ in the Anharmonic Oscillator with $|\alpha|=2$ the center of mass and the maximum of the semiclassical $w_\alpha$ differ after an average phase space rotation of $\pi/2$. This illustrates that the finite width of the Wigner functions (trace) can have a non-negligible effect on the prediction of wave packet center trajectories. \\

\section{Semiclassical ansatz for wave packet evolution including Wigner function evolution}
As argued above, the semiclassical wave-packet Lagrangian approach 
gives a worse prediction of $\braket{\hat{x}}(t)$ and $\braket{\hat{p}}(t)$ for initially (almost) coherent wave packets than the semiclassical evolution of $w_\alpha(x,p) $.  Furthermore, it cannot make a prediction on the evolution of $\Delta x$ and $\Delta p$.  
This motivates a different ansatz for the study of wave packets in a multiband system by semiclassical means: We suggest to construct a localized  wave packet as the eigenstate $\ket{\Psi ^\nu_0}$ of the symbol $H(x,p)$ multiplied by $\ket{\alpha}$ the coherent state of the Harmonic oscillator
\begin{equation}
    \ket{\psi^\nu}=\ket{\alpha}\ket{\Psi_0^\nu(x_\alpha ,p_\alpha)},
\end{equation}
with frozen arguments $x_\alpha$ and $p_\alpha$ in $\ket{\Psi_0^\nu}$.  Under the condition that $\ket{\psi^\nu}$ is approximately in the range of the adiabatic projector $\hat{\pi}^\nu$ associated with the band $e_0^{(\nu)}$ 
\begin{equation}
    \hat{\pi}^\nu \ket{\psi^\nu}=\ket{\psi^\nu}+\mathcal{O}(\varepsilon^2),
\end{equation}
one can approximate the evolution of any slow operators $\hat{o}$ via 
    \begin{equation}
    \begin{aligned}
                    \braket{\hat{o}}(t)=&\int\int\mathrm{d} \lambda \omega_\alpha(x^{\phi_{-t}^\nu},p^{\phi_{-t}^\nu}) o(x,p)+\mathcal{O}(\varepsilon^2)
    \end{aligned}
\end{equation}
where
\begin{equation}
   \omega_\alpha(x,p) =\frac{w_\alpha(x,p)}{1+\varepsilon F^\nu_{x,p}}\\
\end{equation}
with $\phi^\nu$ the semiclassical flow associated with the adiabatic subspace $\mathcal{H}^\nu_\text{adiabatic}$ and $\mathrm{d}\lambda=\mathrm{d}x\mathrm{d}p(1+\varepsilon F^\nu_{xp})$ the invariant phase space measure. Using this ansatz, both wave packet oscillation and deformation can be captured in a semiclassical analysis including Wigner function evolution.

%% file: chapters/Conclusion.tex
\chapter*{Conclusion}
\addcontentsline{toc}{chapter}{Conclusion}
\section*{Results}
In this Master's thesis, the semiclassical limit of wave dynamics via the approximate evolution of the Wigner function  was studied in oscillator systems with different levels of complexity: the Harmonic Oscillator (one-band), the Anharmonic Oscillator (one-band) and the Dirac Oscillator (two-band). 
For the Quantum Harmonic Oscillator,  it was shown that the semiclassical Wigner function evolution coincides with the exact quantum evolution and how the classical limit can be obtained for coherent states $\ket{\alpha}$, which is well known in the literature.  \\
For the  Anharmonic Oscillator and the Dirac Oscillator, the quality of the semiclassical prediction was evaluated by comparison with the exact solutions.  It was observed that the semiclassical analysis can capture the oscillation and deformation of  initially localized wave packets. But it fails to describe the later (fractional) revival of the initial wave packet in the Anharmonic Oscillator and partial  relocalization of wave packets in the Dirac Oscillator.  \\
A precise analysis of the conditions for validity of the semiclassical approximations was performed for both systems.
For the Anharmonic Oscillator, a parameter $\epsilon$ was identified that controls the semiclassical limit for the initial evolution of initially coherent states. When a (fractional) revival of the initial state occurs, the assumption  necessary for identifying $\epsilon$  becomes invalid and $\epsilon$ loses its control over the semiclassical limit. 
For the Dirac Oscillator,  the control parameter of the semiclassical approximation was obtained as a rescaled commutator  $[\hat{x},\hat{p}]\rightarrow\mathrm{i}\varepsilon$.  It was observed that the validity of the semiclassical approximation is also controlled by the form of the Wigner function, which  can lead to an effectively higher band gap. 

The 2 branches of the spectrum of the Dirac Oscillator were associated with the adiabatic subspaces corresponding to the bands $e_0^\pm$. The spectral flow state was associated to the upper or lower branch depending on the parameter $\mu$.  The fact that the semiclassical analysis only applies to states belonging to the same adiabatic subspace  required the construction of a `single-branch almost coherent state` in order to study the dynamics of a localized wave packet in the semiclassical limit. 
 
The effect of the semiclassical first-order corrections terms on the quality of the predictions was studied explicitly. As expected, neglecting the Moyal-term $\braket{H_1}$ for an initially coherent state in the Anharmonic Oscillator decreases the quality of the prediction. For the Dirac Oscillator on the other hand, including the Littlejohn-Flynn term and the Berry curvature gives a clearly better prediction only if one looks at the $\ket{\uparrow}$ and $\ket{\downarrow}$ polarizations separately, but not necessarily for total expectation values. 

The  global dynamics of initially localized wave packets was analyzed  in the Anharmonic Oscillator and Dirac Oscillator using both the exact and approximate semiclassical evolution of the Wigner function.  For both systems, the simple analytical form of the semiclassical flow helped to characterize how the oscillation period, $T_{\text{oscill}}$, and the typical decay time, $\tau_{\text{dec}}$, of the initial evolution of the wave packet centers $\braket{\hat{x}}$ and $\braket{\hat{p}}$ depends on the control parameters. It also helped to understand the initial growth of wave packet width $\Delta x$ and $\Delta p$. \\
In the last chapter, it was shown that a more commonly used Lagrangian ansatz for semiclassical analysis of wave packet centers fails to reproduce behavior that is related to the finite width of a wave packet. This motivated the proposal of an alternative ansatz that includes the semiclassical Wigner function evolution.  
\section*{Perspectives}
In a future work, it would be desirable to test the theoretical predictions for the almost coherent state of the Dirac Oscillator in a numerical or experimental study. A suitable choice could be a numerical implementation of a hexagonal tight-binding crystal with spatially varying opposite on site energy $\Delta(x)$. To make the more realistic ansatz of a Gaussian wave packet in $y$-direction, one would need to include more spatial dimensions in the semiclassical analysis. Furthermore, working in higher spatial dimensions allows the study of nontrivial position-position and momentum-momentum components of the Berry curvature.\\
It would also be interesting to study an `Anharmonic Dirac Oscillator`~\cite{BUCK1981atom-phonon-coupling} ~ 
\begin{equation*}
    \hat{\mathcal{H}}=\left(\begin{array}{cc}
\mu & f(\hat{a}^\dagger\hat{a})  \, \hat{a}^{\dagger} \\
f(\hat{a}^\dagger\,\hat{a})  \hat{a} & -\mu
\end{array}\right)
\end{equation*}
where all the different semiclassical first order corrections are present. Separately neglecting each correction term could distinctly illustrate each contribution to the semiclassical prediction. \\
Another perspective is the study of systems with more than two bands. The geophysical shallow water model for instance is a three-band systems, for which the manifestation of the Berry curvature and the Littlejohn-Flynn-term in wave packet dynamics was shown in~\cite{Perez_2021} by comparing a numerical simulation to the wave packet center dynamics obtained via the semiclassical Lagrangian. The study of this model would also allow to investigate how the semiclassical approximation is controlled in systems that do not involve $\hbar$. A next step would be to test if the proposed semiclassical ansatz, that includes Wigner function evolution, can describe wave packet deformation in the numerical simulations. \\

%% file: literature.bib
@book{Zachos-QMPS:OverviewwithPapers,
author = {Zachos, Cosmas and Fairlie, David and Curtright, Thomas},
year = {2005},
month = {01},
pages = {},
title = {Quantum Mechanics in Phase Space. An Overview with Selected Papers},
isbn = {978-981-238-384-6},
doi = {10.1142/5287}
}

@article{Zachos-QuantumMechanicsinPS,
author = {Curtright, Thomas and Zachos, Cosmas},
year = {2012},
month = {01},
pages = {37-46},
title = {Quantum Mechanics in Phase Space},
volume = {1},
journal = {Asia-Pacific Physics Newsletter},
doi = {10.1142/S2251158X12000069}
}

@article{Case-WFandWTforpedestrinas,
author = {Case, William},
year = {2008},
month = {10},
pages = {},
title = {Wigner functions and Weyl transforms for pedestrians},
volume = {76},
journal = {American Journal of Physics - AMER J PHYS},
doi = {10.1119/1.2957889}
}

@misc{chaturvedi-PSdescriptionsoperators,
      title={Phase-space descriptions of operators and the Wigner distribution in quantum mechanics I. A Dirac inspired view}, 
      author={S. Chaturvedi and E. Ercolessi and G. Marmo and G. Morandi and N. Mukunda and R. Simon},
      year={2005},
      eprint={quant-ph/0507053},
      archivePrefix={arXiv},
      primaryClass={quant-ph},
      url={https://arxiv.org/abs/quant-ph/0507053}, 
}

@article{Wigner-QuantumCorrections,
  title = {On the Quantum Correction For Thermodynamic Equilibrium},
  author = {Wigner, E.},
  journal = {Phys. Rev.},
  volume = {40},
  issue = {5},
  pages = {749--759},
  numpages = {0},
  year = {1932},
  month = {Jun},
  publisher = {American Physical Society},
  doi = {10.1103/PhysRev.40.749},
  url = {https://link.aps.org/doi/10.1103/PhysRev.40.749}
}

@article{Weyl1927,
  author    = {Hermann Weyl},
  title     = {Quantenmechanik und Gruppentheorie},
  journal   = {Zeitschrift für Physik},
  volume    = {46},
  number    = {1},
  pages     = {1--46},
  year      = {1927},
  month     = nov,
  issn      = {0044-3328},
  doi       = {10.1007/BF02055756},
  url       = {https://doi.org/10.1007/BF02055756}
}

@article{vNeumann1931,
  author    = {von Neumann, J.},
  title     = {Die Eindeutigkeit der Schrödingerschen Operatoren},
  journal   = {Mathematische Annalen},
  volume    = {104},
  number    = {1},
  pages     = {570--578},
  year      = {1931},
  month     = dec,
  doi       = {10.1007/BF01457956},
  url       = {https://doi.org/10.1007/BF01457956},
  issn      = {1432-1807}
}

@article{GROENEWOLD1946405,
title = {On the principles of elementary quantum mechanics},
journal = {Physica},
volume = {12},
number = {7},
pages = {405-460},
year = {1946},
issn = {0031-8914},
doi = {https://doi.org/10.1016/S0031-8914(46)80059-4},
url = {https://www.sciencedirect.com/science/article/pii/S0031891446800594},
author = {H.J. Groenewold},
}

@article{Moyal_1949, 
title={Quantum mechanics as a statistical theory},
volume={45},
DOI={10.1017/S0305004100000487}, 
number={1}, 
journal={Mathematical Proceedings of the Cambridge Philosophical Society}, 
author={Moyal, J. E.},
year={1949}, 
pages={99–124}}

@book{BIPM2025,
  author    = {{Bureau International des Poids et Mesures}},
  title     = {Le Système international d'unités / The International System of Units},
  edition   = {9},
  year      = {2025},
  publisher = {Bureau International des Poids et Mesures},
  note      = {Brochure},
  doi       = {10.59161/AUEZ1291},
}

@book{Arnold1989,
  author    = {V. I. Arnold},
  title     = {Mathematical Methods of Classical Mechanics},
  series    = {Graduate Texts in Mathematics},
  volume    = {60},
  edition   = {2},
  publisher = {Springer},
  address   = {New York, NY},
  year      = {1989},
  doi       = {10.1007/978-1-4757-2063-1},
  isbn      = {978-0-387-96890-2},
  note      = {Originally published by Nauka, Moscow, 1974}
}

@article{SundaramNiu,
author = {Sundaram, Ganesh and Niu, Qian},
year = {1999},
month = {07},
pages = {},
title = {Wave-packet dynamics in slowly perturbed crystals: Gradient corrections and Berry-phase effects},
volume = {59},
journal = {Physical Review B},
doi = {10.1103/PhysRevB.59.14915}
}

@phdthesis{perez:tel-03922294,
  TITLE = {{Topological waves in geophysical and astrophysical fluids}},
  AUTHOR = {Perez, Nicolas},
  URL = {https://theses.hal.science/tel-03922294},
  NUMBER = {2022ENSL0021},
  SCHOOL = {{Ecole normale sup{\'e}rieure de lyon - ENS LYON}},
  YEAR = {2022},
  MONTH = Sep,
  TYPE = {Theses},
  HAL_ID = {tel-03922294},
  HAL_VERSION = {v1},
}

@article{Price-BerryCurvature-OpticalLattice-NiuAnsatz,
  title = {Mapping the Berry curvature from semiclassical dynamics in optical lattices},
  author = {Price, H. M. and Cooper, N. R.},
  journal = {Phys. Rev. A},
  volume = {85},
  issue = {3},
  pages = {033620},
  numpages = {12},
  year = {2012},
  month = {Mar},
  publisher = {American Physical Society},
  doi = {10.1103/PhysRevA.85.033620},
  url = {https://link.aps.org/doi/10.1103/PhysRevA.85.033620}
}

@article{wimmerExperimentalMeasurementBerry2017ExampleNiu,
  title = {Experimental Measurement of the {{Berry}} Curvature from Anomalous Transport},
  author = {Wimmer, Martin and Price, Hannah M. and Carusotto, Iacopo and Peschel, Ulf},
  date = {2017-06},
  journaltitle = {Nature Physics},
  shortjournal = {Nature Phys},
  volume = {13},
  number = {6},
  pages = {545--550},
  issn = {1745-2473, 1745-2481},
  doi = {10.1038/nphys4050},
  url = {https://www.nature.com/articles/nphys4050},
  urldate = {2025-10-17},
  langid = {english},
  }

@article{Perez_2021,
   title={Manifestation of the Berry curvature in geophysical ray tracing},
   volume={477},
   ISSN={1471-2946},
   url={http://dx.doi.org/10.1098/rspa.2020.0844},
   DOI={10.1098/rspa.2020.0844},
   number={2248},
   journal={Proceedings of the Royal Society A: Mathematical, Physical and Engineering Sciences},
   publisher={The Royal Society},
   author={Perez, N. and Delplace, P. and Venaille, A.},
   year={2021},
   month=apr }

@phdthesis{Luneau:phd,
  TITLE = {{Topological coupling between slow and fast quantum systems}},
  AUTHOR = {Luneau, Jacquelin},
  URL = {https://theses.hal.science/tel-04351529},
  NUMBER = {2023ENSL0053},
  SCHOOL = {{Ecole normale sup{\'e}rieure de lyon - ENS LYON}},
  YEAR = {2023},
  MONTH = Sep,
  TYPE = {Theses},
  HAL_ID = {tel-04351529},
  HAL_VERSION = {v1},
}

@article{Stiepan_2012,
   title={Semiclassical Approximations for Hamiltonians with Operator-Valued Symbols},
   volume={320},
   ISSN={1432-0916},
   url={http://dx.doi.org/10.1007/s00220-012-1650-5},
   DOI={10.1007/s00220-012-1650-5},
   number={3},
   journal={Communications in Mathematical Physics},
   publisher={Springer Science and Business Media LLC},
   author={Stiepan, Hans-Michael and Teufel, Stefan},
   year={2012},
   month=dec, pages={821–849} }

@book{Teufel2003-AdiabaticPTQM,
  author    = {Stefan Teufel},
  title     = {Adiabatic Perturbation Theory in Quantum Dynamics},
  series    = {Lecture Notes in Mathematics},
  volume    = {1821},
  publisher = {Springer},
  address   = {Berlin, Heidelberg},
  year      = {2003},
  doi       = {10.1007/b13355},
  isbn      = {978-3-540-40723-2},
}

@misc{Lein2010,
  author       = {Max Lein},
  title        = {Quantization and Semiclassics},
  year         = {2010},
  eprint       = {1009.0444},
  archivePrefix= {arXiv},
  primaryClass = {math-ph},
  note         = {Lecture notes, 145 pages},
  doi       = {10.48550/arXiv.1009.0444},
}

@article{Berry1984,
  author    = {Michael V. Berry},
  title     = {Quantal phase factors accompanying adiabatic changes},
  journal   = {Proceedings of the Royal Society of London. Series A, Mathematical and Physical Sciences},
  volume    = {392},
  number    = {1802},
  pages     = {45--57},
  year      = {1984},
  doi       = {10.1098/rspa.1984.0023},
  url       = {https://doi.org/10.1098/rspa.1984.0023}
}

@article{Littlejohn1992,
  title = {Semiclassical theory of spin-orbit coupling},
  author = {Littlejohn, Robert G. and Flynn, William G.},
  journal = {Phys. Rev. A},
  volume = {45},
  issue = {11},
  pages = {7697--7717},
  numpages = {0},
  year = {1992},
  month = {Jun},
  publisher = {American Physical Society},
  doi = {10.1103/PhysRevA.45.7697},
  url = {https://link.aps.org/doi/10.1103/PhysRevA.45.7697}
}

@book{Munster2020,
  author    = {Gernot Münster},
  title     = {Quantentheorie},
  publisher = {De Gruyter Oldenbourg},
  address   = {Berlin, Boston},
  year      = {2020},
  doi       = {10.1515/9783110479966},
  isbn      = {9783110479959}
}

@article{Mostowski,
author = {Mostowski, Jan and Pietraszewicz, Joanna},
year = {2021},
month = {04},
pages = {},
title = {Wigner Function for Harmonic Oscillator and The Classical Limit},
doi = {10.48550/arXiv.2104.06638}
}

@article{banerjiNonlinearWavePacket2001,
  title = {Non-Linear Wave Packet Dynamics of Coherent States},
  author = {Banerji, J},
  date = {2001-02},
  journaltitle = {Pramana - J Phys},
  volume = {56},
  number = {2--3},
  pages = {267--280},
  issn = {0304-4289, 0973-7111},
  doi = {10.1007/s12043-001-0123-8},
  url = {http://link.springer.com/10.1007/s12043-001-0123-8},
  urldate = {2025-10-05}
}

@article{milburnQuantumClassicalLiouville1986,
  title = {Quantum and Classical {{Liouville}} Dynamics of the Anharmonic Oscillator},
  author = {Milburn, G. J.},
  date = {1986-01-01},
  journaltitle = {Phys. Rev. A},
  volume = {33},
  number = {1},
  pages = {674--685},
  issn = {0556-2791},
  doi = {10.1103/PhysRevA.33.674},
  url = {https://link.aps.org/doi/10.1103/PhysRevA.33.674},
  urldate = {2025-10-05},
  langid = {english}
}

@article{Polkovnikov_PSRoQD_2010,
   title={Phase space representation of quantum dynamics},
   volume={325},
   ISSN={0003-4916},
   url={http://dx.doi.org/10.1016/j.aop.2010.02.006},
   DOI={10.1016/j.aop.2010.02.006},
   number={8},
   journal={Annals of Physics},
   publisher={Elsevier BV},
   author={Polkovnikov, Anatoli},
   year={2010},
   month=aug, pages={1790–1852} }

@article{proppDecoherenceLimitsCost,
  title = {Decoherence limiting the cost to simulate an anharmonic oscillator},
  author = {Propp, Tzula B. and Ray, Sayonee and DeBrota, John B. and Albash, Tameem and Deutsch, Ivan H.},
  journal = {Phys. Rev. A},
  volume = {108},
  issue = {6},
  pages = {062219},
  numpages = {21},
  year = {2023},
  month = {Dec},
  publisher = {American Physical Society},
  doi = {10.1103/PhysRevA.108.062219},
  url = {https://link.aps.org/doi/10.1103/PhysRevA.108.062219}
}

@article{rohithFractionalRevivalsSuperposed2014,
  title = {Fractional Revivals of Superposed Coherent States},
  author = {Rohith, M. and Sudheesh, C.},
  date = {2014-02-28},
  journaltitle = {J. Phys. B: At. Mol. Opt. Phys.},
  volume = {47},
  number = {4},
  eprint = {1309.0104},
  eprinttype = {arXiv},
  eprintclass = {quant-ph},
  pages = {045504},
  issn = {0953-4075, 1361-6455},
  doi = {10.1088/0953-4075/47/4/045504},
  url = {http://arxiv.org/abs/1309.0104},
  urldate = {2025-10-05}
}

@article{stobinskaWignerFunctionEvolution2008,
  title = {Wigner {{Function Evolution}} of {{Quantum States}} in {{Presence}} of {{Self-Kerr Interaction}}},
  author = {Stobińska, Magdalena and Milburn, G. J. and Wódkiewicz, Krzysztof},
  date = {2008-07-09},
  journaltitle = {Phys. Rev. A},
  volume = {78},
  number = {1},
  eprint = {quant-ph/0605166},
  eprinttype = {arXiv},
  pages = {013810},
  issn = {1050-2947, 1094-1622},
  doi = {10.1103/PhysRevA.78.013810},
  url = {http://arxiv.org/abs/quant-ph/0605166},
  urldate = {2025-10-05}
}

@article{sudheeshManifestationsWavePacket2004,
  title = {Manifestations of Wave Packet Revivals in the Moments of Observables},
  author = {Sudheesh, C. and Lakshmibala, S. and Balakrishnan, V.},
  date = {2004-08},
  journaltitle = {Physics Letters A},
  volume = {329},
  number = {1--2},
  pages = {14--21},
  issn = {03759601},
  doi = {10.1016/j.physleta.2004.06.085},
  url = {https://linkinghub.elsevier.com/retrieve/pii/S0375960104009132},
  urldate = {2025-10-05},
}

@phdthesis{ReagorCricuitQED2016,
  author       = {Matthew James Reagor},
  title        = {Superconducting Cavities for Circuit Quantum Electrodynamics},
  school       = {Yale University},
  year         = {2016},
  type         = {Doctoral dissertation},
  address      = {New Haven, CT},
  note         = {A dissertation presented to the Faculty of the Graduate School of Yale University in candidacy for the degree of Doctor of Philosophy. ProQuest ID: 10160874},
}

@article{drummondQuantumTheoryOptical1980,
  title = {Quantum Theory of Optical Bistability. {{I}}. {{Nonlinear}} Polarisability Model},
  author = {Drummond, P D and Walls, D F},
  date = {1980-02-01},
  journaltitle = {Journal of Physics A: Mathematical and General},
  shortjournal = {J. Phys. A: Math. Gen.},
  volume = {13},
  number = {2},
  pages = {725--741},
  issn = {0305-4470, 1361-6447},
  doi = {10.1088/0305-4470/13/2/034},
  url = {https://iopscience.iop.org/article/10.1088/0305-4470/13/2/034},
  urldate = {2025-10-05}
}

@article{greinerCollapseRevivalBEC2002,
  title = {Collapse and Revival of the Matter Wave Field of a {{Bose}}–{{Einstein}} Condensate},
  author = {Greiner, Markus and Mandel, Olaf and Hänsch, Theodor W. and Bloch, Immanuel},
  date = {2002-09},
  journaltitle = {Nature},
  shortjournal = {Nature},
  volume = {419},
  number = {6902},
  pages = {51--54},
  issn = {0028-0836, 1476-4687},
  doi = {10.1038/nature00968},
  url = {https://www.nature.com/articles/nature00968},
  urldate = {2025-10-05},
  langid = {english}
}

@article{moshinsky_dirac_1989,
	title = {The Dirac oscillator},
	volume = {22},
	issn = {0305-4470, 1361-6447},
	url = {https://iopscience.iop.org/article/10.1088/0305-4470/22/17/002},
	doi = {10.1088/0305-4470/22/17/002},
	pages = {L817--L819},
	number = {17},
	journaltitle = {J. Phys. A: Math. Gen.},
	author = {Moshinsky, M and Szczepaniak, A},
	urldate = {2025-10-02},
	date = {1989-09-07},
	langid = {english},
}

@inproceedings{sadurni_dirac-moshinsky_2011,
	title = {The Dirac-Moshinsky Oscillator: Theory and Applications},
	url = {http://arxiv.org/abs/1101.3011},
	doi = {10.1063/1.3575536},
	shorttitle = {The Dirac-Moshinsky Oscillator},
	pages = {249--290},
	author = {Sadurni, Emerson},
	urldate = {2025-10-02},
	date = {2011},
	langid = {english},
	eprinttype = {arxiv},
	eprint = {1101.3011 [quant-ph]}
}

@inproceedings{torres_dirac-moshinsky_2010,
	title = {The Dirac-Moshinsky oscillator coupled to an external field and its connection to quantum optics},
	url = {http://arxiv.org/abs/1010.5229},
	doi = {10.1063/1.3537860},
	pages = {301--312},
	author = {Torres, Juan Mauricio and Sadurni, Emerson and Seligman, Thomas H.},
	urldate = {2025-10-02},
	date = {2010},
	langid = {english},
	eprinttype = {arxiv},
	eprint = {1010.5229 [quant-ph]}
}

@article{NogamiCoherentStateDO,
author = {Nogami, Y. and Toyama, F. M.},
title = {Coherent state of the Dirac oscillator},
journal = {Canadian Journal of Physics},
volume = {74},
number = {3-4},
pages = {114-121},
year = {1996},
doi = {10.1139/p96-018},
URL = {https://doi.org/10.1139/p96-018},
eprint = {https://doi.org/10.1139/p96-018}
}

@article{Ghosh2021QuasicoherentstateDO,
author = {P. Ghosh and P. Roy},
title = {Quasi coherent state of the Dirac oscillator},
journal = {Journal of Modern Optics},
volume = {68},
number = {1},
pages = {56--62},
year = {2021},
publisher = {Taylor \& Francis},
doi = {10.1080/09500340.2021.1876261},
URL = {https://doi.org/10.1080/09500340.2021.1876261
},
eprint = {   
        https://doi.org/10.1080/09500340.2021.1876261
}
}

@misc{sanchez-jaynes-cummings-in-PSQM_2025,
	title = {The Jaynes-Cummings model in Phase Space Quantum Mechanics},
	url = {http://arxiv.org/abs/2506.23386},
	doi = {10.48550/arXiv.2506.23386},
	number = {{arXiv}:2506.23386},
	publisher = {{arXiv}},
	author = {Sanchez-Cordova, Mar and Berra-Montiel, Jasel and Molgado, Alberto},
	urldate = {2025-10-02},
	date = {2025-06-29},
	langid = {english},
	eprinttype = {arxiv},
	eprint = {2506.23386 [quant-ph]},
}

@article{BUCK1981atom-phonon-coupling,
title = {Exactly soluble model of atom-phonon coupling showing periodic decay and revival},
journal = {Physics Letters A},
volume = {81},
number = {2},
pages = {132-135},
year = {1981},
issn = {0375-9601},
doi = {https://doi.org/10.1016/0375-9601(81)90042-6},
url = {https://www.sciencedirect.com/science/article/pii/0375960181900426},
author = {B. Buck and C.V. Sukumar}
}

@article{franco-MicrowaveDO_2013,
	title = {First experimental realization of the Dirac oscillator},
	volume = {111},
	issn = {0031-9007, 1079-7114},
	url = {http://arxiv.org/abs/1306.2204},
	doi = {10.1103/PhysRevLett.111.170405},
	pages = {170405},
	number = {17},
	journaltitle = {Phys. Rev. Lett.},
	author = {Franco-Villafañe, J. A. and Sadurní, E. and Barkhofen, S. and Kuhl, U. and Mortessagne, F. and Seligman, T. H.},
	urldate = {2025-10-02},
	date = {2013-10-24},
	langid = {english},
	eprinttype = {arxiv},
	eprint = {1306.2204 [cond-mat]},
}

@article{bermudez-Relativistic_Landau_Levels_2007,
	title = {Mesoscopic Superposition States in Relativistic Landau Levels},
	volume = {99},
	rights = {http://link.aps.org/licenses/aps-default-license},
	issn = {0031-9007, 1079-7114},
	url = {https://link.aps.org/doi/10.1103/PhysRevLett.99.123602},
	doi = {10.1103/PhysRevLett.99.123602},
	pages = {123602},
	number = {12},
	journaltitle = {Phys. Rev. Lett.},
	author = {Bermudez, A. and Martin-Delgado, M. A. and Solano, E.},
	urldate = {2025-10-02},
	date = {2007-09-18},
	langid = {english}
}

@article{longhi_photonicDO_2010,
	title = {Photonic realization of the relativistic Dirac oscillator},
	volume = {35},
	issn = {0146-9592, 1539-4794},
	url = {http://arxiv.org/abs/1009.0159},
	doi = {10.1364/OL.35.001302},
	pages = {1302},
	number = {8},
	journaltitle = {Opt. Lett.},
	author = {Longhi, Stefano},
	urldate = {2025-10-02},
	date = {2010-04-15},
	langid = {english},
	eprinttype = {arxiv},
	eprint = {1009.0159 [physics]}
}

@article{raghu_analogs_2008,
	title = {Analogs of quantum Hall effect edge states in photonic crystals},
	volume = {78},
	issn = {1050-2947, 1094-1622},
	url = {http://arxiv.org/abs/cond-mat/0602501},
	doi = {10.1103/PhysRevA.78.033834},
	pages = {033834},
	number = {3},
	journaltitle = {Phys. Rev. A},
	author = {Raghu, S. and Haldane, F. D. M.},
	urldate = {2025-10-02},
	date = {2008-09-23},
	langid = {english},
	eprinttype = {arxiv},
	eprint = {cond-mat/0602501},
	}

@article{xiao_valley-contrasting_2007,
	title = {Valley-Contrasting Physics in Graphene: Magnetic Moment and Topological Transport},
	volume = {99},
	rights = {http://link.aps.org/licenses/aps-default-license},
	issn = {0031-9007, 1079-7114},
	url = {https://link.aps.org/doi/10.1103/PhysRevLett.99.236809},
	doi = {10.1103/PhysRevLett.99.236809},
	shorttitle = {Valley-Contrasting Physics in Graphene},
	pages = {236809},
	number = {23},
	journaltitle = {Phys. Rev. Lett.},
	author = {Xiao, Di and Yao, Wang and Niu, Qian},
	urldate = {2025-10-02},
	date = {2007-12-07},
	langid = {english},
}

@Article{LectNotesDelplace,
	title={{Berry-Chern monopoles and spectral flows}},
	author={Pierre Delplace},
	journal={SciPost Phys. Lect. Notes},
	pages={39},
	year={2022},
	publisher={SciPost},
	doi={10.21468/SciPostPhysLectNotes.39},
	url={https://scipost.org/10.21468/SciPostPhysLectNotes.39}
}

@article{Haldane,
  title = {Model for a Quantum Hall Effect without Landau Levels: Condensed-Matter Realization of the "Parity Anomaly"},
  author = {Haldane, F. D. M.},
  journal = {Phys. Rev. Lett.},
  volume = {61},
  issue = {18},
  pages = {2015--2018},
  numpages = {0},
  year = {1988},
  month = {Oct},
  publisher = {American Physical Society},
  doi = {10.1103/PhysRevLett.61.2015},
  url = {https://link.aps.org/doi/10.1103/PhysRevLett.61.2015}
}

@article{jaynes_JC_model_1963,
	title = {Comparison of quantum and semiclassical radiation theories with application to the beam maser},
	volume = {51},
	rights = {https://ieeexplore.ieee.org/Xplorehelp/downloads/license-information/{IEEE}.html},
	issn = {0018-9219},
	url = {http://ieeexplore.ieee.org/document/1443594/},
	doi = {10.1109/PROC.1963.1664},
	pages = {89--109},
	number = {1},
	journaltitle = {Proc. {IEEE}},
	author = {Jaynes, E.T. and Cummings, F.W.},
	urldate = {2025-10-02},
	date = {1963}
}

@article{son_chiral_anomaly_Weylmetals_2013,
	title = {Chiral Anomaly and Classical Negative Magnetoresistance of Weyl Metals},
	volume = {88},
	issn = {1098-0121, 1550-235X},
	url = {http://arxiv.org/abs/1206.1627},
	doi = {10.1103/PhysRevB.88.104412},
	pages = {104412},
	number = {10},
	journaltitle = {Phys. Rev. B},
	author = {Son, D. T. and Spivak, B. Z.},
	urldate = {2025-10-02},
	date = {2013-09-13},
	eprinttype = {arxiv},
	eprint = {1206.1627 [cond-mat]},
}

@article{yang_Weyl_semimetal_2011,
	title = {Quantum Hall effects in a Weyl Semi-Metal: possible application in pyrochlore Iridates},
	volume = {84},
	issn = {1098-0121, 1550-235X},
	url = {http://arxiv.org/abs/1105.2353},
	doi = {10.1103/PhysRevB.84.075129},
	shorttitle = {Quantum Hall effects in a Weyl Semi-Metal},
	pages = {075129},
	number = {7},
	journaltitle = {Phys. Rev. B},
	author = {Yang, Kai-Yu and Lu, Yuan-Ming and Ran, Ying},
	urldate = {2025-10-02},
	date = {2011-08-09},
	eprinttype = {arxiv},
	eprint = {1105.2353 [cond-mat]},
}

@article{dodonovSqueezedRewiew2002,
  title = {`{{Nonclassical}}' States in Quantum Optics: A `squeezed' Review of the First 75 Years},
  shorttitle = {`{{Nonclassical}}' States in Quantum Optics},
  author = {Dodonov, V V},
  date = {2002-02-01},
  journaltitle = {Journal of Optics B: Quantum and Semiclassical Optics},
  shortjournal = {J. Opt. B: Quantum Semiclass. Opt.},
  volume = {4},
  number = {1},
  pages = {R1-R33},
  issn = {1464-4266, 1741-3575},
  doi = {10.1088/1464-4266/4/1/201},
  url = {https://iopscience.iop.org/article/10.1088/1464-4266/4/1/201},
  urldate = {2025-10-08},
}

@article{leibfriedTwoIonWavePacket,
  title = {Experimental Demonstration of a Robust, High-Fidelity Geometric Two Ion-Qubit Phase Gate},
  author = {Leibfried, D. and DeMarco, B. and Meyer, V. and Lucas, D. and Barrett, M. and Britton, J. and Itano, W. M. and Jelenković, B. and Langer, C. and Rosenband, T. and Wineland, D. J.},
  date = {2003-03-01},
  journaltitle = {Nature},
  shortjournal = {Nature},
  volume = {422},
  number = {6930},
  pages = {412--415},
  issn = {1476-4687},
  doi = {10.1038/nature01492},
  url = {https://doi.org/10.1038/nature01492},
}

@article{Feldmann-CoherentDynamacisWavePackets,
  title = {Coherent dynamics of excitonic wave packets},
  author = {Feldmann, J. and Meier, T. and von Plessen, G. and Koch, M. and G\"obel, E. O. and Thomas, P. and Bacher, G. and Hartmann, C. and Schweizer, H. and Sch\"afer, W. and Nickel, H.},
  journal = {Phys. Rev. Lett.},
  volume = {70},
  issue = {20},
  pages = {3027--3030},
  numpages = {0},
  year = {1993},
  month = {May},
  publisher = {American Physical Society},
  doi = {10.1103/PhysRevLett.70.3027},
  url = {https://link.aps.org/doi/10.1103/PhysRevLett.70.3027}
}

@article{maternyWavePacketDynamics2000,
  title = {Wave Packet Dynamics in Different Electronic States Investigated by Femtosecond Time-Resolved Four-Wave-Mixing Spectroscopy},
  author = {Materny, A. and Chen, T. and Schmitt, M. and Siebert, T. and Vierheilig, A. and Engel, V. and Kiefer, W.},
  date = {2000-09},
  journaltitle = {Applied Physics B},
  shortjournal = {Appl Phys B},
  volume = {71},
  number = {3},
  pages = {299--317},
  issn = {0946-2171, 1432-0649},
  doi = {10.1007/s003400000343},
  url = {http://link.springer.com/10.1007/s003400000343},
}

@misc{WeissteinAssociatedLaguerrePolynomial,
  author       = {Weisstein, Eric W.},
  title        = {Associated Laguerre Polynomial},
  howpublished = {MathWorld--A Wolfram Web Resource},
  year         = {1999--2025},
  note         = {Last updated: September 19, 2025. September 21, 2025},
  url          = {https://mathworld.wolfram.com/AssociatedLaguerrePolynomial.html}
}

@misc{WeissteinLaguerrePolynomial,
  author       = {Weisstein, Eric W.},
  title        = {Laguerre Polynomial},
  howpublished = {MathWorld--A Wolfram Web Resource},
  note         = {Last updated: September 19, 2025. Accessed: September 21, 2025},
  url          = {https://mathworld.wolfram.com/LaguerrePolynomial.html}
}

@article{Feldheim1942,
  author    = {Ervin Feldheim},
  title     = {Relations entre les polyn\^omes de Jacobi, Laguerre et Hermite},
  journal   = {Acta Mathematica},
  volume    = {75},
  number    = {1},
  pages     = {117--138},
  year      = {1942},
  doi       = {10.1007/BF02404102},
  url       = {https://doi.org/10.1007/BF02404102} ,
  issn      = {1871-2509}
}
